\documentclass[%
aps,
twocolumn,
superscriptaddress,
amsmath,amssymb,
pra,
floatfix,
]{revtex4-2}
\usepackage{url}
\usepackage{graphicx}
\usepackage{placeins}
\usepackage{dcolumn}
\usepackage{bm}
\usepackage{hyperref}

\usepackage{float}
\usepackage{physics}
\usepackage{dsfont}
\usepackage{tikz,pgfplots}
\usepackage{enumerate}
\usepackage{subfigure}
\usepackage{bbold}
\usepackage{makecell}
\usepackage{array}
\makeatletter
\newcommand{\thickhline}{%
    \noalign {\ifnum 0=`}\fi \hrule height 1pt
    \futurelet \reserved@a \@xhline
}
\newcolumntype{"}{@{\hskip\tabcolsep\vrule width 1pt\hskip\tabcolsep}}
\makeatother
\usepackage{multirow}
\usepackage{tabu}
\usepackage{textcomp,booktabs}
\usepackage{colortbl}
\usepackage[T1]{fontenc}
\definecolor{mygray}{gray}{0.9}
\definecolor{mypink}{rgb}{0.99,0.91,0.95}
\definecolor{mycyan}{cmyk}{0.3,0,0,0}

\newcommand{\id}{\text{id}}

\newcommand{\mT}{\mathcal{T}}
\newcommand{\mE}{\mathcal{E}}

\newcommand{\mN}{\mathcal{N}}

\newcommand{\mS}{\mathcal{S}}
\newcommand{\mU}{\mathcal{U}}
\newcommand{\T}{\mathbf{T}}

\definecolor{KB1}{rgb}{0.4,0.3,0.9}
\definecolor{KB2}{rgb}{0.9,0.3,0.9}

\definecolor{daxColor}{HTML}{900C3F}

\makeatletter
\newcommand*{\rom}[1]{\expandafter\@slowromancap\romannumeral #1@}
\makeatother

\usepackage{amsthm}
\newtheorem*{thm*}{Theorem}

\usepackage{tcolorbox}
\tcbuselibrary{theorems}

\newtcbtheorem[number within=section, use counter*=thm]{mythm}{Theorem}%
{colback=red!5,colframe=red!35!red,fonttitle=\bfseries}{thm}

\usepackage{tcolorbox}
\tcbuselibrary{theorems}

\newtcbtheorem[number within=section, use counter*=thm]{mylem}{Lemma}%
{colback=green!5,colframe=green!35!black,fonttitle=\bfseries}{lem}

\usepackage{tcolorbox}
\tcbuselibrary{theorems}

\newtcbtheorem[number within=section, use counter*=thm]{mydef}{Definition}%
{colback=yellow!5,colframe=yellow!70!black,fonttitle=\bfseries}{def}

\usepackage{tcolorbox}
\tcbuselibrary{theorems}

\newtcbtheorem[number within=section, use counter*=thm]{mycor}{Corollary}%
{colback=blue!5,colframe=blue!70!black,fonttitle=\bfseries}{cor}

\usepackage{tcolorbox}
\tcbuselibrary{theorems}

\newtcbtheorem[number within=section, use counter*=thm]{myprop}{Proposition}%
{colback=pink!5,colframe=pink!70!black,fonttitle=\bfseries}{prop}

\pgfplotsset{compat=1.18}
\begin{document}


\title{Fundamental Limitations on Communication over a Quantum Network}

\author{Junjing~Xing}
\address{College of Intelligent Systems Science and Engineering, Harbin Engineering University, Harbin,
Heilongjiang 150001, People's Republic of China}
\author{Tianfeng~Feng}
\address{State Key Laboratory of Optoelectronic Materials and Technologies and School of Physics, 
Sun Yat-sen University, Guangzhou, 
Guangdong 510275, People's Republic of China}
\author{Zhaobing~Fan}
\email{fanzhaobing@hrbeu.edu.cn}
\address{College of Intelligent Systems Science and Engineering, Harbin Engineering University, Harbin,
Heilongjiang 150001, People's Republic of China}
\author{Haitao~Ma}
\email{hmamath@hrbeu.edu.cn}
\address{College of Mathematics Science, Harbin Engineering University, Harbin,
Heilongjiang 150001, People's Republic of China}
\author{Kishor~Bharti}
\email{kishor.bharti1@gmail.com}
\address{Institute of High Performance Computing (IHPC), Agency for Science, Technology and Research (A*STAR), 1 Fusionopolis Way, \#16-16 Connexis, Singapore 138632, Republic of Singapore}
\author{Dax~Enshan~Koh}
\email{dax\_koh@ihpc.a-star.edu.sg}
\address{Institute of High Performance Computing (IHPC), Agency for Science, Technology and Research (A*STAR), 1 Fusionopolis Way, \#16-16 Connexis, Singapore 138632, Republic of Singapore}
\author{Yunlong~Xiao}
\email{mathxiao123@gmail.com}
\address{Institute of High Performance Computing (IHPC), Agency for Science, Technology and Research (A*STAR), 1 Fusionopolis Way, \#16-16 Connexis, Singapore 138632, Republic of Singapore}

\date{\today}
             
\begin{abstract} 
Entanglement, a fundamental feature of quantum mechanics, has long been recognized as a valuable resource in enabling secure communications and surpassing classical limits. However, previous research has primarily concentrated on static entangled states generated at a single point in time, overlooking the crucial role of the quantum dynamics responsible for creating such states. Here, we propose a framework for investigating entanglement across multiple time points, termed temporal entanglement, and demonstrate that the performance of a quantum network in transmitting information is inherently dependent on its temporal entanglement. Through case studies, we showcase the capabilities of our framework in enhancing conventional quantum teleportation and achieving exponential performance growth in the protocol of quantum repeaters. Additionally, our framework effectively doubles the communication distance in certain noise models. Our results address the longstanding question surrounding temporal entanglement within non-Markovian processes and its impact on quantum communication, thereby pushing the frontiers of quantum information science.
\end{abstract}

\maketitle

\begin{figure}
  \centering
\includegraphics[width=0.48\textwidth]{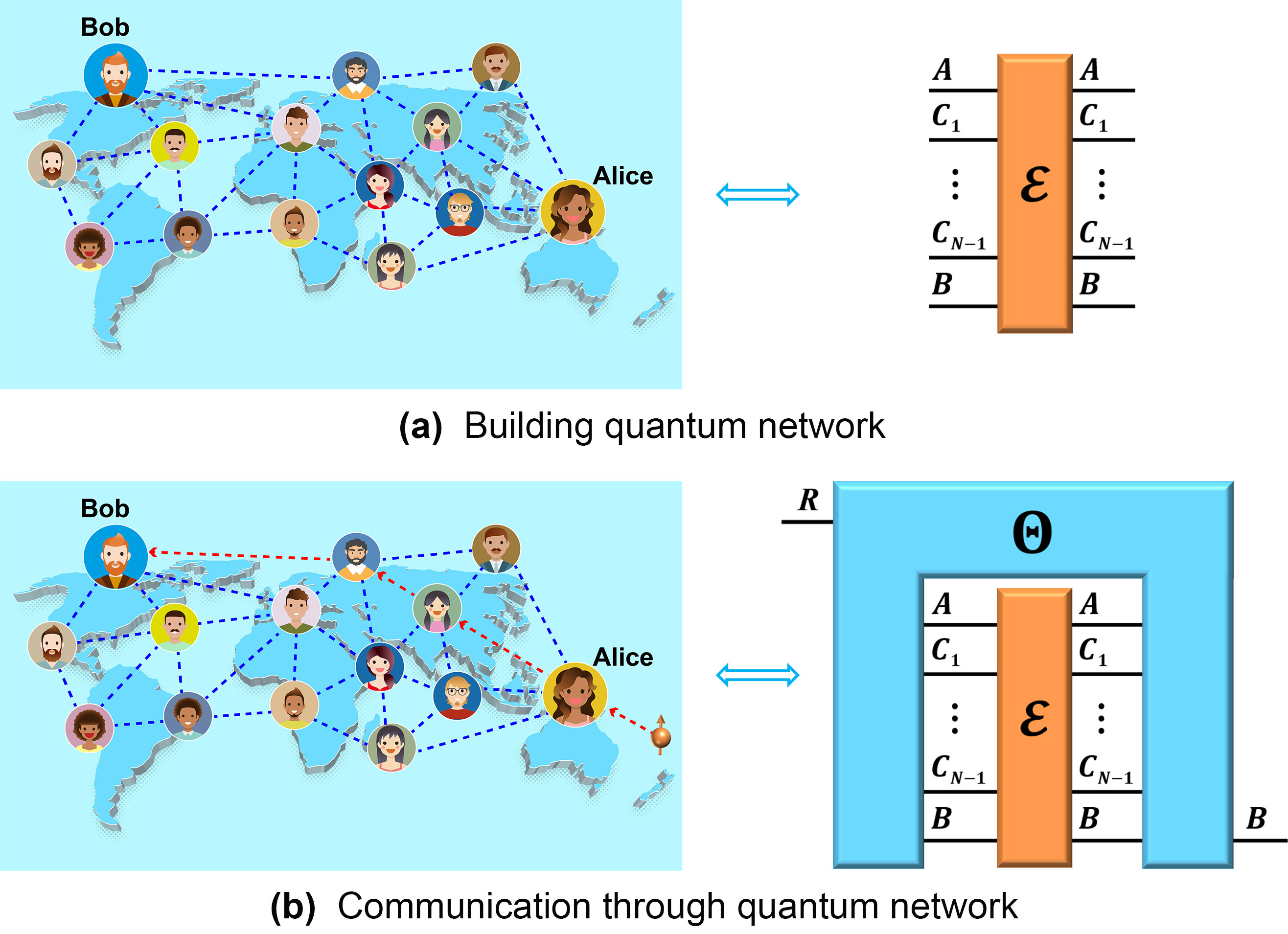}\\
   \caption{\textbf{Quantum Network:} a schematic diagram of single-round quantum network communication. The physical quantum network connects the sender Alice ($A$), the receiver Bob ($B$), and third-party agents ($C_k$ with $k=1, \ldots, N-1$), and is created using a multi-input and multi-output quantum channel $\mE$ (a), highlighted in orange. System $R$ encodes the quantum information, while the blue superchannel $\Theta$ represents its subsequent transmission within the network (b). 
   }
  \label{fig:Network-single-round}
\end{figure}

\noindent \textbf{Introduction}-- The advent of modern network technology has catalyzed a monumental transformation in our daily lives, reshaping the way we communicate and access information. From instant messaging to social media platforms, networks have revolutionized our interactions with one another, enabling real-time connectivity across the globe. The impact of networks on our daily lives is undeniable, and it continues to shape the way we work, learn, and play in an ever-evolving digital landscape. Notwithstanding these remarkable achievements, our current communication infrastructure is built on classical information processing techniques that are constrained by physical limitations. In stark contrast, global quantum network (Fig.~\ref{fig:Network-single-round})~\cite{Kimble2008,Pirandola2016,Simon2017,doi:10.1126/science.aam9288} holds the promise of an unprecedented leap forward in communication capabilities and security by capitalizing on a unique resource of quantum mechanics -- quantum entanglement~\cite{RevModPhys.81.865}.

Despite considerable progress in investigating entanglement and its effectiveness in quantum communication, most studies have predominantly focused on a static view of entanglement at specific time points. However, the intrinsic nature of quantum systems defies fixation to any particular moment, embracing an unceasing odyssey of incessant evolution. In quantum information science, the evolution of quantum systems finds its expression through the framework of quantum channels~\cite{nielsen_chuang_2010,wilde_2013,watrous_2018}, encompassing diverse processes such as the operations of polarization beam splitters, wave plates, and single photon detectors in photonic systems. Mirroring the dynamic evolution inherent in quantum systems, quantum channels also undergo temporal transformations that shape their behavior over time~\cite{PhysRevA.80.022339,PhysRevLett.101.060401,PhysRevA.84.012311}. To encapsulate their evolution, quantum superchannels~\cite{Chiribella_2008} emerges as a powerful tool. While a quantum channel can be regarded as a single quantum process, a quantum superchannel incorporates a sequential combination of two processes: pre-processing and post-processing procedures, with the integration of quantum memory serving as a connecting bridge. This formulation empowers us to explore the fundamental nature of quantum dynamics spanning multiple time points, where quantum channels are interconnected through the orchestration of quantum memory. These quantum dynamics underpin a myriad of protocols, including quantum causal inference~\cite{ried2015quantum,causal-NC,xiao2023quantum}, quantum-enhanced agents~\cite{Mile2012,thompson2017using,PhysRevX.12.011007,Wu2023}, quantum network communication~\cite{Pant2019,PRXQuantum.1.020317,PRXQuantum.2.017002,Hermans2022,doi:10.1116/5.0051881,doi:10.1116/5.0092069,doi:10.1116/5.0118569,fang2022quantum,azuma2022quantum}, distributed quantum computing~\cite{7562346,10.1145/3233188.3233224,8910635}, and non-Markovian open systems~\cite{PhysRevA.97.012127,PhysRevLett.122.140401,PRXQuantum.2.030201}. Hence, there is a pressing need for a more comprehensive understanding of entanglement across multiple temporal instances and its impact on quantum communications.

Guided by this dynamic perspective, we embark on an illuminating journey into quantum teleportation~\cite{PhysRevLett.69.2881,PhysRevLett.70.1895,PhysRevA.54.3824,Zeilinger1997,PhysRevLett.80.1121}, where the remarkable efficiency in surpassing classical communication limitations stems from the entangling bipartite channel shared between the sender and receiver. Real-world quantum communication is susceptible to environmental factors that degrade its performance. Free space encounters atmospheric thinness and turbulence~\cite{Pan2017,Diamanti2017,Xia_2018,Dai2020,PhysRevLett.128.170501,RevModPhys.94.035001}, while fiber channels face temperature fluctuations, Kerr non-linearity, and reference frame disparities~\cite{Marcikic2003,Landry07,Valivarthi2016,Grosshans2016}. In response to these issues, conventional strategies focus on post-processing through maximizing local operations and classical communication (LOCC)~\cite{PhysRevA.60.1888}. Yet, the general manipulation of quantum channels encompasses both pre-processing and post-processing. This prompts fundamental questions: Which approach, pre-processing or post-processing, demonstrates superior efficacy in unlocking the communication capabilities of noisy bipartite channels? Can superchannel-assisted protocols outperform conventional scenarios without incurring additional resource costs? Additionally, what is the ultimate performance limit achievable by leveraging superchannels? Similar inquiries apply to quantum repeaters~\cite{PhysRevLett.81.5932}. Looking ahead to general quantum networks, wherein multiple rounds of communication occur among the sender, receiver, and intermediate agents, an intriguing query surfaces: What is the optimal performance for transmitting quantum information? This model assumes profound significance in the domains of distributed and cloud-based quantum computing~\cite{PhysRevA.94.032329,Ku2020,PhysRevA.105.042610,Ma2022}. Nevertheless, the answer to this crucial question remains elusive.

\noindent \textbf{Results}-- This work answers a diverse array of questions, converging upon a central theme: the intricate interplay of quantum dynamics across multiple temporal instances and their impact on the performance of quantum communication. Our investigation unveils the driving force behind these dynamics -- temporal entanglement,  the evolution of entanglement over time. Through this lens, we make a series of interesting discoveries: (\romannumeral1) For quantum teleportation, we showcase the supremacy of pre-processing techniques over all post-processing approaches under specific noise models. This emphatically highlights the indispensable role of pre-processing in enhancing quantum communication vis-\`a-vis conventional protocols. (\romannumeral2) For quantum repeaters, we uncover the astonishing ability of pre-processing techniques to effectively double the communication distance in select scenarios. Moreover, under specific noise models, the power of pre-processing becomes even more remarkable, resulting in exponential performance growth as the number of connection nodes increases. (\romannumeral3) For general quantum networks, we formulate their performance limit in terms of temporal entanglement, resolving a long-standing open question since the inception of quantum information science. Our results not only establish a fundamental limitation for quantum network communication, igniting a captivating path for the future development of quantum communication technology, but also introduce a practical and experimentally viable approach to enhance existing quantum communication technologies through pre-processing techniques.

\noindent \textbf{Temporal Entanglement}-- Quantum entanglement -- the emergence of non-local correlations among multiple systems -- marks a definitive departure from classical physics~\cite{RevModPhys.81.865}. This holistic property of compound systems has unlocked avenues for quantum cryptography~\cite{PhysRevLett.67.661,Ekert1992,PhysRevLett.77.2818}, communication~\cite{Zhao2004,Jin2010,Ma2012,Takeda2013,Pirandola2015,Sun2016}, and computing~\cite{PhysRevLett.86.5188,10.5555/2011492.2011495,PhysRevA.68.022312,Briegel2009,biamonte2020entanglement}. Although investigations of entanglement have grown significantly in recent years, most studies focus on entanglement within a single quantum process, including static entanglement of states~\cite{PhysRevLett.76.722,10.1063/1.3483717,PhysRevLett.106.130503,6353584,PhysRevLett.119.180506} and dynamical entanglement of channels~\cite{6556948,PhysRevLett.125.040502,PhysRevA.107.012429,PhysRevLett.125.180505,PhysRevA.103.062422}. While such works have yielded valuable insights, a comprehensive understanding of the dynamics of molecules and biochemical systems~\cite{Lambert2013}, distributed and cloud-based quantum computing, as well as quantum network communication, demands a theory of temporal entanglement that can describe non-local correlations associated with non-Markovian quantum processes across various time points.

\begin{figure}
  \centering
\includegraphics[width=0.48\textwidth]{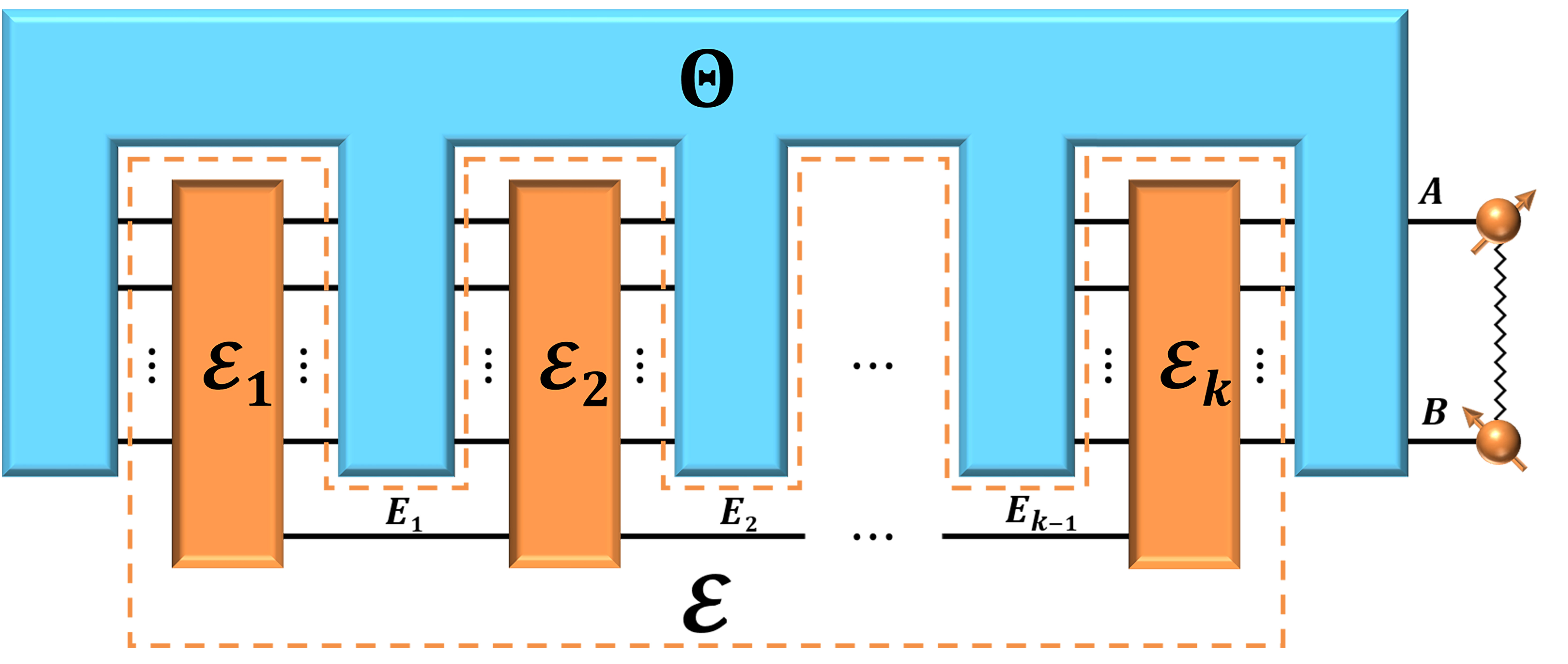}\\
   \caption{\textbf{Process of distilling static entanglement from a non-Markovian process:} 
   For a non-Markovian quantum process $\mE$ (depicted in orange) consisting of $k$ multipartite quantum channels, namely $\mE:= \mE_k\circ\cdots\circ\mE_1$, a free manipulation $\Theta$ (shown in blue) can be applied to transform it into a static state shared between systems $A$ and $B$. The fidelity $\Tr[\Theta(\mE)\cdot \phi^+_d]$ quantifies the similarity between the output state $\Theta(\mE)$ and a $d$-dimensional maximally entangled state $\phi^+_d$. By maximizing over all possible free manipulations, we obtain the $d$-fidelity of $\mS$ distillation $F_{\max}^{d,\mS}(\mE):= \max_{\Theta\,\,\mS\text{-free}} \Tr[\Theta(\mE)\cdot \phi^+_d]$. It represents an upper bound on the amount of static entanglement that can be obtained from $\mE$ through $\mS$-free manipulation.
   }
  \label{fig:ds}
\end{figure}

Investigating temporal entanglement encoded in non-Markovian processes is a challenge, with far-reaching implications for quantum information processing, computation, and metrology~\cite{PhysRevLett.123.110501,PhysRevLett.127.060501,PhysRevLett.130.070803}. Concurrently, the rapid development of quantum resource theory as a powerful tool for understanding and manipulating diverse quantum phenomena has opened up new avenues for tackling this challenge~\cite{COECKE201659,RevModPhys.91.025001,8678741,Bartosz2021NC,PhysRevLett.127.060402}. Here, riding the waves of quantum resource theories, we explore the nature and properties of temporal entanglement in quantum systems. The object of research is a non-Markovian process $\mE$~\cite{PhysRevA.86.010102,PhysRevLett.114.090402,PhysRevLett.120.040405,LI20181,White2020} comprising $k$ multipartite quantum channels $\mE_i$ ($1\leqslant i\leqslant k$) that are interconnected via quantum memory, i.e., $\mE:= \mE_k\circ\cdots\circ\mE_1$ (colored orange in Fig.~\ref{fig:ds}). 

Free manipulation of $\mE$ is a variable concept that depends on the restrictions imposed in a physical laboratory. For example, LOCC, which involves communication between parties and is considered free in theoretical investigations of entanglement theory, may not be feasible in practical experiments due to the impractically large number of communication rounds it may require, coupled with the exponentially growing amount of communication bits in each round. In such cases, experimentalists may choose to use a modified set of free operations denoted as $\text{LOCC}_{1}(\text{poly}(d))$. This approach involves local operations and one-way classical communication, with the communication complexity polynomially dependent on the dimension ($d:= d_R$) of the message system $R$, followed by a coarse-graining of quantum instruments~\cite{Chitambar2014}. 
By employing these operations, the operational complexity is reduced, making the experiment more feasible in real-world scenarios. Returning to the transformation of a non-Markovian process, we define a manipulation $\Theta$ (colored blue in Fig.~\ref{fig:ds}) as being $\mS$-free if it can be expressed as a sequence of quantum channels, with each channel belonging to the set $\mS$. As technology continues to progress, it opens up new possibilities for performing increasingly complex and powerful operations, expanding the feasible set of manipulations. To ensure that the operations performed by $\mS$ are at least as powerful as those achievable in laboratories, it is essential to assume that $\text{LOCC}_{1}(\text{poly}(d))\subset\mS$. We remark that while \emph{convexity} and \emph{closedness} are desirable mathematical properties of free manipulations, we do not require them in our analysis as they are not necessary and cannot be justified a priori.

Our goal is to evaluate the temporal entanglement encoded in a non-Markovian process $\mE$, and explore its relevance in quantum network communication. To accomplish this, we introduce the $d$-fidelity of $\mS$ distillation, denoted as $F_{\max}^{d,\mS}(\mE)$, which quantifies the maximum amount of static entanglement that can be extracted and distributed among designated parties through $\mS$-free manipulations (Fig.~\ref{fig:ds}). We further demonstrate that this quantity plays a crucial role in determining the performance of quantum network communication. By emphasizing the broadest form of non-Markovian processes, our exploration of temporal entanglement naturally encompasses previous studies~\cite{brukner2004quantum,Milz_2018,PhysRevA.98.012328,Zych2019,PhysRevResearch.3.023028,10.21468/SciPostPhys.10.6.141,PhysRevLett.128.220401}, assimilating them as specific instances. A more detailed resource-theoretic framework is presented in~\cite[Sec.~\rom{2}]{SM}.

\begin{figure*}
  \centering
\includegraphics[width=1\textwidth]{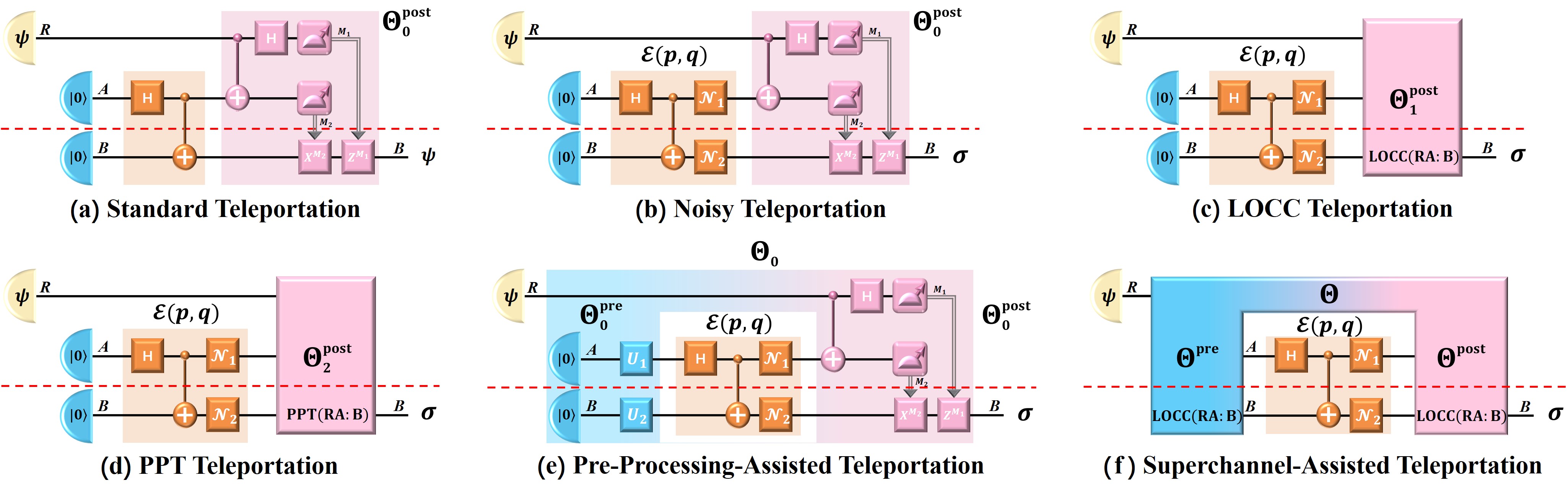}\\
   \caption{\textbf{Teleportation protocols:}
  (a) standard teleportation, where the quantum information $\psi_R$ is transmitted from sender Alice to receiver Bob. Alice has access to systems $RA$, while Bob has access to system $B$. The protocol employs a standardized LOCC operation $\Theta^{\text{post}}_{0}$, which is graphically represented by the pink box in the diagram.
  (b) noisy teleportation, where the entangling gates are followed by channels $\mN_1(p)$ and $\mN_2(q)$, resulting in a noisy entangling gate $\mE(p, q)$.
  (c) LOCC teleportation, which maximizes all possible LOCC operations after applying the noisy entangling gate $\mE(p,q)$.
  (d) PPT teleportation, where the set of allowable operations following the application of the noisy entangling gate $\mE(p,q)$ has been extended to include all PPT channels.
  (e) pre-processing-assisted teleportation, where a special pre-processing protocol $\Theta^{\text{pre}}_0:= \mU_1(\ketbra{0}{0})_A\otimes\mU_2(\ketbra{0}{0})_B$ has been considered, followed by the noisy entangling gate $\mE(p,q)$ and standardized LOCC operation $\Theta^{\text{post}}_0$.
  (f) superchannel-assisted teleportation, in which a LOCC superchannel $\Theta$ has transformed the entangling gate $\mE(p, q)$ into a teleportation channel $\Theta(\mE(p, q))_{R\to B}$. 
  We remark that protocol f does not require any additional resources compared to protocol c, as the extra pre-processing involved in protocol f falls under the umbrella of LOCC operations, which are considered free. These diagrams distinguish between systems belonging to Alice, which are located above the dashed red line, and those belonging to Bob, which are situated below the dashed line.
  }
  \label{fig:qt}
\end{figure*}

\noindent \textbf{Quantum Teleportation}-- Teleportation is a fundamental process in quantum information science that allows the secure transfer of quantum information between distant parties without physically transmitting the information carrier~\cite{Georgescu2022}. It involves four essential components: First, the message to be transmitted is encoded in a quantum state $\psi$ on system $R$; Second, a pre-processing step prepares two ancillary qubits in the state $\ket{0}$ for both the sender Alice and the receiver Bob; Third, an entangling gate that creates a quantum correlation between the sender and receiver; Finally, a post-processing step $\Theta^{\text{post}}_0$ involves standard Bell measurements on the sender's system followed by a local unitary operation on the receiver's system. Fig.~\ref{fig:qt}(a) visualizes the physical implementation of standard quantum teleportation. 

Achieving perfect information transmission requires the creation of a maximally entangled state between the sender and receiver. While this can be accomplished theoretically, practical implementations are challenged by the presence of noise and decoherence, which can significantly impact the reliability and accuracy of the communication. Before analyzing a general noisy model, let us examine a specific scenario in which local noise follows the gate that creates maximally entangled state. We denote the resulting bipartite channel shared between Alice and Bob as $\mE(p, q)$ (Fig.~\ref{fig:qt}(b)), with $p$ and $q$ representing the noise parameters acting on the sender's and receiver's systems, respectively. To fully exploit the capabilities of noisy entangling gate $\mE(p, q)$ and optimize the performance of teleportation, the conventional approach~\cite{PhysRevA.60.1888} involves maximizing over all possible LOCC post-processing protocols where the implementation of the noisy entangling gate is followed by a LOCC operation shared between the sender Alice and the receiver Bob (Fig.~\ref{fig:qt}(c)). This approach encompasses $\Theta^{\text{post}}_0$ as a specific case, thus enabling a more robust and efficient implementation of quantum teleportation. It is natural to wonder whether the conventional approach can achieve optimal performance given a noisy entangling gate? 

Our analysis indicates that there is still room for improvement in the performance of noisy quantum teleportation protocols, and that new strategies can be developed to achieve better results. The temporal entanglement shared between Alice and Bob, which is carried by the bipartite channel $\mE(p, q)$, is the \emph{bona fide} quantum resource in teleportation. To fully exploit the capabilities of noisy entangling gates, we consider the most general way of manipulating quantum channels, 
namely, by using quantum superchannels. This concept involves pre-processing and post-processing operations connected through a quantum memory. In the conventional approach to quantum teleportation, the maximization over all possible LOCC post-processing operations is typically performed. However, it is important to recognize that this approach neglects the potential benefits of maximizing all LOCC for pre-processing as well. By incorporating LOCC superchannel-assisted teleportation protocols (Fig.~\ref{fig:qt}(f)), we can achieve a better performance of quantum teleportation in the presence of noise, without requiring additional resources. 

\begin{figure}
  \centering
\includegraphics[width=0.48\textwidth]{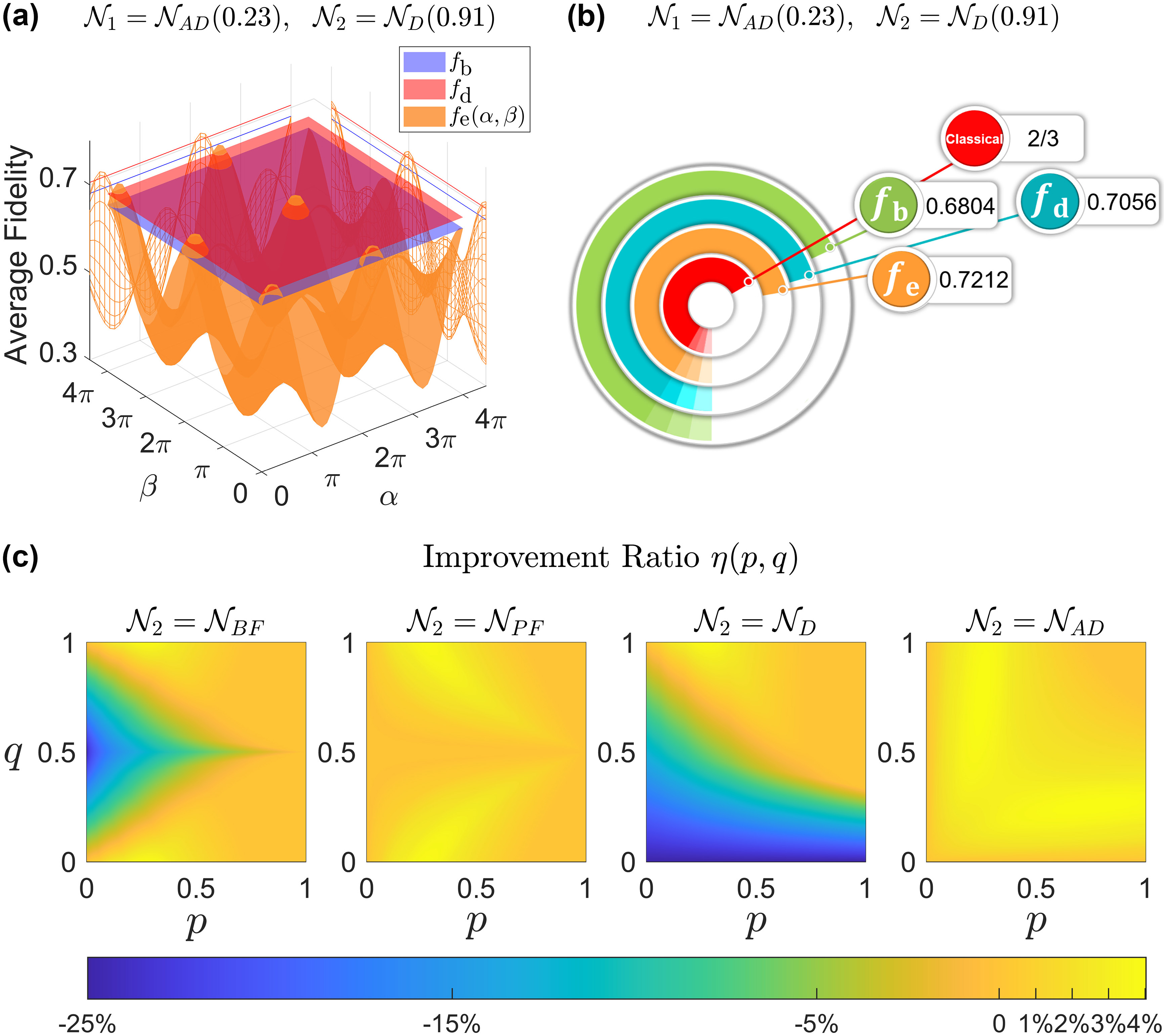}\\
   \caption{\textbf{Performance comparison of different teleportation protocols:} 
   In order to evaluate the effectiveness of a quantum channel $\mE$ in transmitting quantum information, we employ the average fidelity, $f(\mE):= \int d\psi\Tr[\psi\cdot\mE(\psi)]$, where the integral is taken over the Haar measure. We consider the noise channels shown in Fig.~\ref{fig:qt} with $\mN_1=\mN_{AD}(p=0.23)$ and $\mN_2=\mN_{D}(q=0.91)$ being amplitude damping and depolarizing channels. The pre-processing, given by $U_1(\alpha):= R_{y}(\alpha)$ and $U_2(\beta):= R_{y}(\beta)$, are rotations along the $y$-axis with angles $\alpha$ and $\beta$. The performances of protocols b, c, d, e, and f are denoted as $f_{\text{b}}$, $f_{\text{c}}$, $f_{\text{d}}$, $f_{\text{e}}(\alpha, \beta)$, and $f_{\text{f}}$, satisfying $f_{\text{b}}\leqslant f_{\text{c}}\leqslant f_{\text{d}}$ and $f_{\text{e}}(\alpha, \beta)\leqslant f_{\text{f}}$. Figure (a) demonstrates a performance comparison between $f_{\text{b}}$, $f_{\text{d}}$, and $f_{\text{e}}(\alpha, \beta)$. We maximize the performance of protocol e by optimizing over all $R_y$ gates in the pre-processing step, resulting in $f_{\text{e}}:=\max_{\alpha, \beta} f_e(\alpha, \beta)$ with $f_{\text{e}}(\alpha, \beta)\leqslant f_{\text{e}}\leqslant f_{\text{f}}$. Figure (b) displays the advantage of performance $f_e$ over $f_{\text{b}}$, $f_{\text{d}}$, and the classical limit. Figure (c) illustrates the improvement ratio $\eta(p,q):=(f_{\text{e}}- f_{\text{d}})/f_{\text{d}}$ for various noisy models, where we set $\mN_1= \mN_{AD}$ and choose $\mN_2$ from bit flip $\mN_{BF}$, phase flip $\mN_{PF}$, depolarizing $\mN_{D}$, and amplitude damping $\mN_{AD}$ channels. 
   }
  \label{fig:tel-num}
\end{figure}

While it is evident that the teleportation performance of the superchannel-assisted protocols in Fig.~\ref{fig:qt}(f) outperforms that of the conventional approach in Fig.~\ref{fig:qt}(c), comparing the two can be challenging as both require the maximization over the set of LOCC operations, whose mathematical structure is difficult to characterize. Nevertheless, we propose a new pre-processing-assisted teleportation protocol that incorporates local unitaries as a pre-processing step, followed by the standard post-processing step $\Theta^{\text{post}}_0$ (Fig.~\ref{fig:qt}(e)); this protocol serves as a lower bound for the performance of superchannel-assisted protocol (Fig.~\ref{fig:qt}(f)). At the same time, we propose a positive-partial-transpose (PPT) teleportation protocol that optimizes the post-processing step over all possible completely-PPT-preserving operations (Fig.~\ref{fig:qt}(d)); this protocol functions as an upper bound for the performance of conventional protocol (Fig.~\ref{fig:qt}(c)). We find noisy entangling gates that the protocol in Fig.~\ref{fig:qt}(e) outperforms Fig.~\ref{fig:qt}(d), providing a clear distinction between the protocols shown in Fig.~\ref{fig:qt}(f) and Fig.~\ref{fig:qt}(c). In particular, we conducted a numerical experiment where system $A$ underwent an amplitude damping channel with a noise parameter $p= 0.23$, and system $B$ experienced a depolarizing channel with a noise parameter $q= 0.91$. To assess the impact of pre-processing on the performance under different parameter regimes, we present the results of a comprehensive analysis in Fig.~\ref{fig:tel-num}(a). Upon maximizing over all parameters of pre-processing, we present a comparison between protocols of Fig.~\ref{fig:qt}(e) and Fig.~\ref{fig:qt}(d) in Fig.~\ref{fig:tel-num}(b). Our results unequivocally demonstrate that, in this case, the performance of PPT teleportation, despite employing entangling operations, is inferior to pre-processing-assisted teleportation that relies solely on pre-processing operations. Finally, we present the improvement ratio for various noise models and parameters in Fig.~\ref{fig:tel-num}(c), highlighting the efficacy of pre-processing-assisted teleportation. 

In actual experiments, the magnitude of ambient noise frequencies is often considerably lower compared to the frequency of photon distribution. This particular characteristic allows for efficient learning of these noises within a defined time interval. As a result, the implementation of pre-processing techniques becomes not only theoretically viable but also practically feasible. While the findings presented here provide valuable insights into several noisy models, it is important to note that Fig.~\ref{fig:tel-num} represents only a small fraction of the potential models that 
can be explored. As such, we urge interested readers to consult \cite[Sec.~\rom{3}]{SM} for a thorough account of our numerical experiments and additional theoretical results that may further inform future investigations.

The characterization of the ultimate capabilities of a bipartite channel $\mE(p, q)$ in the context of quantum teleportation under all possible $S$-free superchannels is a fundamental question in quantum information theory. While previous discussions have focused on specific choices of free operations, a comprehensive understanding of the fundamental limitation remains elusive. Our framework of quantum network communication, which accounts for non-Markovian processes and encompasses channels as special cases, will address this problem and provide a unified approach to investigating the role of temporal entanglement in quantum communication.

\noindent \textbf{Quantum Repeater}-- Quantum communication, despite its numerous advantages over classical communication, encounters a fundamental obstacle when transmitting information over long distances~\cite{PhysRevLett.81.3563}: the likelihood of errors increases as the length of the channel connecting the parties grows. This is especially true when using photonic quantum communication over optical fibers, as both photon absorption and depolarization tend to increase exponentially with the length of the fiber. As a result, longer quantum communications experience significant signal quality degradation, leading to increased error rates and reduced efficiency and reliability. A promising approach to overcome this challenge involves the incorporation of quantum repeaters~\cite{PhysRevLett.81.5932}. These repeaters can be built with auxiliary particles at intermediate connection points, enabling the transmission of quantum information over longer distances~\cite{Duan2001,Yuan2008,PhysRevA.79.032325,PhysRevLett.104.180503,RevModPhys.83.33,Munro2012,PhysRevLett.112.250501,7010905,Azuma2015,PhysRevLett.117.210501,Pirandola2017,PhysRevLett.120.030503,Lukin2020,PhysRevX.10.021071,PRXQuantum.1.010301,Duan2021,Liorni_2021,Jiang2021,Sharman2021quantumrepeaters,PhysRevLett.127.220502,Wang2023,PhysRevApplied.19.024070}. By distributing entangled pairs of particles, quantum repeaters can help to maintain the integrity of the transmitted information and ensure that the communication remains reliable, efficient, and secure, even over long distances.

Entanglement swapping~\cite{PhysRevLett.71.4287,PhysRevLett.78.3031,PhysRevLett.80.3891} is a critical technical tool that enables quantum repeaters by creating a new, entangled pair between remote locations. This process bypasses the need for direct transmission, and entangling gates are valuable resources within this setup. The construction of a quantum repeater can be seen as a multipartite quantum channel shared between sender, receiver, and intermediate connection points. Since entanglement swapping is a post-processing step, a free superchannel should be the most general approach to manipulating the multipartite quantum channel that underlies quantum repeaters. Motivated by this, we investigate the potential of pre-processing-assisted quantum repeater protocols and their ability to enhance the performance of quantum communication. The behavior of superchannel-assisted repeaters remains a topic for investigation in our forthcoming framework for quantum networks.

\begin{figure}
  \centering
\includegraphics[width=0.48\textwidth]{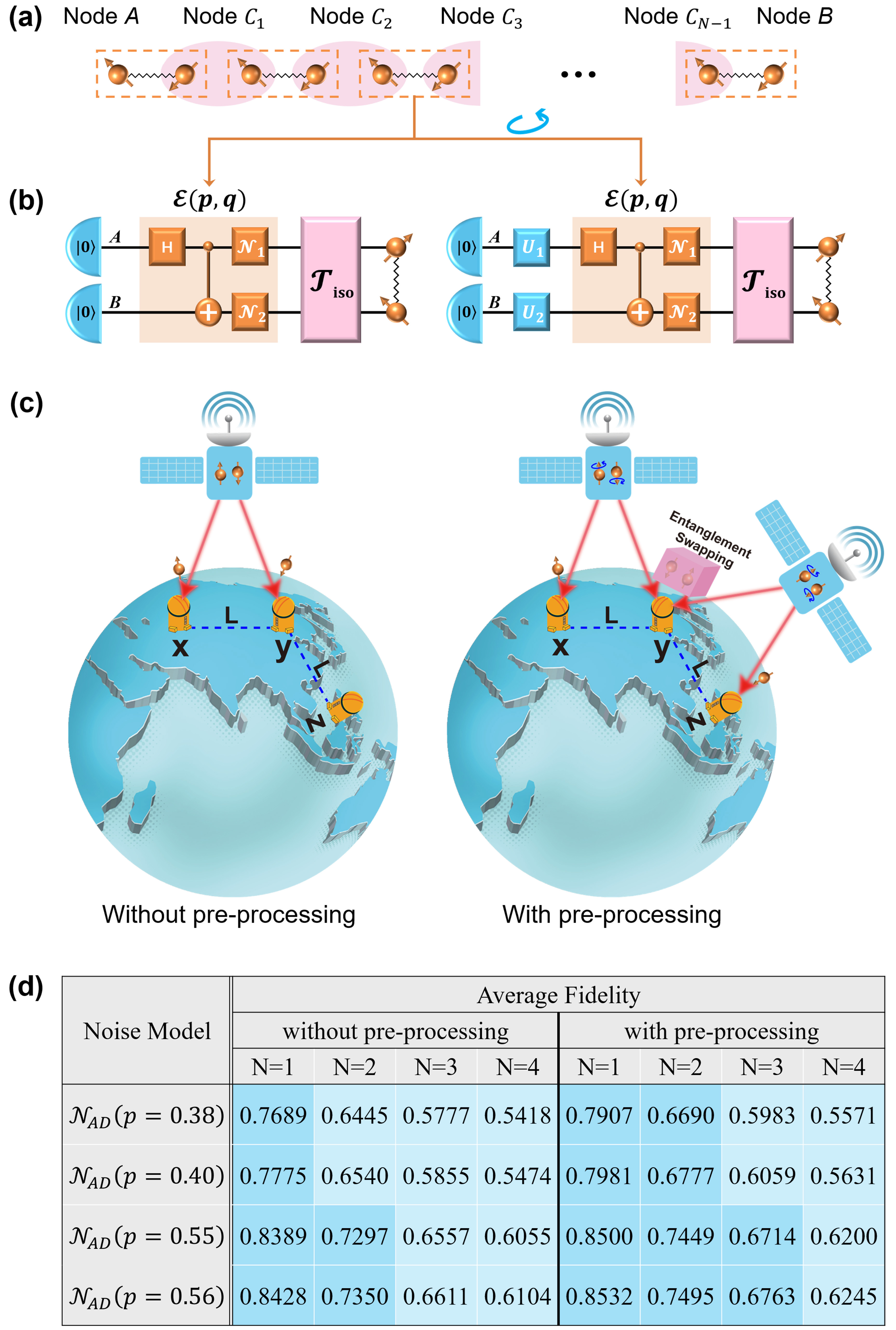}\\
   \caption{\textbf{Enhancing quantum repeaters:} 
   Figure (a) illustrates a quantum repeater schematic where distant nodes $A$ and $B$ are divided into $N$ segments. $N$ pairs of qubits are created and shared between nodes. Each intermediate node undergoes entanglement swapping. 
   Figure (b) depicts the creation of noise singlet shared between nodes. We use the same model, i.e., $\mE(p, q)$, discussed previously in teleportation (Fig.~\ref{fig:qt}), followed by isotropic twirling $\mT_{\text{iso}}(\cdot):=\int dU \ U\otimes\overline{U}(\cdot) U^{\dag}\otimes U^{\T}$~\cite{PhysRevA.59.4206}.
   Figure (c) shows how communication distance can be doubled using pre-processing techniques. Assuming a satellite distributes a noise singlet, with $\mN_1$ as the amplitude damping channel with $p=0.38$ and $\mN_2$ as the noiseless channel, to ground stations labeled $x$ and $y$, separated by a distance $L$. While information transmission from $x$ to an additional ground station $z$ located at the same distance $L$ from $y$ via entanglement swapping cannot exceed the classical limit, as $0.6445<2/3$, pre-processing can be used to surpass the limit without additional resource consumption.
   Figure (d) analyzes the fidelity of a state obtained through $N$ copies of noise singlet under entanglement swapping, and compare their performance with and without pre-processing. The amplitude damping channel $\mN_1$ has a noisy parameter chosen from $\{0.38, 0.40, 0.55, 0.56\}$, while the identity channel $\mN_2=\id$ represents noiseless evolution. The average fidelity is indicated by blue shading, with dark blue representing performance that exceeds the classical limit of $2/3$, and light blue indicating cases where the performance falls below this limit.
   }
  \label{fig:rp}
\end{figure}

In the conventional protocol of quantum repeaters, a communication channel between two distant nodes $A$ and $B$ is divided into $N$ segments, and $N$ pairs of qubits are created, with each pair being shared between a connection point and a node (Fig.~\ref{fig:rp}(a)). Let us now demonstrate the benefits of using pre-processing -- that is, how pre-processing applied prior to multipartite quantum channel that creates quantum repeaters can enhance their performance. Our research focuses on the use of noisy singlets, as illustrated in Fig.~\ref{fig:rp}(b). We compare the performance of a conventional quantum repeater, where entanglement swapping is performed directly on the noisy singlets, to that of a pre-processing-assisted quantum repeater, where pre-processing is applied before entangling gate that creates quantum repeater. The only difference between the noisy singlets in Fig.~\ref{fig:rp}(b) is the presence of pre-processing. The pre-processing assisted protocol outperforms the conventional approach, as it maximizes over all possible pre-processings, while the conventional approach is simply obtained by setting the local unitary in pre-processing to the identity. To evaluate the effectiveness,
we computed the improvement ratio $\zeta(p, q)$, which represents the percentage increase in performance attained through the use of pre-processing (see~\cite[Sec.~\rom{4}A]{SM} for a formal definition). By connecting $N$ adjacent pairs through entanglement swapping, we can generate a new pair with fidelity $1/4+3\left((4F_1-1)/3\right)^N/4$ with $F_1$ being the entanglement fidelity of noisy singlet shown in the left subfigure of Fig.~\ref{fig:rp}(b). Assuming that the maximization over pre-processing has been taken into account, the entanglement fidelity of the resulting state can be expressed as $1/4+3\left(1+\zeta(p, q)\right)^N\left((4F_1-1)/3\right)^N/4$, displaying an exponential growth, compared with the one without pre-processing, as the number of $N$ increases.

Next, we will transition to conducting numerical experiments to further investigate this topic. Specifically, we examine the case with a single noisy gate: $\mN_1$ is the amplitude damping channel with noisy parameter $p=0.38$, while at the same time $\mN_2$ is the noiseless channel. Assume that such a state can be used to teleport quantum information with length $L$. Our results demonstrate that without pre-processing, only one copy of such a state can beat the classical limit $2/3$, resulting in a communication distance of $L$. However, with pre-processing, even the connection of two copies, with a communication distance of $2L$, can outperform the classical bound of communication. Fig.~\ref{fig:rp}(c) portrays a schematic diagram of the doubling of communication distance, which highlights the significant potential of pre-processing techniques for global quantum communication. We also observe a similar doubling effect in the case of $p=0.4$, where the use of pre-processing techniques doubles the communication distance in transmitting quantum information. 
However, the data in the last two rows of Fig.~\ref{fig:rp}(d) reveals that doubling the communication distance is not always feasible. Nevertheless, the pre-processing techniques enable the extension of connective copies of states in the quantum network of Fig.~\ref{fig:rp}(a) from $2$ to $3$ for different noise parameters. A more comprehensive analysis has been included in~\cite[Sec.~\rom{4}A]{SM}. 

\noindent \textbf{Quantum Network}-- The development of a reliable quantum network capable of transmitting information over long distances is an indispensable step towards the advancement of quantum communication and computation. The quantum repeater has emerged as a promising approach for achieving this goal. However, the assumption of identical Bell measurements on independent and identically distributed (i.i.d.) quantum states in each round, which is commonly used in this approach, cannot be justified a priori despite its analytical convenience. In reality, fluctuations in error rates can occur due to variations in nodes or environmental factors, necessitating the introduction of an adaptive communication protocol that can dynamically adjust its behavior in response to such changes. The presence of imperfect devices and transmission channels further exacerbates this problem, highlighting the need for continuous monitoring and adjustment of the network to ensure reliable and efficient quantum communication. In classical wireless communication, adaptive communication protocols~\cite{10.1145/313451.313529,10.1145/1322263.1322295,10.1145/1754414.1754423} have proven effective in optimizing communication parameters to account for variations in network conditions. Extending this approach to the quantum realm, we propose a framework called \emph{adaptive quantum communication protocols}. The framework comprises multiple rounds of quantum communications, connected by quantum memory, which enables the optimization of communication parameters based on the current state of the network. This allows for the adaptation of communication parameters to changing conditions, ensuring that communication remains reliable and efficient even in the presence of noise and other environmental factors.

Single-round quantum network communications serve as fundamental building blocks within adaptive quantum communication protocols. Such a network, as depicted in Fig.~\ref{fig:Network-single-round}(a), is constructed using a multipartite quantum channel $\mE$, responsible for generating and maintaining entanglement among the sender Alice, the receiver Bob, and any intermediary agents involved. Notably, both quantum teleportation and repeaters can be regarded as specific instances of this general description. To fully unfold the potential of $\mE$, it is essential to employ a $\mS$-free manipulation $\Theta$, as illustrated in Fig.~\ref{fig:Network-single-round}(b). The term ``free'' in this context refers to the experimental accessibility of manipulating $\mE$ in a laboratory setting, enabling the creation of a communication channel $\Theta(\mE)$ between sender Alice and receiver Bob. 

\begin{figure}
    \centering   \includegraphics[width=0.48\textwidth]{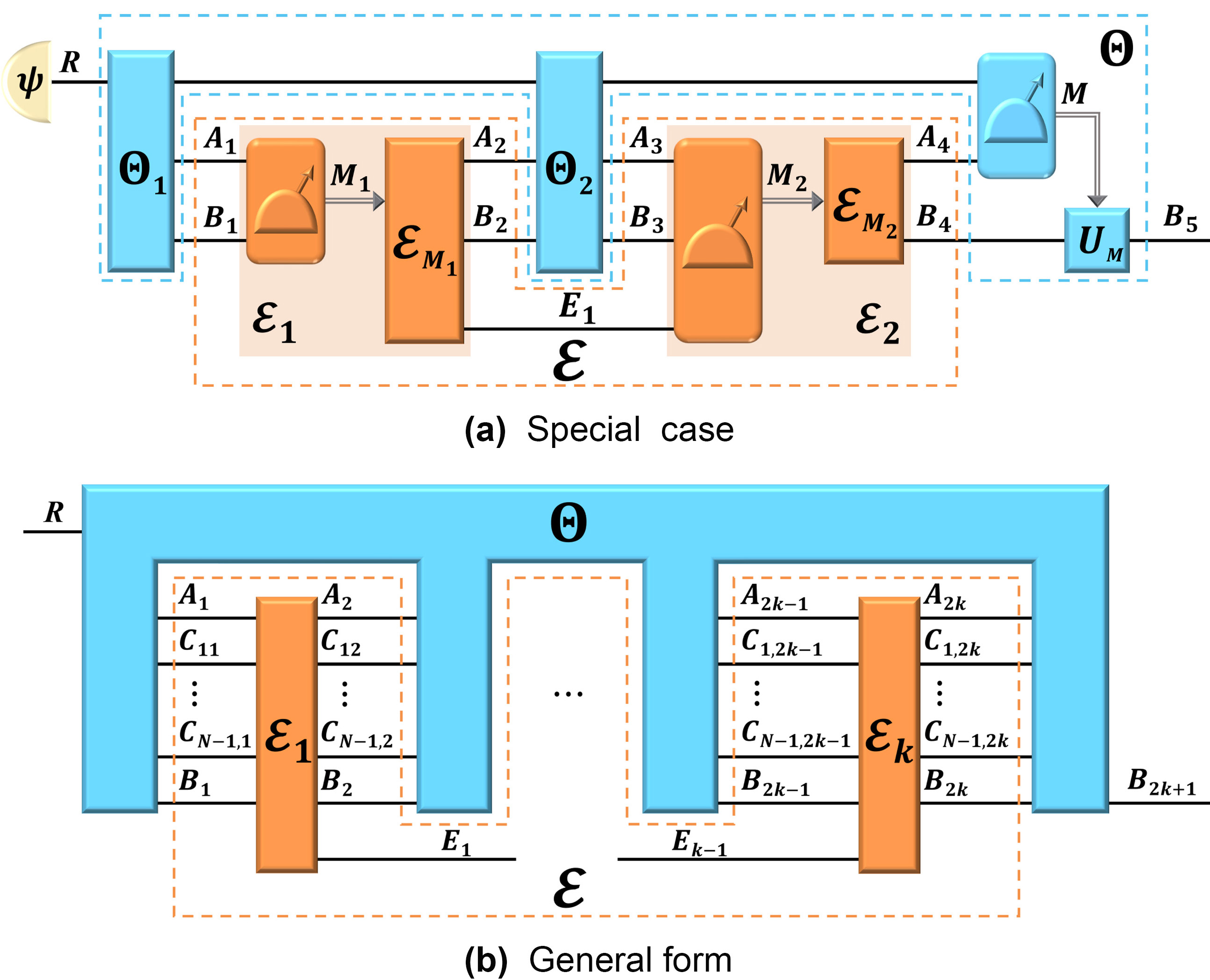}
    \caption{\textbf{Adaptive quantum communication protocols:} 
    Figure (a) shows an example of adaptive quantum communication protocol. The protocol uses multipartite operations $\mE_1$ and $\mE_2$ implemented via a positive operator-valued measure (POVM) and a retrieving quantum channel.
    Figure (b) presents the general adaptive quantum communication protocol $\mE$ in orange, which can be converted into a communication channel between the message and the receiver's system using free manipulation $\Theta$ in blue.
    }
    \label{fig:net}
\end{figure}

For a protocol to be truly adaptive, it should incorporate multiple rounds of quantum network communications linked by memory. Fig.~\ref{fig:net}(a) provides a concrete example of adaptive quantum communication protocol, which helps to develop a more intuitive understanding. This protocol involves two resourceful bipartite operations, denoted as $\mE_1$ and $\mE_2$, each comprising a learning machine implemented through a joint POVM on multiple systems, and a retrieving device realized using a quantum channel. The communication between these operations is dynamically adjusted based on changing conditions of the channel, thanks to the use of a memory-assisted quantum feedback control scheme. This dynamic adjustment allows the protocol to respond to variations in network conditions, such as the presence of intermediate nodes between sender and receiver. While this special scheme does not account for intermediate nodes, the inclusion of such nodes can greatly enhance the long-distance quantum communication capabilities of the protocol. 

An adaptive quantum communication protocol is formulated by composing finite rounds of multipartite quantum channels, denoted by $\mE_1$ through $\mE_k$, resulting in the overall non-Markovian process $\mE:= \mE_k\circ\cdots\circ\mE_1$. This shared resource, $\mE$, is used by Alice, Bob, and intermediaries over multiple time points. To ensure secure communication, they apply a $\mS$-free manipulation $\Theta$, resulting in channel $\Theta(\mE)$ between the message and receiver systems, as illustrated in Fig.~\ref{fig:net}(b). The $d$-fidelity of $\mS$ distillation can be used to quantify the temporal entanglement associated with an adaptive quantum communication protocol. At the same time, the communication performance of the resulting channel $\Theta(\mE)$ can be characterized by the optimization $f_{\max}(\mE):= \max_{\Theta\,\,\mS\text{-free}} f(\Theta(\mE))$, where $f$ is the average fidelity. The following theorem shows that the performance is entirely determined by the amount of temporal entanglement that can be generated and utilized by the adaptive quantum communication protocol.

\begin{thm*}  
An adaptive quantum communication protocol $\mE$ can be converted into a communication channel $\Theta(\mE)$, shared between the message and receiver systems (Fig.~\ref{fig:net}(b)). This transformation utilizes a $\mS$-free manipulation $\Theta$, involving a series of quantum channels, each belonging to the set $\mS$, wherein $\mS$ encompasses $\text{LOCC}_{1}(\text{poly}(d))$ as a subset. The optimal performance $f_{\max}$, as measured by the average fidelity, under all possible $\mS$-free manipulations is given by
\begin{align}\label{eq:all-in-one}
f_{\max}(\mE)
=
\frac{d F_{\max}^{d,\mS}(\mE)+ 1}{d+1},
\end{align}
where $F_{\max}^{d,\mS}(\mE)$ stands for the $d$-fidelity of $\mS$ distillation of $\mE$ (Fig.~\ref{fig:ds}), and $d$ corresponds to the dimension of the message system. 
\end{thm*}

When $\mE$ adopts the specific form of $\mE(p, q)$, Eq.~\ref{eq:all-in-one} delineates the fundamental limits of teleportation considered in Fig.~\ref{fig:qt}. In the context of repeaters, where nodes are interconnected by implementing $\mE_1\otimes\cdots\otimes\mE_N$, the optimal performance thereof is governed by $F_{\max}^{d,\mS}(\mE_1\otimes\cdots\otimes\mE_N)$. If $\mE$ is regarded as a bipartite state subject to LOCC, our theorem encompasses and extends the earlier seminal findings presented in Ref.~\cite{PhysRevA.60.1888}.

The generality of our theorem arises from two aspects. Firstly, unlike many existing models, our model does not rely on any specific assumptions or structures about the protocol, making it applicable to a wide range of communication scenarios, including those with different numbers of communication rounds or spatial correlations shared between quantum systems. The only constraint is the principle of causality~\cite{PhysRevLett.87.170405,RevModPhys.76.93,Horodecki2019}, which ensures that information cannot be transmitted from the future to the past. This allows our theorem to be readily adapted to arbitrary quantum causal networks. Secondly, our approach does not assume any specific manipulations of the adaptive quantum communication protocols or make any assumptions about their mathematical properties. This lack of constraints enables our theorem to provide a more comprehensive and flexible framework for understanding and analyzing adaptive quantum communication protocols under different quantum resource theories.

\noindent \textbf{Discussion}-- In this work, we demonstrate that the performance of a quantum communication protocol is solely determined by its temporal entanglement. Therefore, the effectiveness of communication is contingent upon the manipulation of temporal entanglement. In quantum teleportation, the temporal entanglement is carried by a bipartite channel between two time points, and its manipulation is characterized by a superchannel. Building upon this consideration, we propose a novel superchannel-assisted teleportation protocol. Our numerical experiments show that even a subprotocol with pre-processing can surpass the conventional approach. In contrast to catalyst-assisted protocols~\cite{PhysRevLett.83.3566,PhysRevA.71.042319,PhysRevA.103.022403,PhysRevX.11.011061,PhysRevLett.126.150502,PhysRevLett.127.080502,PhysRevLett.127.150503,PhysRevLett.128.240501,PhysRevLett.129.120506}, our method stands out by eliminating the need for ancillary systems, resulting in a more streamlined and practical approach for experimental implementation. In quantum repeaters, temporal entanglement is carried by tensors of bipartite channels shared between connection points, and their manipulation is also characterized by the framework of superchannels. Theoretical investigation has further shown that the pre-processing techniques can lead to an exponential growth in performance with respect to the number of connection points, offering a promising approach to achieving high-fidelity quantum communication over long distances. In addition, numerical experiments have demonstrated the advantages conferred by doubling the communication distance within specific scenarios, providing further evidence for the effectiveness of pre-processing techniques in quantum repeaters.

Adaptive quantum communication protocols are a key component in the construction of general quantum networks, which enable the iterative exchange of quantum information between parties. Among these protocols, quantum memory plays a critical role in facilitating communication, allowing for the storage and retrieval of quantum states that can be accessed and manipulated by the parties involved. The performance of these protocols in transmitting quantum information is determined by the temporal entanglement of the network. Understanding the relationship between quantum network communication and temporal entanglement can help improve the performance of non-Markovian communication protocols in practice. Specifically, more temporal entanglement leads to better performance in communication. 

The need for further investigation is apparent, and it begs the question: how can we calculate the optimal performance of an adaptive quantum communication protocol? This problem depends on two factors: the set of free operations that are allowed in a given physical laboratory, and how optimization can be performed under these free operations~\cite{PhysRevLett.123.070502,PhysRevA.104.022401,Berta2022,PhysRevA.107.012428,doi:10.1116/5.0135467}. Solving such questions requires a more detailed discussion, which we leave for future works. Finally, it is essential to note that, advances in quantum communication protocols have implications beyond quantum information processing. For example, quantum networks can improve the accuracy and resolution of clocks and sensors~\cite{Komar2014,Guo2020,Liu2021,Beloy2021,Malia2022}, secure communication, and enable new forms of distributed and cloud-based computing. Therefore, further research in this area is critical for advancing both fundamental science and technology.


\section*{Acknowledgments}
We would like to thank Varun Narasimhachar, Davit Aghamalyan, Rahul Arvind, and Mile Gu for fruitful discussions. This research is supported by the National Research Foundation, Singapore and A*STAR under its Quantum Engineering Programme (NRF2021-QEP2-02-P03), and by A*STAR under its Central Research Funds. 


\bibliography{Bib}
\end{document}



\title{Fundamental Limitations on Communication over a Quantum Network\\Supplemental Material}

\author{Junjing~Xing}
\address{College of Intelligent Systems Science and Engineering, Harbin Engineering University, Harbin,
Heilongjiang 150001, People's Republic of China}
\author{Tianfeng~Feng}
\address{State Key Laboratory of Optoelectronic Materials and Technologies and School of Physics, 
Sun Yat-sen University, Guangzhou, 
Guangdong 510275, People's Republic of China}
\author{Zhaobing~Fan}
\email{fanzhaobing@hrbeu.edu.cn}
\address{College of Intelligent Systems Science and Engineering, Harbin Engineering University, Harbin,
Heilongjiang 150001, People's Republic of China}
\author{Haitao~Ma}
\email{hmamath@hrbeu.edu.cn}
\address{College of Mathematics Science, Harbin Engineering University, Harbin,
Heilongjiang 150001, People's Republic of China}
\author{Kishor~Bharti}
\email{kishor.bharti1@gmail.com}
\address{Institute of High Performance Computing (IHPC), Agency for Science, Technology and Research (A*STAR), 1 Fusionopolis Way, \#16-16 Connexis, Singapore 138632, Republic of Singapore}
\author{Dax~Enshan~Koh}
\email{dax\_koh@ihpc.a-star.edu.sg}
\address{Institute of High Performance Computing (IHPC), Agency for Science, Technology and Research (A*STAR), 1 Fusionopolis Way, \#16-16 Connexis, Singapore 138632, Republic of Singapore}
\author{Yunlong~Xiao}
\email{mathxiao123@gmail.com}
\address{Institute of High Performance Computing (IHPC), Agency for Science, Technology and Research (A*STAR), 1 Fusionopolis Way, \#16-16 Connexis, Singapore 138632, Republic of Singapore}

\date{\today}
             
\begin{abstract} 
Quantum communication is a rapidly advancing and highly promising field that has the potential to revolutionize information transmission by providing unprecedented levels of security and efficiency. In this work, we explore the intrinsic relationship between the entanglement of quantum dynamics and the performance of quantum communication, from quantum teleportation and quantum repeaters to general quantum networks. To ensure a comprehensive and rigorous treatment of the topic, we provide this supplemental material that offers a detailed examination and rigorous proof of the results presented in the main text. This supplementary material is designed to be explicit and self-contained, with a focus on providing a thorough understanding of the underlying principles and mechanisms that drive quantum communication. Importantly, the supplementary material provides an extensive array of supplementary data, encompassing not only the results expounded upon in the main text but also other pertinent information essential for a comprehensive understanding of the communication over a quantum network. 
\end{abstract}

\maketitle
\tableofcontents

\newpage


\section{\label{sec:QD}Quantum Dynamics}
In this section, we lay the foundation for the results presented in this work by introducing our notations and investigating the physical properties of a quantum channel through its Choi-Jamio\l kowski operator (refer to Def.~\ref{def:CJ}). We also formulate the sequential composition of channels using the link product (refer to Def.~\ref{def:LP}) in Subsec.~\ref{subsec:QC}. After a bit of these concepts, we delve into two important quantifiers of a quantum channel: the average fidelity (refer to Def.~\ref{def:AF}) and the entanglement fidelity (refer to Def.~\ref{def:EF}). By employing the link product in Lem.~\ref{lem:AFEF} of Subsec.~\ref{subsec:QQC}, we establish their intrinsic connection. In addition, we discuss the most general way of manipulating a quantum channel -- quantum superchannel via the language of non-signalling maps (refer to Def.~\ref{def:NS}) in Subsec.~\ref{subsec:QSC}. Finally, we address the broadest framework of non-Markovian quantum processes, which we refer to as quantum circuit fragments in Subsec.~\ref{subsec:Fragment}. Our aim in this section is to provide a clear and concise understanding of the concepts and terminologies that will be used throughout the rest of the work.


\subsection{\label{subsec:QC}Quantum Channels: From Choi-Jamio\l kowski Isomorphism to Link Product}

Just as Lego bricks can be assembled and connected in countless ways to construct various objects, including castles, machinery, vehicles, Millennium Falcon, and even robots, quantum channels are the fundamental building blocks of the circuit model for quantum computing and information processing. These channels play a critical role in numerous aspects of quantum theory, including the preparation of quantum states, implementation of quantum measurements, handling of noise from system-environment interactions, and execution of quantum gates in a circuit. Essentially, the concept of quantum channels is integral to the whole field of quantum theory, acting as a foundational tool that enables scientists to push the boundaries of what is possible in the realm of quantum information processing and beyond. 
In this subsection, we will briefly review the theory of quantum channels, which forms the backbone of this work. The main mathematical toolkit employed here is known as Choi-Jamio\l kowski (CJ) isomorphism~\cite{JAMIOLKOWSKI1972275,CHOI1975285} in literature. While we provide an overview of the concept and the corresponding mathematical framework, readers seeking a more in-depth understanding can refer to Refs.~\cite{nielsen_chuang_2010,wilde_2013,watrous_2018} and the cited references therein, which offer comprehensive insights into the topic.

Generally speaking, a linear map $\mE_{A\to B}$ from linear operators to linear operators is called a quantum channel if it satisfies the following conditions: 
\begin{itemize}
\item {\bf Complete Positivity (CP)};
\item {\bf Trace Preservation (TP)}.
\end{itemize}
Here, complete positivity means that by applying $\mE_{A\to B}$ to part of a quantum state $\rho_{A^{'}A}$, the resultant state is still positive semidefinite, namely $\id_{A^{'}}\otimes\mE_{A\to B}(\rho_{A^{'}A})\geqslant0$. Meanwhile, as indicated by the name, the trace-preserving property guarantees that $\Tr[\mE(\rho)]= \Tr[\rho]$ for all $\rho$. A powerful tool for investigating the mathematical properties of quantum channels is the so-called Choi-Jamio\l kowski (CJ) isomorphism. Formally, it is defined as

\begin{mydef}
{Choi-Jamio\l kowski Isomorphism~\cite{JAMIOLKOWSKI1972275,CHOI1975285}}{CJ}
Given a quantum channel $\mathcal{E}_{A\to B}$, there is an isomorphism between map $\mathcal{E}_{A\to B}$ and matrix $J^{\mE}_{AB}$ (known as the Choi-Jamio\l kowski operator of channel $\mathcal{E}_{A\to B}$). Mathematically, $J^{\mE}_{AB}$ is defined as
\begin{align}\label{eq:cj}
J^{\mathcal{E}}_{AB}
:=
\id_{A}\otimes\mathcal{E}_{A^{'}\to B} (\Gamma_{AA^{'}}),
\end{align}
where $\Gamma_{AA^{'}}:= \ketbra{\Gamma}{\Gamma}_{AA^{'}}$, and $\ket{\Gamma}_{AA^{'}}:= \sum_{i}\ket{ii}_{AA^{'}}$ stands for the unnormalized maximally entangled state (UMES) with $A^{'}$ being a replica of system $A$, and $\{\ket{i}\}$ being an orthonormal basis on $A$. For maximally entangled state (MES) $\phi^{+}_{AA^{'}}$, we define the Choi-Jamio\l kowski state of channel $\mathcal{E}_{A\to B}$ as
\begin{align}\label{eq:cj-state}
\rho^{\mE}_{AB}
:=
\id_{A}\otimes\mathcal{E}_{A^{'}\to B} (\phi^{+}_{AA^{'}}),
\end{align}
where $\phi^{+}_{AA^{'}}:= \ketbra{\phi^{+}}{\phi^{+}}_{AA^{'}}$, and $\ket{\phi^{+}}_{AA^{'}}:= (1/\sqrt{d_A})\sum_{i}\ket{ii}_{AA^{'}}$ with $d_{A}$ being the dimension of system $A$. It is now straightforward to check that $\rho^{\mE}_{AB}= J^{\mathcal{E}}_{AB}/d_A$.
\end{mydef}

\begin{figure}[h]
\includegraphics[width=0.75\textwidth]{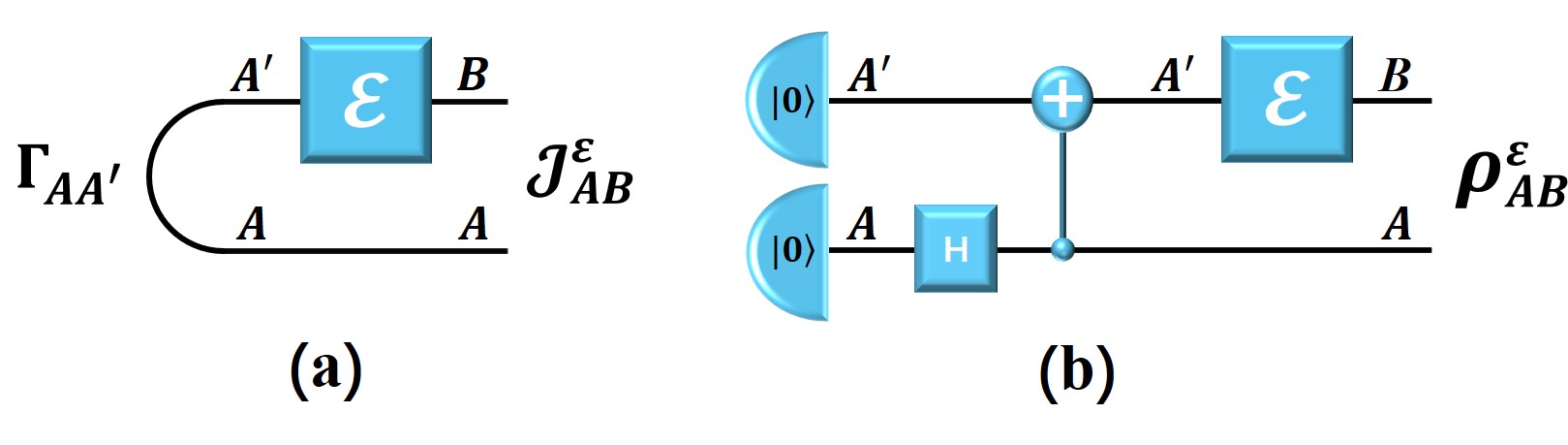}
\caption{(color online) Generations of the Choi-Jamio\l kowski operator $J^{\mE}_{AB}$ (see Eq.~\ref{eq:cj}) and state $\rho^{\mE}_{AB}$  (see Eq.~\ref{eq:cj-state}). In (a), $\Gamma_{AA^{'}}$ stands for $\ketbra{\Gamma}{\Gamma}_{AA^{'}}$ with $\ket{\Gamma}_{AA^{'}}:=\sum_{i}\ket{ii}_{AA^{'}}$. When the unnormalized operator $\Gamma$ has been replaced by the maximally entangled state $\phi^+$, we obtain the Choi-Jamio\l kowski state $\rho^{\mE}_{AB}$, shown in (b).}
\label{fig:cj}
\end{figure} 

\noindent
See Fig.~\ref{fig:cj} for an illustration of Eq.~\ref{eq:cj} and Eq.~\ref{eq:cj-state}. 

\begin{remark}
In the fields of quantum information, computing, and condensed matter physics, conventional diagrammatic techniques, such as tensor networks~\cite{doi:10.1080/14789940801912366,ORUS2014117,Bridgeman_2017,RevModPhys.93.045003} and ZX-calculus~\cite{10.1007/978-3-540-70583-3_25,Coecke_2011,coecke_kissinger_2017,https://doi.org/10.48550/arxiv.2012.13966,coecke2023basic}, are widely used. Although our graphical representation bears similarity to these existing techniques, there exists a key difference in our approach. Specifically, in the conventional techniques, a cup symbol (i.e., $\subset$) represents a vector $\ket{\Gamma}$, whereas in our work, it represents the unnormalized maximally entangled state $\Gamma$. Similarly, in the conventional approach, a cap symbol (i.e., $\supset$) represents the effect $\bra{\Gamma}$. However, in our approach, the cup acting on an operator $M$ represents the trace of $\Gamma\cdot M$, denoted as $\Tr[\Gamma\cdot M]$. Finally, we note that the \emph{yanking equations}, which relate the graphical notations in ZX-calculus, also hold in our graphical language.
\end{remark}

\begin{remark}\label{remark:effect}
The Choi-Jamio\l kowski operator for quantum effects can be defined by substituting the quantum channel in Eq.~\ref{eq:cj} with a completely positive and trace non-increasing map (CPTNI). This mathematical tool is crucial in understanding the impact of quantum measurement and probabilistic protocols in quantum information processing.
\end{remark}

Equipped with Choi-Jamio\l kowski isomorphism, we can now translate the properties of completely positive and trace-preserving into the language of Choi-Jamio\l kowski operators. More precisely, a linear map $\mathcal{E}_{A\to B}$ is completely positive if and only if its Choi-Jamio\l kowski operator $J^{\mathcal{E}}_{AB}$ is positive semidefinite, i.e., $J^{\mathcal{E}}_{AB}\geqslant0$. Furthermore, such a map is trace-preserving if and only if $J^{\mathcal{E}}_{AB}$ satisfies $\Tr_{B}[J^{\mathcal{E}}_{AB}]= \1_{A}$. To conclude, given a linear operator $J^{\mathcal{E}}_{AB}$, it represents a quantum channel from system $A$ to system $B$ if both $J^{\mathcal{E}}_{AB}\geqslant0$ and $\Tr_{B}[J^{\mathcal{E}}_{AB}]= \1_{A}$ are satisfied. 

A natural question that arises is how to determine the Choi-Jamio\l kowski operator of the sequential composition of two quantum channels, such as $\mE_{A\to B}$ and $\mF_{B\to C}$, denoted by $(\mF\circ\mE)_{A\to C}$, in terms of $J^\mE_{AB}$ and $J^\mF_{BC}$. To tackle this question, we can utilize a useful tool called the \emph{link product} $\star$. This product allows us to efficiently compute the Choi-Jamio\l kowski operator of the composite channel and is a powerful technique for analyzing the properties of quantum channels.

\begin{mydef}
{Link Product~\cite{PhysRevA.80.022339,PhysRevLett.101.060401,PhysRevA.84.012311}}{LP}
Given two Hermitian operators $M$ and $N$ with $X$ being the intersection of the systems on which M and N act, their link product, denoted by $M\star N$, is defined as 
  \begin{equation}
      M \star N:= \Tr_X[M^{\T_X}\cdot N],
  \end{equation}
indicating that the link between $M$ and $N$, namely system $X$, has been ``swallowed'' by the product $\star$. Here $\Tr_X$ and $\T_X$ represent the partial trace and partial transpose over the common system $X$ respectively, and the identity operator $\1$ has been ignored.
\end{mydef}

\begin{figure}[h]
    \centering
\includegraphics[width=0.85\textwidth]{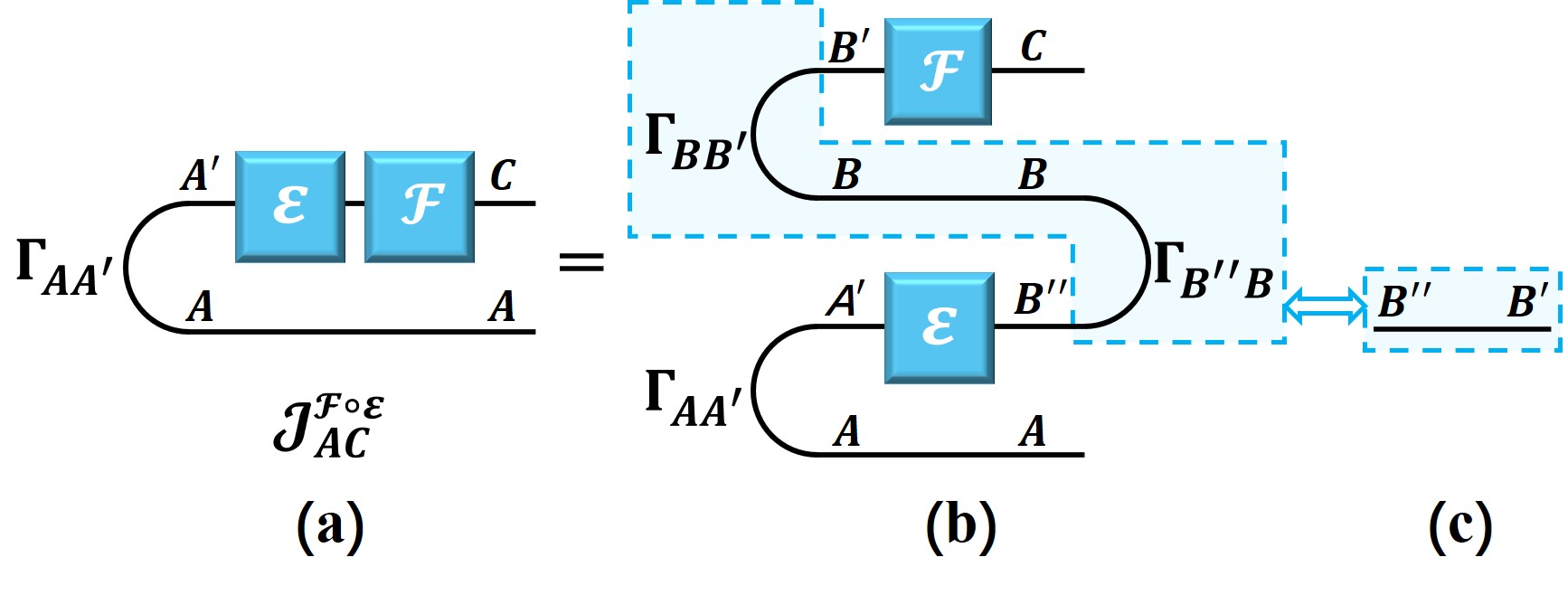}
\caption{(Color online) Physical realization (up to a scalar) of the Choi-Jamio\l kowski operator $J^{\mF \circ\mE}_{AC}$ (see Eq.~\ref{eq:c-lp} for detail). Here the cup (i.e. $\subset$) stands for the preparation of $\Gamma_{AA^{'}}$, and the cap (i.e. $\supset$) represents the effect of measuring the operator with respect to $\bra{\Gamma}$. In particular, (a) illustrates the Choi-Jamio\l kowski operator of channel $\mF\circ\mE$. In (b), we re-derive the Choi-Jamio\l kowski operator $J^{\mF\circ\mE}$ by using the yanking equation, which states that the snake line shown in blue dashed box of (b) is equal to the line drawn in blue dashed box of (c).
}
\label{fig:lp}
\end{figure}

\noindent With the help of the link product $\star$, we can now address the question posed earlier regarding the Choi-Jamio\l kowski operator of the composite channel $(\mF\circ\mE)_{A\to C}$ in terms of $J^\mE_{AB}$ and $J^\mF_{BC}$. In this context, system $B$ serves as the link between $J^\mE_{AB}$ and $J^\mF_{BC}$, and we can obtain the resultant Choi-Jamio\l kowski operator $J^{\mE\circ\mF}_{AC}$ as follows
\begin{align}\label{eq:c-lp}
J^{\mF\circ\mE}_{AC}
=
J^\mE_{AB}\star J^\mF_{BC}
=
\Tr_{B}
[
\left((J^\mE_{AB})^{\T_B}\otimes\1_{C}\right)
\cdot
\left(\1_{A}\otimes J^\mF_{BC}\right)
].
\end{align}
A graphical visualization of link product appeared in Eq.~\ref{eq:c-lp} is demonstrated in Fig.~\ref{fig:lp}, where the equivalence between 

\noindent
Fig.~\ref{fig:lp}(a) and Fig.~\ref{fig:lp}(b) is obtained by using the yanking equation~\cite{10.1007/978-3-540-70583-3_25,Coecke_2011,coecke_kissinger_2017,https://doi.org/10.48550/arxiv.2012.13966}; that is
\begin{align}\label{eq:c-lp-pt}
\Tr_{B}
[
\left((J^\mE_{AB})^{\T_B}\otimes\1_{C}\right)
\cdot
\left(\1_{A}\otimes J^\mF_{BC}\right)
]
=
\Tr_{BB^{''}}[
\left(J^\mE_{AB^{''}}\otimes J^{\mF}_{BC}\right)
\cdot
(\1_{A}\otimes\Gamma_{B^{''}B}\otimes\1_{C})
].
\end{align}
The definition of the link product (e.g., Eq.~\ref{eq:c-lp}) involves the partial transpose $\T_{B}$, which is not completely positive and may seem unphysical. However, our Eq.~\ref{eq:c-lp-pt} and Fig.~\ref{fig:lp} demonstrate that the partial transpose can be obtained by implementing Bell state measurements (up to a scalar factor). This highlights the fact that seemingly unphysical operations like $\T_{B}$ can have practical interpretation in quantum information processing.

Let us discuss two specific scenarios to illustrate the usefulness of the link product $\star$ (see Def.~\ref{def:LP}): First, the link product helps us to demonstrate the action of quantum channel in terms of its Choi-Jamio\l kowski operator. More precisely, given a quantum state $\rho$ acting on system $A$, the quantum channel $\mE_{A\to B}$ will transform it into
\begin{align}\label{eq:rho-e}
\mE(\rho)= \rho\star J^{\mE}_{AB},
\end{align}
where $J^{\mE}_{AB}$ is the Choi-Jamio\l kowski operator of channel $\mE$; Second, the \emph{Born rule} can be rewritten as a link product. In particular, denote by $p_x$ the probability distribution obtained by measuring $\rho$ with respect to $0\leqslant M_x\leqslant\1$, then we have
\begin{align}\label{eq:rho-m}
p_x=
\rho\star M_x^{\T}.
\end{align}
Here $M_x^{\T}$ stands for the Choi-Jamio\l kowski operator of effect $M_x$ (see Remark~\ref{remark:effect}).

\begin{remark}\label{remark:lk}
The link product $\star$ is commutative under SWAP~\cite{Chiribella_2016}. Take operators $M_{AB}$ and $N_{BC}$ for instance, we have 
\begin{align}
M_{AB}\star N_{BC}= 
\text{SWAP}_{A\leftrightarrow C}(N_{BC}\star M_{AB}).
\end{align}
Once the order of operations is determined, we can disregard the need for the SWAP operation and treat the resulting link product as a commutative operation. However, it is important to note that without the inclusion of the SWAP operation, the property of commutativity no longer holds. For convenience of presentation, in this work, we will fix the order of systems to make the link product commutative. Let's take Eq.~\ref{eq:c-lp} for example, when the lexicographical order is considered, we have
\begin{align}
J^{\mF\circ\mE}_{AC}
=
J^\mE_{AB}\star J^\mF_{BC}
=
J^\mF_{BC}\star J^\mE_{AB}.
\end{align}
Similarly, Eqs.~\ref{eq:rho-e} and \ref{eq:rho-m} can be rewritten as $\mE(\rho)= J^{\mE}_{AB}\star\rho$ and $p_x= M_x^{\T}\star\rho$ respectively. Furthermore, it is worth noting that in addition to commutativity, the link product $\star$ is also associative. This property allows for the grouping of multiple Choi-Jamio\l kowski operators without changing the outcome of their product and is a useful feature in the analysis of complex network structures.
\end{remark}


\subsection{\label{subsec:QQC}How to Quantify a Quantum Channel: Twist It} 
While there is a variety of ways to quantify quantum channels, good channel quantifiers must have operational interpretations in some quantum information processing or quantum computation tasks. In this subsection, we review two such channel quantifiers, namely the average fidelity $f$ and the entanglement fidelity $F$. Specifically, the average fidelity $f(\mE)$ characterizes the performance of a quantum channel $\mE$ in transmitting quantum information, and meanwhile the entanglement fidelity $F(\mE)$ measures how $\mE$ preserves the quantum correlations between distant parties. Despite the obvious differences in mathematical expression and physical interpretation, the average fidelity and the entanglement fidelity enjoy a one-to-one correspondence, which was originally proved by M. Horodecki, P. Horodecki, and R. Horodecki in Ref.~\cite{PhysRevA.60.1888}. Here we provide an alternative proof of the one-to-one correspondence between average fidelity and entanglement fidelity, showing a different perspective on these channel quantifiers. Our mathematical toolkit includes the link product (see Def.~\ref{def:LP}) and twirling.

Let us start with the first channel quantifier: the average fidelity. For an input state $\psi_A$ acting on system $A$, the quantum channel $\mE_{A\to B}$ will transform it into the state $\mE(\psi)$, and we can use the \emph{Uhlmann fidelity} $F_{\text{U}}$~\cite{UHLMANN1976273,doi:10.1080/09500349414552171} to measure the similarity between the input and output states of channel $\mE_{A\to B}$, i.e., 
\begin{align}\label{eq:FU}
F_{\text{U}}(\psi, \mE(\psi))= \bra{\psi}\mE(\psi)\ket{\psi}= \Tr[\psi\cdot\mE(\psi)].
\end{align}
In this work, we have adopted the notational convention that $\psi:=\ketbra{\psi}{\psi}$. Now the consideration of similarity between the input and output states over all possible $\psi$ leads to the average fidelity $f(\mE)$.

\begin{mydef}
{Average Fidelity}{AF}
The average fidelity $f(\mE_{A\to B})$ of a quantum channel $\mE_{A\to B}$ is defined by
\begin{align}\label{eq:af}
f(\mE_{A\to B})
:=
\int d\psi F_{\text{U}}(\psi_{B}, \mE_{A\to B}(\psi_{A}))
=
\int d\psi \Tr[\psi_{B}\cdot\mE_{A\to B}(\psi_{A})],
\end{align}
where the integral is over the \emph{Haar measure}, satisfying $\int d\psi=1$. Notation $F_{\text{U}}$ stands for the Uhlmann fidelity. 
\end{mydef}

The second channel quantifier considered in this work is the so-called entanglement fidelity. Applying the quantum channel $\mE_{A^{'}\to B}$ to part of the maximally entangled state $\phi^{+}_{AA^{'}}$ leads to the output state $\id_{A}\otimes\mE_{A^{'}\to B}(\phi^{+}_{AA^{'}})$, whose difference with the maximally entangled state implies the entanglement fidelity $F(\mE)$.

\begin{mydef}
{Entanglement Fidelity}{EF}
The entanglement fidelity $F(\mE_{A\to B})$ of a quantum channel $\mE_{A\to B}$ is defined by
\begin{align}\label{eq:ef}
F(\mE_{A\to B})
:=
F_{\text{U}}(\phi^{+}_{AB}, \mE_{A^{'}\to B}(\phi^{+}_{AA^{'}}))
=
\Tr[\phi^{+}_{AB}\cdot\left(\id_{A}\otimes\mE_{A^{'}\to B}(\phi^{+}_{AA^{'}})\right)],
\end{align}
where $F_{\text{U}}$ stands for the Uhlmann fidelity, with $\phi^{+}$ being the maximally entangled state. 
\end{mydef}

\begin{remark}
Since the systems of quantum channels are usually clear from the context, the notations of average fidelity $f(\mE_{A\to B})$ (see Eq.~\ref{eq:af}) and entanglement fidelity $F(\mE_{A\to B})$ (see Eq.~\ref{eq:ef}) will be abbreviated to $f(\mE)$ and $F(\mE)$ respectively, and the maximally entangled state will be shortened to $\phi^{+}$. Sometimes we will also ignore the identity channel $\id$. Adopting these conventions, the cumbersome expressions can be simplified. For example, Eqs.~\ref{eq:af} and \ref{eq:ef} can be rewritten as $f(\mE)= \int d\psi \Tr[\psi\cdot\mE(\psi)]$ and $F(\mE)= \Tr[\phi^{+}\cdot\mE(\phi^{+})]$ respectively. 
\end{remark}

Symmetry is a fundamental concept in nature and is of great importance in quantum information processing and computing. Twirling is a powerful technique in quantum information theory that enables the analysis of quantum systems by exploiting their symmetries. Specifically, twirling transforms a quantum state in a symmetric manner, preserving its properties under certain operations. This technique has numerous applications in quantum information theory, such as the analysis of entanglement~\cite{959270}, quantum error correction~\cite{PhysRevA.88.012314,cai2019constructing}, and communication protocols~\cite{PhysRevA.60.1888}. In this context, we can explore the symmetry of quantum dynamics by applying different types of twirling. One such type is channel twirling, denoted by $\mT_{\text{ch}}$.

\begin{mydef}
{Channel Twirling}{CT}
Given a quantum channel $\mE$, the channel twirling of $\mE$ is defined by
\begin{align}\label{eq:ct}
\mT_{\text{ch}}(\mE)
:=
\int dU \,\mU^{\dag}\circ\mE\circ\mU,
\end{align}
where the integral is over the \emph{Haar measure}, with $\mU(\rho):= U\,\rho\, U^{\dag}$ and $\mU^{\dag}(\rho):= U^{\dag}\,\rho\, U$. Notation $\circ$ stands for the composition of channels, and $^{\dag}$ represents the \emph{Hermitian adjoint} of operators.
\end{mydef}

\noindent It is worth mentioning that both the average fidelity $f(\mE)$ (see Def.~\ref{def:AF}) and entanglement fidelity $F(\mE)$ (see Def.~\ref{def:EF}) are invariant under channel twirling $\mT_{\text{ch}}$. To make our work self-contained, we will provide the formal statements about these properties, including their proofs.

\begin{mylem}
{Invariance of Average Fidelity under Channel Twirling}{IAF}
The average fidelity $f(\mE)$ (see Eq.~\ref{eq:af}) of a quantum channel $\mE$ is invariant under channel twirling $\mT_{\text{ch}}$ (see Eq.~\ref{eq:ct}); that is
\begin{align}\label{eq:iaf}
f(\mT_{\text{ch}}(\mE))
=
f(\mE).
\end{align}
\end{mylem}

\begin{proof}
According to Def.~\ref{def:AF} and Def.~\ref{def:CT}, we have 
\begin{align}
f(\mT_{\text{ch}}(\mE))
&=
\int d\psi \Tr[\psi\cdot\mT_{\text{ch}}(\mE)(\psi)]\\
&=
\int d\psi\int dU\,
\Tr[\psi\cdot\mU^{\dag}\circ\mE\circ\mU(\psi)]\\
&=
\int d\psi\int dU\,
 \Tr[\mU(\psi)\cdot\mE(\mU(\psi))]\\
&=
\int dU f(\mE)\\
&=
f(\mE),
\end{align}
which completes the proof. 
\end{proof}

\begin{mylem}
{Invariance of Entanglement Fidelity under Channel Twirling}{IEF}
The entanglement fidelity $F(\mE)$ (see Eq.~\ref{eq:ef}) of a quantum channel $\mE$ is invariant under channel twirling $\mT_{\text{ch}}$ (see Eq.~\ref{eq:ct}); that is
\begin{align}\label{eq:ief}
F(\mT_{\text{ch}}(\mE))
=
F(\mE).
\end{align}
\end{mylem}

\begin{proof}
Note that $\overline{U}\otimes U \ket{\phi^{+}}=\ket{\phi^{+}}$, where $\overline{U}$ stands for the complex conjugate of $U$. Hence, we obtain 
\begin{align}
F(\mT_{\text{ch}}(\mE))
&=
\Tr[\phi^{+}\cdot\mT_{\text{ch}}(\mE)(\phi^{+})]\\
&=
\Tr[\phi^{+}\cdot\left(\mU^{\dag}\circ\mE\circ\mU(\phi^{+})\right)]\\
&=
\Tr[
\left(\1\otimes U\,\phi^{+}\,\1\otimes U^{\dag}\right)
\cdot
\left(U^{\T}\otimes\1\,\mE(\phi^{+})\,\overline{U}\otimes\1\right)
]\\
&=
\Tr[
\left(\overline{U}\otimes U\,\phi^{+}\,U^{\T}\otimes U^{\dag}\right)
\cdot
\mE(\phi^{+})
]\\
&=
\Tr[
\phi^{+}
\cdot
\mE(\phi^{+})
]\\
&=
F(\mE),
\end{align}
which completes the proof of the lemma.
\end{proof}
Another type of twirling is the isotropic twirling $\mT_{\text{iso}}$~\cite{PhysRevA.59.4206}, which was originally introduced by M. Horodecki and P. Horodecki in formulating the reduction criterion of separability.

\begin{mydef}
{Isotropic Twirling}{IT}
Given a bipartite state $\rho$, the isotropic twirling of $\rho$ is defined by
\begin{align}\label{eq:it}
\mT_{\text{iso}}(\rho)
:=
\int dU \,
U\otimes\overline{U}\,\rho\, U^{\dag}\otimes U^{\T},
\end{align}
where the integral is over the \emph{Haar measure}.
\end{mydef}

\begin{figure}[h]
    \centering
    \includegraphics[width=0.55\textwidth]{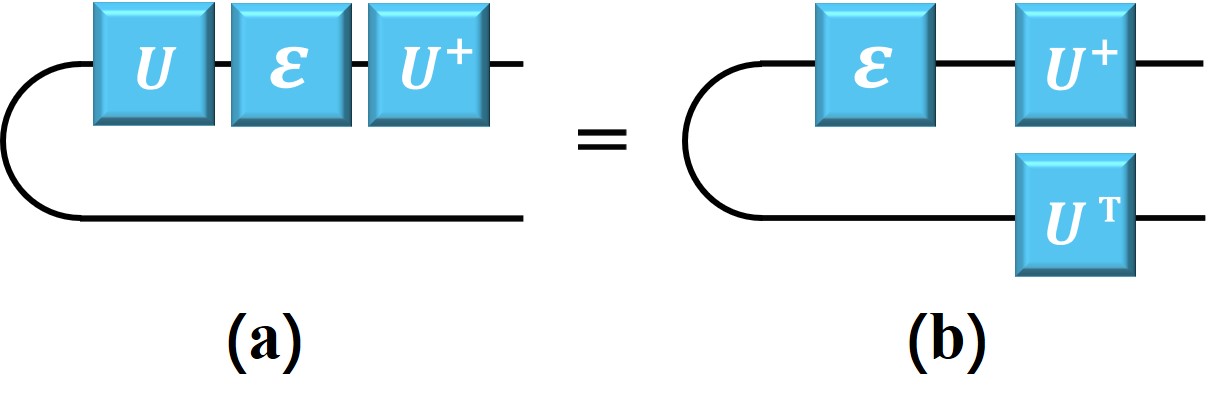}
    \caption{(Color online) Equivalence between the Choi-Jamio\l kowski operators of $J^{\mU^{\dag}\circ\mE\circ\mU}$ in (a) and $U^{\T}\otimes U^{\dag}\,J^{\mE}\, \overline{U}\otimes U$ in (b), which leads to the equation $J^{\mT_{\text{ch}}(\mE)}= \mT_{\text{iso}}(J^{\mE})$ as required.}
    \label{fig:tc}
\end{figure}

\begin{remark}\label{rem:ct-it}
Despite their differences in expressions, channel twirling $\mT_{\text{ch}}$ (see Eq.~\ref{eq:ct}) and isotropic twirling $\mT_{\text{iso}}$ (see Eq.~\ref{eq:it}) are closely related with each other. In particular, the Choi-Jamio\l kowski operator $J^{\mT_{\text{ch}}(\mE)}$ of channel $\mT_{\text{ch}}(\mE)$ is equal to $\mT_{\mathrm{iso}}(J^{\mE})$, obtained by implementing the isotropic twirling $\mT_{\text{iso}}$ to the Choi-Jamio\l kowski operator of channel $\mE$. A graphic proof of above statement, i.e., $J^{\mT_{\text{ch}}(\mE)}= \mT_{\text{iso}}(J^{\mE})$, is given in Fig.~\ref{fig:tc}. 
\end{remark}

The isotropic twirling $\mT_{\text{iso}}$ is indeed a quantum channel, and thus its Choi-Jamio\l kowski operator $J^{\mT_{\text{iso}}}$ completely characterizes the mathematical properties of $\mT_{\text{iso}}$. For more related results, we refer the interested readers to the work of E. M. Rains~\cite{959270}, which also contains the prototype of link product (see Def.~\ref{def:LP}).

\begin{mylem}
{Choi-Jamio\l kowski Operator of Isotropic Twirling~\cite{959270}}{CJIT}
Assume the isotropic twirling $\mT_{\text{iso}}$ is a map from systems $A_1\otimes B_1$ to $A_2\otimes B_2$, with $d:= \dim A_1= \dim A_2= \dim B_1= \dim B_2$, then its Choi-Jamio\l kowski operator is given by
\begin{align}\label{eq:cjit}
J^{\mT_{\text{iso}}}_{A_1B_1A_2B_2}
=
\phi^{+}_{A_1B_1}\otimes\phi^{+}_{A_2B_2}
+
\frac{1}{d^2-1} 
(\1_{A_1B_1}- \phi^{+}_{A_1B_1}) 
\otimes 
(\1_{A_2B_2}- \phi^{+}_{A_2B_2})
,
\end{align}
where $\phi^{+}$ is the maximally entangled state. By using the conventional abbreviations $A:= A_1A_2$, $B:= B_1B_2$, $\mathfrak{1}:= A_1B_1$, and $\mathfrak{2}:= A_2B_2$, Eq.~\ref{eq:cjit} can be simplified to $J^{\mT_{\text{iso}}}_{AB}= \phi^{+}_{\mathfrak{1}}\otimes\phi^{+}_{\mathfrak{2}}+(\1_{\mathfrak{1}}- \phi^{+}_{\mathfrak{1}})\otimes(\1_{\mathfrak{2}}- \phi^{+}_{\mathfrak{2}})/(d^2-1)$.
\end{mylem}

We now wish to show that the output state of isotropic twirling $\mT_{\text{iso}}: \mathfrak{1}\to \mathfrak{2}$ is always a noisy singlet, i.e. in the form of $p\phi^{+}_{\mathfrak{2}}+(1-p)\1_{\mathfrak{2}}/d^2$. To do so, let us consider the link product between a given state $\rho_{\mathfrak{1}}$ acting on system $\mathfrak{1}$ and the Choi-Jamio\l kowski operator of $\mT_{\text{iso}}$. Written out explicitly, we have

\begin{align}
\mT_{\text{iso}}(\rho_{\mathfrak{1}})
&=
J^{\mT_{\text{iso}}}_{AB}\star \rho_{\mathfrak{1}}\\
&=
\left(\phi^{+}_{\mathfrak{1}}\otimes\phi^{+}_{\mathfrak{2}}+\frac{1}{d^2-1}(\1_{\mathfrak{1}}- \phi^{+}_{\mathfrak{1}})\otimes(\1_{\mathfrak{2}}- \phi^{+}_{\mathfrak{2}})\right)
\star\rho_{\mathfrak{1}}\\
&=
\Tr[\phi^{+}_{\mathfrak{1}}\cdot\rho_{\mathfrak{1}}]\phi^{+}_{\mathfrak{2}}
+
\frac{(1-\Tr[\phi^{+}_{\mathfrak{1}}\cdot\rho_{\mathfrak{1}}])}{d^2-1}(\1_{\mathfrak{2}}- \phi^{+}_{\mathfrak{2}})
\\
&=
\frac{\Tr[\phi^{+}_{\mathfrak{1}}\cdot\rho_{\mathfrak{1}}]d^2-1}{d^2-1}
\phi^{+}_{\mathfrak{2}}
+
\frac{(1-\Tr[\phi^{+}_{\mathfrak{1}}\cdot\rho_{\mathfrak{1}}])d^2}{d^2-1}
\frac{\1_{\mathfrak{2}}}{d^2},\label{eq:iso-4}
\end{align}
implying that

\begin{mythm}
{Noisy Singlet}{NS}
Assume the isotropic twirling $\mT_{\text{iso}}$ is a map from systems $A_1\otimes B_1$ to $A_2\otimes B_2$ (i.e., $A_1B_1\to A_2B_2$), with $d:= \dim A_1= \dim A_2= \dim B_1= \dim B_2$, and $\rho_{\mathfrak{1}}$ being an input state acting on system $A_1\otimes B_1$, then the output state under $\mT_{\text{iso}}$ is a noisy singlet, i.e.,
\begin{align}\label{eq:noisy_singlet}
p\phi^{+}_{\mathfrak{2}}+(1-p)\frac{\1_{\mathfrak{2}}}{d^2},
\end{align}
with the parameter $p$ being characterized by
\begin{align}
p=\frac{\Tr[\phi^{+}_{\mathfrak{1}}\cdot\rho_{\mathfrak{1}}]d^2-1}{d^2-1}.
\end{align}
Here we have adopted the notational conventions that $\mathfrak{1}:= A_1B_1$, and $\mathfrak{2}:= A_2B_2$.
\end{mythm}

Let us turn our attention to the output channel of channel twirling $\mT_{\text{ch}}: A_2\to A_1, B_1\to B_2$. For any input channel $\mE: A_1\to B_1$, the Choi-Jamio\l kowski operator of output channel $\mT_{\text{ch}}(\mE)$ is then given by
\begin{align}
J^{\mT_{\text{ch}}(\mE)}_{\mathfrak{2}}
&=
\mT_{\text{iso}}(J^{\mE}_{\mathfrak{1}})\label{eq:ch-1}\\
&=
\left(\phi^{+}_{\mathfrak{1}}\otimes\phi^{+}_{\mathfrak{2}}+\frac{1}{d^2-1}(\1_{\mathfrak{1}}- \phi^{+}_{\mathfrak{1}})\otimes(\1_{\mathfrak{2}}- \phi^{+}_{\mathfrak{2}})\right)
\star J^{\mE}_{\mathfrak{1}}\label{eq:ch-2}\\
&=
\Tr[\phi^{+}_{\mathfrak{1}}\cdot J^{\mE}_{\mathfrak{1}}]\phi^{+}_{\mathfrak{2}}
+
\frac{(d-\Tr[\phi^{+}_{\mathfrak{1}}\cdot J^{\mE}_{\mathfrak{1}}])}{d^2-1}(\1_{\mathfrak{2}}- \phi^{+}_{\mathfrak{2}})
\label{eq:ch-3}
\\
&=
\frac{\Tr[\phi^{+}_{\mathfrak{1}}\cdot J^{\mE}_{\mathfrak{1}}]d^2-d}{d(d^2-1)}
\Gamma_{\mathfrak{2}}
+
\frac{(d-\Tr[\phi^{+}_{\mathfrak{1}}\cdot J^{\mE}_{\mathfrak{1}}])d}{d^2-1}
\frac{\1_{\mathfrak{2}}}{d},
\end{align}
where the first equality (i.e., Eq.~\ref{eq:ch-1}) has been investigated in Remark.~\ref{rem:ct-it}, the second equality (i.e., Eq.~\ref{eq:ch-2}) is a result of Lem.~\ref{lem:CJIT}, and the third equality (i.e., Eq.~\ref{eq:ch-3}) follows from the property of trace-preservation; namely,
\begin{align}
\1_{\mathfrak{1}}\star J^{\mE}_{\mathfrak{1}}= \Tr_{A_1}[\1_{A_1}]= d.
\end{align}
Thanks to the Choi-Jamio\l kowski isomorphism, we know that $\Gamma_{\mathfrak{2}}$ represents the Choi-Jamio\l kowski operator of the noiseless quantum channel $\id_{A_2\to B_2}$. Meanwhile, $\1_{\mathfrak{2}}/d$ is the Choi-Jamio\l kowski operator of the replacement channel $\1_{B_2}/d\otimes\Tr_{A_2}$. Written in full, that is
\begin{align}
\mT_{\text{ch}}(\mE)
=
\left(
\frac{\Tr[\phi^{+}_{\mathfrak{1}}\cdot J^{\mE}_{\mathfrak{1}}]d^2-d}{d(d^2-1)}
\right)
\id_{A_2\to B_2}
+
\left(
\frac{(d-\Tr[\phi^{+}_{\mathfrak{1}}\cdot J^{\mE}_{\mathfrak{1}}])d}{d^2-1}
\right)
\frac{\1_{B_2}}{d}\otimes\Tr_{A_2}.
\end{align}
To conclude, we have the following theorem for channel twirling $J^{\mT_{\text{ch}}}$.

\begin{mythm}
{Depolarizing Channel}{DC}
Given a channel $\mE: A_1\to B_1$, the channel twirling $\mT_{\text{ch}}: A_2\to A_1, B_1\to B_2$ will transform it into a depolarizing channel, i.e., 
\begin{align}\label{eq:dc-1}
\mT_{\text{ch}}(\mE)
=
p\,\id_{A_2\to B_2}+(1-p)\,\frac{\1_{B_2}}{d}\otimes\Tr_{A_2},
\end{align}
where the parameter $p$ is given by
\begin{align}\label{eq:dc-2}
p
=
\frac{\Tr[\phi^{+}_{\mathfrak{1}}\cdot J^{\mE}_{\mathfrak{1}}]d^2-d}{d(d^2-1)},
\end{align}
with $d:= \dim A_1= \dim A_2= \dim B_1= \dim B_2$.
\end{mythm}

Before the end of this subsection, let us return to the average fidelity $f(\mE)$ (see Def.~\ref{def:AF}) and the entanglement fidelity $F(\mE)$ (see Def.~\ref{def:EF}). In Ref.~\cite{PhysRevA.60.1888}, M. Horodecki, P. Horodecki, and R. Horodecki find the essential connection between $f(\mE)$ and $F(\mE)$, in particular that

\begin{mylem}
{Connection between Average Fidelity and Entanglement Fidelity~\cite{PhysRevA.60.1888}}{AFEF}
Given a quantum channel $\mE: A\to B$, its average fidelity $f(\mE)$ (see Def.~\ref{def:AF}) and entanglement fidelity $F(\mE)$ (see Def.~\ref{def:EF}) satisfy the following equation,
\begin{align}\label{eq:afef}
f(\mE)
=
\frac{dF(\mE)+1}{d+1},
\end{align}
with $d:= \dim A= \dim B$.
\end{mylem}

\begin{remark}
There are several ways to prove Lem.~\ref{lem:AFEF}. The original proof was given by M. Horodecki, P. Horodecki, and R. Horodecki in Ref.~\cite{PhysRevA.60.1888}, which had later been simplified by M. A. Nielsen in Ref.~\cite{NIELSEN2002249}. Here we will provide an alternative proof based on the Choi-Jamio\l kowski operator of isotropic twirling $\mT_{\text{iso}}$ (see Lem.~\ref{lem:CJIT}) and the link product $\star$ (see Def.~\ref{def:LP}). In particular, our method relies heavily on Thm.~\ref{thm:DC} presented in this subsection, whose proof is completed by using the Choi-Jamio\l kowski operator of isotropic twirling and the link product. Hence, we conclude that our proof of Lem.~\ref{lem:AFEF} is based on $J^{\mT_{\text{iso}}}$ and $\star$. Besides its fundamental importance in understanding channel quantifiers, Lem.~\ref{lem:AFEF} is also a key to characterizing general quantum teleportation~\cite{PhysRevA.60.1888}.
\end{remark}

\begin{proof}
To prove this claim, we start with the input state $\psi$ and consider the corresponding output under quantum channel $\mT_{\text{ch}}(\mE)$. Thanks to Eq.~\ref{eq:dc-1} of Thm.~\ref{thm:DC}, we immediately obtain 
\begin{align}
\mT_{\text{ch}}(\mE)(\psi)
=
p\,\psi
+
(1-p)\,\frac{\1}{d},
\end{align}
with $p$ given by Eq.~\ref{eq:dc-2}. Thus, by using Lem.~\ref{lem:IAF}, we see that
\begin{align}\label{eq:afef-2}
f(\mE)
=
f(\mT_{\text{ch}}(\mE))
=
p+ \frac{1-p}{d}
=
\frac{p(d-1)+1}{d}.
\end{align}
On the other hand, the entanglement fidelity $F(\mE)$ of channel $\mE$ is equal to $\Tr[\phi^{+}_{\mathfrak{1}}\cdot J^{\mE}_{\mathfrak{1}}]/d$, implying that
\begin{align}\label{eq:afef-3}
p=\frac{
F(\mE)d^2-1
}{
d^2-1
}.
\end{align}
Combining Eq.~\ref{eq:afef-2} with Eq.~\ref{eq:afef-3} leads to Eq.~\ref{eq:afef}, as desired.
\end{proof}


\subsection{\label{subsec:QSC}Quantum Superchannels: Pre-Processing, Post-Processing, and Non-Signalling}

Exploring quantum channels primarily involves manipulating them in specific contexts. For instance, when dealing with a noisy quantum channel, determining its capacity to transmit information through the use of encoder and decoder leads to the development of quantum information theory -- a rapidly evolving and dynamic field that seeks to unleash the extraordinary capabilities of quantum mechanics in revolutionizing information processing. Mathematically, the sender's encoder and the receiver's decoder can be represented as a bipartite quantum operation known as a \emph{superchannel}~\cite{Chiribella_2008}. This concept encapsulates the most general way of manipulating a quantum channel, including the channel twirling $\mT_{\text{ch}}$ (see Def.~\ref{def:CT}) discussed in Subsec.~\ref{subsec:QQC}. Quantum superchannels also constitute a fundamental tool for understanding dynamical resources and the associated resource measures~\cite{8678741,Bartosz2021NC,PhysRevLett.127.060402}. In this subsection, we provide a rigorous introduction to the basic concepts of quantum superchannels and investigate their properties using the powerful Choi-Jamio\l kowski isomorphism (see Def.~\ref{def:CJ}).

A quantum superchannel is formed by combining two quantum channels, which can be implemented in sequence and connected through a quantum memory in accordance with the principle of causality in the theory of relativity~\cite{PhysRevLett.87.170405,RevModPhys.76.93,Horodecki2019}. This causal structure is formally known as \emph{non-signalling} (NS) between the parties involved in quantum coding protocols. A comprehensive review of non-signalling codes is beyond the scope of this discussion. Interested readers are referred to Ref.~\cite{7115934} and the references therein for a more detailed introduction. In the context of causal boxes, the causality property of quantum superchannels is referred to as \emph{semicausality}~\cite{10.1007/978-94-017-0990-3_45,PhysRevA.64.052309,T.Eggeling_2002}. The use of the Choi-Jamio\l kowski isomorphism has proven to be a powerful tool in analyzing the properties of semicausal maps and their relation to \emph{semilocalizability}, as discussed in previous works~\cite{PhysRevA.74.012305}. Now let us start with the definition of non-signalling.

\begin{mydef}
{Non-Signalling}{NS}
A bipartite channel $\Theta_{A_1B_1\to A_2B_2}$ is non-signalling from $B:=B_1B_2$ to $A:=A_1A_2$, denoted as $B\not\to A$, if
\begin{align}\label{eq:ns-ba}
\Tr_{B_2}\circ\,\Theta_{A_1B_1\to A_2B_2}
=
\Tr_{B_1}\otimes\,\mE_{A_1\to A_2},
\end{align}
holds for some quantum channel $\mE_{A_1\to A_2}$. Similarly, a bipartite channel $\Theta_{A_1B_1\to A_2B_2}$ is non-signalling (NS) from $A\to B$, denoted as $A\not\to B$, if 
\begin{align}\label{eq:ns-ab}
\Tr_{A_2}\circ\,\Theta_{A_1B_1\to A_2B_2}
=
\Tr_{A_1}\otimes\,\mF_{B_1\to B_2},
\end{align}
holds for some quantum channel $\mF_{B_1\to B_2}$. A map $\Theta$ is non-signalling if it is both $A\not\to B$ and $B\not\to A$.
\end{mydef}

\begin{figure}[h]
    \centering
    \includegraphics[width=0.9\textwidth]{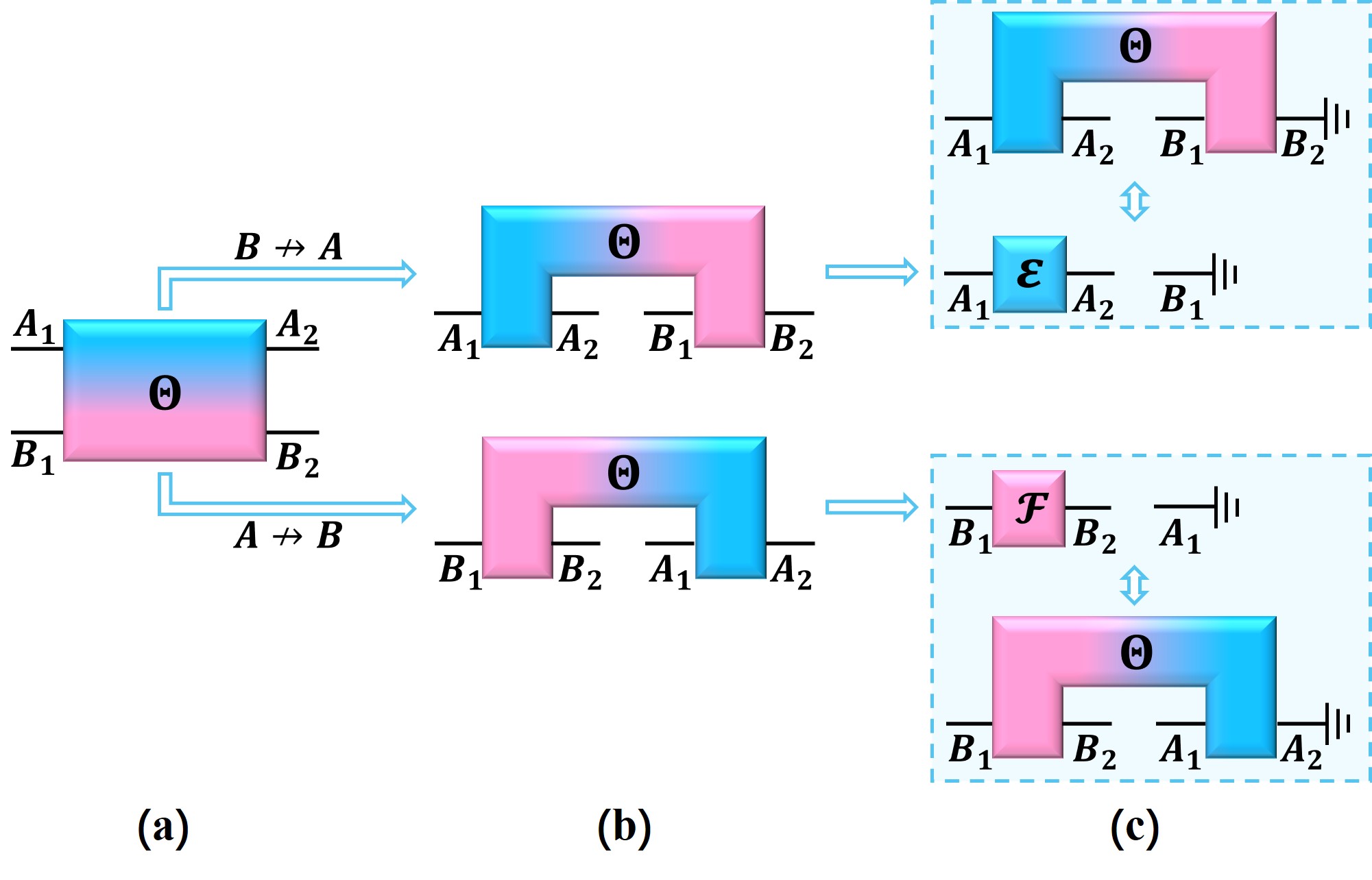}
    \caption{(Color online) Demonstration of non-signalling bipartite channel $\Theta$. A general bipartite quantum channel $\Theta_{A_1B_1\to A_2B_2}$ is depicted in (a), where systems $A_1$ and $A_2$ belongs to Alice, systems $B_1$ and $B_2$ belongs to Bob. If the bipartite channel $\Theta$ is non-signalling from $B$ to $A$, i.e., Bob's action cannot affect Alice, we arrive at the top of (b). In other words, tracing out system $B_2$ will lead to the form of $\mE_{A_1\to A_2}\otimes\Tr_{B_1}$ (see Eq.~\ref{eq:ns-ba}) for some quantum channel $\mE$ on Alice's side, shown in the top box of (c). On the other hand, if the bipartite channel $\Theta$ is non-signalling from $A$ to $B$, i.e. Alice's behavior won't influence Bob, the bipartite channel $\Theta$ can be transformed into the form illustrated in the bottom of (b). In this case, discarding system $A_2$ will imply the decomposition of bipartite operation between Alice and Bob; that is $\Tr_{A_1}\otimes\mF_{B_1\to B_2}$ (see Eq.~\ref{eq:ns-ab}) with $\mF$ being a quantum acting on Bob's side, painted in the bottom box of (c).
    }
    \label{fig:ns}
\end{figure}

\noindent
Physically, non-signalling from $B$ to $A$ means that no action on system $B$ can cause any detectable effect on system $A$. A visualization of non-signalling bipartite channels is provided in Fig.~\ref{fig:ns}.

\begin{figure}[h]
    \centering
    \includegraphics[width=1\textwidth]{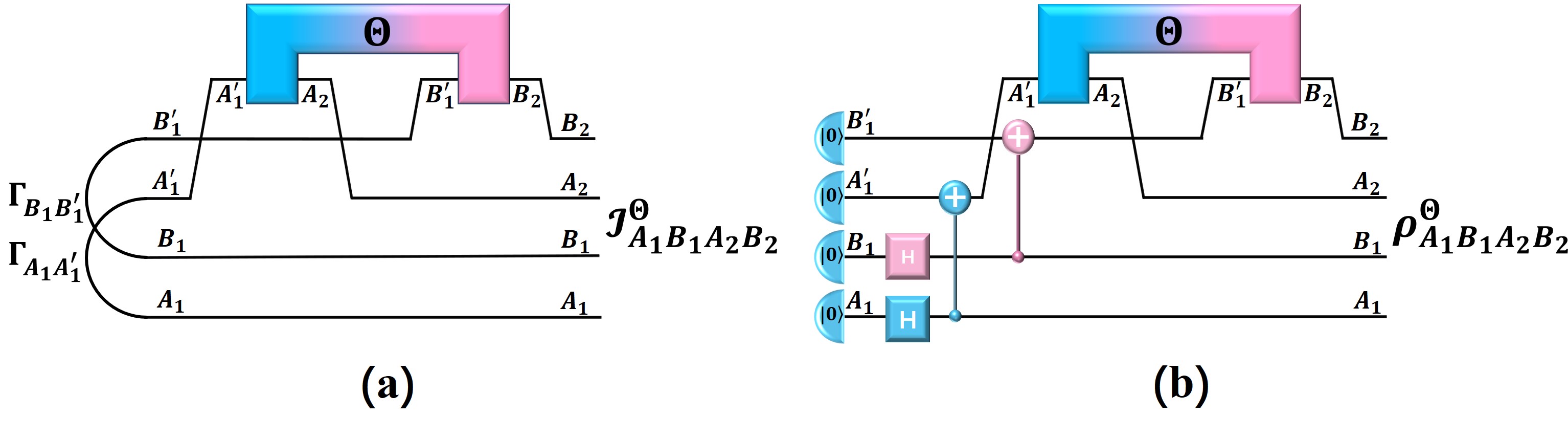}
    \caption{
    (Color online) Pictorial representation of the Choi-Jamio\l kowski operator $J^{\Theta}_{A_1B_1\to A_2B_2}$ (see Eq.~\ref{eq:sc-cj}) for superchannel $\Theta_{A_1B_1\to A_2B_2}$ that is non-signalling from $B$ to $A$ (a), and circuit realization of normalized Choi-Jamio\l kowski state $\rho^{\Theta}_{A_1B_1\to A_2B_2}$ (b). Here $\rho^{\Theta}_{A_1B_1\to A_2B_2}= J^{\Theta}_{A_1B_1\to A_2B_2}/d_{A_1}d_{B_1}$.
    }
    \label{fig:sc-cj}
\end{figure}

Given a quantum channel $\mE_{A_2\to B_1}$, the general quantum supermap $\Theta_{A_1B_1\to A_2B_2}$ that maps $\mE_{A_2\to B_1}$ to a channel $\Theta(\mE)_{A_1\to B_2}$, called superchannel, satisfies the following conditions~\cite{Chiribella_2008}:

\begin{itemize}
\item {\bf Completely Positivity (CP)};
\item {\bf Trace Preservation (TP)};
\item {\bf Non-Signalling (NS) from $B$ to $A$}.
\end{itemize}
Similar to the construction of Choi-Jamio\l kowski operator for a quantum channel (see Def.~\ref{def:CJ} and Fig.~\ref{fig:cj}), we can introduce the Choi-Jamio\l kowski operator for superchannels. In particular, the Choi-Jamio\l kowski operator for superchannel $\Theta_{A_1B_1\to A_2B_2}$ is defined as
\begin{align}\label{eq:sc-cj}
J^{\Theta}_{A_1B_1A_2B_2}
:=
\id_{A_1}\otimes\Theta_{A^{'}_1B^{'}_1\to A_2B_2}\otimes\id_{B_1} 
(\Gamma_{A_1A^{'}_1}\otimes\Gamma_{B_1B^{'}_1}),
\end{align}
where $\Gamma_{A_1A^{'}_1}$ and $\Gamma_{B_1B^{'}_1}$ are unnormalized maximally entangled states acting on systems $A_1A^{'}_1$ and $B_1B^{'}_1$ respectively. An illustration of Eq.~\ref{eq:sc-cj} is given in Fig.~\ref{fig:sc-cj}(a), including the corresponding quantum circuit realization (see Fig.~\ref{fig:sc-cj}(b)).  

Using the language of Choi-Jamio\l kowski isomorphism, we say that a quantum supermap $\Theta_{A_1B_1\to A_2B_2}$ is completely
positive if and only if its Choi-Jamio\l kowski operator $J^{\Theta}_{A_1B_1A_2B_2}$ is positive semidefinite, namely $J^{\Theta}_{A_1B_1A_2B_2}\geqslant0$. It is said to be trace-preserving if and only if $\Tr_{B_1B_2}J^{\Theta}_{A_1B_1A_2B_2}= \1_{A_1A_2}$. Finally, such a supermap is non-signalling from $B$ to $A$ if and only if its Choi-Jamio\l kowski operator $J^{\Theta}_{A_1B_1A_2B_2}$ satisfies 

\begin{align}
\Tr_{B_2}[J^{\Theta}_{A_1B_1A_2B_2}]
=
J^{\mE}_{A_1A_2}\otimes\1_{B_1},
\end{align}
where $J^{\mE}_{A_1A_2}$ stands for the Choi-Jamio\l kowski operator of some quantum channel $\mE$ from $A_1$ to $A_2$. 

\begin{figure}[h]
    \centering
    \includegraphics[width=0.7\textwidth]{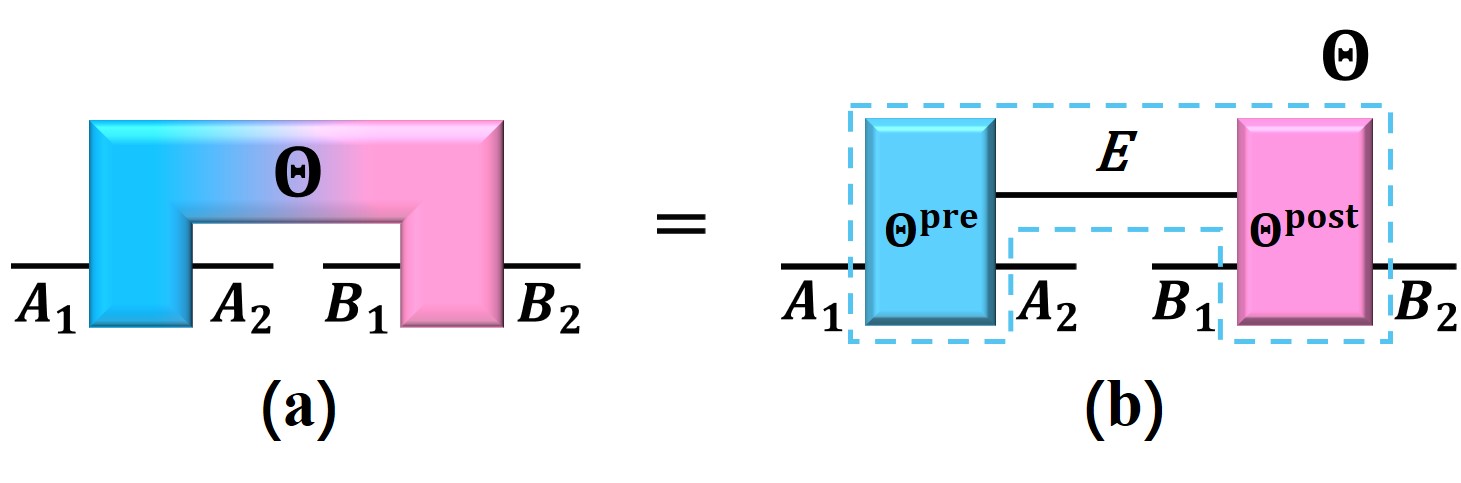}
    \caption{
    (Color online) Decomposition of superchannel $\Theta_{A_1B_1\to A_2B_2}$ (see Eq.~\ref{eq:sc-dec}). In other words, superchannel $\Theta$ (a) can be realized by preparing a pre-processing $\Theta^{\text{pre}}_{A_1\to A_2E}$ and a post-processing $\Theta^{\text{post}}_{B_1E\to B_2}$, which are connected by a hidden memory system $E$ (b).
    }
    \label{fig:FigSM06}
\end{figure}

Equivalently, a superchannel $\Theta_{A_1B_1\to A_2B_2}$ can also be viewed as the sequential composition of two quantum channels; that is
\begin{align}\label{eq:sc-dec}
\Theta_{A_1B_1\to A_2B_2}
=
\Theta^{\text{post}}_{B_1E\to B_2}\circ\Theta^{\text{pre}}_{A_1\to A_2E},
\end{align}
where $E$ is a memory system between $\Theta^{\text{pre}}$ and $\Theta^{\text{post}}$ (see Fig.~\ref{fig:FigSM06}). 
Here, quantum channels $\Theta^{\text{pre}}$ and $\Theta^{\text{post}}$ are known as the pre-processing and post-processing of $\Theta$ respectively. Denote their Choi-Jamio\l kowski operators as $J^{\text{pre}}$ and $J^{\text{post}}$, then it is straightforward to write down the Choi-Jamio\l kowski operator of superchannel $\Theta$ by using link product $\star$ (see Def.~\ref{def:LP}), i.e., 

\begin{align}\label{eq:sc-cj-lp}
J^{\Theta}_{A_1B_1A_2B_2}
=
J^{\text{pre}}_{A_1A_2}\star J^{\text{post}}_{B_1B_2}.
\end{align}
For any quantum channel $\mE_{A_2\to B_1}$, the Choi-Jamio\l kowski operator of resultant channel $\Theta(\mE)_{A_1\to B_2}$ under $\Theta_{A_1B_1\to A_2B_2}$ is characterized by
\begin{align}
J^{\Theta(\mE)}_{A_1B_2}
=
J^{\Theta}_{A_1B_1A_2B_2}\star J^{\mE}_{A_2B_1}
=
\Tr_{A_2B_1}[J^{\Theta}_{A_1B_1A_2B_2}
\cdot
\left(
(J^{\mE}_{A_2B_1})^{\T}\otimes\1_{A_1B_2}
\right)].
\end{align}
Then, for an input state $\rho_{A_1}$, the output state of channel $\Theta(\mE)_{A_1\to B_2}$ reads
\begin{align}
\Theta(\mE)_{A_1\to B_2}(\rho_{A_1})
=
J^{\Theta(\mE)}_{A_1B_2}\star\rho_{A_1}
=
J^{\Theta}_{A_1B_1A_2B_2}\star J^{\mE}_{A_2B_1}\star\rho_{A_1}
=
\Tr_{A_2B_1}[J^{\Theta}_{A_1B_1A_2B_2}
\cdot
\left(
\rho_{A_1}^{\T}\otimes
(J^{\mE}_{A_2B_1})^{\T}\otimes\1_{B_2}
\right)].
\end{align}
The formula presented above is well-defined due to the commutativity and associativity properties of the link product $\star$, as noted in Remark~\ref{remark:lk}.


\subsection{\label{subsec:Fragment}Quantum Circuit Fragments: Framework of Non-Markovian Dynamics}

Quantum channels are a central concept in quantum mechanics, characterizing the time-evolution of quantum systems between two different time points (see Fig.~\ref{fig:cj}). The most general way of manipulating quantum channels is through the use of quantum superchannels, which describe the non-Markovian dynamics of four time points (see Fig.~\ref{fig:sc-cj}). These superchannels capture the complex interplay between system and memory, allowing for a more comprehensive understanding of quantum dynamics. A notable example is the superchannel $\Theta_{A_1B_1\to A_2B_2}$, shown in Fig.~\ref{fig:FigSM06}(b). This superchannel captures the system-memory dynamics from $A_1$ to $A_2$ and from $B_1$ to $B_2$, via the composition of two channels, i.e., $\Theta^{\text{pre}}_{A_1\to A_2E}$ and $\Theta^{\text{post}}_{B_1E\to B_2}$. The study of non-Markovian quantum processes that involve multiple time points is critical for various fields, such as distributed quantum computing~\cite{7562346,10.1145/3233188.3233224,8910635}, cloud-based quantum computing~\cite{PhysRevA.94.032329,Ku2020,PhysRevA.105.042610,Ma2022}, quantum network communication~\cite{Kimble2008,Pirandola2016,Simon2017,doi:10.1126/science.aam9288,Pant2019,PRXQuantum.1.020317,PRXQuantum.2.017002,Hermans2022,doi:10.1116/5.0051881,doi:10.1116/5.0092069,doi:10.1116/5.0118569,fang2022quantum,azuma2022quantum}, quantum-enhanced agents~\cite{Mile2012,thompson2017using,PhysRevX.12.011007,Wu2023}, quantum metrology~\cite{PhysRevLett.123.110501,PhysRevLett.127.060501,PhysRevLett.130.070803}, open quantum systems~\cite{PhysRevA.97.012127,PhysRevLett.122.140401,PRXQuantum.2.030201}, quantum clocks and sensors~\cite{Komar2014,Guo2020,Liu2021,Beloy2021,Malia2022}. Furthermore, it has significant implications for emerging areas such as quantum biology~\cite{Lambert2013}, making it a topic of utmost importance. Generally, non-Markovian quantum processes offer a window into the interplay between system and environment, allowing us to explore the rich landscape of quantum phenomena that arise from it. In this subsection, we introduce the concept of \emph{quantum circuit fragment} as a general framework for understanding these processes. The corresponding Choi-Jamio\l kowski operator, which characterizes the quantum circuit fragment, is known as the \emph{quantum comb}, and has been extensively investigated in literature~\cite{PhysRevA.80.022339,PhysRevLett.101.060401,PhysRevA.84.012311}.

\begin{figure}[h]
    \centering
\includegraphics[width=1\textwidth]{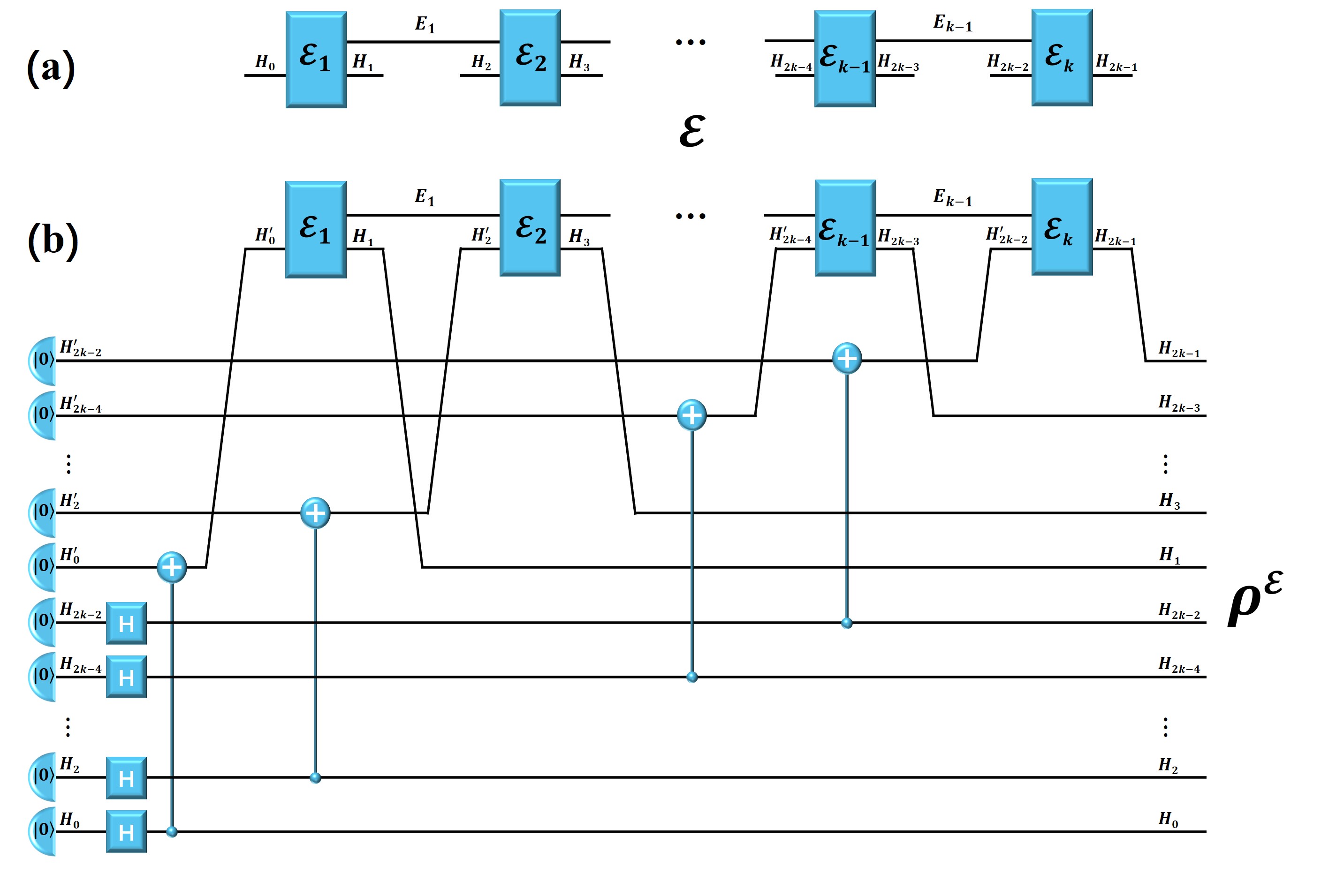}
    \caption{(Color online) The quantum circuit fragment $\mE$ (a) and the implementation of its Choi-Jamio\l kowski state $\rho^{\mE}$ (b). Here $\mE$ is constituted of $k$ sequential quantum channels $\mE_1, \mE_2, \ldots, \mE_k$ (see Eq.~\ref{eq:qcf}).}
    \label{fig:FigSM07}
\end{figure}

Consider the collection of all quantum circuit fragments consisting of $k$ sequential quantum channels, denoted as $\mathfrak{F}_{k}$. It then follows that for any given point-to-point quantum channel $\mE$, we have
\begin{align}
\mE\in\mathfrak{F}_{1}.
\end{align}
Here, the set $\mathfrak{F}_{1}$ represents the exact set of all quantum channels, encompassing the full range of physical operations that can be used to manipulate quantum states. Notably, both the set of quantum states, denoted by $\mathfrak{S}_{\text{State}}$, and the set of positive operator-valued measures (POVMs), denoted by $\mathfrak{S}_{\text{POVM}}$, are subsets of $\mathfrak{F}_{1}$~\cite{wilde_2013}. Writing everything out explicitly, we have
\begin{align}
\mathfrak{S}_{\text{State}}\cup\mathfrak{S}_{\text{POVM}}
\subset
\mathfrak{F}_{1}
=
\mathfrak{S}_{\text{Channel}},
\end{align}

\noindent 
where $\mathfrak{S}_{\text{Channel}}$ stands for the collection of all quantum channels. The set $\mathfrak{F}_{2}$, on the other hand, refers to the collection of all quantum superchannels. In the context of quantum causal inference~\cite{ried2015quantum,causal-NC,xiao2023quantum}, the set of causal maps under investigation is denoted as $\mathfrak{S}_{\text{CMap}}$. By examining the relationship between these two sets, it becomes evident that $\mathfrak{S}_{\text{CMap}}$ is naturally embedded within $\mathfrak{F}_{2}$, as all causal maps can be realized as special cases of superchannels.
\begin{align}
\mathfrak{S}_{\text{CMap}}
\subset
\mathfrak{F}_{2}
=
\mathfrak{S}_{\text{SChannel}}.
\end{align}
Here $\mathfrak{S}_{\text{SChannel}}$ is the set of all quantum superchannels. Let us now turn our attention to the case of a general set $\mathfrak{F}_{k}$, which comprises elements that consist of $k$ sequential quantum channels. Consider a sequence of quantum channels ${\mE_1, \mE_2, \ldots, \mE_k}$ that belongs to $\mathfrak{F}_{1}$ and possesses the following defining characteristics:
\begin{align}
&\mE_1:\mH_0\to\mH_1E_1,\\
&\mE_i:\mH_{2i-2}E_{i-1}\to\mH_{2i-1}E_i,
\,\,\,\,\,\,\,\, 
2\leqslant i\leqslant k-1\\
&\mE_k:\mH_{2k-2}E_{k-1}\to\mH_{2k-1}.
\end{align}
In this context, it is important to note that intermediate channels typically have both an environment input system and an environment output system. In contrast, the first channel in the sequence generally lacks an environment input system, while the final channel lacks an environment output system. In our analysis of quantum circuits, we adopt the term ``quantum circuit fragment''~\cite{xiao2023quantum} to refer to the overall dynamics of the circuit, which is composed of a sequence of $k$ quantum channels. In full, a quantum circuit fragment can be expressed as the composite of $k$ quantum channels, denoted as
\begin{align}\label{eq:qcf}
\mE:=\mE_k\circ\mE_{k-1}\circ\cdots\circ\mE_2\circ\mE_1.
\end{align}
Quantum circuit fragment (see Eq.~\ref{eq:qcf}) offers us the most general way of demonstrating quantum dynamics of non-Markovian processes~\cite{PhysRevA.86.010102,PhysRevLett.114.090402,PhysRevLett.120.040405,LI20181,White2020}, including the time evolution of quantum systems under environment coupling and designing and operating quantum devices in quantum computing. The corresponding Choi-Jamio\l kowski operator is called quantum comb~\cite{PhysRevA.80.022339,PhysRevLett.101.060401,PhysRevA.84.012311}. See Fig.~\ref{fig:FigSM07} for an illustration of the dynamics of quantum circuit fragment $\mE$ introduced in Eq.~\ref{eq:qcf} and its Choi-Jamio\l kowski state $\rho^{\mE}$. 
Denote the Choi-Jamio\l kowski operator of channel $\mE_{i}$ ($1\leqslant i\leqslant k$) as $J^{\mE_{i}}$, then the quantum comb $J^{\mE}$ can be re-expressed as the link product between channels, namely
\begin{align}
J^{\mE}=
J^{\mE_{1}}\star J^{\mE_{2}}
\star\cdots\star
J^{\mE_{k-1}}\star J^{\mE_{k}}.
\end{align}

The process of measuring a quantum circuit fragment is referred to as an \emph{interactive measurement}, which is an essential tool for witnessing non-Markovianity and inferring causality. While the topic of interactive measurements is complex and important, a full discussion of this topic is beyond the scope of this work. We therefore leave the exploration of interactive measurements to future research endeavors~\cite{xiao2023quantum}. Nevertheless, it is worth noting that interactive measurements represent a promising avenue for advancing our understanding of quantum systems over multiple time points. By leveraging the principles of interactive measurement, we can gain valuable insights into the behavior of quantum systems and develop new approaches to solving complex problems in various fields.


\section{\label{sec:TE}Temporal Entanglement}

In this section, we present a comprehensive framework of temporal entanglement from a quantum resource theoretical perspective, extending the entanglement theory from states and channels to the most general form of quantum circuit fragments (refer to Subsec.~\ref{subsec:Fragment}). In Subsec.~\ref{subsec:ECF}, we demonstrate the framework of quantum resource theory and examine various options for free morphism in the context of temporal entanglement. Additionally, we introduce the fidelity of quantum circuit fragments (refer to Def.~\ref{def:FD}) concerning various free morphisms in Subsec.~\ref{subsec:CFD}. It is essential to develop this framework to investigate the concepts of quantum teleportation, quantum repeaters, and communication in quantum networks thoroughly. The connection between temporal entanglement and quantum network communication lies at the heart of our research, and we will unravel this relationship in greater detail in the following sections.


\subsection{\label{subsec:ECF}Entangled Circuit Fragments: Quantum Correlations across Space and Time}

Quantum entanglement is one of the most remarkable and puzzling features of quantum mechanics, where two or more particles become intrinsically connected in such a way that the state of one particle instantaneously affects the state of the other, even when they are separated by significant distances. This extraordinary phenomenon has emerged as a key resource in various fields~\cite{RevModPhys.81.865}, ranging from communication~\cite{Zhao2004,Jin2010,Ma2012,Takeda2013,Pirandola2015,Sun2016}, cryptography~\cite{PhysRevLett.67.661,Ekert1992,PhysRevLett.77.2818}, to computation~\cite{PhysRevLett.86.5188,10.5555/2011492.2011495,PhysRevA.68.022312,Briegel2009,biamonte2020entanglement}. The study of entangled states, which represent a form of quantum entanglement at a given moment, has become a major focus of research since its inception, owing to their indispensable role in understanding the fundamentals of quantum mechanics and their potential for advancing a wide range of quantum technologies. As such, entangled states have been the subject of numerous research articles in the literature. In the realm of physical systems, change is the only constant. As famously stated by Chuck Palahniuk in Fight Club, ``nothing is static, everything is evolving''. Recent progress has extended the theory of entanglement~\cite{PhysRevLett.76.722,10.1063/1.3483717,PhysRevLett.106.130503,6353584,PhysRevLett.119.180506} to encompass quantum channels~\cite{6556948,PhysRevLett.125.040502,PhysRevA.107.012429,PhysRevLett.125.180505,PhysRevA.103.062422}, enabling a comprehensive investigation into the intricate entanglement of dynamical processes. Both point-to-point and bipartite channels have been explored, offering a multifaceted approach to understanding quantum dynamics. Notably, there is no fundamental difference between quantum states and quantum channels, as both are elements of $\mathfrak{F}_{1}$, i.e., quantum circuit fragments with a single process. Despite these advancements, the generic entanglement theory of quantum dynamics over multiple time points~\cite{brukner2004quantum,Milz_2018,PhysRevA.98.012328,Zych2019,PhysRevResearch.3.023028,10.21468/SciPostPhys.10.6.141,PhysRevLett.128.220401}, represented by the quantum circuit fragment $\mathfrak{F}_{k}$ (see Subsec.~\ref{subsec:Fragment}), is still absent. Deeper insights into the intricate interactions of quantum systems, particularly as they evolve over time, hold immense potential to illuminate fundamental theoretical concepts and practical applications. The continued expansion of our understanding of quantum dynamics over multiple time points will undoubtedly shape the future of quantum computing and technology. In order to tackle this issue, our attention is directed towards examining temporal entanglement - that is, the entanglement of general quantum circuit fragments - in this subsection. Through our analysis, we aim to develop a quantum resource theory that aligns with our findings, obtained through a meticulous analysis. 

Quantum resource theory is a vibrant and rapidly developing field at the forefront of modern quantum theories. At its core, the theory focuses on free states or channels, which can be prepared or manipulated without the need for any additional resources~\cite{COECKE201659,RevModPhys.91.025001}. On the contrary, the use of resources can provide quantum advantages in a wide range of tasks, including channel (or phase) discrimination~\cite{PhysRevLett.102.250501,PhysRevA.93.042107,PhysRevLett.116.150502,PhysRevLett.120.230504,PhysRevLett.122.140402} and exclusion~\cite{PhysRevLett.125.110401,PhysRevLett.125.110402}. Entangled states~\cite{RevModPhys.81.865} and coherent states~\cite{RevModPhys.89.041003} are among the most important and widely studied quantum resources in quantum resource theory. They represent distinct types of resources that can be used to perform a variety of tasks with advantages that cannot be achieved using classical resources alone. For example, entangled states can be used to perform tasks such as quantum teleportation, superdense coding, and quantum cryptography, while coherent states can be used in applications such as quantum metrology and quantum computing. In order to streamline the nomenclature of quantum resource theories and clarify the distinction between the subject of study and its manipulation, we adopt the language of category theory~\cite{mac2013categories}. Specifically, we refer to the subject of study as \emph{free objects} and the corresponding physical manipulation of these objects as \emph{free morphisms} in this work. For instance, in the conventional resource theory of entanglement, the SWAP operation serves as a morphism. On the other hand, in the investigation of dynamical entanglement, the SWAP becomes an object. In summary, a quantum resource theory $\mR$ is a framework that allows researchers to study the properties and behaviors of quantum resources in a rigorous and systematic way. It is constituted of three fundamental components that are essential to its functioning:
\begin{itemize}
\item {\bf Set of Free Objects $\text{ob}(\mR)$:} These are quantum circuit fragments that can be prepared or manipulated without the need for any additional resources. The free objects are the starting point for understanding the properties of more complex quantum systems over multiple time points, such as entangled or coherent non-Markovian quantum processes;
\item {\bf Set of Free Morphisms $\text{hom}(\mR)$:} These are the physical transformations that can be applied to the free objects without creating new resources. Free morphisms play a crucial role in the manipulation and transformation of quantum resources, and they provide a framework for understanding the ways in which quantum circuit fragments can be transformed and manipulated;
\item {\bf Golden Rule $\Theta(\text{ob}(\mR))\subset\text{ob}(\mR), \forall\,\Theta\in\text{hom}(\mR)$:} This is a fundamental principle that guides the manipulation and transformation of quantum resources. It states that in any given quantum resource theory, the set of free objects should be closed under the action of the free morphisms.
\end{itemize}
A prominent example is the study of conventional resource theory of entanglement for quantum states, where $\text{ob}(\mR)$ is the set of all separable states. Meanwhile, set $\text{hom}(\mR)$ is the collection of all local operations and classical communication (LOCC). The use of free morphisms has been instrumental in establishing the resource theory of entanglement of quantum states. By considering the action of free morphisms, it is possible to generate all separable states using local operations and classical communication. This means that the entire resource theory of entanglement can be built upon the framework of free morphisms. In practical laboratory settings, there are physical limitations on the types of operations that can be performed, and as a result, the set of LOCC can be divided into different classes~\cite{Chitambar2014}: 
\begin{itemize}
\item {\bf Local Operations ($\text{LO}$):} quantum operations that can be performed on individual subsystems of a composite quantum system. These operations can be performed independently by each party without any communication between them~\cite{PhysRevX.8.031020}. A typical example is the local unitary (LU) operations~\cite{PhysRevA.58.1833,PhysRevLett.104.020504,PhysRevA.82.032121}, which form a special class of local operations, namely $\text{LU}\subset\text{LO}$, that involve applying a unitary transformation to one or more subsystems of a composite quantum system. These operations are commonly used in the context of entangled quantum states to manipulate the entanglement between subsystems;
\item {\bf $\text{LOCC}_{1}(\text{poly}(d))$:} local quantum operations and one-way classical communication with respect to a fixed party, followed by a coarse-graining map. Here, the communication complexity is polynomially dependent on an integral parameter $d$. This parameter $d$ represents a measure of the dimensionality or complexity of the quantum system under consideration, such as the dimension of the message system. By incorporating the concept of $\text{LOCC}_{1}(\text{poly}(d))$, the communication complexity is effectively reduced, making it more feasible to implement in real-world scenarios.
\item {\bf $\text{LOCC}_{1}$:} local quantum operations and one-way classical communication with respect to a fixed party, followed by a coarse-graining map. $\text{LOCC}_{1}$ is a specific subclass of $\text{LOCC}$ operations that has been extensively studied in the field of quantum information theory. It has been shown to be strictly less powerful than the set of all $\text{LOCC}$ operations, which means that there are entangled states that cannot be transformed into one another using $\text{LOCC}_{1}$ operations but can be transformed using more general $\text{LOCC}$ operations;
\item {\bf $\text{LOCC}_{k}$:} quantum channels that are $\text{LOCC}$-linked -- obtaining by implementing $\text{LOCC}_{1}$ and coarse-graining maps -- to a element of $\text{LOCC}_{k-1}$, where $k\geq 2$. $\text{LOCC}_{k}$ operations are a generalization of $\text{LOCC}_{1}$ and provide a way to perform more complex entanglement manipulation tasks. However, as $k$ increases, the set of $\text{LOCC}_{k}$ operations becomes more powerful and less well-understood;
\item {\bf $\text{LOCC}_{\mathds{N}}$:} a element of $\text{LOCC}_{k}$ for some $k\in\mathds{N}$;
\item {\bf $\text{LOCC}$:} quantum channel $\mE$ is a $\text{LOCC}$ channel if there exists a sequence of quantum channels $ \{\mE_1, \mE_2, \ldots\}$ satisfying (\romannumeral1) each of them belongs to the set $\text{LOCC}_{\mathds{N}}$, (\romannumeral2) $\mE_k$ is $\text{LOCC}$-linked to $\mE_{k-1}$, and (\romannumeral3) $\mE_k^{'}$, obtained by implementing coarse-graining maps on $\mE_k$, converges to $\mE$. $\text{LOCC}$ operations are important in quantum information theory because they are considered to be the most basic and physically realistic class of operations that can be performed on entangled quantum systems. They are also used as a benchmark for the study of several important tasks, such as entanglement distillation, entanglement cost, and quantum teleportation;  
\item {\bf $\overline{\text{LOCC}_{\mathds{N}}}$:} topological closure of $\text{LOCC}_{\mathds{N}}$.
\end{itemize}
The formal definitions of coarse-graining maps, $\text{LOCC}$-link, and convergence of channels can be found in Ref.~\cite{Chitambar2014}, which is a comprehensive reference on $\text{LOCC}$. These concepts are important for studying the behavior of entangled quantum systems under local operations and classical communication. In addition to these concepts, there are many other related operations in entanglement theory that are of interest to researchers. These include:
\begin{itemize}
\item {\bf Separable Operations ($\text{SEP}$):} quantum channels that can be expressed using a product form of \emph{Kraus operators}. It is noteworthy that when considering two bipartite pure quantum states, the existence of a separable operation capable of transforming one into the other implies the existence of a $\text{LOCC}$ operation that can achieve the same result. Specifically, Gheorghiu and Griffiths have demonstrated that any transformation of a pure state using separable operations satisfies the same necessary and sufficient conditions as an $\text{LOCC}$ transformation~\cite{PhysRevA.78.020304}. It should be noted, however, that this equivalence does not hold for mixed states, as illustrated by Chitambar and Duan in Ref.~\cite{PhysRevLett.103.110502}. In fact, every $\text{LOCC}$ operation can be expressed in terms of a Kraus operator-sum representation, using product Kraus operators only. However, the converse is not necessarily true;
\item {\bf Separability Preserving Operations ($\text{SEPP}$):} quantum channels that preserve the separability of states. To further elaborate, $\text{SEPP}$ is a set of operations that are non-entangling, meaning they do not create any entanglement from separable states. The study of $\text{SEPP}$, especially its approximate versions, is important in the context of entanglement manipulation, as it provides a framework for the reversible manipulation of entanglement. A deeper exploration of approximate $\text{SEPP}$ can be found in various literature references, including Ref.~\cite{Brandao2008,Brandao2010,berta2022gap,Lami2023}, which discuss the generalization of quantum Stein’s lemma and the second law of entanglement manipulation. These references provide detailed explanations and mathematical formulations of the properties and applications of $\text{SEPP}$, and can be a valuable resource for those interested in studying this topic;
\item {\bf Positive-Partial-Transpose Preserving Operations ($\text{PPT}$):} quantum channels that completely preserve the positivity of the partial transpose of states. More precisely, any quantum state that has a positive partial transpose will remain positive under a $\text{PPT}$ operation. This property distinguishes $\text{PPT}$ operations from other types of quantum operations, such as $\text{LOCC}$ operations, because they can create bound entangled states that cannot be created using only $\text{LOCC}$ operations. One of the key advantages of using $\text{PPT}$ operations over $\text{LOCC}$ operations is that PPT operations are generally easier to characterize. This is because the set of $\text{PPT}$ operations is a convex set that can be described by using semidefinite programming (SDP) techniques~\cite{PhysRevA.60.179,PhysRevA.63.019902,PhysRevLett.90.027901}, whereas the set of $\text{LOCC}$ operations is a non-convex set that is much harder to handle.
\end{itemize}
The models under consideration follow a hierarchical structure, with each model being more general and encompassing than the previous one. This hierarchy allows for a systematic study of the different models and their relationships to one another in entanglement theory. Specifically, the models considered in this work can be organized in the following hierarchy, starting from the most restrictive:
\begin{align}\label{eq:hi}
\text{LO}\subsetneq\text{LOCC}_{1}(\text{poly}(d))\subset\text{LOCC}_{1}
\subsetneq\text{LOCC}_{k}\subsetneq\text{LOCC}_{k+1}\subsetneq\text{LOCC}_{\mathds{N}}\subsetneq\text{LOCC}\subsetneq\overline{\text{LOCC}_{\mathds{N}}}\subsetneq\text{SEP}\subsetneq\text{SEPP}\subsetneq\text{PPT},
\end{align}
for any $k>1$. 

\begin{remark}
While $\text{LOCC}_{1}(\text{poly}(d))$ is clearly a subset of $\text{LOCC}_{1}$, encompassing operations with a narrower range of communication complexities, it is still unclear whether this inclusion is strict. This ambiguity arises from the possibility that there are scenarios where $\text{LOCC}_{1}(\text{poly}(d))$ and $\text{LOCC}_{1}$ result in the same set of achievable transformations. To gain a comprehensive understanding of the relationship between $\text{LOCC}_{1}(\text{poly}(d))$ and $\text{LOCC}_{1}$, more thorough examination and in-depth analysis are required.
\end{remark}

We will now expand the scope of these definitions to encompass quantum circuit fragments -- the carrier of temporal entanglement. The first new and important concept is called the $\mS$-simulable quantum circuit fragment.

\begin{mydef}
{$\mS$-Simulable Quantum Circuit Fragment}{Simulable}
A quantum circuit fragment $\mE\in\mathfrak{F}_{k}$ can be referred to as a $\mS$-simulable quantum circuit fragment if it can be decomposed into a series of quantum channels $\{\mE_1, \mE_2, \ldots, \mE_k\}$, where each $\mE_i\in\mS$. The set $\mS$ is chosen from $\{\text{LO}, \text{LOCC}_{1}(\text{poly}(d)), \text{LOCC}_{k}, \text{LOCC}_{\mathds{N}}, \text{LOCC}, \overline{\text{LOCC}_{\mathds{N}}}, \text{SEP}, \text{SEPP}, \text{PPT}\}$. We denote the set of all $\mS$-simulable quantum circuit fragments as $\mathfrak{F}_{k}(\mS)$.
\end{mydef}

\begin{remark}
When examining $\text{LOCC}_{1}$-simulable quantum circuit fragments, namely $\mathfrak{F}_{k}(\text{LOCC}_{1})$, we impose the restriction that all of its components are $\text{LOCC}_{1}$ with respect to the same party. 
\end{remark}

Having introduced $\mS$-simulable quantum circuit fragments in Def.~\ref{def:Simulable}, we can now present the formal definition of an entangled circuit fragment, which captures the essence of temporal entanglement over multiple time points.

\begin{mydef}
{Entangled Circuit Fragment}{ECF}
We refer to a quantum circuit fragment $\mE$ as an entangled circuit fragment if it does not belong to the set $\mE\notin\mathfrak{F}_{k}(\text{SEPP})$.
\end{mydef}

\begin{remark}
The significance of Def.~\ref{def:ECF} lies in its ability to extend the concept of entanglement to its most general form, which includes not just quantum states and channels but also quantum circuit fragments. In doing so, it provides a crucial framework for characterizing and understanding entanglement in the context of non-Markovian quantum processes. In particular, by introducing 
\begin{align}
\text{Sta}(\mS):= \{\rho\,|\,\rho\in\mathfrak{F}_{1}(\mS), \Tr[\rho]=1\},
\end{align}
with $\mS\in\{\text{LO}, \text{LOCC}_{k}, \text{LOCC}_{\mathds{N}}, \text{LOCC}, \overline{\text{LOCC}_{\mathds{N}}}, \text{SEP}, \text{SEPP}, \text{PPT}\}$, the collection of separable states is completely captured by
\begin{align}
\text{Sta}(\text{LOCC})=\text{Sta}(\text{SEP})=\text{Sta}(\text{SEPP}).
\end{align}
If a state $\rho\notin\text{Sta}(\text{SEPP})$, it is entangled. Therefore, an entangled state can be regarded as a specific instance of an entangled quantum circuit fragment, which consists of a single quantum process.
\end{remark}

While we trigger the investigation of entangled quantum circuit fragments , or directly temporal entanglement, our objective of this work is only to explore their connection to quantum network communication, including quantum teleportation and quantum repeater-based protocols. We will not dive into the complete framework here, but our upcoming works will present a more comprehensive understanding of temporal entanglement.


\subsection{\label{subsec:CFD} Fidelity of Circuit Fragment Distillation: Quantifying Temporal Entanglement}

Entanglement distillation is a critical component of quantum information processing, involving the conversion of a large number of weakly entangled quantum states into a smaller number of highly entangled ones. This transformation is essential for numerous quantum communication and computation applications, as the strength of entanglement between parties constrains their operational effectiveness. In this subsection, we will delve into the topic of entanglement distillation for quantum circuit fragments, and introduce the notion of $k$-fidelity of $\mS$ distillation.

With a focus on practicality, our attention in this work is directed towards single-shot entanglement distillation of quantum circuit fragments. More precisely, we are interested in transforming a single instance of quantum circuit fragment $\mE\in\mathfrak{F}_{k}$ into a state that closely approximates the maximally entangled state $\phi^{+}_{d}$ through the application of operation $\Theta\in\mathfrak{F}_{k+1}(\mS)$ with $\mS\in\{\text{LO}, \text{LOCC}_{k}, \text{LOCC}_{\mathds{N}}, \text{LOCC}, \overline{\text{LOCC}_{\mathds{N}}}, \text{SEP}, \text{SEPP}, \text{PPT}\}$; that is

\begin{mydef}
{$d$-Fidelity of $\mS$ Distillation}{FD}
The $d$-fidelity $F_{d, \mS}(\mE)$ of a quantum circuit fragment $\mE\in\mathfrak{F}_{k}$ under operation $\Theta\in\mathfrak{F}_{k+1}(\mS)$ is defined by
\begin{align}\label{eq:FD}
F_{d, \mS}(\mE)
:=
\max_{\Theta\in\mathfrak{F}_{k+1}(\mS)} F_{\text{U}}(\Theta(\mE), \phi^{+}_{d})
=
\max_{\Theta\in\mathfrak{F}_{k+1}(\mS)}
\Tr[\Theta(\mE)\cdot\phi^{+}_{d}],
\end{align}
where $F_{\text{U}}$ stands for the Uhlmann fidelity, with $\phi^{+}_{d}$ being the $d$-dimensional maximally entangled state. 
\end{mydef}

The $d$-Fidelity of $\mS$ distillation offers a quantitative assessment of how closely the output channel -- $\Theta(\mE)$ -- resembles the ideal, maximally entangled state $\phi^{+}_{d}$ under operations $\Theta\in\mathfrak{F}_{k+1}(\mS)$, characterizing their similarity through the overlap between $\Theta(\mE)$ and $\phi^{+}_{d}$. The functional $F_{d, \mS}(\mE)$ is a generalization of the concept of single-shot entanglement distillation, where the goal is to generate states that are as close as possible to a maximally entangled state under LOCC from initial object, which can be a quantum state or a quantum channel. When $\mE$ is a quantum state and the set of allowed operations are LOCC, i.e., $\mS= \text{LOCC}$, then $F_{d, \mS}(\mE)$ recovers the concept of single-shot entanglement distillation under LOCC. On the other hand, when the set of operations that can be applied to state $\mE$ are PPT channels, namely $\mS= \text{PPT}$, then $F_{d, \mS}(\mE)$ gives the fidelity of $d$-state PPT distillation~\cite{959270}. Moreover, when $\mE$ is a quantum channel and $\Theta$ represents a superchannel, the function $F_{d, \mS}(\mE)$ captures the concept of single-shot entanglement distillation of dynamical entanglement~\cite{6556948,PhysRevLett.125.040502,PhysRevA.107.012429,PhysRevLett.125.180505,PhysRevA.103.062422}.

\begin{remark}
Here, $F_{d, \mS}(\mE)$ is not an entanglement monotone as it is not faithful, i.e., $F_{d, \mS}(\mE)>0$ if and only if $\mE$ is an entangled circuit fragment (see Def.~\ref{def:ECF}). But, it is still monotonically non-increasing under free morphism $\Xi\in\mathfrak{F}_{k+1}(\mS)$. That is, given a quantum circuit fragment $\mE$, we have $F_{d, \mS}(\mE)\geqslant F_{d, \mS}(\Xi(\mE))$ holds for all $\Xi\in\mathfrak{F}_{k+1}(\mS)$.
\end{remark}

This subsection introduced the basic idea of $d$-Fidelity of $\mS$ distillation $F_{d, \mS}(\mE)$ and explored its importance in single-shot entanglement distillation. However, the comprehensive scope of temporal entanglement extends far beyond what was discussed here. This raises many questions, such as: How can we formulate the standard entanglement cost and distillable entanglement in the asymptotic case of quantum circuit fragments under $\mathfrak{F}_{k}(\text{LOCC})$? Is it possible to solve the entanglement cost of a quantum circuit fragment and find a quantity that unifies both $\kappa$ entanglement~\cite{PhysRevLett.125.040502} for states and max-logarithmic negativity~\cite{PhysRevLett.125.180505} for bipartite channels in the case of $\mathfrak{F}_{k}(\text{PPT})$? These questions exceed the boundaries of this work and will be presented in our upcoming works.


\section{\label{sec:QC}Quantum Communications}

\begin{figure}[h]
  \centering
\includegraphics[width=1\textwidth]{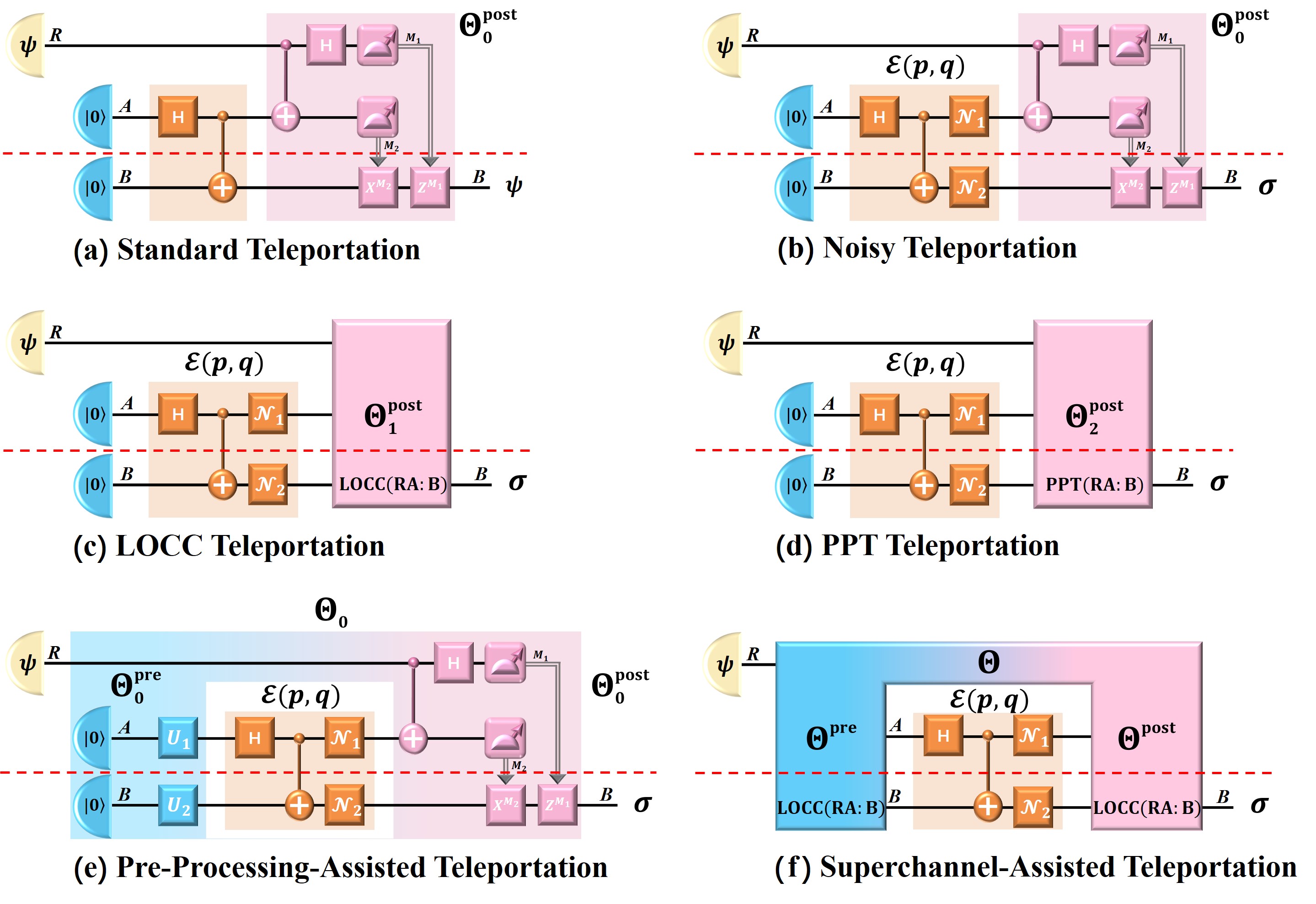}\\
  \caption{(Color online) Quantum teleportation protocols: 
  (a) standard teleportation, a fundamental process in quantum information theory where a quantum state is transmitted from one location to another. The protocol involves the sharing of a maximally entangled state between two parties, Alice and Bob, who are situated at distinct locations. The quantum information that needs to be transmitted from Alice to Bob is stored in the state $\psi_R$ that acts on system $R$. Alice has access to systems $RA$, while Bob has access to system $B$. The protocol is enabled by implementing a standardized operation, which is represented by the pink box in the diagram and denoted as $\Theta^{\text{post}}_{0}\in\mathfrak{F}_{1}(\text{LOCC}(RA:B))$; 
  (b) noisy teleportation, where the entangling gates are followed by channels $\mN_1(p)$ and $\mN_2(q)$ (see Eq.~\ref{eq:N1N2}), resulting in a noisy entangling gate $\mE(p, q)$ (see Eq.~\ref{eq:EG}). Despite these noises, the transmission of quantum information $\psi_R$ remains through $\Theta^{\text{post}}_0$; 
  (c) LOCC teleportation~\cite{PhysRevA.60.1888}. Rather than relying on a standardized operation $\Theta^{\text{post}}_0$ executed by remote parties, a more versatile $\text{LOCC}$ channel $\Theta_1^{\text{post}}\in\mathfrak{F}_{1}(\text{LOCC}(RA:B))$ has been utilized to facilitate the transmission of quantum information between the composite systems $RA$ and $B$. After optimizing over all possible $\text{LOCC}$ channels, this protocol demonstrates superior performance compared to the previous protocol (b); 
  (d) PPT teleportation. The set of allowable operations has been expanded to include all PPT channels, which means that the LOCC channel $\Theta^{\text{post}}_1$ has been replaced by a PPT channel $\Theta^{\text{post}}_2\in\mathfrak{F}_{1}(\text{PPT}(RA:B))$. As a result, the performance of quantum state teleportation has been further enhanced; 
  (e) pre-processing-assisted teleportation. When the entangling gate (represented by the brown box) is seen as a shared dynamical resource between Alice and Bob, there is no justification for limiting consideration solely to the implementation of post-processing. In this case, we consider a special pre-processing protocol $\Theta^{\text{pre}}_0:= \mU_1(\ketbra{0}{0})_A\otimes\mU_2(\ketbra{0}{0})_B$, followed by the standard teleportation protocol $\Theta^{\text{post}}_0$; 
  (f) superchannel-assisted teleportation, achieved by implementing a superchannel $\Theta\in\mathfrak{F}_{2}(\text{LOCC})$. Here, gate $\mE(p, q)$ has been transformed into channel $\Theta(\mE(p, q))_{R\to B}$. This consideration leads to superior performance compared to the case with only post-processing, i.e., protocol (c);
  From protocols (b) to (f), the output is represented by $\sigma$. These diagrams distinguish between systems belonging to Alice, which are located above the dashed red line, and those belonging to Bob, which are situated below the dashed line.
  }
  \label{fig:qt}
\end{figure}

In this section, we will provide a fresh perspective on quantum teleportation by focusing on the role of resourceful quantum channels in the protocol. Instead of simply considering the entanglement of a static state, we will examine the crucial role of the entanglement-generating channel that connects the sender and receiver. To explore the optimal performance of a bipartite channel in quantum teleportation, we will introduce a new superchannel-assisted teleportation protocol (shown in Fig.~\ref{fig:qt}(f)) in Subsec.~\ref{subsec:QT}. In the subsequent Subsec.~\ref{subsec:FB}, we will dive deeper into the general mathematical framework that underlies these protocols. By doing so, we will gain a more complete understanding of the principles that govern a variety of quantum teleportation protocols (illustrated in Fig.~\ref{fig:qt}) and the role of entanglement-generating channel in superchannel-assisted quantum teleportation. Finally, in Subsec.~\ref{subsec:NE}, we will demonstrate the superior performance of our newly proposed teleportation protocol compared to conventional methods in certain noise models. This result highlights the practical significance of our approach and its potential for real-world applications.


\subsection{\label{subsec:QT}Quantum Teleportation: A Channel-Theoretic Perspective}

At the heart of quantum communication is the ability to securely transfer information between parties~\cite{Georgescu2022}. This is crucial for a variety of applications, from quantum cryptography to quantum computing. The most widely recognized and effective method for achieving secure transfer of quantum information is through the use of quantum teleportation~\cite{PhysRevLett.69.2881,PhysRevLett.70.1895,PhysRevA.54.3824,Zeilinger1997,PhysRevLett.80.1121}. This protocol, which was first introduced by Bennett et al. in Ref.~\cite{PhysRevLett.70.1895}, enables the transfer of quantum states from one location to another, without physically transporting the particles that carry the information. Its effectiveness has been experimentally demonstrated in Refs.~\cite{Zeilinger1997,PhysRevLett.80.1121}. Recently, efforts to build a global-scale quantum network have resulted in the successful ground-to-satellite teleportation over distances of up to $1, 400$ kilometers as reported in Ref.~\cite{Pan2017}. To facilitate a thorough exploration of related topics, we highly recommend consulting Ref.~\cite{Diamanti2017,Xia_2018,Dai2020,PhysRevLett.128.170501,RevModPhys.94.035001}. Theoretically, the transfer of any quantum information from Alice to Bob can be achieved perfectly through the use of a maximally entangled state shared between them (see Fig.~\ref{fig:qt}(a)). However, the physical implementation of teleportation is subject to various sources of quantum noise resulting from interactions between the system and its environment (see Fig.~\ref{fig:qt}(b))~\cite{Diamanti2017,Xia_2018,Dai2020,PhysRevLett.128.170501,RevModPhys.94.035001,Marcikic2003,Landry07,Valivarthi2016,Grosshans2016}. The conventional approach to combating the noise is to replace the standard teleportation protocol (i.e., $\Theta_0^{\text{post}}$ of Fig.~\ref{fig:qt}(b)) with a general LOCC protocol (i.e., $\Theta_1^{\text{post}}$ of Fig.~\ref{fig:qt}(c))~\cite{PhysRevA.60.1888}, as $\Theta_0^{\text{post}}$ is only a special case of LOCC. The power of teleportation originates from the entangling gate (brown boxes in Fig.~\ref{fig:qt}) between different parties. One might then wonder: does the conventional approach (see Fig.~\ref{fig:qt}(c)) fully leverage the capabilities of the entangling gate?

Regrettably, the answer to the question posed above is negative. There is concrete evidence that the use of a superchannel-based protocol is demonstrably superior to one that relies solely on post-processing techniques. To provide a more concrete illustration, we demonstrate that there exist examples where a straightforward application of local unitary operations in the pre-processing stage (see Fig.~\ref{fig:qt}(e)) can outperform any possible PPT-post-processing technique (see Fig.~\ref{fig:qt}(d)), which naturally encompasses all LOCC-post-processing protocols as a subset (see Fig.~\ref{fig:qt}(c)). 

To provide a comprehensive explanation, we will evaluate each of the protocols depicted in Fig.~\ref{fig:qt} systematically, along with their corresponding performance.

\begin{itemize}
\item {\bf Protocol a: Standard Teleportation (see Fig.~\ref{fig:qt}(a)).} The mechanism of teleportation can be traced back to classical textbooks, such as Refs.~\cite{nielsen_chuang_2010,wilde_2013,watrous_2018}. Here, we visually demonstrate the standard teleportation through a diagram in Fig.~\ref{fig:id_channel}.
\begin{figure}[h]
  \centering
\includegraphics[width=1\textwidth]{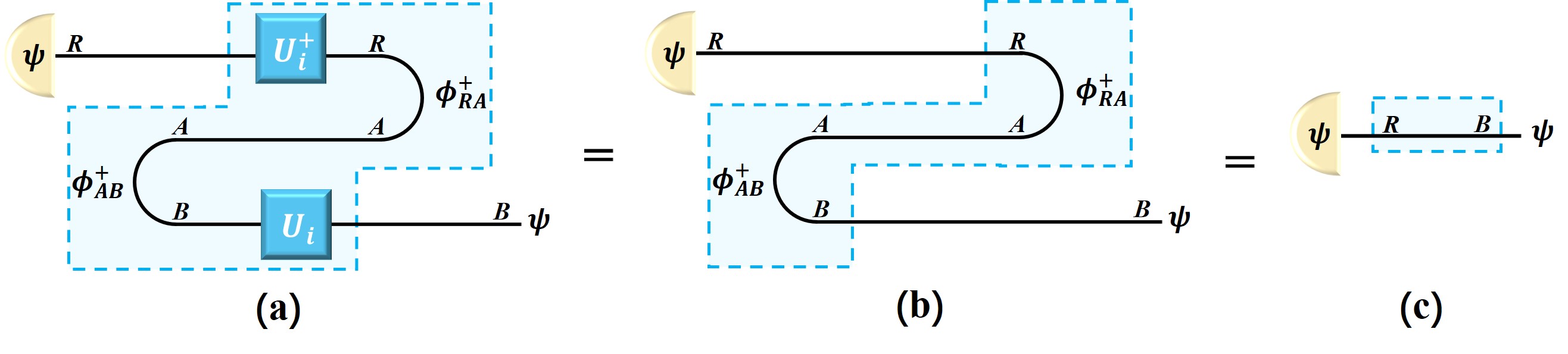}
\caption{(Color online) Graphical illustration of standard teleportation procedure (see Fig.~\ref{fig:qt}(a)). Here, $(U_i\otimes\1\ket{\phi^{+}})^{\dag}$ stands for the standard Bell measurements with $U_0=\1$, $U_1=Z$, $U_2=X$ and $U_3=XZ$. Thanks to $U\otimes\1\ket{\phi^{+}}= \1\otimes U^{\T}\ket{\phi^{+}}$, we have (a) = (b). By further using the yanking equation, we obtain the equivalence between (b) and (c). The technique employed here is akin to that illustrated in Fig.~\ref{fig:lp}.
}
\label{fig:id_channel}
\end{figure} 
\item {\bf Protocol b: Noisy Teleportation (see Fig.~\ref{fig:qt}(b)).} In this case, the entangling gate is subject to local noise, which affects its performance. Denote by $\mN_1$ and $\mN_2$ the local noise affecting systems $A$ and $B$, respectively. In the subsequent analysis, we will make use of the following noise models; that are
\begin{align}
\mN_1, \mN_2\in \{\text{Bit Flip}\,\,\,\mN_{\text{BF}}, \text{Phase Flip}\,\,\,\mN_{\text{PF}}, \text{Depolarizing}\,\,\,\mN_{\text{D}}, \text{Amplitude Damping}\,\,\,\mN_{\text{AD}}\}.
\end{align}

As a fundamental type of error in quantum computing and communication, the bit flip channel $\mN_{\text{BF}}$ can cause a qubit state to flip from $0$ to $1$ or from $1$ to $0$ with a certain probability, which is typically described by a single parameter known as the error probability. This type of error can arise from a variety of noise sources, such as thermal fluctuations and electromagnetic interference. The phase flip channel $\mN_{\text{PF}}$, which arises from fluctuations in the local magnetic field, can cause a qubit phase to flip between $|0\rangle$ and $|1\rangle$ states with a certain probability. Specifically, if the qubit is initially in the state $\alpha \ket{0} + \beta \ket{1}$, then the phase flip channel can transform this state to $\alpha \ket{0} - \beta \ket{1}$ with a certain probability. As a combination of the bit flip, phase flip, and the bit-phase flip channels, the quantum depolarizing channel, denoted as $\mN_{\text{D}}$, takes its initial state to a completely mixed state with some probability. The amplitude damping channel $\mN_{\text{AD}}$ is a type of noise that is commonly encountered in various physical systems, such as superconducting qubits and trapped ions. This channel causes a qubit to lose energy and transition to a lower energy state, resulting in a decay of the amplitude of its quantum state. Using the Kraus decomposition $\mN(\cdot)=\sum_i E_i \cdot E^{\dag}_i$, we can characterize these noise models in Tab.~\ref{tab:kraus-nc}. For more details on this topic, see Ref.~\cite{nielsen_chuang_2010}.

\begin{table}[h]
    \centering
    \includegraphics[width=0.8\textwidth]{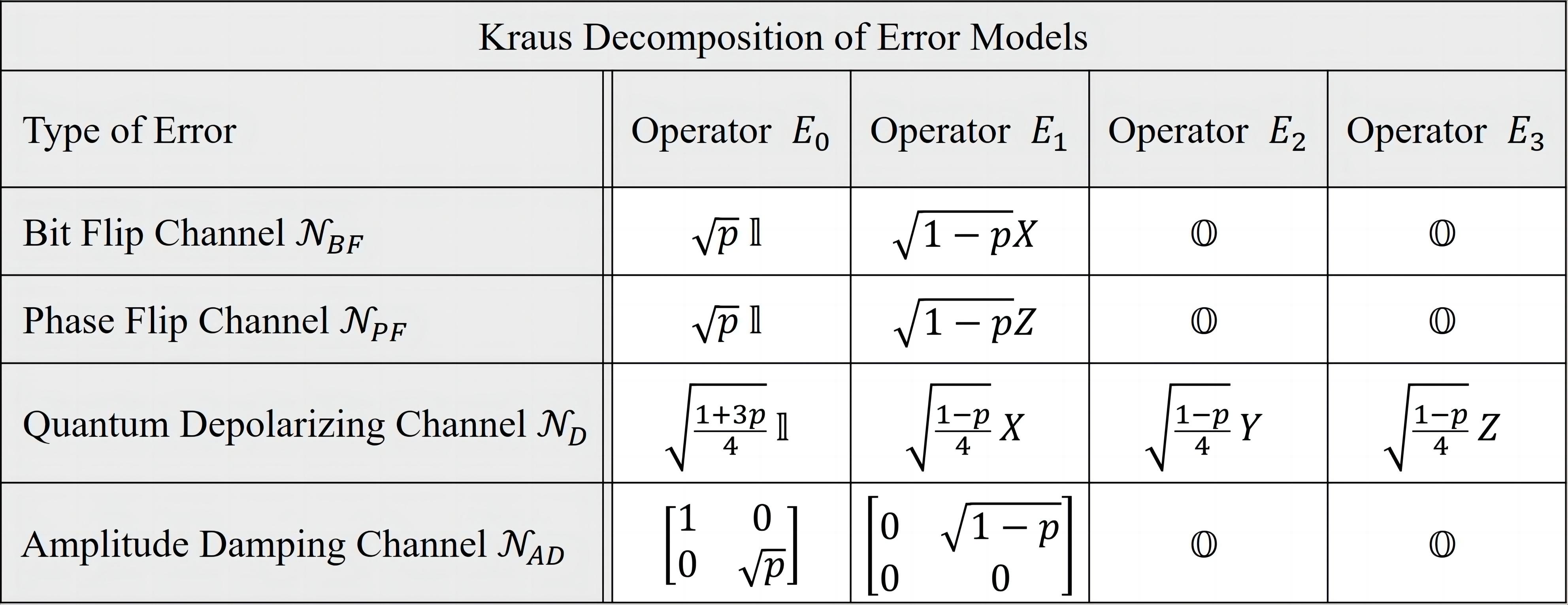}
    \caption{Kraus decomposition of the fundamental noisy channels, including bit flip channel $\mN_{\text{BF}}$, phase flip channel $\mN_{\text{PF}}$, quantum depolarizing channel $\mN_{\text{D}}$, and amplitude damping channel $\mN_{\text{AD}}$. In this context, the symbols $\1$, $X$, $Y$, and $Z$ represent the \emph{Pauli operators}.}
    \label{tab:kraus-nc}
\end{table}

To differentiate the parameters associated with the quantum channels $\mN_1$ and $\mN_2$, we denote their corresponding parameters as $p$ and $q$, respectively. More precisely, we express this distinction as
\begin{align}\label{eq:N1N2}
\mN_1 &= \mN_1(p),\\
\mN_2 &= \mN_2(q).
\end{align}
These notations allow us to clearly identify the parameter associated with each channel, which is essential for analyzing and comparing their effects in noisy quantum teleportation. Having defined $\mN_1$ and $\mN_2$, we can now completely characterize the entangling gate between Alice and Bob as $\mE(p, q)$, colored brown in Fig.~\ref{fig:qt}(b). Expressed mathematically, the noisy entangling gate $\mE(p, q)$ can be represented by
\begin{align}\label{eq:EG}
\mE(p, q)_{AB\to AB}:=
\mN_1(p)_{A\to A}\otimes\mN_2(q)_{B\to B}\circ \text{C}_{\text{NOT}, AB\to AB}\circ \text{H}_{A\to A}\otimes\id_{B\to B}.
\end{align}
Here, we use the symbol $\text{H}$ to refer to the \emph{Hadamard gate}, and $\text{C}_{\text{NOT}}$ to denote the controlled NOT gate (also known as the controlled-X gate). In this scenario, the resultant channel $\mF_{\text{b}}(p, q)$ from system $R$ to $B$ takes the form of
\begin{align}\label{eq:channel-fb}
\mF_{\text{b}}(p, q)_{R\to B}
:=
\Theta^{\text{post}}_{0, RAB\to B}\circ\mE(p, q)_{AB\to AB}(\ketbra{0}{0}_{A}\otimes\ketbra{0}{0}_{B}).
\end{align}
The subscript `b' in $\mF$ indicates that the channel from $R$ to $B$ is obtained using protocol b. In order to measure and evaluate the performance of protocol b, we use the metric of average fidelity $f$, which is defined in Def.~\ref{def:AF}. This metric provides a quantitative measure of how accurately the teleported state $\mF_{\text{b}}(p, q)(\psi)$ matches the original state $\psi$ that was being teleported. In particular, the average fidelity $f$ of channel $\mF_{\text{b}}(p, q)$ is given by
\begin{align}
f(\mF_{\text{b}}(p, q))=
\int d\psi F_{\text{U}}\left(\psi, \mF_{\text{b}}(p, q)(\psi)\right)
=
\int d\psi \Tr[\psi\cdot\left(\mF_{\text{b}}(p, q)(\psi)\right)].
\end{align}
To simplify notation, we can use the abbreviation $f_{\text{b}}$ to refer to the performance of protocol b, i.e.,
\begin{align}\label{eq:fb}
f_{\text{b}}:= f(\mF_{\text{b}}(p, q)).
\end{align}
Using Lem.~\ref{lem:AFEF}, we can prove that $f_b$ can be expressed as
\begin{align}\label{eq:fidelity-fb}
f_{\text{b}}=
\frac{2F(\mF_{\text{b}}(p, q))+ 1}{3}.
\end{align}
Here, $F(\mF_{\text{b}}(p, q))$ represents the entanglement fidelity (defined in Def.~\ref{def:EF}) of the channel $\mF_{\text{b}}(p, q)$.
\item {\bf Protocol c: LOCC Teleportation (see Fig.~\ref{fig:qt}(c)).} It is worth noting that the standard teleportation protocol $\Theta^{\text{post}}_{0}$ involves performing Bell measurements and applying local unitary operations based on the measurement outcomes. Such an operation can be understood as a specific instance of a LOCC channel. Given this connection, it is reasonable to consider whether optimizing over all possible LOCC channels could potentially improve the performance of teleportation. Formally, if we replace the standardized operation $\Theta^{\text{post}}_{0}$, as depicted in Fig.~\ref{fig:qt}(b), with a general LOCC channel $\Theta^{\text{post}}_{1}\in\text{LOCC}(RA:B)$, the resultant channel $\mF_{\text{c}}(\Theta^{\text{post}}_{1}, p, q)$ from the sender to the receiver can be expressed as
\begin{align}\label{eq:channel-fc}
\mF_{\text{c}}(\Theta^{\text{post}}_{1}, p, q)_{R\to B}:=
\Theta^{\text{post}}_{1, RAB\to B}\circ\mE(p, q)_{AB\to AB}(\ketbra{0}{0}_{A}\otimes\ketbra{0}{0}_{B}).
\end{align}
While both $\mF_{b}(p, q)$ (see Eq.~\ref{eq:channel-fb}) and $\mF_{\text{c}}(\Theta^{\text{post}}_{1}, p, q)$ (see Eq.~\ref{eq:channel-fc}) are influenced by parameters $p$ and $q$, $\mF_{\text{c}}(\Theta^{\text{post}}_{1}, p, q)$ is more complex as it also takes into account the variable of LOCC channel $\Theta^{\text{post}}_{1}$. The inclusion of $\Theta^{\text{post}}_{1}$ in $\mF_{\text{c}}(\Theta^{\text{post}}_{1}, p, q)$ adds an extra layer of complexity that enhances the performance of teleportation. The expression for the average fidelity $f$ (see Eq.~\ref{eq:af}) of $\mF_{\text{c}}(\Theta^{\text{post}}_{1}, p, q)$ is
\begin{align}
f(\mF_{\text{c}}(\Theta^{\text{post}}_{1}, p, q))
=
\int d\psi F_{\text{U}}\left(\psi, \mF_{\text{c}}(\Theta^{\text{post}}_{1}, p, q)(\psi)\right)
=
\int d\psi \Tr[\psi\cdot\mF_{\text{c}}(\Theta^{\text{post}}_{1}, p, q)(\psi)].
\end{align}
Now, if we maximize over all possible $\Theta^{\text{post}}_{1}\in\text{LOCC}(RA:B)$, we get
\begin{align}\label{eq:fc}
f_{\text{c}}
:=
\max_{\Theta^{\text{post}}_{1}\in\text{LOCC}(RA:B)}
f(\mF_{\text{c}}(\Theta^{\text{post}}_{1}, p, q)).
\end{align}
Because $\Theta^{\text{post}}_{0}$ is a LOCC channel and we have optimized over all possible choices of $\Theta^{\text{post}}_{1}$, the resulting fidelity of $f_{\text{c}}$ is guaranteed to be greater than or equal to that of $f_{\text{b}}$, i.e., 
\begin{align}\label{eq:ine-cb}
f_{\text{c}}\geqslant f_{\text{b}}.
\end{align}
The fidelity $f_{\text{c}}$ represents the maximum achievable fidelity that can be obtained via conventional teleportation. However, the computation of the quantity $f_{\text{c}}$ remains elusive to us in general. This difficulty  lies in the fact that the mathematical representation of LOCC operations is often hard to specify.
\item {\bf Protocol d: PPT Teleportation (see Fig.~\ref{fig:qt}(d)).}
Expanding the class of permissible post-processing operations beyond just LOCC channels can lead to an improvement in establishing an upper bound for the fidelity $f_{\text{c}}$ (see Eq.~\ref{eq:fc}). We can achieve this by incorporating positive partial transpose (PPT) channels into our analysis. Specifically, we consider the use of a PPT channel $\Theta^{\text{post}}_{2}\in\text{PPT}(RA:B)$ for post-processing operations. In this scenario, the resulting channel $\mF_{\text{d}}(\Theta^{\text{post}}_{2}, p, q)$ from the sender to the receiver can be fully characterized by
\begin{align}\label{eq:channel-fd}
\mF_{\text{d}}(\Theta^{\text{post}}_{2}, p, q)_{R\to B}:=
\Theta^{\text{post}}_{2, RAB\to B}\circ\mE(p, q)_{AB\to AB}(\ketbra{0}{0}_{A}\otimes\ketbra{0}{0}_{B}).
\end{align}
The average fidelity $f$ (see Eq.~\ref{eq:af}) of channel $\mF_{\text{d}}(\Theta^{\text{post}}_{2}, p, q)$ is now given by
\begin{align}
f(\mF_{\text{d}}(\Theta^{\text{post}}_{2}, p, q))
=
\int d\psi F_{\text{U}}\left(\psi, \mF_{\text{d}}(\Theta^{\text{post}}_{2}, p, q)(\psi)\right)
=
\int d\psi \Tr[\psi\cdot
\left(\mF_{\text{d}}(\Theta^{\text{post}}_{2}, p, q)(\psi)\right)].
\end{align}
Taking the maximum over all PPT channel $\Theta^{\text{post}}_{2}$, we arrive at the following fidelity.
\begin{align}\label{eq:fd}
f_{\text{d}}
:=
\max_{\Theta^{\text{post}}_{2}\in\text{PPT}(RA:B)}
f(\mF_{\text{d}}(\Theta^{\text{post}}_{2}, p, q)).
\end{align}
As LOCC channels represent a subset of PPT channels (as shown in Eq.~\ref{eq:hi}), we can derive an inequality between the fidelities $f_{\text{c}}$ and $f_{\text{d}}$ as follows
\begin{align}\label{eq:ine-dc}
f_{\text{d}}\geqslant f_{\text{c}}.
\end{align}
The fidelity $f_{\text{d}}$ is the maximum achievable fidelity of teleportation that can be obtained through post-processing using PPT channels.
\item {\bf Protocol e: Pre-Processing-Assisted Teleportation (see Fig.~\ref{fig:qt}(e)).} Traditionally, post-processing, such as $\Theta^{\text{post}}_{0}$ in protocols a (see Fig.~\ref{fig:qt}(a)) and b (see Fig.~\ref{fig:qt}(b)), $\Theta^{\text{post}}_{1}$ in protocol c (see Fig.~\ref{fig:qt}(c)), and $\Theta^{\text{post}}_{2}$ in protocol d (see Fig.~\ref{fig:qt}(d)), has been the primary focus for mitigating noise in teleportation protocols. However, incorporating pre-processing techniques alongside post-processing can further enhance the fidelity and reliability of teleportation. There are two primary factors that motivate the pre-processing-assisted quantum teleportation, as outlined below: (\romannumeral1) First, in the realm of quantum communication theory, particularly in the investigation of quantum channel capacity, reliable transmission of information requires both encoding and decoding. Pre-processing serves as a means of encoding the information, while post-processing functions as a decoding mechanism. Therefore, incorporating both pre-processing and post-processing techniques is crucial for achieving efficient and accurate quantum teleportation. (\romannumeral2) Second, when viewed through the lens of quantum resource theory~\cite{RevModPhys.91.025001}, specifically the resource theory of dynamical entanglement~\cite{6556948,PhysRevLett.125.040502,PhysRevA.107.012429,PhysRevLett.125.180505,PhysRevA.103.062422}, the shared resource between Alice and Bob is the bipartite channel $\mE(p, q)$ (brown boxes in Fig.~\ref{fig:qt}(b)-(f)). As such, it is imperative to evaluate its maximum performance under all feasible free morphisms, namely the set of all LOCC superchannels $\mathfrak{F}_{2}(\text{LOCC}(RA:B))$ (for further information, please refer to Def.~\ref{def:Simulable} in Subsec.~\ref{subsec:ECF}). Additionally, the preparation of the initial state $\ketbra{0}{0}_{A}\otimes\ketbra{0}{0}_{B}$ is merely a specific instance of free pre-processing. Therefore, it is not necessary to exclusively focus on this particular choice, as any free pre-processing can be employed instead. In this scenario, we examine a specific selection of free pre-processing $\Theta^{\text{pre}}_{0}$ 
\begin{align}\label{eq:theta-pre-0}
\Theta^{\text{pre}}_{0}:=
\mU_1(\ketbra{0}{0})_{A}\otimes\mU_2(\ketbra{0}{0})_{B}
\end{align}
and demonstrate that even with this particular choice, the resulting fidelity cannot be attained by any protocol that employs solely LOCC post-processing. In Eq.~\ref{eq:theta-pre-0}, $\mU_1$ and $\mU_2$ are local unitary gates satisfying $\mU_i(\cdot)= U_i \cdot U_i^{\dag}$ with $i\in\{1, 2\}$. It is worth noting that any unitary operator $U$ acting on a qubit can be written in terms of a global phase shift, a rotation operator about the $y$-axis, and two rotation operators about the $z$-axis~\cite{nielsen_chuang_2010}. More precisely, we have
\begin{align}
U=e^{i\alpha} \left[
  \begin{array}{cc}
    e^{-i\beta/2} & 0  \\
    0 & e^{i\beta/2}\\
  \end{array}
\right]
\left[
  \begin{array}{cc}
    \cos \frac{\gamma}{2} & -\sin \frac{\gamma}{2}  \\
    \sin \frac{\gamma}{2} & \cos \frac{\gamma}{2}\\
  \end{array}
\right]
 \left[
  \begin{array}{cc}
    e^{-i\delta/2} & 0  \\
    0 & e^{i\delta/2}\\
  \end{array}
\right].
\end{align}
Here, $\alpha$, $\beta$, $\gamma$, and $\delta$ are real-valued parameters. To facilitate our analysis and numerical simulations, we simplify the discussion by focusing on specific unitary operations, which has the following form:
\begin{align}\label{eq:u12}
U_1(\alpha):=
\left[
  \begin{array}{cc}
    \cos \frac{\alpha}{2} & -\sin \frac{\alpha}{2}  \\
    \sin \frac{\alpha}{2} & \cos \frac{\alpha}{2}\\
  \end{array}
\right],\quad
U_2(\beta):=
\left[
  \begin{array}{cc}
    \cos \frac{\beta}{2} & -\sin \frac{\beta}{2}  \\
    \sin \frac{\beta}{2} & \cos \frac{\beta}{2}\\
  \end{array}
\right].
\end{align}
With the pre-processing step taken into account, the overall effect of the protocol is to establish a quantum channel, denoted by $\mF_{\text{e}}(\Theta^{\text{pre}}_{0}(\alpha, \beta), p, q)$ or simply $\mF_{\text{e}}(\alpha, \beta, p, q)$, from the sender to the receiver. This channel can be expressed as follows
\begin{align}\label{eq:channel-fe}
\mF_{\text{e}}(\alpha, \beta, p, q)_{R\to B}
:=
&\Theta^{\text{post}}_{0, RAB\to B}\circ\mE(p, q)_{AB\to AB}\circ\Theta^{\text{pre}}_{0, \mathbb{C}\to AB}\\
=
&\Theta^{\text{post}}_{0, RAB\to B}\circ\mE(p, q)_{AB\to AB}
\left(\left(U_1(\alpha)\ketbra{0}{0} U_1^{\dag}(\alpha)\right)_A\otimes
\left(U_2(\beta)\ketbra{0}{0} U_2^{\dag}(\beta)\right)_B\right),
\end{align}
where the unitary operators $U_1(\alpha)$ and $U_2(\beta)$ are defined in Eq.~\ref{eq:u12}. Mathematically, the average fidelity $f$ (see Eq.~\ref{eq:af}) of the resulting channel $\mF_{\text{e}}(\alpha, \beta, p, q)$ is expressed as
\begin{align}
f(\mF_{\text{e}}(\alpha, \beta, p, q))
=
\int d\psi F_{\text{U}}\left(\psi, \mF_{\text{e}}(\alpha, \beta, p, q)(\psi)\right)
=
\int d\psi \Tr[\psi\cdot
\left(
\mF_{\text{e}}(\alpha, \beta, p, q)(\psi)
\right)].
\end{align}
By maximizing over all possible values of $\alpha$ and $\beta$, we arrive at the following fidelity expression.
\begin{align}\label{eq:fe}
f_{\text{e}}
:=
\max_{\alpha, \beta}f(\mF_{\text{e}}(\alpha, \beta, p, q)).
\end{align}
As $\mF_{\text{e}}(\alpha, \beta, p, q)$ (see Eq.~\ref{eq:channel-fe}) equals $\mF_{\text{b}}(p, q)$ (see Eq.~\ref{eq:channel-fb}) when $\alpha=0$ and $\beta=0$ (see Eq.~\ref{eq:u12}), it follows that
\begin{align}\label{eq:ine-eb}
f_{\text{e}}\geqslant f_{\text{b}},
\end{align}
where $f_{\text{b}}$ is defined in Eq.~\ref{eq:fb}. Although protocol e has the ability to mitigate the effects of noise and improve the fidelity of teleportation, its interplay with $f_{\text{c}}$ (see Eq.~\ref{eq:fc}) and $f_{\text{d}}$ (see Eq.~\ref{eq:fd}) is currently unknown.
\item {\bf Protocol f: Superchannel-Assisted Teleportation (see Fig.~\ref{fig:qt}(f)).} A quantum superchannel is the most comprehensive method for manipulating a quantum channel, allowing the transformation of a quantum channel into another quantum channel by enabling the channel to have memory and transmit information across multiple time steps. This feature makes quantum superchannels a crucial component in the development of efficient and reliable quantum communication protocols and quantum computing algorithms. To mitigate the noise that occurs in $\mE(p, q)$, this protocol utilizes a quantum superchannel $\Theta= \Theta^{\text{post}}\circ\Theta^{\text{pre}}\in\mathfrak{F}_{2}(\text{LOCC}(RA:B))$. This superchannel consists of both pre-processing $\Theta^{\text{pre}}$ and post-processing $\Theta^{\text{post}}$ channels, which are both local operations and classical communication channels. Now, the output channel $\mF_{\text{f}}(\Theta, p, q)$ that transmits information from $R$ to $B$ reads
\begin{align}\label{eq:channel-ff}
\mF_{\text{f}}(\Theta, p, q)_{R\to B}:=
\Theta\left(\mE(p, q)\right)=\Theta^{\text{post}}\circ\mE(p, q)\circ\Theta^{\text{pre}}.
\end{align}
In this case, the average fidelity $f$ (see Eq.~\ref{eq:af}) of the output channel $\mF_{\text{f}}(\Theta, p, q)$ is calculated as
\begin{align}
f(\mF_{\text{f}}(\Theta, p, q))
=
\int d\psi F_{\text{U}}\left(\psi, \mF_{\text{f}}(\Theta, p, q) (\psi)\right)
=
\int d\psi \Tr[\psi\cdot
\left(
\mF_{\text{f}}(\Theta, p, q)(\psi)
\right)].
\end{align}
Upon maximizing over all LOCC superchannels, we obtain the following expression for fidelity.
\begin{align}\label{eq:ff}
f_{\text{f}}
:=
\max_{\Theta\in\mathfrak{F}_{2}(\text{LOCC}(RA:B))}
f(\mF_{\text{f}}(\Theta, p, q)).
\end{align}
The subscript `f' is used in average fidelity $f$ to signify that the fidelity is determined using protocol f. Protocol e considers a specific LOCC superchannel $\Theta_0:=\Theta^{\text{post}}_{0}\circ\Theta^{\text{pre}}_{0}$ (see Fig.~\ref{fig:qt}(e)), which is only a special case of the broader class of LOCC superchannels. The fidelity $f_{\text{f}}$ is obtained by maximizing over all LOCC superchannels, and therefore captures the optimal performance of the teleportation protocol. As a result, we can establish the following relationship between $f_{\text{e}}$ (see Eq.~\ref{eq:fe}) and $f_{\text{f}}$ (see Eq.~\ref{eq:ff}).
\begin{align}\label{eq:ine-fe}
f_{\text{f}}\geqslant f_{\text{e}}.
\end{align}
Additionally, it is worth mentioning that protocol c, as shown in Fig.~\ref{fig:qt}(c), implements a special form of LOCC superchannel. This further illuminates the connection between the two protocols, which can be summarized as follows
\begin{align}\label{eq:ine-fc}
f_{\text{f}}\geqslant f_{\text{c}}.
\end{align}
The protocol f discussed here provides the most general method for manipulating the bipartite channel shared between the sender and receiver, and effectively leverages the resources associated with the noisy entangling gate $\mE(p, q)$. Therefore, if LOCC operations are considered to be free, the average fidelity achieved by $f_{\text{f}}$ represents the highest performance achievable in practical scenarios.
\end{itemize}

\begin{figure}[h]
    \centering
    \includegraphics[width=0.3\textwidth]{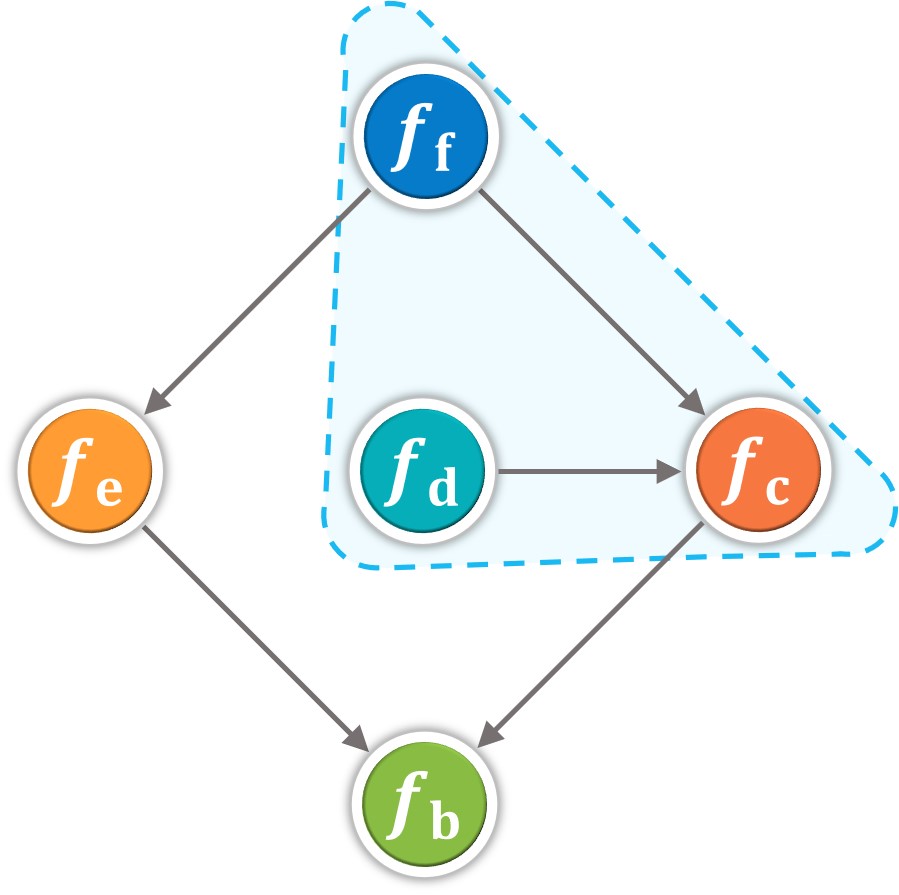}
    \caption{(Color online) Hasse diagram for the average fidelities. Here the arrows connecting the nodes indicate the direction of the partial ordering between the quantities. Specifically, an arrow pointing from node $f_x$ to node $f_y$ indicates that the quantity $f_x$ is greater than or equal to $f_y$, with $x, y\in \{\text{b}, \text{c}, \text{d}, \text{e}, \text{f}\}$. Computing the fidelities, as shown in the blue dashed box, involves optimizing over a set of allowable operations, making it a difficult problem to solve.}
    \label{fig:poset}
\end{figure}

To assess the relative performance of each protocol (see Fig.~\ref{fig:qt}), we will compare their average fidelity (see Eq.~\ref{eq:af}). From the results presented in Eq.~\ref{eq:ine-cb} and Eq.~\ref{eq:ine-dc}, it is evident that the fidelity of protocol d is superior to those of protocols c and b.
\begin{align}\label{eq:ine-dcb}
f_{\text{d}}\geqslant f_{\text{c}}\geqslant f_{\text{b}}.
\end{align}
Similarly, Eq.~\ref{eq:ine-eb} and Eq.~\ref{eq:ine-fe} reveal that the fidelity of protocol f is higher than that of protocols e and b.
\begin{align}\label{eq:ine-feb}
f_{\text{f}}\geqslant f_{\text{e}}\geqslant f_{\text{b}}.
\end{align}
However, the relationship between the average fidelity of protocol e and those of protocols d and c remains ambiguous, and requires further examination. The inequality chains described above, i.e., Eq.~\ref{eq:ine-dcb} and Eq.~\ref{eq:ine-feb}, hold for any noisy bipartite channel $\mE(p, q)$ (brown boxes in Fig.~\ref{fig:qt}(b)-(f)), meaning that they apply to a broad range of noise models. Therefore, these bounds provide a general framework for evaluating the performance of quantum teleportation protocols in the presence of noise. Mathematically, the average fidelities associated with different protocols form a partially ordered set (poset), which can be visualized using a \emph{Hasse diagram} such as Fig.~\ref{fig:poset}. In this poset, the nodes represent different  average fidelities, and there is a directed edge from node $f_x$ to node $f_y$ if and only if $f_y$ is less than or equal to $f_x$ ($x, y\in \{\text{b}, \text{c}, \text{d}, \text{e}, \text{f}\}$); that is
\begin{align}
    f_x\to f_y
    \quad\iff\quad
    f_x\geqslant f_y,
    \quad
    x, y\in \{\text{b}, \text{c}, \text{d}, \text{e}, \text{f}\}.
\end{align}

In the upcoming subsection (i.e., Subsec.~\ref{subsec:NE}), we will further demonstrate that there are specific noise models (in terms of $\mN_1(p)$ and $\mN_2(q)$) for which the fidelity of protocol e is strictly higher than that of protocol d. 
\begin{align}\label{eq:ine-ed}
f_{\text{e}}> f_{\text{d}}.
\end{align}
Building on previous analysis, we can now establish the following hierarchy in this case.
\begin{align}\label{eq:ine-fedcb}
f_{\text{f}}\geqslant f_{\text{e}}> 
f_{\text{d}}\geqslant f_{\text{c}}\geqslant f_{\text{b}}.
\end{align}
This analysis highlights the importance of investigating the performance of quantum teleportation protocols with pre-processing, as the effectiveness of a simply LOCC pre-processing-assisted protocol like e (see Fig.~\ref{fig:qt}(e)) cannot be achieved even by allowing all possible PPT post-processing operations (see Fig.~\ref{fig:qt}(d)). Therefore, these results emphasize the necessity of investigating the potential benefits of superchannel techniques in the development of robust quantum communication protocols, such as the protocol f illustrated in Fig.~\ref{fig:qt}(f).


\subsection{\label{subsec:FB}Fidelity Benchmarking: The Role of Temporal Entanglement}

To assess the quality of the teleportation protocols, we typically use the average fidelity, which measures the similarity between the input state and the teleported state. However, computing the average fidelity can be a challenging task, especially for high-dimensional states. To simplify the discussion, we can roughly categorize the fidelities considered in Subsec.~\ref{subsec:QT} into two types: (\romannumeral1) The first type corresponds to protocols with standard teleportation operation $\Theta^{\text{post}}_{0}$, such as $f_{\text{e}}$ (see Eq.~\ref{eq:fe}) and $f_{\text{b}}$ (see Eq.~\ref{eq:fb}), which are determined solely by the properties of the quantum channel and the measurements; (\romannumeral2) The second type, on the other hand, requires the maximization of fidelity over a set of allowed operations or protocols, such as $f_{\text{f}}$ (see Eq.~\ref{eq:ff}), $f_{\text{d}}$ (see Eq.~\ref{eq:fd}), and $f_{\text{c}}$ (see Eq.~\ref{eq:fc}). This optimization allows us to find the maximum fidelity achievable using the given set of operations and protocols, and therefore provides a way to compare the performance of different quantum teleportation schemes. In this subsection, we will explore the process of calculating these fidelities and prepare ourselves for upcoming numerical experiments. Through this discussion, we will gain a deeper understanding of the mathematical underpinnings of the fidelities and develop the necessary tools to confidently analyze our experimental results.

\begin{figure}[h]
    \centering
\includegraphics[width=0.45\textwidth]{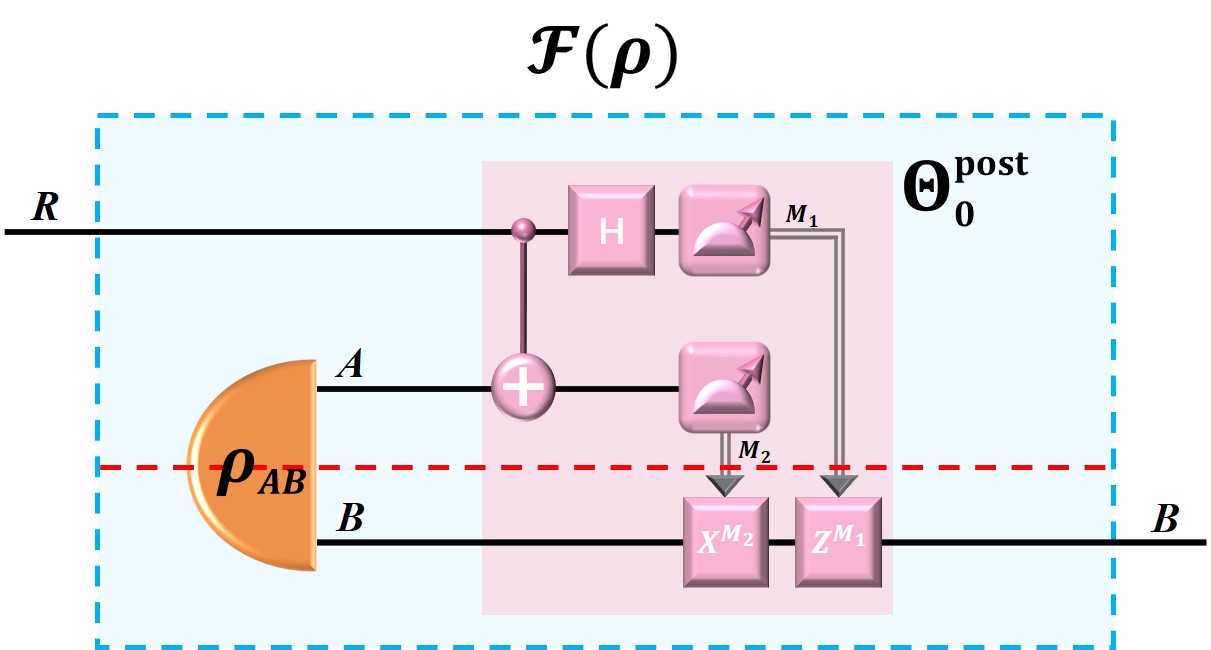}
    \caption{(Color online) Teleportation channel $\mF(\rho)$ (see Eq.~\ref{eq:channel-rho}), which is generated by first preparing a bipartite state $\rho_{AB}$ and subsequently implementing a standard teleportation operation $\Theta^{\text{post}}_{0}$. The resulting channel $\mF(\rho)$ can be used to transmit quantum states over distances.}
    \label{fig:channel-rho}
\end{figure}

Let's begin our discussion of teleportation protocols b and e. It is worth noting that both protocols share a common feature (see Fig.~\ref{fig:qt}(b) and (e)): their post-processing step is based on the standard teleportation operation $\Theta^{\text{post}}_{0}$. In other words, after the measurements are performed, the classical information is sent from the receiver to the sender, who can then apply the appropriate correction operations to reconstruct the teleported state. This use of a common post-processing operation makes it easier to compute the performance of the two protocols. To begin, let us refer to the state of the system $AB$ before post-processing $\Theta^{\text{post}}_{0}$ as $\rho_{AB}$. It is important to understand the relationship between this state and the teleportation channel $\mF(\rho)$ that is generated by $\rho_{AB}$. For a visual representation, consult the accompanying Fig.~\ref{fig:channel-rho}.
Here channel $\mF(\rho)$ is given by
\begin{align}\label{eq:channel-rho}
\mF(\rho)_{R\to B}:=\Theta^{\text{post}}_{0, RAB\to B}(\rho_{AB}).
\end{align}
The average fidelity of the teleportation channel $\mF(\rho)$ is a key metric for assessing its performance in quantum teleportation. However, its calculation is notoriously challenging. The problem can be simplified by borrowing the insights from Lemma~\ref{lem:AFEF}, which enables us to transform the calculation of average fidelity $f$ into the calculation of entanglement fidelity $F$. Although the calculation of entanglement fidelity is still a formidable task, we present a new lemma that establishes a direct connection between the entanglement fidelity of $\mF(\rho)$ and the fidelity of the state $\rho$ shared between the sender and receiver. Our proof is based on a novel diagrammatic approach that offers a fresh perspective on the problem and provides valuable intuition for understanding the underlying physics.

\begin{mylem}
{Connection between Entanglement Fidelity and Uhlmann fidelity}{EFUF}
When a state $\rho$ on system $AB$ is subjected to the standard teleportation operation $\Theta^{\text{post}}_{0}$, as illustrated in Fig.~\ref{fig:channel-rho}, the resulting entanglement fidelity $F(\mF(\rho))$ of the generated channel $\mF(\rho)$ (see Eq.~\ref{eq:channel-rho}) can be expressed as the Uhlmann fidelity $F_{\text{U}}(\rho, \phi^{+})$ between $\rho$ and the maximally entangled state $\phi^{+}$. More specifically, this is given by
\begin{align}
F(\mF(\rho))= F_{\text{U}}(\rho, \phi^{+})= \Tr[\rho\cdot\phi^+].
\end{align}
Above equation simplifies the calculation of entanglement fidelity $F$ for the teleportation channel $\mF(\rho)$. By providing a closed-form expression for the quantity, our equation allows for efficient and accurate evaluation of the performance of the channel, which is essential for assessing its suitability for practical applications in quantum information processing.
\end{mylem}

\begin{proof}
Our analysis is geared towards qubits for the sake of clarity and simplicity. In order to illustrate our findings, we have opted for a pictorial method. Specifically, we have depicted the entanglement fidelity $F(\mF(\rho))$ as a diagram, as shown in Fig.~\ref{fig:EFUF}.

\begin{figure}[h]
    \centering
\includegraphics[width=1\textwidth]{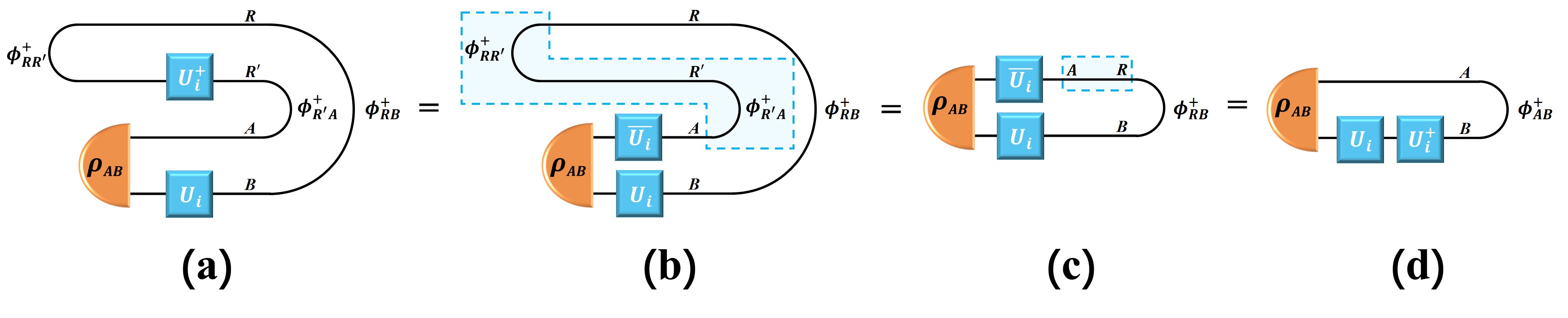}
    \caption{(Color online) A graphical representation of the entanglement fidelity $F$ (see Def.~\ref{def:EF}) of channel $\mF(\rho)$ (see Eq.~\ref{eq:channel-rho}).}
    \label{fig:EFUF}
\end{figure}

Given the Bell measurement's equal probability of all outcomes at $1/4$ in this scenario, we will narrow our focus to the measurement $\1\otimes U_{i}\ket{\phi^{+}}$. The mathematical characteristic of the maximally entangled state permits us to equate Fig.~\ref{fig:EFUF}(a) to Fig.~\ref{fig:EFUF}(b) and Fig.~\ref{fig:EFUF}(c) to Fig.~\ref{fig:EFUF}(d). In addition, Fig.~\ref{fig:EFUF}(b) and Fig.~\ref{fig:EFUF}(c) are demonstrated to be identical through the application of the yanking equation. The correspondence between Fig.~\ref{fig:EFUF}(d) and the Uhlmann fidelity $F_{\text{U}}(\rho, \phi^{+})$ is apparent, thereby concluding the proof.
\end{proof}

\begin{remark}
While our proof in Lemma~\ref{lem:EFUF} focuses on qubits, it should be noted that our approach is applicable beyond this specific scenario. In fact, one can substitute the Pauli operators with the Heisenberg-Weyl operators to extend our method to systems of any finite dimension. Moreover, while we present many of our results with a pictorial representation in the qubit case, it is important to emphasize that our findings hold for arbitrary finite-dimensional systems.
\end{remark}

To analyze the properties of the quantum teleportation protocols b and e, let us denote the state of the system $AB$ before post-processing $\Theta^{\text{post}}_{0}$ as 
\begin{align}
\rho_{\text{b}}
&:=
\mN_1(p)_{A\to A}\otimes\mN_2(q)_{B\to B}\circ \text{C}_{\text{NOT}, AB\to AB}\circ \text{H}_{A\to A}\otimes\id_{B\to B}
(\ketbra{0}{0}_{A}\otimes\ketbra{0}{0}_{B}),\label{eq:rho-b}\\
\rho_{\text{e}}(\alpha, \beta)
&:=
\mN_1(p)_{A\to A}\otimes\mN_2(q)_{B\to B}\circ \text{C}_{\text{NOT}, AB\to AB}\circ \text{H}_{A\to A}\otimes\id_{B\to B}\notag\\
&\quad\quad\quad\quad\quad\quad\quad\quad\quad\quad\quad\quad\quad\quad\quad\quad\quad\quad\quad\quad\quad\quad\quad\quad
\left(\left(U_1(\alpha)\ketbra{0}{0} U_1^{\dag}(\alpha)\right)_A\otimes
\left(U_2(\beta)\ketbra{0}{0} U_2^{\dag}(\beta)\right)_B\right).\label{eq:rho-e-ab}
\end{align}
Lemma~\ref{lem:EFUF} provides a key result for evaluating the entanglement fidelity of the teleportation channel $\mF(\rho_{\text{b}})= \mF_{\text{b}}(p, q)$ (see Eq.~\ref{eq:channel-fb}) and $\mF(\rho_{\text{e}}(\alpha, \beta))= \mF_{\text{e}}(\alpha, \beta, p, q)$ (see Eq.~\ref{eq:channel-fe}). Specifically, the entanglement fidelity can be expressed in a compact form, as shown in the following equation
\begin{align}
F(\mF(\rho_{\text{b}}))
&=
\Tr[\rho_{\text{b}}\cdot\phi^+],\\
F(\mF(\rho_{\text{e}}(\alpha, \beta)))
&=
\Tr[\rho_{\text{e}}(\alpha, \beta)\cdot\phi^+].
\end{align}
Drawing upon the insights presented in Lemma~\ref{lem:AFEF}, we can obtain a concise expression for the average fidelity of protocol b. This result is fundamental for assessing the efficacy of the noisy teleportation protocol in transmitting quantum states from the sender to the receiver.

\begin{mythm}
{Performance of Protocol b}{protocol-b}
The expression for the average fidelity of protocol b, i.e., $f_{\text{b}}$ (see Eq.~\ref{eq:fb}), is given by
\begin{align}
f_{\text{b}}=
\frac{2\Tr[\rho_{\text{b}}\cdot\phi^+]+ 1}{3},
\end{align}
where $\rho_{\text{b}}$ is defined in Eq.~\ref{eq:rho-b}.
\end{mythm}

\noindent Protocol e can also be evaluated in terms of average fidelity using Lemma~\ref{lem:AFEF}, similar to protocol b. However, the calculation of the average fidelity for protocol e is more complex compared to that for protocol b. This is because we need to maximize over all possible values of the pre-processing parameters $\alpha$ and $\beta$, which introduces additional computational complexity.

\begin{mythm}
{Performance of Protocol e}{protocol-e}
The average fidelity of channel protocol e, i.e., $f_{\text{e}}$ (see Eq.~\ref{eq:fe}), can be calculated using the formula
\begin{align}
f_{\text{e}}=
\frac{2\left(\max\limits_{\alpha, \beta}\left\{\Tr[\rho_{\text{e}}(\alpha, \beta)\cdot\phi^+]\right\}\right)+ 1}{3}.
\end{align}
Here, $\rho_{\text{e}}(\alpha, \beta)$ is given by Eq.~\ref{eq:rho-e-ab}.
\end{mythm}

Moving on to more complex problems, we will now examine the performance of protocols c, d, and f. Our analysis will begin with protocols c and d, which share the same state of system $AB$ before post-processing (the region highlighted in pink of Fig.~\ref{fig:qt}), namely $\rho_{\text{b}}$ (see Eq.~\ref{eq:rho-b}). To determine the efficacy of protocols c and d, we can use the lemma that follows, which provides us with a path forward.

\begin{figure}[h]
    \centering
    \includegraphics[width=0.4\textwidth]{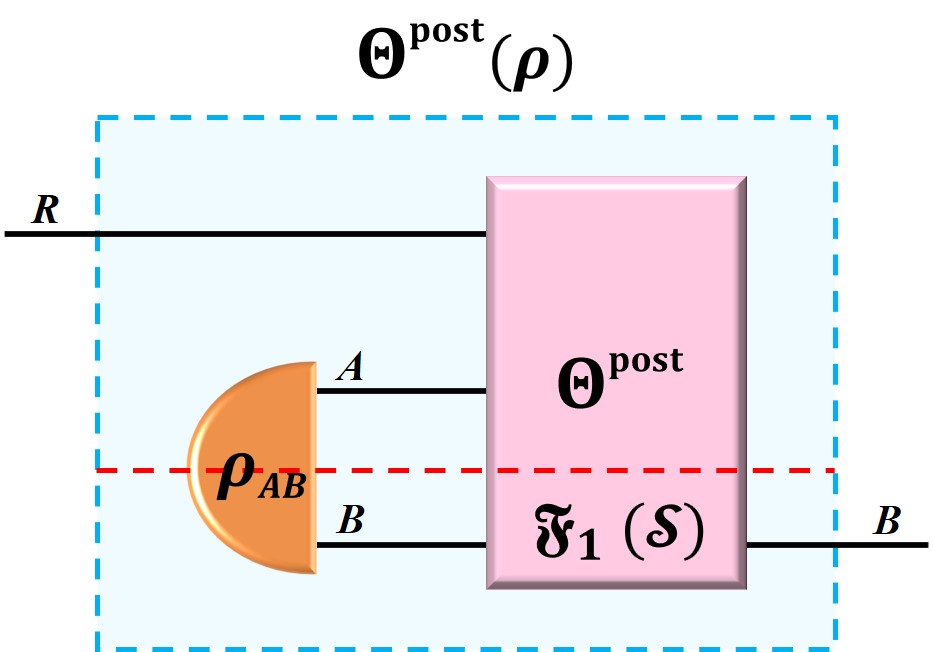}
    \caption{(Color online) Teleportation channel $\Theta^{\text{post}}(\rho)$, which is created by first preparing an initial state $\rho_{AB}$, and then subjected to post-processing operations $\Theta^{\text{post}}_{RAB\to B}\in\mathfrak{F}_{1}(\mS)$. The post-processing operation $\Theta^{\text{post}}_{RAB\to B}$ is designed to transform the static state $\rho$ into a dynamical resource, which can be used to teleport quantum information from message system $R$ to receiver's system $B$.}
    \label{fig:channel-S}
\end{figure}

\begin{mylem}
{Connection between Average Fidelity and Entanglement Distillation}{AFED}
When a bipartite quantum state $\rho$ on system $AB$, with dimension $\dim A= \dim B= d$, undergoes a post-processing $\Theta^{\text{post}}_{RAB\to B}\in\mathfrak{F}_{1}(\mS)$, where $\mS$ is a set of allowed operations that can be one of the following $\{\text{LOCC}_{1}(\text{poly}(d)), \text{LOCC}_{k}, \text{LOCC}_{\mathds{N}}, \text{LOCC}, \overline{\text{LOCC}_{\mathds{N}}}, \text{SEP}, \text{SEPP}, \text{PPT}\}$, then the maximal average fidelity of the resulting channel $\Theta^{\text{post}}(\rho)$, as demonstrated in Fig.~\ref{fig:channel-S}, is completely characterized by the $d$-fidelity of $\mS$ distillation, which is defined in Def.~\ref{def:FD}. In other words, we can determine the maximal average fidelity as
\begin{align}\label{eq:state-fidelity-lem}
\max_{\Theta^{\text{post}}\in\mathfrak{F}_{1}(\mS)}f(\Theta^{\text{post}}(\rho))
=
\frac{d F_{d, \mS}(\rho)+ 1}{d+1},
\end{align}
where $\mathfrak{F}_{1}(\mS)$ is defined as the collection of all free channels, as specified in Def.~\ref{def:Simulable}.
\end{mylem}

\begin{remark}
In Eq.~\ref{eq:state-fidelity-lem}, $F_{d, \mS}(\rho)$, appeared on the right-hand side, is a specific instance of the more general $d$-fidelity of $\mS$ distillation $F_{d, \mS}(\mE)$ (defined in Def.~\ref{def:FD}). To develop a more complete understanding of the expression, we unpack the formula for $F_{d, \mS}(\rho)$. 
\begin{align}\label{eq:FD-rho}
F_{d, \mS}(\rho)
:= 
\max_{\Omega\in\mathfrak{F}_{1}(\mS)} F_{\text{U}}(\Omega(\rho), \phi^{+}_{d})
=
\max_{\Omega\in\mathfrak{F}_{1}(\mS)}
\Tr[\Omega(\rho)\cdot\phi^{+}_{d}].
\end{align}
Since the input of $F_{d, \mS}$ is a state, the corresponding free morphism simplifies to the case of a superchannel without pre-processing, which is equivalent to a quantum channel.
\end{remark}

\begin{proof}
The proof will start by showing that the left-hand side is smaller than or equal to the right-hand side. Assuming that the maximum of the left-hand side is achieved by some quantum channel $\Xi\in\mathfrak{F}_{1}(\mS)$, we have
\begin{align}\label{eq:state-fidelity-lem-left}
\max_{\Theta^{\text{post}}\in\mathfrak{F}_{1}(\mS)}f(\Theta^{\text{post}}(\rho))
=
f(\Xi(\rho))
=
\frac{d F(\Xi(\rho))+ 1}{d+1}
=
\frac{d \Tr[\Xi(\rho)(\phi^{+}_{d})\cdot\phi^{+}_{d}]+ 1}{d+1}
\leqslant
\frac{d F_{d, \mS}(\rho)+ 1}{d+1}.
\end{align}
The second equation is derived from Lem.~\ref{lem:AFEF}, the third equation is based on the definition of entanglement fidelity (see Def.~\ref{def:EF}), and the inequality is a direct consequence of Eq.~\ref{eq:FD-rho}. Without loss of generality, we can assume that both $\rho$ and $\phi^{+}_{d}$ act on the system $AB$. We can then apply a local operation to transform $\Xi(\rho)(\phi^{+}_{d})$ from system $RB$ to $AB$. For a visual illustration, see Fig.~\ref{fig:state-fidelity-lem}(a). Since local operations (i.e., $\text{LO}$) belong to the set of free morphisms $\mathfrak{F}_{1}(\mS)$, the overall operation is still a free channel in $\mathfrak{F}_{1}(\mS)$.

\begin{figure}[h]
    \centering
    \includegraphics[width=1\textwidth]{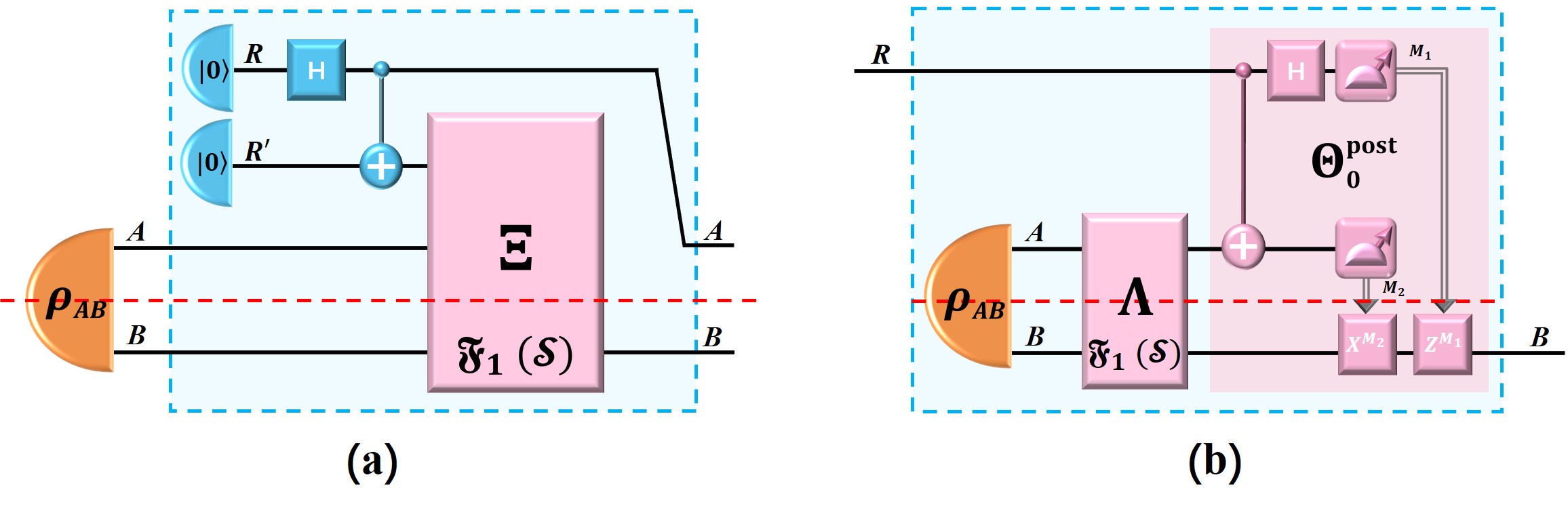}
    \caption{(Color online) Visual demonstration of the proof of Lem.~\ref{lem:AFED}. Figure (a) illustrates the role of $\Xi$ behind Equation~\ref{eq:state-fidelity-lem-left}, which is a key element of our proof. On the other hand, Figure (b) depicts the construction of a teleportation channel based on $\Lambda$ (see Eq.~\ref{eq:state-fidelity-lem-right}). In this diagram, the systems located above the dashed red line belong to Alice, while the systems below the dashed line belong to Bob. While we have focused on qubit cases in this figure for the sake of simplicity, it is worth noting that our proof holds for systems of arbitrary finite dimension $d$.}
    \label{fig:state-fidelity-lem}
\end{figure}

Next, we will show the inverse direction of the proof by demonstrating that the right-hand side is smaller than or equal to the left-hand side. In this part of the proof, let us assume that the right-hand side value of $F_{d, \mS}(\rho)$ is achieved by a particular quantum channel $\Lambda\in\mathfrak{F}_{1}(\mS)$. Then, we have
\begin{align}\label{eq:state-fidelity-lem-FD}
F_{d, \mS}(\rho)
=
\Tr[\Lambda(\rho)\cdot\phi^{+}_{d}]
=
F(\mF(\Lambda(\rho)))
=
F(\Theta^{\text{post}}_{0}\circ\Lambda(\rho)),
\end{align}
where $\Theta^{\text{post}}_{0}$ is the standard teleportation operation (see the pink box of Fig.~\ref{fig:qt}(a)), this implies
\begin{align}\label{eq:state-fidelity-lem-right}
\frac{d F_{d, \mS}(\rho)+ 1}{d+1}
=
\frac{d F(\Theta^{\text{post}}_{0}\circ\Lambda(\rho))+ 1}{d+1}
=
f(\Theta^{\text{post}}_{0}\circ\Lambda(\rho))
\leqslant
\max_{\Theta^{\text{post}}\in\mathfrak{F}_{1}(\mS)}f(\Theta^{\text{post}}(\rho)).
\end{align}
Since $\Lambda\in\mathfrak{F}_{1}(\mS)$ and $\Theta^{\text{post}}_{0}\in\text{LOCC}_{1}(\text{poly}(d))\subset\mathfrak{F}_{1}(\mS)$, we have $\Theta^{\text{post}}_{0}\circ\Lambda\in\mathfrak{F}_{1}(\mS)$. An illustration of the teleportation channel $\Theta^{\text{post}}_{0}\circ\Lambda(\rho)$ is provided in Fig.~\ref{fig:state-fidelity-lem}(b). The second equation of Eq.~\ref{eq:state-fidelity-lem-FD} is obtained using Lem.~\ref{lem:EFUF}, which allows us to express the $d$-fidelity of $\mS$ distillation $F_{d, \mS}(\rho)$ with respect to the entanglement fidelity of channel $\Theta^{\text{post}}_{0}\circ\Lambda(\rho)$. Combining Eq.~\ref{eq:state-fidelity-lem-left} with Eq.~\ref{eq:state-fidelity-lem-right}, we can conclude that the two sides are equal, which completes the proof.
\end{proof}

\begin{remark}
We make two remarks about Lem.~\ref{lem:AFED}. Firstly, it generalizes the main theorem presented in Ref.~\cite{PhysRevA.60.1888}, by extending its scope beyond the LOCC operations considered in that work. Specifically, while LOCC might require an impractically large number of communication rounds, our lemma applies to a broader range of operations, including $\text{LOCC}_{k}$, making it more practical. Secondly, the proof of our lemma demonstrates that our results are applicable to any set $\mS$ that includes $\text{LOCC}_{1}(\text{poly}(d))$, i.e., $\text{LOCC}_{1}(\text{poly}(d))\subset\mS$. This means that the inequality in Eq.~\ref{eq:state-fidelity-lem} still holds for a broader range of operations beyond those explicitly considered in the lemma. This observation highlights the generality and versatility of our approach, and its potential applications to various teleportation scenarios.
\end{remark}

Equipped with Lem.~\ref{lem:AFED}, we can now express the performance of protocols c and d in a more precise and general way. Specifically, we can apply the inequality in Eq.~\ref{eq:state-fidelity-lem} to obtain the average fidelity of teleportation channel produced by these protocols. This allows us to better understand the effects of noise and imperfections on the entanglement, and to optimize the protocols for practical applications. First, let's consider the performance of protocol c. Written out explicitly, we have

\begin{mythm}
{Performance of Protocol c}{protocol-c}
To calculate the average fidelity of protocol c, denoted by $f_{\text{c}}$ (as defined in Eq.~\ref{eq:fc}), we can use the following expression
\begin{align}\label{eq:per-pro-c}
f_{\text{c}}=
\frac{2\left(\max\limits_{\Omega\in\text{LOCC}(A:B)}
\Tr[\Omega(\rho_{\text{b}})\cdot\phi^+]\right)+ 1}{3},
\end{align}
where $\rho_{\text{b}}$ is defined in Eq.~\ref{eq:rho-b}.
\end{mythm}

Although Eq.~\ref{eq:per-pro-c} provides a formula for the average fidelity of protocol c, it is still difficult to compute due to the complexity of the mathematical structure of LOCC. This limitation has motivated the development of protocol d, which offers a computable solution to the problem.

\begin{mythm}
{Performance of Protocol d}{protocol-d}
To determine the average fidelity of protocol d, which is denoted by $f_{\text{d}}$ as defined in Eq.~\ref{eq:fd}, we can utilize the following expression
\begin{align}\label{eq:per-pro-d}
f_{\text{d}}=
\frac{2\left(\max\limits_{\Omega\in\text{PPT}(A:B)}
\Tr[\Omega(\rho_{\text{b}})\cdot\phi^+]\right)+ 1}{3},
\end{align}
where $\rho_{\text{b}}$ is defined in Eq.~\ref{eq:rho-b}.
\end{mythm}

Thanks to the mathematical property of positive partial transpose (PPT) operations, the optimization term in the right-hand side of Eq.~\ref{eq:per-pro-d} now takes a tractable semi-definite programming (SDP) form~\cite{doi:10.1137/1038003,boyd_vandenberghe_2004}. This enables us to efficiently optimize the fidelity and make performance comparisons between protocol d and other teleportation protocols. Through a deeper examination of the aforementioned optimization, we can simplify it to~\cite{959270}
\begin{align}
\max\limits_{\Omega\in\text{PPT}(A:B)}
\Tr[\Omega(\rho_{\text{b}})\cdot\phi^+]
=
&\max\,\,\,\Tr[\rho_{\text{b}}\cdot X_{AB}],\notag\\
&\,\,\text{s.t.}\,\,\,\,\,\,\,\0\leqslant X_{AB}\leqslant\1,\,\,\,
-\frac{1}{2}\1\leqslant X_{AB}^{\T_B}\leqslant\frac{1}{2}\1,
\end{align}
where $\T_B$ denotes the partial transpose with respect to system $B$. Writing everything out explicitly, we have

\begin{mycor}
{Performance of Protocol d Simplified}{protocol-d-sim}
To determine the average fidelity of protocol d, which is denoted by $f_{\text{d}}$ as defined in Eq.~\ref{eq:fd}, we can utilize the following expression
\begin{align}\label{eq:per-pro-d-sim}
f_{\text{d}}=
\frac{2\max\left\{\Tr[\rho_{\text{b}}\cdot X_{AB}]
\,\bigg|\,
\0\leqslant X_{AB}\leqslant\1,\,
-\frac{1}{2}\1\leqslant X_{AB}^{\T_B}\leqslant\frac{1}{2}\1
\right\}+ 1}{3},
\end{align}
where $\rho_{\text{b}}$ is defined in Eq.~\ref{eq:rho-b}.
\end{mycor}

The implementation of a quantum superchannel significantly increases the complexity of calculating the average fidelity for protocol f (see Fig.~\ref{fig:qt}(f)). Fortunately, temporal entanglement provides a natural framework for simplifying this calculation. By exploiting the properties of temporal entanglement, we can simplify the calculation of the average fidelity and improve the efficiency of quantum teleportation. Here, we will delve into the details of how temporal entanglement can be leveraged to enhance the performance of quantum teleportation and to achieve higher fidelities. Our starting point will be a generalized version of Lem.~\ref{lem:AFED}.

\begin{figure}[h]
    \centering
    \includegraphics[width=0.4\textwidth]{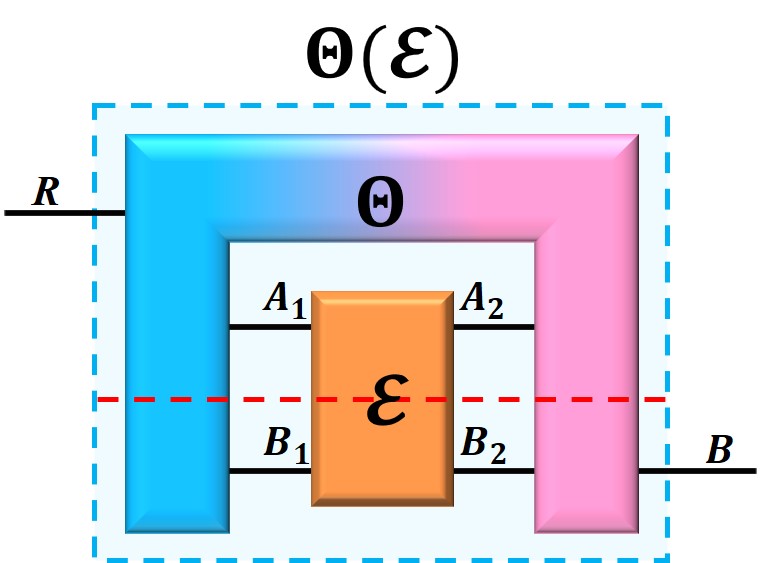}
    \caption{(Color online) Teleportation channel $\Theta(\mE)$ used for transmitting quantum information between distant systems. The channel is obtained by applying a superchannel $\Theta\in\mathfrak{F}_{2}(\mS)$ to a bipartite channel $\mE$. The bipartite channel plays a crucial role in generating the essential entanglement required for teleportation, while the superchannel applies a series of designed operations between the sender, receiver, and message systems to guarantee the reliable and accurate transmission of quantum states.}
    \label{fig:channel-C}
\end{figure}

\begin{mylem}
{Channel Version of Lem.~\ref{lem:AFED}}{AFED-channel}
Consider a bipartite quantum channel $\mE_{A_1B_1\to A_2B_2}$ with dimensions $\dim A_1= \dim A_2= \dim B_1= \dim B_2= d$. Let us apply a superchannel $\Theta_{RA_2B_2\to A_1B_1B}$ from the set $\mathfrak{F}_{2}(\mS)$ to the channel $\mE$, where $\mS$ is a set of permissible operations from $\{\text{LOCC}_{1}(\text{poly}(d)), \text{LOCC}_{k}, \text{LOCC}_{\mathds{N}}, \text{LOCC}, \overline{\text{LOCC}_{\mathds{N}}}, \text{SEP}, \text{SEPP}, \text{PPT}\}$. We can then characterize the maximum achievable average fidelity of the resulting channel $\Theta(\mE)$ (see Fig.~\ref{fig:channel-C}) using the $d$-fidelity of $\mS$ distillation, defined in Def.~\ref{def:FD}. More precisely, the connection between the maximum achievable average fidelity under free superchannels and the temporal entanglement can be formally characterized by the following equation
\begin{align}\label{eq:channel-fidelity-lem}
\max_{\Theta\in\mathfrak{F}_{2}(\mS)}f(\Theta(\mE))
=
\frac{d F_{d, \mS}(\mE)+ 1}{d+1},
\end{align}
where $\mathfrak{F}_{2}(\mS)$ is defined as the collection of all free superchannels, as specified in Def.~\ref{def:Simulable}.
\end{mylem}

\begin{proof}
We will prove the equation by establishing an upper bound for the left-hand side and a lower bound for the right-hand side. Firstly, we will show that the left-hand side is bounded above by the right-hand side. Assuming that there exists a quantum superchannel $\Xi\in\mathfrak{F}_{2}(\mS)$ that maximizes the left-hand side of Eq.~\ref{eq:channel-fidelity-lem}, we see that
\begin{align}\label{eq:channel-fidelity-lem-left}
\max_{\Theta\in\mathfrak{F}_{2}(\mS)}f(\Theta(\mE))
=
f(\Xi(\mE))
=
\frac{d F(\Xi(\mE))+ 1}{d+1}
=
\frac{d \Tr[\Xi(\mE)(\phi^{+}_{d})\cdot\phi^{+}_{d}]+ 1}{d+1}
\leqslant
\frac{d F_{d, \mS}(\mE)+ 1}{d+1}.
\end{align}
We can derive the second equation of Eq.~\ref{eq:channel-fidelity-lem-left} directly from Lemma~\ref{lem:AFEF}. To obtain the third equation of Eq.~\ref{eq:channel-fidelity-lem-left}, we use the definition of entanglement fidelity for a quantum channel, as given in Def.~\ref{def:EF}. We note that the inequality holds because we can prepare a local $\phi^+_d$ state on Alice's side, followed by the free superchannel $\Xi\in\mathfrak{F}_{2}(\mS)$. After the interaction, we can apply the local noiseless channel $\id_{R\to A}\in\text{LO}$. Therefore, we can view the overall effect of $\Xi(\cdot)(\phi^{+}_{d})$ as a superchannel that belongs to $\mathfrak{F}_{2}(\mS)$. Fig.~\ref{fig:channel-fidelity-lem} provides a visual representation of the action of $\Xi(\cdot)(\phi^{+}_{d})$.

\begin{figure}[h]
    \centering
    \includegraphics[width=1\textwidth]{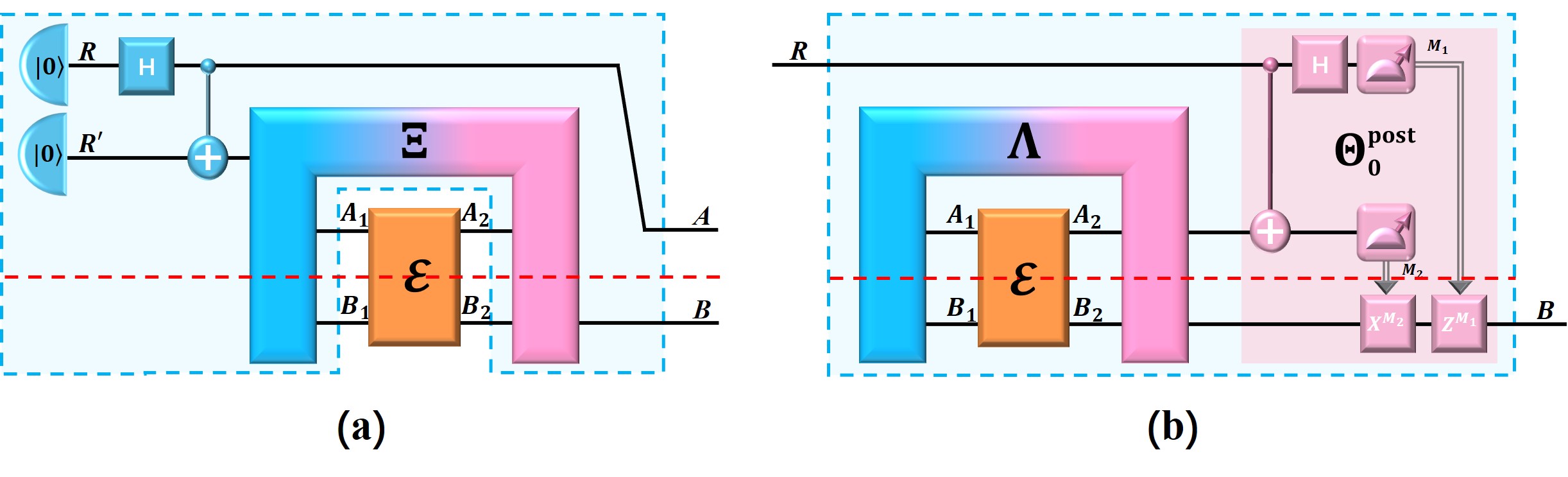}
    \caption{(Color online) Visual demonstration of the proof of Lem.~\ref{lem:AFED-channel}.
    Figure (a) shows how the superchannel $\Xi$ supports Eq.~\ref{eq:channel-fidelity-lem-left}, which is a crucial part of our proof. Meanwhile, figure (b) illustrates how the teleportation channel is constructed using $\Lambda$, as described in Eq.~\ref{eq:channel-fidelity-lem-right}.
    The dashed red line in the diagram serves as a visual separator between the systems belonging to Alice, which are positioned above the line, and the systems belonging to Bob, which are located below the line. Although we have illustrated qubit cases in this figure for simplicity, it is important to emphasize that our proof of Lem.~\ref{lem:AFED-channel} applies to systems of any finite dimension $d$.}
    \label{fig:channel-fidelity-lem}
\end{figure}

Secondly, we will show that the left-hand side is bounded below by the right-hand side. Assuming that a particular quantum channel $\Lambda\in\mathfrak{F}_{2}(\mS)$ achieves the right-hand side value of $F_{d, \mS}(\mE)$, then we can express $F_{d, \mS}(\mE)$ equivalently in the following form
\begin{align}\label{eq:channel-fidelity-lem-FD}
F_{d, \mS}(\mE)
=
\Tr[\Lambda(\mE)\cdot\phi^{+}_{d}]
=
F(\mF(\Lambda(\mE)))
=
F(\Theta^{\text{post}}_{0}\circ\Lambda(\mE)),
\end{align}
where the second equation of Eq.~\ref{eq:channel-fidelity-lem-FD} is a direct consequence of Lem.~\ref{lem:EFUF}, and the last equation of Eq.~\ref{eq:channel-fidelity-lem-FD} comes from Eq.~\ref{eq:channel-rho}, see also Fig.~\ref{fig:channel-rho} for an illustration. It is worth noting that in this context, $\Lambda(\mE)$ denotes a quantum state that operates on the composite system $AB$, while $\Theta^{\text{post}}_{0}\circ\Lambda(\mE)$ represents a quantum channel that maps the message system $R$ to Bob's system $B$. This immediately implies that
\begin{align}\label{eq:channel-fidelity-lem-right}
\frac{d F_{d, \mS}(\mE)+ 1}{d+1}
=
\frac{d F(\Theta^{\text{post}}_{0}\circ\Lambda(\mE))+ 1}{d+1}
=
f(\Theta^{\text{post}}_{0}\circ\Lambda(\mE))
\leqslant
\max_{\Theta\in\mathfrak{F}_{2}(\mS)}f(\Theta(\mE)).
\end{align}
Since the standard teleportation operation $\Theta^{\text{post}}_{0}$ can be realized by using only local operations and one-way classical communication with the communication complexity polynomially dependent on the dimension of the message system, it is an element of the set $\text{LOCC}_{1}(\text{poly}(d))$. When we compose the superchannel $\Lambda$ with the channel $\Theta^{\text{post}}_{0}$, the resulting channel still belongs to the set $\mathfrak{F}_{2}(\mS)$ for any $\mS\in\{\text{LOCC}_{1}(\text{poly}(d)), \text{LOCC}_{k}, \text{LOCC}_{\mathds{N}}, \text{LOCC}, \overline{\text{LOCC}_{\mathds{N}}}, \text{SEP}, \text{SEPP}, \text{PPT}\}$, which leads to the inequality of Eq.~\ref{eq:channel-fidelity-lem-right}.
Finally, by combining Eq.~\ref{eq:channel-fidelity-lem-left} with Eq.~\ref{eq:channel-fidelity-lem-right}, we have shown that the two sides of Eq.~\ref{eq:channel-fidelity-lem} are equal, as required.
\end{proof}

While entanglement monotones have been extensively studied in the context of bipartite channels, their relevance to other quantum information tasks remains uncertain. To address this open question, we investigate the role of bipartite channel, viewed as a carrier of  temporal entanglement, in quantum teleportation and show that it provides unique advantages. Specifically, Lem.~\ref{lem:AFED-channel} demonstrates that the temporal entanglement of a bipartite channel is crucial for determining the optimal performance of communication protocols that involve encoding, decoding, and quantum memory in quantum teleportation. Furthermore, we find that protocol f consistently outperforms the conventional protocol c when viewed as a function of the quantum channel, and that even stricter protocols exist (see Subsec.~\ref{subsec:NE}). Our proof of Lem.~\ref{lem:AFED-channel} also demonstrates that Eq.~\ref{eq:channel-fidelity-lem} applies to any set that includes $\mathfrak{F}_{2}(\text{LOCC}_{1}(\text{poly}(d)))$, namely $\text{LOCC}_{1}(\text{poly}(d))\subset\mS$. One of the key implications of Lem.~\ref{lem:AFED-channel} is that we can express the average fidelity of protocol f in a different way, by using the entanglement fidelity that corresponds to it. To be more precise, we can obtain

\begin{mythm}
{Performance of Protocol f}{protocol-f}
In order to determine the average fidelity of protocol f, which is denoted as $f_{\text{f}}$ and defined in Eq.~\ref{eq:ff}, we can apply the following expression
\begin{align}\label{eq:per-pro-f}
f_{\text{f}}=
\frac{2\left(\max\limits_{\Theta\in\mathfrak{F}_{2}(\text{LOCC})}
\Tr[\Theta(\mE(p, q))\cdot\phi^+]\right)+ 1}{3}.
\end{align}
In this context, $\Theta$ refers to a specific type of LOCC superchannel, i.e., $\mathfrak{F}_{2}(\text{LOCC})$, that maps the quantum channel $\mE(p, q)$ to a bipartite state.
\end{mythm}

Protocol f (see Fig.~\ref{fig:qt}(f)) is designed to fully exploit the resources of bipartite channel $\mE(p, q)$ in quantum communication, taking advantage of the generality of the superchannel formalism for manipulating quantum channels. We can further express the performance of protocol $f$ in terms of entanglement fidelity, as demonstrated in Thm.~\ref{thm:protocol-f}. However, the evaluation of this performance remains a formidable challenge due to the intricate structure of the LOCC-superchannel $\mathfrak{F}_{2}(\text{LOCC})$, which is even more complex than that of a LOCC channel. Unfortunately, now we are unable to provide a systematic method for computing it. The exploration of such a method will require the development of new techniques and insights, and we leave it as an open problem for future researches. Although we are unable to present a quantitative evaluation of the performance of protocol f in this work, we can nonetheless demonstrate a clear advantage of protocol f over the conventional teleportation protocol c (see Fig.~\ref{fig:qt}(c)), which relies solely on post-processing techniques. In the upcoming subsection, we will demonstrate that protocol f achieves a higher entanglement fidelity and, as a result, a higher average fidelity than protocol c for a broad range of noisy models. This superiority underscores the potential of superchannel techniques in expanding the capabilities of quantum communication beyond the confines of conventional teleportation protocols.


\subsection{\label{subsec:NE}Numerical Experiments: Dancing with Data}

In Subsec.~\ref{subsec:FB}, we have simplified the average fidelity of the protocols introduced in Subsec.~\ref{subsec:QT}. Building upon these results, we have conducted numerical experiments to compare the performance of two protocols, e and d, shown in Fig.~\ref{fig:qt}(e) and Fig.~\ref{fig:qt}(d), respectively. Protocol e employs LO pre-processing $\Theta^{\text{pre}}_{0}$ and $\text{LOCC}_{1}(\text{poly}(d))$ post-processing $\Theta^{\text{post}}_{0}$, while protocol d uses PPT post-processing. Our results show that, in some cases, protocol e outperforms protocol d, with the average fidelity $f_{\text{e}}$ being strictly greater than $f_{\text{d}}$, which immediately implies that $f_{\text{f}}>f_{\text{c}}$. This demonstrates the necessity of implementing LOCC pre-processing in quantum teleportation. In this subsection, we will provide details of our numerical experiments and discuss the implications of our findings. 

Before delving into our analysis, we will introduce a set of noise models that will be considered in this subsection. These noise models are listed in Tab.~\ref{tab:noise-models} and will serve as the foundation for our subsequent discussions. For a more detailed discussion of the noise models we are using in our analysis, please refer to the introduction of protocol b in Subsec.~\ref{subsec:QT}. There, we provide a comprehensive overview of the noise models and their associated parameters, which are central to our analysis of the protocol.

\begin{table}[h]
    \centering
    \includegraphics[width=0.95\textwidth]{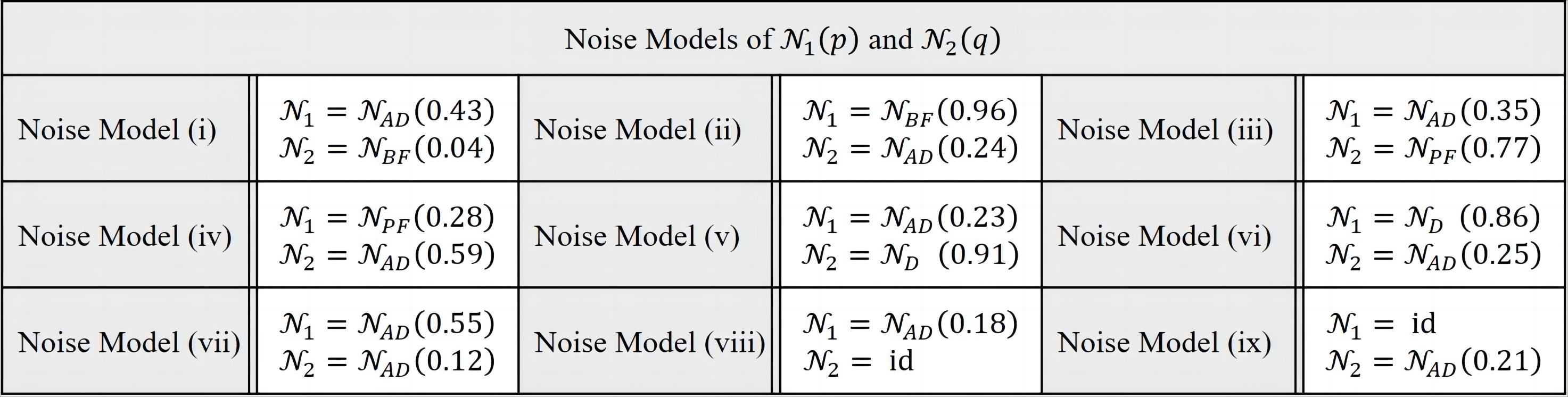}
    \caption{List of noise models $\mN_1(p)$ and $\mN_2(q)$ considered in the bipartite channel $\mE(p, q)$ (see Eq.~\ref{eq:EG}). Here, both $\mN_1(p)$ and $\mN_2(q)$ are noisy channels that are controlled by a single parameter. Abbreviations BF, PF, D, and AD stand for bit flip, phase flip, depolarizing, and amplitude damping channels, which are different types of noise that can affect the transmission of quantum information over the channel. To better understand these noise types, their Kraus are provided in Tab.~\ref{tab:kraus-nc}.}
    \label{tab:noise-models}
\end{table}

To better understand the impact of the noisy models on the performance of protocol e, we have defined a fine-grained average fidelity metric $f_{\text{e}}(\alpha, \beta)$ for some fixed $p$ and $q$; that is
\begin{align}\label{eq:fe-ab}
f_{\text{e}}(\alpha, \beta):=
f(\mF_{\text{e}}(\alpha, \beta, p, q)),
\end{align}
where the channel $\mF_{\text{e}}(\alpha, \beta, p, q)$ is defined in Eq.~\ref{eq:channel-fe}. The quantity $f_{\text{e}}(\alpha, \beta)$ allows us to measure the average fidelity of the protocol e for each noisy model, providing a more comprehensive understanding of how these models shape the behavior of the protocol. Such insights are essential for optimizing the protocol and ensuring its effectiveness in a variety of conditions. Lastly, it is worth noting that the definition of $f_{\text{e}}$ (see Eq.~\ref{eq:fe}) involves maximizing the fidelity with respect to all parameters $\alpha$ and $\beta$. Written in full, that is
\begin{align}
f_{\text{e}}= \max_{\alpha, \beta} f_{\text{e}}(\alpha, \beta).  
\end{align}

\begin{figure}
    \centering
    \includegraphics[width=0.9\textwidth]{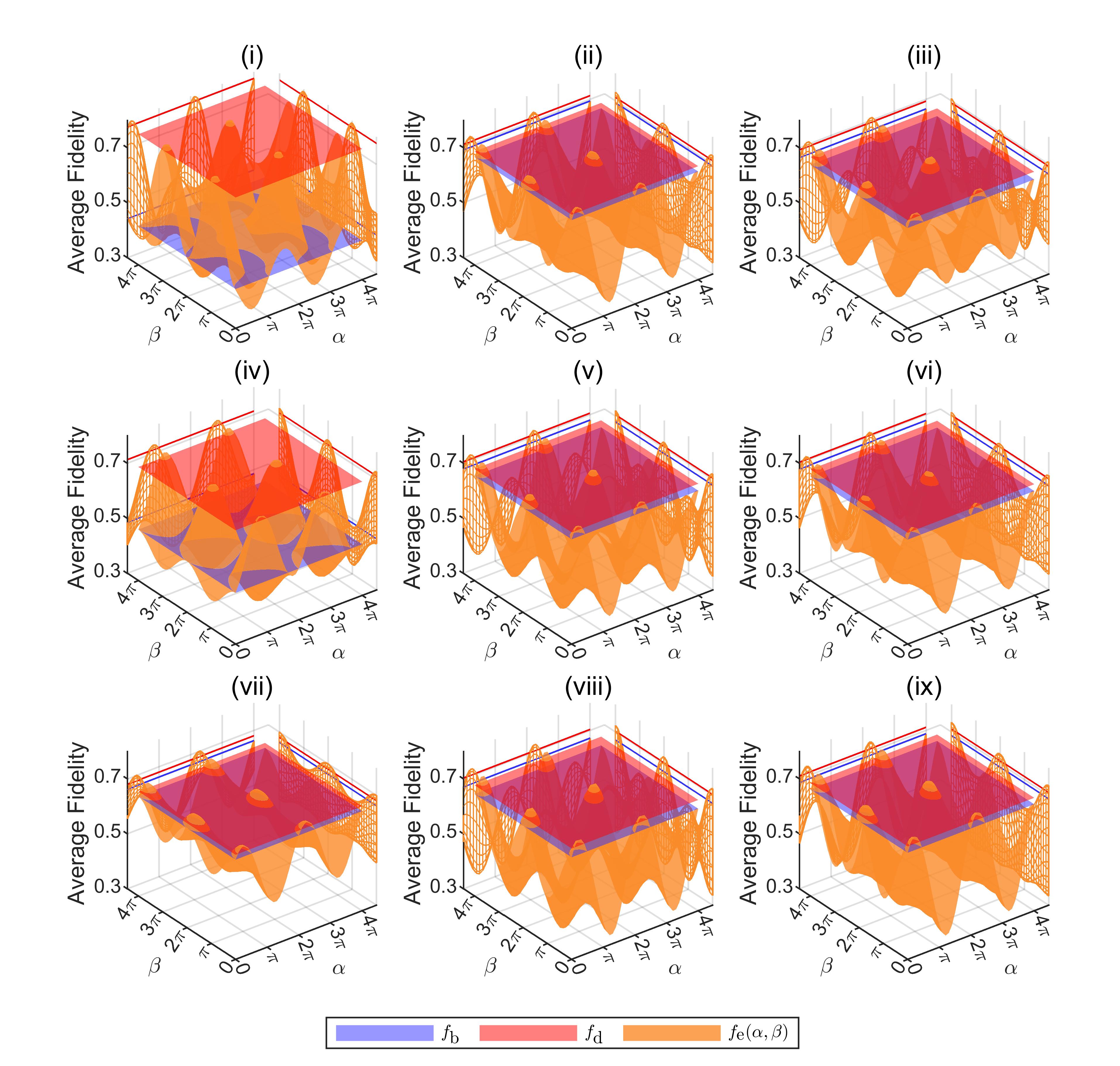}
    \caption{(Color online) Numerical demonstration of average  fidelity $f_{\text{b}}$ (blue), $f_{\text{d}}$ (red) and $f_{\text{e}}(\alpha, \beta)$ (orange) with respect to noise models (\romannumeral1) - (\romannumeral9) considered in Tab.~\ref{tab:noise-models}. In particular, $f_{\text{b}}$ (see Eq.~\ref{eq:fb}) stands for the performance of noisy teleportation (see Fig.~\ref{fig:qt}(b)), $f_{\text{d}}$ (see Eq.~\ref{eq:fd}) represents the performance of PPT teleportation (see Fig.~\ref{fig:qt}(d)) and $f_{\text{e}}(\alpha, \beta)$ (see Eq.~\ref{eq:fe-ab}) denotes the performance of channel $\mF_{\text{e}}(\alpha, \beta, p, q)$ (see Eq.~\ref{eq:channel-fe}) simulated by pre-processing-assisted teleportation (see Fig.~\ref{fig:qt}(e)).}
    \label{fig:f1}
\end{figure}

\begin{figure}
    \centering
\includegraphics[width=1\textwidth]{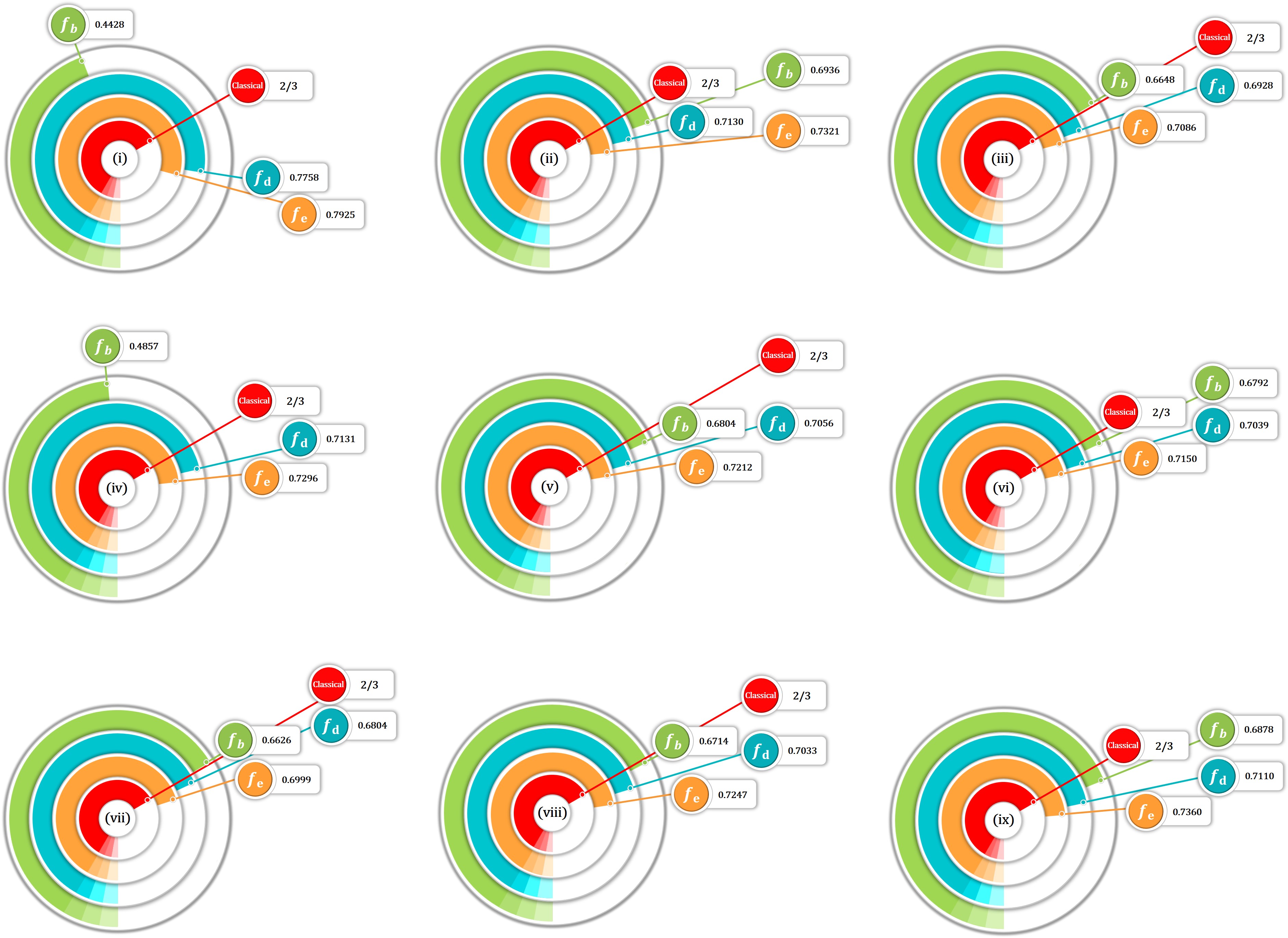}
    \caption{(Color online) Visual comparison between the average fidelity $f_{\text{b}}$, $f_{\text{d}}$, and $f_{\text{e}}$. Each circle in the diagram contains a lowercase Roman numeral at its center, which represents the specific noise model being considered in Tab.~\ref{tab:noise-models}. The red concentric circles in this diagram represent the classical limit of teleportation, while the orange circles indicate the performance of the pre-processing-assisted quantum teleportation protocol (see Fig.~\ref{fig:qt}(e)). This protocol outperforms both the blue circles of the PPT teleportation (see Fig.~\ref{fig:qt}(d)) and the green circles of noisy teleportation (see Fig.~\ref{fig:qt}(b)).}
    \label{fig:f2}
\end{figure}

We present numerical demonstrations of the average fidelity of three different teleportation protocols: $f_{\text{b}}$ (blue), $f_{\text{d}}$ (red), and $f_{\text{e}}(\alpha, \beta)$ (orange), under different noise models (\romannumeral1) - (\romannumeral9) listed in Tab.~\ref{tab:noise-models}. The fidelity values were obtained by using the theoretical results presented in Subsec.~\ref{subsec:FB}. The fidelity $f_{\text{b}}$ represents the performance of noisy teleportation, where the input state is transferred from Alice to Bob via an entangled channel subject to the effects of noise as shown in Fig.~\ref{fig:qt}(b). The fidelity $f_{\text{d}}$ represents the performance of PPT teleportation, where the noisy entangled channel $\mE(p, q)$ is followed by PPT operation, as shown in Fig.~\ref{fig:qt}(d). The fidelity $f_{\text{e}}(\alpha, \beta)$ represents the performance of channel $\mF_{\text{e}}(\alpha, \beta, p, q)$, which is simulated by pre-processing-assisted teleportation, as shown in Fig.~\ref{fig:qt}(e). For each noise model, we calculate the average fidelity of these protocols and illustrate the corresponding results in Fig.~\ref{fig:f1}. On the other hand, Fig.~\ref{fig:f2} provides an alternative comparison between $f_{\text{b}}$, $f_{\text{d}}$, and $f_{\text{e}}$ (not $f_{\text{e}}(\alpha, \beta)$), which demonstrates that protocol e outperforms PPT teleportation and hence all other free-post-processing-assisted teleportation protocols.

\begin{figure}
    \centering
    \includegraphics[width=0.95\textwidth]{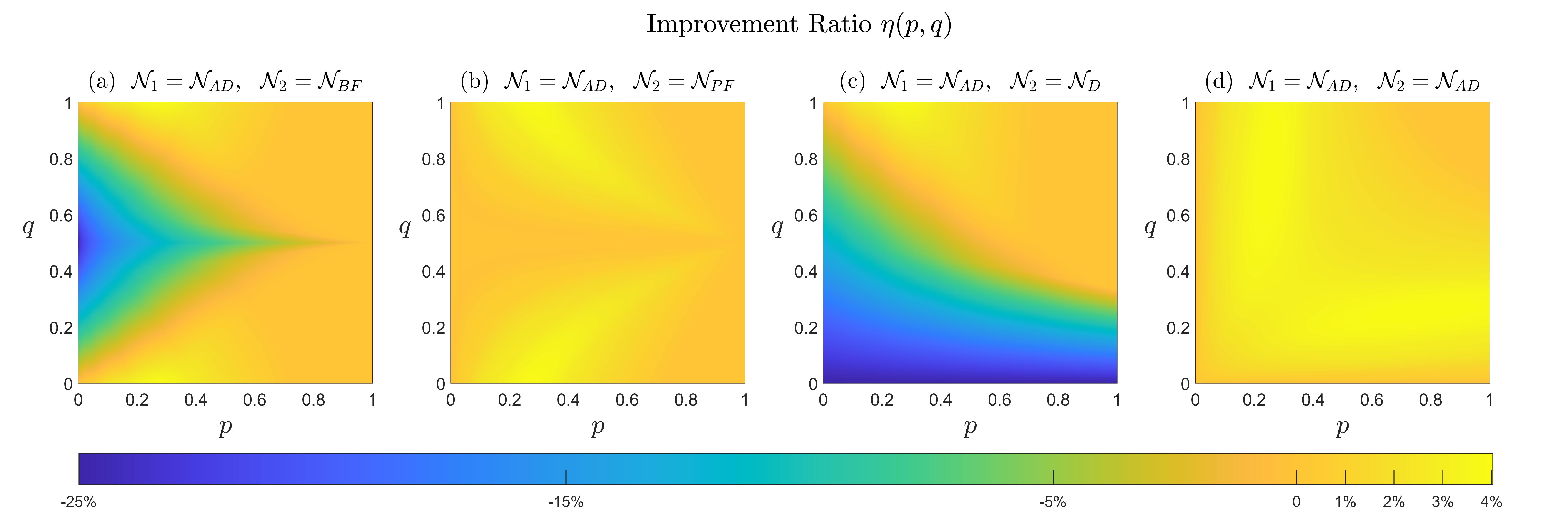}
    \caption{(Color online) Numerical demonstration of the improvement ratio $\eta(p, q)$, as defined in Eq.~\ref{eq:ratio}, across various combinations of noise models. These include the amplitude damping channel $\mN_{\text{AD}}$ and bit flip channel $\mN_{\text{BF}}$, amplitude damping channel $\mN_{\text{AD}}$ and phase flip channel $\mN_{\text{PF}}$, amplitude damping channel $\mN_{\text{AD}}$ and quantum depolarizing channel $\mN_{\text{D}}$, as well as two amplitude damping channels.}
    \label{fig:ratio}
\end{figure}

To assess the advantage of using pre-processing-assisted protocols, we have also calculated the improvement ratio, formally defined as
\begin{align}\label{eq:ratio}
\eta(p, q):= \frac{f_{\text{e}}- f_{\text{d}}}{f_{\text{d}}}.
\end{align}
Here, we have not assumed any specific noise model in our calculation of the improvement ratio, which makes the ratio a function of the parameters $p$ and $q$. We use protocol d as a reference point, as computing the average fidelity for LOCC protocols is generally challenging due to the complex mathematical structure of LOCC. Fig.~\ref{fig:ratio} displays the improvement ratio for various combinations of noise channels. Fig.~\ref{fig:ratio} clearly shows that while pre-processing-assisted teleportation protocols are powerful tools for combating noise, they cannot always outperform post-processing-assisted protocols. In fact, these two types of protocols are complementary to each other. However, in some cases, the effect of pre-processing cannot be achieved even with entangling post-processing, such as PPT operations. These numerical observations highlight the importance of considering protocol f, which combining both pre-processing and post-processing techniques, when designing practical quantum communication protocols.

\begin{remark}
To better illustrate the benefits of pre-processing (i.e., protocol e) in Fig.~\ref{fig:ratio}, we have examined various combinations of noise models. However, to truly observe the advantages of pre-processing, it appears that at least one of the noise channels should be an amplitude damping channel (see Eq.~\ref{eq:EG} or brown boxes in Fig.~\ref{fig:qt}). This observation is based on numerical data, and we currently lack a theoretical explanation for this behavior. Further investigation is needed to gain a deeper understanding of the underlying principles that govern the effectiveness of pre-processing in mitigating the detrimental effects of noise.
\end{remark}

\begin{remark}
When evaluating the performance of protocols d and e, it can be observed that the impact of noise $\mN_1(p)$ and $\mN_2(q)$ is symmetric. Since the SWAP operation is part of both the SEPP and PPT operations (see Eq.~\ref{eq:hi}), the maximum quantity $\max_{\Omega\in\mathrm{PPT}(A:B)}\Tr[\Omega(\rho_{\text{b}})\cdot\phi^+]$ (see Thm.~\ref{thm:protocol-d}) and consequently
the performance of protocol d remains symmetrical even when there is an exchange of $\mN_1(p)$ and $\mN_2(q)$, i.e.,
\begin{align}
    &\max_{\Omega\in\mathrm{PPT}(A:B)}
    \Tr[\Omega\left(
    \mN_1(p)_{A\to A}\otimes\mN_2(q)_{B\to B}\circ \text{C}_{\text{NOT}, AB\to AB}\circ 
    \text{H}_{A\to A}\otimes\id_{B\to B}
    (\ketbra{0}{0}_{A}\otimes\ketbra{0}{0}_{B})
    \right)\cdot\phi^+]\\
    =
    &\max_{\Omega\in\mathrm{PPT}(A:B)}
    \Tr[\Omega\left(
    \mN_2(q)_{A\to A}\otimes\mN_1(p)_{B\to B}\circ \text{C}_{\text{NOT}, AB\to AB}\circ 
    \text{H}_{A\to A}\otimes\id_{B\to B}
    (\ketbra{0}{0}_{A}\otimes\ketbra{0}{0}_{B})
    \right)\cdot\phi^+].
\end{align}
Protocol e exhibits symmetrical behavior in terms of the quantity $\max_{\alpha, \beta}\{\Tr[\rho_{\text{e}}(\alpha, \beta)\cdot\phi^+]\}$ (see Thm.~\ref{thm:protocol-e}), even if the noise sources $\mN_1(p)$ and $\mN_2(q)$ are exchanged, as we have maximized over all possible values of $\alpha$ and $\beta$, and these parameters exhibit symmetry. Hence, the ratio observed when combining noise sources $\mN_1= \mN_{\text{AD}}$ and $\mN_2= \mN_{\text{BF}}$ is symmetric to the ratio observed when switching the noise sources as $\mN_2= \mN_{\text{AD}}$ and $\mN_1= \mN_{\text{BF}}$. Therefore, it is unnecessary to gather data for the case where $\mN_1= \mN_{\text{BF}}$ and $\mN_2= \mN_{\text{AD}}$, as the figure exhibits symmetry with respect to the line $p=q$. Similar symmetrical results can be obtained for other combinations of noise models. This observation simplifies the data collection processes and highlights the fundamental relationships between the different noise models and their impact on the performance of quantum communication protocols.
\end{remark}

\begin{remark}
To generate Fig.~\ref{fig:ratio} with a parameter resolution of $0.001$, we executed our program on a standard server equipped with a CPU (Intel(R) Xeon(R) Platinum 8280*2, 56 cores, 2.70GHz clock), GPU (NVIDIA Quadro RTX 5000*4, 3072-core, 16GB memory), and RAM (1TB memory). The program was run in $84$ concurrent processes, taking approximately $471.54$ hours in total. Each sub-figure took $118.18$ hours (Fig.~\ref{fig:ratio}(a)), $115.99$ hours (Fig.~\ref{fig:ratio}(b)), $118.32$ hours (Fig.~\ref{fig:ratio}(c)), and $119.05$ hours (Fig.~\ref{fig:ratio}(d)), respectively. We calculated the running time using the tic and toc commands in Matlab.
\end{remark}

\begin{remark}
In recent years, there has been growing interest in exploring the use of catalysts to enhance the performance of quantum information tasks~\cite{PhysRevLett.83.3566,PhysRevA.71.042319,PhysRevA.103.022403,PhysRevX.11.011061,PhysRevLett.126.150502,PhysRevLett.127.080502,PhysRevLett.127.150503,PhysRevLett.128.240501,PhysRevLett.129.120506}, including the task of quantum teleportation~\cite{PhysRevLett.127.080502}. While several studies have shown promising results in this area, it's important to note that there are some practical limitations to consider. Specifically, the advantages gained from using catalysts may not always be applicable or feasible in real-world scenarios: (\romannumeral1) Firstly, the catalytic protocols usually utilize an infinite-dimensional catalyst. (\romannumeral2) Secondly, it should be noted that only the marginal of the system corresponds to the original catalyst. Let's direct our attention to catalytic teleportation to delve into further details. Suppose an initial system at state $\rho_{AB}$ is shared between Alice and Bob, which has been utilized to establish a teleportation channel (see Fig.~\ref{fig:channel-S}). In order to enhance the efficiency of teleportation, an additional system $\omega_{A^{'} B^{'}}$ shared between Alice and Bob has been introduced, referred to as the catalyst. By applying an LOCC transformation $\Theta^{\mathrm{post}}: ABA^{'} B^{'}\to ABA^{'} B^{'}$, the final state is obtained, with the marginals $\sigma_{AB}$ and $\omega_{A^{'} B^{'}}$. 
\begin{align}
    \Tr_{A^{'} B^{'}}\Theta^{\mathrm{post}}(\rho\otimes\omega)
    &=
    \sigma,\label{eq:cata-s}\\
    \Tr_{AB}\Theta^{\text{post}}(\rho\otimes\omega)
    &=
    \omega.\label{eq:cata-c}
\end{align}
Notably, the fidelity of $\sigma_{AB}$ in teleporting a state is higher compared to the initial state $\rho_{AB}$. However, the dimension of $\omega_{A^{'} B^{'}}$ is infinite in existing protocols. Importantly, as per Eqs.~\ref{eq:cata-s} and \ref{eq:cata-c}, tracing out systems $A^{'} B^{'}$ is necessary to obtain $\sigma$ and achieve better teleportation performance, unless 
\begin{align}
    \Theta^{\text{post}}(\rho\otimes\omega)= \sigma\otimes\omega,
\end{align}
holds exactly, even if they are arbitrarily close. But, once the catalyst system has been discarded, it is impossible to obtain $\omega$ again, preventing its reuse in subsequent protocols and thus rendering it inconsistent with the definition of a catalyst. Unlike catalytic quantum teleportation, our protocol e achieves improvements without requiring ancillary systems (see Fig.~\ref{fig:qt}(e)), making it more experimentally feasible.
\end{remark}

In summary, teleportation protocols rely on the bipartite channel shared between parties. If we consider only the static state shared between the sender and the receiver as a resource, the use of pre-processing is limited, resulting in the maximal fidelity of $f_{\text{c}}$ achieved by LOCC post-processing. However, viewing the protocol from a channel perspective allows for suitable LOCC pre-processing that can improve teleportation to a fidelity of $f_{\text{f}}$ without consuming additional resources. There are many examples that demonstrate the improvement in performance when pre-processing is implemented in the presence of noises. In comparison to protocols that do not employ pre-processing, these examples, including Figs.~\ref{fig:f1}, \ref{fig:f2}, and \ref{fig:ratio}, show that pre-processing can enhance the quality of information transmission, reduce the effects of noise and decoherence, and increase the overall fidelity of the teleportation protocol. To achieve a thorough understanding of the capabilities and limitations of quantum teleportation, it is imperative to explore protocols from a channel perspective. By leveraging superchannel-assisted protocols, we can reach the ultimate limits in quantum communication. Thus, gaining a deeper understanding of the underlying dynamical resource, namely the bipartite channel, that enables these protocols can provide valuable insights into the fundamental nature of quantum information and its transmission. Such insights can drive the development of novel quantum communication protocols that exploit the full potential of quantum channels, ultimately leading to significant advances in the field of quantum information processing.


\section{\label{sec:QN}Quantum Network}

Quantum communication is a rapidly evolving field that holds great promise for secure and efficient information transfer. In our previous Sec.~\ref{sec:QC}, we demonstrated the crucial role of entanglement-generating channels in enabling higher fidelity quantum teleportation, compared with the conventional teleportation (protocol c of Fig.~\ref{fig:qt}). Building on this foundation, we now extend our framework to the realm of quantum network communication, which requires more complex resource manipulations. In Subsec.~\ref{subsec:QR}, we employ the pre-processing techniques to enhance the performance of quantum repeaters, a key technology for long-distance quantum communication. Our detailed analysis and numerical experiments across various noise models illustrate the practical feasibility and robustness of our approach. Along the way, in Subsec.~\ref{subsec:AIO}, we introduce the concept of \emph{adaptive quantum communication protocol} that allows for multiple rounds of quantum operations and possible quantum memory, providing a more general framework for quantum network communication. Additionally, our result shows that the optimal performance of an adaptive quantum communication protocol is directly correlated with the temporal entanglement it possesses. Our findings demonstrate that temporal entanglement, introduced in Sec.~\ref{sec:TE}, plays a critical role in transmitting quantum information across a quantum network, underscoring the practical importance of this unique quantum property.


\subsection{\label{subsec:QR}Quantum Repeater: Towards a Global Quantum Communication Network}

Quantum communication over long distances poses a fundamental challenge~\cite{PhysRevLett.81.3563}: the likelihood of errors increases as the length of the channel connecting the parties grows. This is especially true when using photonic quantum communication over optical fibers, as both photon absorption and depolarization tend to increase exponentially with the length of the fiber. As a result, longer quantum communications experience significant signal quality degradation, leading to increased error rates and reduced efficiency and reliability. Fortunately, there is a solution to this challenge: the use of quantum repeaters~\cite{PhysRevLett.81.5932}. These repeaters can be built with auxiliary particles at intermediate connection points, enabling the transmission of quantum information over longer distances~\cite{Duan2001,Yuan2008,PhysRevA.79.032325,PhysRevLett.104.180503,RevModPhys.83.33,Munro2012,PhysRevLett.112.250501,7010905,Azuma2015,PhysRevLett.117.210501,Pirandola2017,PhysRevLett.120.030503,Lukin2020,PhysRevX.10.021071,PRXQuantum.1.010301,Guo2020,Duan2021,Liorni_2021,Jiang2021,Sharman2021quantumrepeaters,PhysRevLett.127.220502,Wang2023,PhysRevApplied.19.024070}. By distributing entangled pairs of particles across the channel, quantum repeaters can help to maintain the integrity of the transmitted information and ensure that the communication remains reliable, efficient, and secure, even over long distances.

One of the key technical tools that enables quantum repeater is entanglement swapping~\cite{PhysRevLett.71.4287,PhysRevLett.78.3031,PhysRevLett.80.3891}, which enables long-distance quantum communication. The process involves the transfer of entanglement from two pairs of particles to create a new, entangled pair between two distant locations. This approach allows for the establishment of entanglement across long distances, bypassing the need for direct transmission, which can be subject to significant signal degradation and other sources of interference. By viewing the bipartite channel that generates entanglement between distant locations as a resource, researchers can leverage entanglement swapping as a post-processing step to extract the maximum benefit from this resource. This process is similar to the role of $\Theta^{\text{post}}_{0}$ in standard quantum teleportation (see Subsec.~\ref{subsec:QT}). In this subsection, we will explore the pre-processing-assisted quantum repeater protocol and its ability to significantly enhance the fidelity of quantum communication. Compared to traditional quantum repeater protocols, which rely solely on entanglement swapping, the pre-processing-assisted protocol offers an exponential increase in fidelity.

\begin{figure}[h]
    \centering
    \includegraphics[width=1\textwidth]{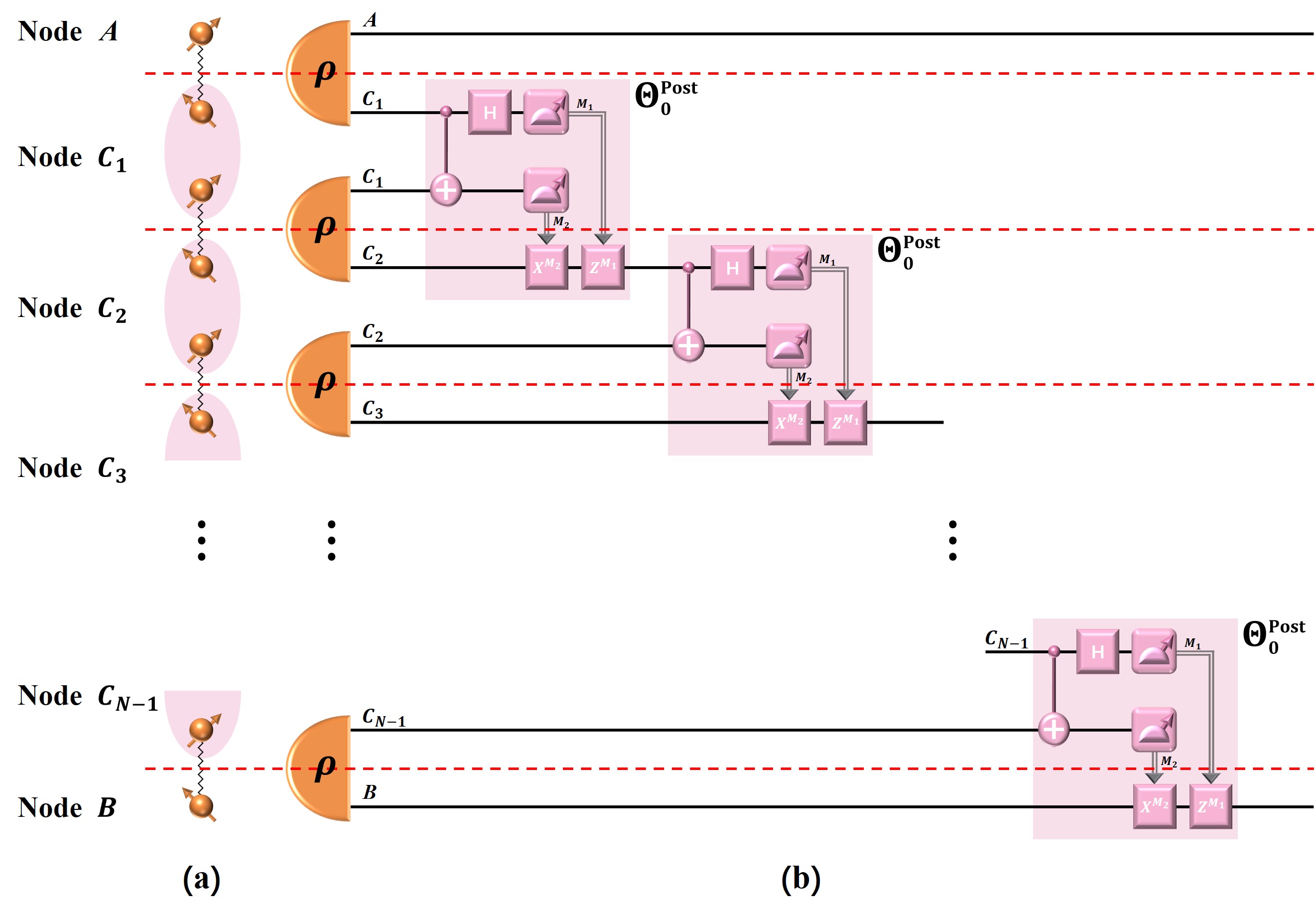}
    \caption{(Color online) Quantum repeaters: establishing an entangled state between system $A$ and $B$ by connecting a sequence of $N$ bipartite states $\rho$ through $N-1$ intermediate nodes from $C_1$ to $C_{N-1}$. Here, we build the connection using standard entanglement swapping technique, whereby Bell measurements are made within the same intermediate node. Figure (a) depicts a schematic diagram of the quantum repeaters. Meanwhile, Figure (b) shows the circuit diagram of the quantum repeaters.}
    \label{fig:repeater}
\end{figure}

To establish a secure communication channel between distant nodes $A$ and $B$, we divide the channel into $N$ segments and create $N$ pairs of qubits, each shared between a connection point and a node. For simplicity, we assume that all $N$ pairs are noisy singlets (see Eq.~\ref{eq:noisy_singlet}) with the same fidelity $F_1$. Using entanglement swapping, we perform Bell measurements on the shared qubits at the intermediate nodes $C_1$ through $C_{N-1}$ and transmit the measurement outcomes between the nodes to effectively teleport the quantum state of the original qubit from node $A$ to node $B$, as illustrated in Fig.~\ref{fig:repeater}. To be more precise, we can assume that the noisy singlet is expressed in the form of Eq.~\ref{eq:noisy_singlet}. Therefore, the relationship between the noisy parameter $p$ and the entanglement fidelity $F_1$ can be characterized by the following formula.
\begin{align}\label{eq:p-to-F}
F_1=\frac{p(d^2-1)+1}{d^2}.
\end{align}
By connecting $N$ neighboring pairs through entanglement swapping, as demonstrated in Fig.~\ref{fig:repeater}, a new pair is obtained with fidelity $F_{N}$. The relationship between $F_{N}$ and $F_1$ is captured by the following lemma.

\begin{mylem}
{Concentrating Noisy Singlets Through Entanglement Swapping}{EF-repeater}
    Using entanglement swapping, as demonstrated in Fig.~\ref{fig:repeater}, $N$ pairs of noisy singlets (see Eq.~\ref{eq:noisy_singlet}) with fidelity $F_1$ can be connected to obtain an entangled state with fidelity $F_N$. The fidelity of this state can be expressed as follows
    \begin{equation}\label{eq:EF-repeater}
    F_N=\frac{1}{4}+\frac{3}{4}\left(\frac{4F_1-1}{3}\right)^N,
    \end{equation}
    which provides a quantitative understanding of the relationship between the fidelity of the initial pairs and the resulting fidelity after connecting them.
\end{mylem}

\begin{remark}
    Some readers may consider Eq.~\ref{eq:EF-repeater} to be a special case of Eq.~$5$ from Ref.~\cite{PhysRevLett.81.5932}, where both the elementary pairs and all operations involved in the connection process are perfect. However, this is not the case. Ref.~\cite{PhysRevLett.81.5932} uses the Werner state (see Ref.~\cite{PhysRevA.40.4277}) as their elementary pair, which is different from the noisy singlet (see Ref.~\cite{PhysRevA.59.4206}) that we investigate in this work. Furthermore, the proof for Eq.~$5$ in Ref.~\cite{PhysRevLett.81.5932} is missing, making our Lem.~\ref{lem:EF-repeater} a more comprehensive reference for readers interested in quantum repeaters. By providing a detailed proof of Lem.~\ref{lem:EF-repeater}, we offer a clear and rigorous understanding of the relationship between the fidelity of the initial pairs and the resulting fidelity after they are connected using entanglement swapping.
\end{remark}

\begin{proof}
We will utilize mathematical induction to demonstrate the validity of this lemma. Given two noisy singlets $\rho= p\phi^{+}+(1-p)\1\otimes\1/4$ defined over systems $AC_1$ and $C_1C_2$, both characterized by the same noise parameter $0\leqslant p\leqslant1$, we can express the resulting overall state as follows
\begin{align}\label{eq:repeater-2copies-decom}
\rho_{AC_1}\otimes\rho_{C_1C_2}
&=
p^2 \left(\phi^{+}_{AC_1}\otimes\phi^{+}_{C_1C_2}\right)
+
p(1-p) \left( \phi^{+}_{AC_1}\otimes\frac{\1_{C_1}}{2} \otimes\frac{\1_{C_2}}{2} \right)
+
p(1-p) \left( \frac{\1_{A}}{2} \otimes\frac{\1_{C_1}}{2}\otimes\phi^{+}_{C_1C_2} \right)
\notag\\
&\quad\quad\quad\quad\quad\quad\quad\quad\quad\quad\quad\quad\quad\quad\quad\quad\quad\quad\quad\quad\quad\quad\quad\quad
+
(1-p)^2 \left( \frac{\1_{A}}{2} \otimes\frac{\1_{C_1}}{2}\otimes\frac{\1_{C_1}}{2}\otimes\frac{\1_{C_2}}{2} \right).
\end{align}
Fig.~\ref{fig:repeater} demonstrates the use of the notation $C_1$ to represent different systems coexisting at the same node. As the distinction between these systems is clear from the context, there should be no confusion arising from this notation. We can decompose the first term on the right-hand side of Eq.~\ref{eq:repeater-2copies-decom} in the following way
\begin{align}
    \ket{\phi^+}_{AC_1}\otimes\ket{\phi^+}_{C_1C_2}
    =
    \frac{1}{2\sqrt{2}}
    \bigg[
    \ket{0}_{A}
    \left(\ket{\phi_{00}}_{C_1C_1}+\ket{\phi_{10}}_{C_1C_1}\right)
    \ket{0}_{C_2}
    &+
    \ket{0}_{A}
    \left(\ket{\phi_{01}}_{C_1C_1}+\ket{\phi_{11}}_{C_1C_1}\right)
    \ket{1}_{C_2}
    \\
    +\ket{1}_{A}
    \left(\ket{\phi_{01}}_{C_1C_1}-\ket{\phi_{11}}_{C_1C_1}\right)
    \ket{0}_{C_2}
    &+
    \ket{1}_{A}
    \left(\ket{\phi_{00}}_{C_1C_1}-\ket{\phi_{10}}_{C_1C_1}\right)
    \ket{1}_{C_2}
    \bigg],
\end{align}
where $\ket{\phi_{ij}}$ stands for the four Bell states, namely $\ket{\phi_{ij}}:= \1\otimes   X^jZ^i\ket{\phi^+}$, with $i,j \in \{0,1\}$. After performing a Bell measurement and obtaining the outcomes $(i, j)$ and applying the unitary operation $X^jZ^i$ to the $C_2$ system, the resulting state of the composite system $AC_2$ becomes a maximally entangled state $\phi^+_{AC_2   }$. Next, we will consider the second component on the right-hand side of Eq.~\ref{eq:repeater-2copies-decom}, which has the following decomposition
\begin{align}
    \phi^{+}_{AC_1}\otimes\frac{\1_{C_1}}{2} \otimes\frac{\1_{C_2}}{2}
    =\frac{1}{8}
    \bigg[
    &\bigg(\ket{0}_{A}(\ket{\phi_{00}}_{C_1C_1}+\ket{\phi_{10}}_{C_1C_1})
    +\ket{1}_{A}(\ket{\phi_{01}}_{C_1C_1}-\ket{\phi_{11}}_{C_1C_1})\bigg)\\
    &\bigg(\bra{0}_{A}(\bra{\phi_{00}}_{C_1C_1}+\bra{\phi_{10}}_{C_1C_1})
    +\bra{1}_{A}(\bra{\phi_{01}}_{C_1C_1}-\bra{\phi_{11}}_{C_1C_1})\bigg)\\
    +&\bigg(\ket{0}_{A}(\ket{\phi_{01}}_{C_1C_1}+\ket{\phi_{11}}_{C_1C_1})
    +\ket{1}_{A}(\ket{\phi_{00}}_{C_1C_1}-\ket{\phi_{10}}_{C_1C_1})\bigg)\\
    &\bigg(\bra{0}_{A}(\bra{\phi_{01}}_{C_1C_1}+\bra{\phi_{11}}_{C_1C_1})+
    \bra{1}_{A}(\bra{\phi_{00}}_{C_1C_1}-\bra{\phi_{10}}_{C_1C_1})\bigg)
    \bigg]\otimes \frac{\1_{C_2}}{2}.
\end{align}
Regardless of which Bell measurement is performed on node $C_1$, the resulting state that acts on system $AC_2$ is always the same: the maximally mixed state $\1_{A}\otimes\1_{C_2}/4$. Our attention will now turn to the third component on the right-hand side of Eq.~\ref{eq:repeater-2copies-decom}.
\begin{align}
    \frac{\1_{A}}{2} \otimes\frac{\1_{C_1}}{2}\otimes\phi^{+}_{C_1C_2}
    =\frac{\1_{A}}{2}\otimes\frac{1}{8}
    \bigg[
    &\bigg((\ket{\phi_{00}}_{C_1C_1}+\ket{\phi_{10}}_{C_1C_1})\ket{0}_{C_2}
    +(\ket{\phi_{01}}_{C_1C_1}+\ket{\phi_{11}}_{C_1C_1})\ket{1}_{C_2}\bigg)
    \\
    &\bigg((\bra{\phi_{00}}_{C_1C_1}+\bra{\phi_{10}}_{C_1C_1})\bra{0}_{C_2}
    +(\bra{\phi_{01}}_{C_1C_1}+\bra{\phi_{11}}_{C_1C_1})\bra{1}_{C_2}\bigg)\\
    +&\bigg((\ket{\phi_{01}}_{C_1C_1}-\ket{\phi_{11}}_{C_1C_1})\ket{0}_{C_2}
    +(\ket{\phi_{00}}_{C_1C_1}-\ket{\phi_{10}}_{C_1C_1})\ket{1}_{C_2}\bigg)\\
    &\bigg((\bra{\phi_{01}}_{C_1C_1}-\bra{\phi_{11}}_{C_1C_1})\bra{0}_{C_2}
    +(\bra{\phi_{00}}_{C_1C_1}-\bra{\phi_{10}}_{C_1C_1})\bra{1}_{C_2}\bigg)
    \bigg].
\end{align}
Similarly to the situation with the second component, the resulting state that acts on system $AC_2$ is always the maximally mixed state, irrespective of which Bell measurement is implemented on node $C_1$. Likewise, the last component on the right-hand side of Eq.~\ref{eq:repeater-2copies-decom} results in the maximally mixed state, meaning that any Bell measurement performed on this component will yield the same output state on system $AC_2$, namely the maximally mixed state. 

In conclusion, only the first component on the right-hand side of Eq.~\ref{eq:repeater-2copies-decom} can produce a maximally entangled state through entanglement swapping, while all other components will only lead to the maximally mixed state. In full, the resulting state of $\rho_{AC_1}\otimes\rho_{C_1C_2}$ under entanglement swapping is given by
\begin{equation}\label{eq:2copies}
  p^2\phi^+_{AC_2} 
  +
  \left(2p(1-p)+(1-p)^2\right)
  \frac{\1_A}{2}\otimes \frac{\1_{C_2}}{2}
  =p^2\phi^+_{AC_2} +\left(1-p^2\right)\frac{\1_A}{2}\otimes \frac{\1_{C_2}}{2}.
\end{equation}
Moving forward, we can analyze the entanglement fidelity of Eq.~\ref{eq:2copies}. Writing everything out explicitly, we have
\begin{equation}
    F_2=\Tr[
    \left(
    p^2\phi^+_{AC_2} +\left(1-p^2\right)\frac{\1_A}{2}\otimes \frac{\1_{C_2}}{2}
    \right)
    \cdot \phi^+_{AC_2}]=\frac{1}{4}+\frac{3}{4}p^2=\frac{1}{4}+\frac{3}{4}\left(\frac{4F_1-1}{3}\right)^2.
\end{equation}
Here the final equation is obtained directly from Eq.~\ref{eq:p-to-F}, which is a consequence of the relationship between the entanglement fidelity and the noise parameter of the noisy singlet.

To establish a general formula for the resulting state under entanglement swapping, we can perform a mathematical induction. Specifically, let us assume that $N-1$ copies of the noisy singlet will yield a state on system $AC_{N-1}$ in the following form
\begin{align}
    p^{N-1}\phi^+_{AC_{N-1}} 
    +\left(1-p^{N-1}\right)\frac{\1_A}{2}\otimes\frac{\1_{C_{N-1}}}{2}.
\end{align}
Under this assumption, the resulting global state over nodes $A$, $C_{N-1}$, and $B$ turns out to be
\begin{align}
    &\left(p^{N-1}\phi^+_{AC_{N-1}} 
    +\left(1-p^{N-1}\right)\frac{\1_A}{2}\otimes\frac{\1_{C_{N-1}}}{2}
    \right)
    \otimes
    \left(
    p\phi^{+}_{C_{N-1}B}+(1-p)\frac{\1_{C_{N-1}}}{2}\otimes\frac{\1_{B}}{2}
    \right)\\
    =
    &p^N
    \left(\phi^+_{AC_{N-1}}\otimes \phi^+_{C_{N-1}B}\right)
    +p^{N-1}(1-p)
    \left(\phi^+_{AC_{N-1}}\otimes\frac{\1_{C_{N-1}}}{2}\otimes\frac{\1_{B}}{2}\right)+ (1-p^{N-1})p\left(\frac{\1_{A}}{2}\otimes \frac{\1_{C_{N-1}}}{2} \otimes \phi^+_{C_{N-1}B}\right)\label{eq:ncopies1}\\
    &\quad\quad\quad\quad\quad\quad\quad
    \quad\quad\quad\quad\quad\quad\quad
    \quad\quad\quad\quad\quad\quad\quad\quad\quad
    +(1-p^{N-1})(1-p)
    \left(
    \frac{\1_{A}}{2}\otimes\frac{\1_{C_{N-1}}}{2}\otimes
    \frac{\1_{C_{N-1}}}{2}\otimes\frac{\1_{B}}{2}
    \right)\label{eq:ncopies2}.
\end{align}
Based on prior discussions, it is known that entanglement swapping can only generate a single maximally entangled state over system $AB$ from the tensor product of two maximally entangled states, i.e., $\phi^+_{AC_{N-1}}\otimes \phi^+_{C_{N-1}B}$. Any other states present in Eq.~\ref{eq:ncopies1} and Eq.~\ref{eq:ncopies2} would result in maximally mixed states after applying entanglement swapping. Thus, after the entanglement swapping operation, the final state on system $AB$ would be
\begin{equation}\label{eq:ncopies}
  p^N \phi^+_{AB} 
  +
  \left(p^{N-1}(1-p)+ (1-p^{N-1})p+ (1-p^{N-1})(1-p)\right)
  \frac{\1_A}{2}\otimes \frac{\1_{B}}{2}
  =p^N\phi^+_{AB} +\left(1-p^N\right)\frac{\1_A}{2}\otimes \frac{\1_{B}}{2}.
\end{equation}
The process of mathematical induction is now complete. To further our analysis, let us examine the entanglement fidelity of Eq.~\ref{eq:ncopies} by expressing the equation in full detail.
\begin{equation}
    F_N=\Tr[
    \left(
    p^N\phi^+_{AB} +
    \left(1-p^N\right)\frac{\1_A}{2}\otimes \frac{\1_{B}}{2}
    \right)
    \cdot \phi^+_{AB}]=\frac{1}{4}+\frac{3}{4}p^N=\frac{1}{4}+\frac{3}{4}\left(\frac{4F_1-1}{3}\right)^N.
\end{equation}
As previously noted, the final equation is a direct consequence of Eq.~\ref{eq:p-to-F}, which establishes the relationship between the entanglement fidelity and the parameter of noisy singlet. This relationship allows us to derive the expression for the entanglement fidelity presented in the last equation. Having established all necessary arguments and demonstrated the validity of Eq.~\ref{eq:EF-repeater}, we conclude that the proof is complete.
\end{proof}

\begin{figure}[h]
    \centering
    \includegraphics[width=0.85\textwidth]{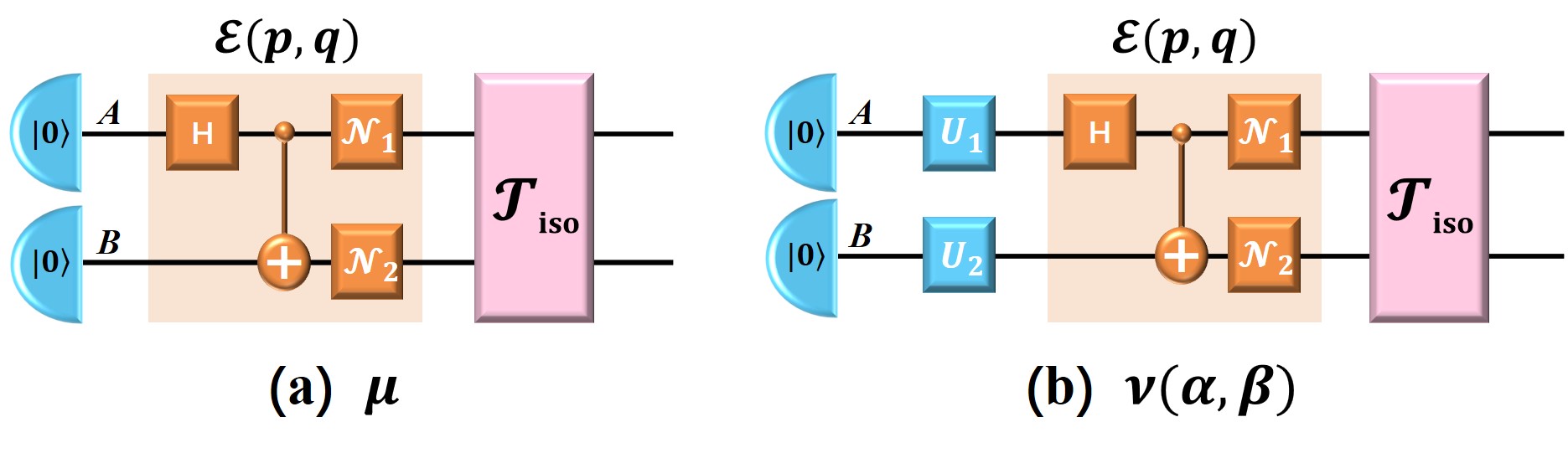}
    \caption{(Color online) Generations of entangled resources. The first one is denoted by $\mu$ (see Eq.~\ref{eq:mu}) and is generated in (a), while the second one is denoted by $\nu(\alpha, \beta)$ (see Eq.~\ref{eq:nu-ab}) and is generated in (b). The entangling gate $\mE(p,q)$ and isotropic twirling $\mT_{\text{iso}}$ are defined in Eq.~\ref{eq:EG} and Eq.~\ref{eq:it} respectively. In (b), the pre-processing step involves two unitary operations, namely $U_1$ and $U_2$, with respective parameters $\alpha$ and $\beta$, as defined in Eq.~\ref{eq:u12}.
    }
    \label{fig:repeater-iso}
\end{figure}

In the previous Lem.~\ref{lem:EF-repeater}, we established a general result for quantum repeaters that employ noisy singlets. We will now focus on a specific preparation of noisy singlet and investigate its performance in more detail. As we discussed in Subsec.~\ref{subsec:QQC} and established through Thm.~\ref{thm:NS}, the output of the isotropic twirling operator $\mT_{\text{iso}}$ always results in a noisy singlet. Leveraging this fact, we can now choose the following state as the initial state to be shared between the nodes in Fig.~\ref{fig:repeater}.
\begin{align}\label{eq:mu}
    \mu:= \mT_{\text{iso}}(\rho_{\text{b}}).
\end{align}
Here, $\rho_{\text{b}}$ is defined in Eq.~\ref{eq:rho-b}, which represents the noisy entangled state without implementing any pre-processing. Due to the property that $F(\mT_{\text{iso}}(\rho))= F(\rho)$ holds for any input state $\rho$, we can conclude that $F(\mu)= F(\rho_{\text{b}})$. Fig.~\ref{fig:repeater-iso}(a) illustrates the process of generating $\mu$. 

Moving forward, let's investigate the initial state with pre-processing, as discussed in protocol e of Subsec.~\ref{subsec:QT}. Mathematically, we can express the new initial state as $\rho_{\text{e}}(\alpha, \beta)$, which is defined in Eq.~\ref{eq:rho-e-ab}. After applying the isotropic twirling operation, we obtain
\begin{align}\label{eq:nu-ab}
    \nu(\alpha, \beta):= \mT_{\text{iso}}(\rho_{\text{e}}(\alpha, \beta)).
\end{align}
The generation of $\nu(\alpha, \beta)$ is depicted in Fig.~\ref{fig:repeater-iso}(b).
In this context, we are concerned with the scenario where the maximization is performed over all parameters $\alpha$ and $\beta$, resulting in the following expression for the entanglement fidelity.
\begin{align}
    F(\nu):= \max\limits_{\alpha, \beta}\Tr[\mT_{\text{iso}}\left(\rho_{\text{e}}(\alpha, \beta)\right)\cdot\phi^+]
    = \max\limits_{\alpha, \beta}\Tr[\rho_{\text{e}}(\alpha, \beta)\cdot\phi^+].
\end{align}
The inclusion of pre-processing can affect the output of bipartite channels, resulting in differences between $\mu$ and $\nu(\alpha, \beta)$. To determine the effectiveness of pre-processing-assisted protocols, we calculated the improvement ratio $\zeta(p, q)$, which represents the percentage increase in performance achieved by using pre-processing.
\begin{align}\label{eq:ratio-network}
    \zeta(p, q):= 
    \frac{\max_{\alpha, \beta} F(\rho_{\text{e}}(\alpha, \beta))- F(\rho_{\text{b}})}{F(\rho_{\text{b}})- \frac{1}{4}}.
\end{align}
The improvement ratio $\zeta(p, q)$ is determined by the characteristics of the noisy entangling channel, namely $\mE(p, q)$ (see Eq.~\ref{eq:EG}). As $p$ and $q$ are the parameters that control the behavior of the bipartite channel, the improvement ratio is inherently dependent on these parameters. Tab.~\ref{tab:ratio-zeta} outlines our analysis of four distinct noise models, where we thoroughly examine their individual impact on the improvement ratio $\zeta(p,q)$.

\begin{table}[h]
    \centering
    \includegraphics[width=0.95\textwidth]{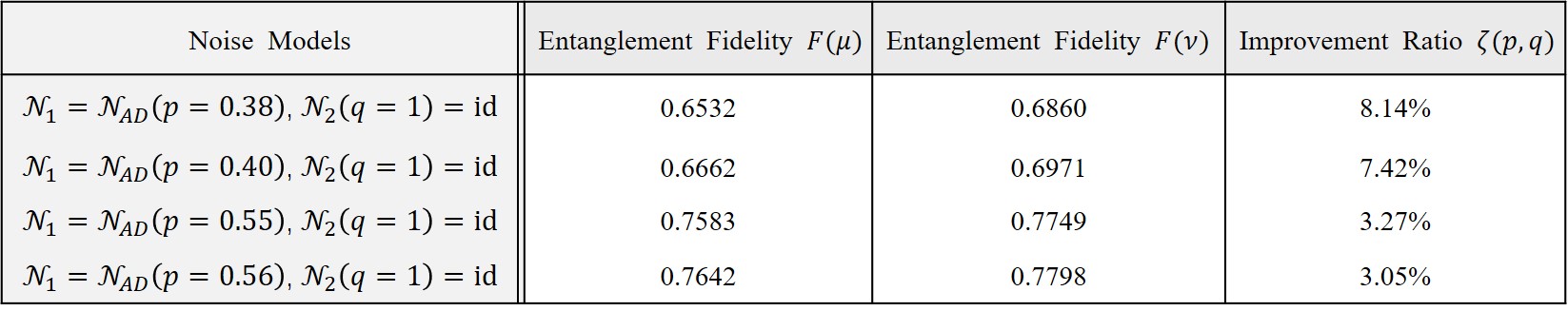}
    \caption{Quantifying the effectiveness of pre-processing techniques: an investigation of improvement ratio $\zeta(p,q)$, defined in Eq.~\ref{eq:ratio-network}, across different noise models. In particular, we analyze the noise models of amplitude damping channel $\mN_1(p)= \mN_{\text{AD}}(p)$ with $p\in\{0.38, 0.40, 0.55, 0.56\}$, while employing the identity channel $\mN_2= \id$ to represent a noiseless evolution of the system (see Fig.~\ref{fig:repeater-iso}). We list the corresponding improvement ratios in the final column.}
    \label{tab:ratio-zeta}
\end{table}

\begin{remark}
    To streamline our discussion, we focus on the noisy singlet with a noise parameter $0<p=\frac{4F_1-1}{3}\leqslant1$ (see Eq.~\ref{eq:p-to-F}). This implies that $F_1$ is strictly greater than $1/4$. As a result, the quantity $\zeta(p,q)$ defined in Eq.~\ref{eq:ratio-network} is guaranteed to be non-negative. In this case, the utilization of pre-processing techniques in the quantum repeater protocol presents a clear advantage in the context of the quantity $\zeta(p,q)$. Specifically, Tab.~\ref{tab:ratio-zeta} shows that the use of pre-processing significantly enhances the performance of the protocol compared to the case without pre-processing, which reflects the effectiveness of the pre-processing-assisted protocol in quantum communication over long distances. These findings underscore the importance of incorporating pre-processing techniques into the design of quantum repeater protocols and highlight the potential for further advancements in the field of quantum communication.
\end{remark}

Our current objective is to build a quantum network using either $\mu$ (see Eq.~\ref{eq:mu}) or $\nu(\alpha, \beta)$ (see Eq.~\ref{eq:nu-ab}) as our entangled resource, and we want to demonstrate the advantages of pre-processing in quantum network communication. To evaluate the performance of these resources, we will measure the fidelity of the resulting state obtained through entanglement swapping, as illustrated in Fig.~\ref{fig:repeater}. Afterward, we will compare their effectiveness in quantum network communication by referring to the following theorem.

\begin{mythm}
{Performance of Quantum Network with and without Pre-Processing}{repeater}
By connecting $N$ adjacent pairs of $\mu$ (as defined in Eq.~\ref{eq:mu}) through entanglement swapping (as depicted in Fig.~\ref{fig:repeater}) in quantum network, we can generate a new pair with fidelity
\begin{align}\label{eq:mu-N}
    \frac{1}{4}+\frac{3}{4}\left(\frac{4F(\mu)-1}{3}\right)^N.
\end{align}
Assuming that the maximization over pre-processing in terms of $U_1(\alpha)\otimes U_2(\beta)$ (as shown in Eq.~\ref{eq:u12}) has been taken into account, the fidelity of the resulting state can be expressed as
\begin{align}\label{eq:nu-N}
    \frac{1}{4}+\frac{3}{4}\left(\frac{4F(\nu)-1}{3}\right)^N
    =
    \frac{1}{4}+\frac{3}{4}\left(1+\zeta(p, q)\right)^N
    \left(\frac{4F(\mu)-1}{3}\right)^N.
\end{align}
Here, the symbol $\zeta(p, q)$, as defined in Eq.~\ref{eq:ratio-network}, represents the improvement ratio achieved by using pre-processing-assisted protocols (see Fig.~\ref{fig:repeater-iso}(b)) compared to standard protocols without pre-processing (see Fig.~\ref{fig:repeater-iso}(a)), measured in terms of the entanglement fidelities obtained.
\end{mythm}

\begin{proof}
    The expressions in Eq.~\ref{eq:mu-N} and the left-hand side of Eq.~\ref{eq:nu-N} are direct consequences of Lem.~\ref{lem:EF-repeater}, which provides a key insight into the performance of the repeater protocol by relating its fidelity to the fidelity of its constituent parts. The proof of the right-hand side of Eq.~\ref{eq:nu-N} is a straightforward application of the definition of the improvement ratio $\zeta(p,q)$ and the property of isotropic twirling $\mT_{\text{iso}}$. Specifically, we can derive the following result by using the invariant properties of isotropic twirling 
    \begin{align}
        F(\mu)&= F(\rho_{\text{b}}),\\
        F(\nu)&= \max\limits_{\alpha, \beta} F(\rho_{\text{e}}(\alpha, \beta)),
    \end{align}
    where $\rho_{\text{b}}$ and $\rho_{\text{e}}(\alpha, \beta)$ are defined in Eq.~\ref{eq:rho-b} and Eq.~\ref{eq:rho-e-ab} respectively. These equations, combined with the definition of the improvement ratio $\zeta(p,q)$, immediately yield the right-hand side of Eq.~\ref{eq:nu-N}. Taken together, these results provide a rigorous mathematical foundation for the performance analysis of the repeater protocol, and demonstrate the power of the improvement ratio concept in characterizing the benefits of using pre-processing-assisted protocols.
\end{proof}

\begin{remark}
    According to the results presented in Thm.~\ref{thm:repeater}, it is evident that the quantum network with pre-processing, which involves generating $\nu(\alpha, \beta)$ at each step, can perform better than the protocol that uses the initial state $\mu$ without pre-processing. Specifically, the resulting entanglement fidelity of the former protocol displays an exponential growth as the number of $N$ increases.
\end{remark}

\begin{figure}
    \centering   \includegraphics[width=1\textwidth]{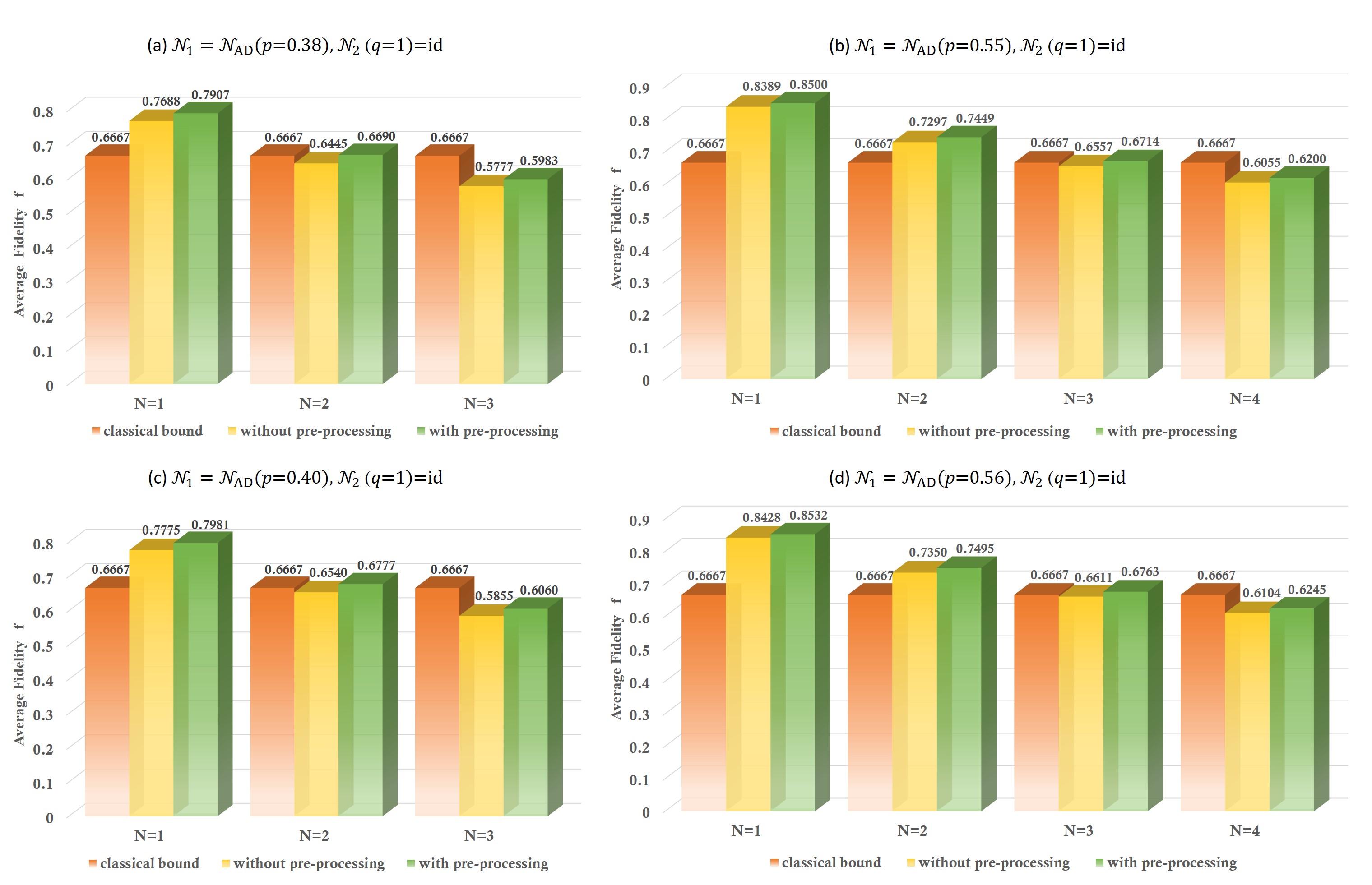}
    \caption{(Color online) Improving quantum network communication with pre-processing techniques. Here, figures (a) and (c) demonstrate a remarkable doubling effect in the communication distance achieved through the use of pre-processing techniques in quantum repeater protocols, shown in Fig.~\ref{fig:repeater}. The comparison between states $\mu= \rho_{\text{iso}}(p)$ (without pre-processing, defined in Eq.~\ref{eq:rho-p}) and $\nu(\alpha, \beta)= \rho_{\text{iso}}(p, \alpha, \beta)$ (with pre-processing, defined in Eq.~\ref{eq:rho-p-ab}) highlights the efficacy of pre-processing techniques in improving the performance of quantum communication systems. On the other hand, for different noise parameters, figures (b) and (d) illustrate that the use of pre-processing techniques extends the connective copies of states in quantum network from $N= 2$ copies to $N= 3$ copies. The generations of states $\rho_{\text{iso}}(p)$ and $\rho_{\text{iso}}(p, \alpha, \beta)$ are depicted in Fig.~\ref{fig:repeater-iso}.
   }
    \label{fig:Network-Twirl-AD}
\end{figure}

Up to this point, our focus has been on examining the theoretical impact of pre-processing in quantum networks that utilize repeaters. Next, we will transition to conducting numerical experiments to further investigate this topic. In the first experiment, we examine the bipartite channel $\mE(p, q)$ (see brown box of Fig.~\ref{fig:repeater-iso}), which only contains a single noisy gate. Specifically, we consider $\mN_1(p)$ as the amplitude damping channel (see Tab.~\ref{tab:kraus-nc}), i.e., $\mN_1(p)= \mN_{\text{AD}}(p)$, while at the same time $\mN_2(q=1)=\id$. In this case, let us denote the state shared between parties as
\begin{align}\label{eq:rho-p}
    \rho_{\text{iso}}(p):= \mT_{\text{iso}, AB\to AB}\circ\mN_{\text{AD}}(p)_{A\to A}\otimes \id_{B\to B}\circ \text{C}_{\text{NOT}, AB\to AB}\circ \text{H}_{A\to A}\otimes\id_{B\to B}
    (\ketbra{0}{0}_{A}\otimes\ketbra{0}{0}_{B}).
\end{align}
By applying the pre-processing $U_1(\alpha)\otimes U_2(\beta)$, as defined in Eq.~\ref{eq:u12}, to the systems, the shared state between the parties becomes
\begin{align}\label{eq:rho-p-ab}
    \rho_{\text{iso}}(p, \alpha, \beta)&:=
    \mT_{\text{iso}, AB\to AB}\circ
    \mN_{\text{AD}}(p)_{A\to A}\otimes \id_{B\to B}\circ \text{C}_{\text{NOT}, AB\to AB}\circ \text{H}_{A\to A}\otimes\id_{B\to B}\notag\\
    &\quad\quad\quad\quad\quad\quad\quad
    \quad\quad\quad\quad\quad\quad\quad
    \quad\quad\quad\quad\quad\quad\quad\quad\quad\quad
    \left(\left(U_1(\alpha)\ketbra{0}{0} U_1^{\dag}(\alpha)\right)_A\otimes
    \left(U_2(\beta)\ketbra{0}{0} U_2^{\dag}(\beta)\right)_B\right).
\end{align}
To fully exploit the potential of the pre-processing technique, we aim to optimize all available values of $\alpha$ and $\beta$ within the state $\rho_{\text{iso}}(p, \alpha, \beta)$.

\begin{figure}[h]
    \centering
    \includegraphics[width=0.61\textwidth]{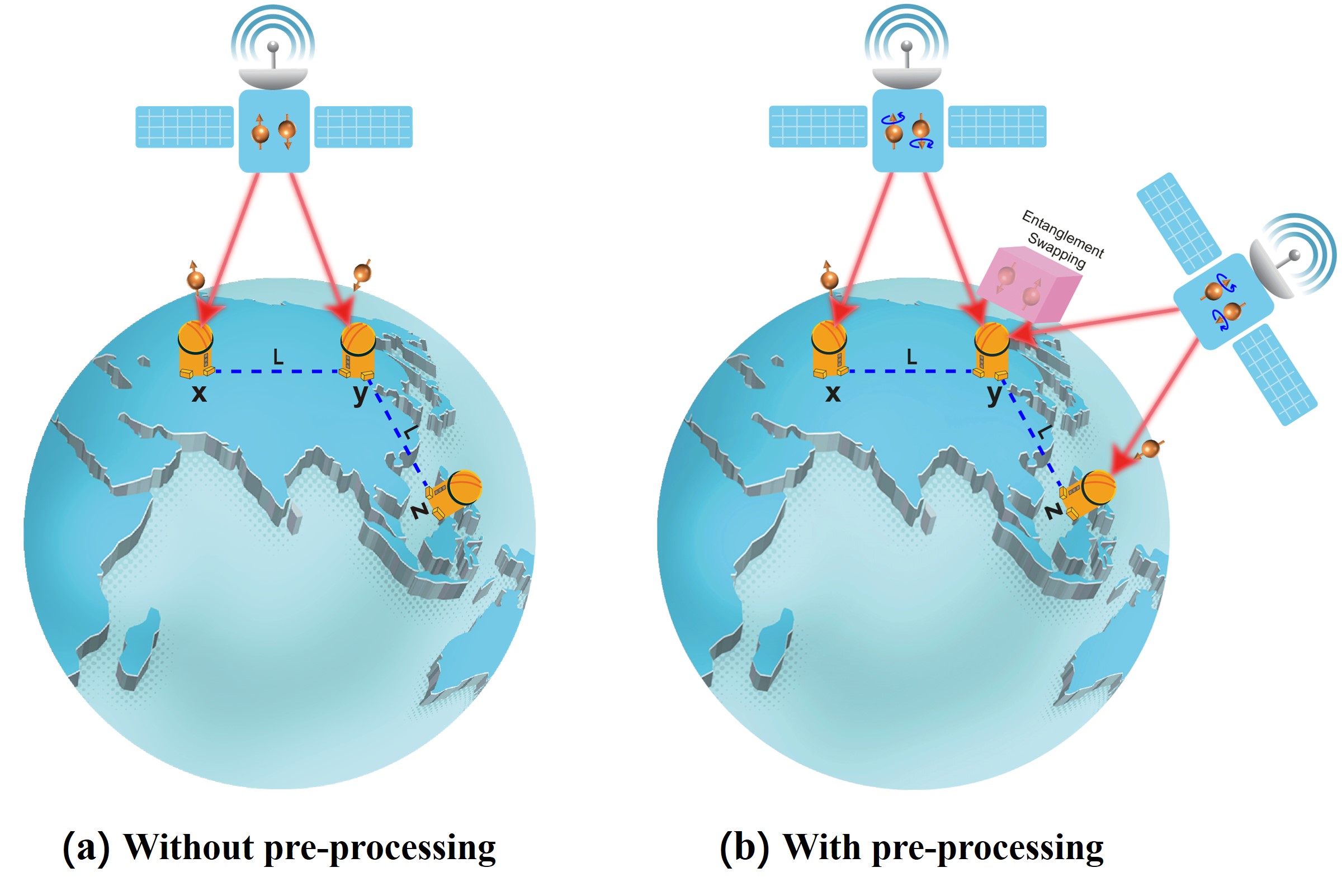}
    \caption{(Color online) Doubling of communication distance by employing pre-processing techniques: Assuming a satellite distributes an entangled state $\rho_{\text{iso}}(0.38)$, as defined in Eq.~\ref{eq:rho-p}, to two ground stations, labeled $x$ and $y$, separated by a distance $L$, quantum communication with a performance superior to classical protocols can be demonstrated, as shown in (a). However, if an additional ground station $z$ exists, located at the same distance $L$ from $y$, information transmission from $x$ to $z$ via entanglement swapping at $y$ cannot surpass the classical limit, as $0.6445<2/3$ (see Fig.~\ref{fig:Network-Twirl-AD}(a)). Nevertheless, through the use of pre-processing techniques, it is possible to beat the classical limit without consuming additional resources, i.e., dynamical entanglement, in this protocol, as illustrated in (b). The phenomenon of doubled quantum communication distance is not limited to the entangled state $\rho_{\text{iso}}(0.38)$, as demonstrated by our results for states $\rho_{\text{iso}}(p)$ with $p=0.4$ (see Fig.\ref{fig:Network-Twirl-AD}(c)) and $\rho(p, \alpha, \beta)$ with $p=0.38$ (see Fig.\ref{fig:Network-AD}(a)) and $p=0.4$ (see Fig.~\ref{fig:Network-AD}(c)).}
    \label{fig:doubling}
\end{figure}

Fig.~\ref{fig:Network-Twirl-AD} presents the results of our first numerical experiment with different parameters $p$, showcasing how the quantum network performs under varying copies in this scenario. In Fig.~\ref{fig:Network-Twirl-AD}(a), we focus on the case, where the noise parameter of $\rho_{\text{iso}}(p)$ is $p=0.38$, and assume that each $\rho_{\text{iso}}(0.38)$ can be used to teleport quantum information with length $L$. Our results demonstrate that without pre-processing, only one copy of $\rho_{\text{iso}}(0.38)$ can beat the classical limit, resulting in a communication distance of $L$. However, with pre-processing, even the connection of two copies, with a communication distance of $2L$, can outperform the classical bound of communication. Fig.~\ref{fig:doubling} portrays a schematic diagram of the doubling of communication distance, which highlights the significant potential of pre-processing techniques for global quantum communication. We also observe a similar doubling effect in the case of $p=0.4$, where the use of pre-processing techniques doubles the communication distance in transmitting quantum information, as shown in Fig.~\ref{fig:Network-Twirl-AD}(c). This finding is particularly noteworthy as it demonstrates the robustness of pre-processing techniques across different noise regimes. In Fig.~\ref{fig:Network-Twirl-AD}(b) and Fig.~\ref{fig:Network-Twirl-AD}(d), we demonstrate that although we cannot double the communication distance, we can extend the connective copies of states in quantum network of Fig.~\ref{fig:repeater} from $2$ copies to $3$ copies. This extension enables the realization of longer quantum communication channels, which is of critical importance in situations where the initial communication distance of each resource shared between parties is limited.

\begin{remark}
Our lemma, denoted as Lem.~\ref{lem:EF-repeater}, establishes a crucial link between the shared noisy singlet states among communication nodes and the overall performance of quantum repeaters. However, the practical application of the commonly employed noisy bipartite channel $\mathcal{E}(p, q)$, including the generations of state $\rho_{\text{b}}$ in Eq.~\ref{eq:rho-b} and state $\rho_{\text{e}}(\alpha, \beta)$ in Eq.~\ref{eq:rho-e-ab}, to the context of quantum repeater protocols is inherently constrained by its inability to generate the desired noisy singlet states. Consequently, a pivotal gap emerges, demanding a novel approach to bridge it. This void is skillfully filled through the ingenious application of isotropic twirling~Def.~\ref{def:IT}, a ubiquitous and efficient tool employed in the domain of entanglement theory. By leveraging isotropic twirling, we accomplish the transformation of any input state into a noisy singlet, as eloquently highlighted by our profound theorem, denoted as Theorem~\ref{thm:NS}. This masterful technique rectifies the previous deficiency while meticulously preserving the delicate degree of entanglement intrinsic to the system. After applying the isotropic twirling $\mT_{\text{iso}}$, we transform the states $\rho_{\text{b}}$ and $\rho_{\text{e}}(\alpha, \beta)$ into the  forms of $\mu:= \mT_{\text{iso}}(\rho_{\text{b}})$ (see Eq.~\ref{eq:mu}) and $\nu(\alpha, \beta):= \mT_{\text{iso}}(\rho_{\text{e}}(\alpha, \beta))$ (see Eq.~\ref{eq:nu-ab}), respectively. The undeniable elegance inherent in the twirling channel empowers us to establish a comprehensive framework for quantifying entanglement between any two nodes within quantum repeaters endowed with distributed entanglement. Moreover, our analysis unearths a profound correlation between entanglement fidelity shared between nodes and the overall performance, enriching our understanding of the intricate interplay between these fundamental concepts.

However, we must confront the formidable challenge posed by the practical implementation of the isotropic twirling $\mT_{\text{iso}}$, as articulated by Eq.~\ref{eq:it}. 
\begin{align}\label{eq:it-cdot}
\mT_{\text{iso}}(\cdot)
:=
\int dU \,
U\otimes\overline{U}\,\cdot\, U^{\dag}\otimes U^{\T}.
\end{align}
This formulation necessitates the intricate computation of an integral over the unitary group, rendering the ideal isotropic twirling an arduous feat requiring an infinite number of unitary operations acting upon the input states. These theoretical reflections inevitably propel us towards a critical inquiry: can twirling channels, imbued with such remarkable potential, be effectively harnessed within practical quantum network scenarios? This question assumes paramount importance given the inevitable constraints imposed by the limited repertoire of operations available within real-world quantum networks. In the ensuing analysis, we embark on a numerical exploration of this concern and reveal a remarkable outcome: even with a finite set of randomly generated unitary operations, we can achieve a high-fidelity approximation of the ideal isotropic twirling. To accomplish this, we introduce the following quantum channel for consideration.
\begin{align}\label{eq:it-app}
\mT_{\text{iso}, \text{app}}(\cdot)
:=
\frac{1}{M}\sum^M_{i=1}
U_i\otimes \overline{U_i}\cdot U^{\dag}_i\otimes U^{\T}_i.
\end{align}
Here $M$ symbolizes the count of randomly generated unitaries taken into account during the analysis. In our numerical analysis, we generate $M$ unitaries using the ``RandomUnitary'' function available in \emph{QETLAB} (Quantum Entanglement Theory LABoratory). Notably, when $M$ approaches infinity, the channel $\mT_{\text{iso}, \text{app}}$ perfectly approximates the ideal isotropic twirling $\mT_{\text{iso}}$. To assess the impact of $\mT_{\text{iso}, \text{app}}$, we employ the Uhlmann fidelity $F_{\text{U}}$ as a metric to quantify the degree of similarity. This is given by
\begin{align}\label{eq:FU_mix}
    F_{\text{U}}(
    \mT_{\text{iso}, \text{app}}(\rho), 
    \mT_{\text{iso}}(\rho)
    )
    =
    \left(\Tr\left[\sqrt{\sqrt{\mT_{\text{iso}, \text{app}}(\rho)}
    \mT_{\text{iso}}(\rho)
    \sqrt{\mT_{\text{iso}, \text{app}}(\rho)}}\right]\right)^2.
\end{align}
In this investigation, we numerically examine two distinct scenarios, namely when $\rho$ corresponds to $\rho_{\text{b}}$ (without pre-processing) or $\rho_{\text{e}}(\alpha, \beta)$ (with pre-processing). Our pivotal query revolves around ascertaining the minimal number of randomly generated unitaries necessary to surpass an exacting performance threshold of $99.5\%$, namely 
\begin{align}
F_{\text{U}}(
\mT_{\text{iso}, \text{app}}(\rho), 
\mT_{\text{iso}}(\rho)
)\geqslant 0.995.
\end{align}
Should it be $M\geqslant 100$, $M\geqslant 1000$, or even $M\geqslant 10000$? In this work, we present compelling evidence that a mere $M\geqslant 20$ random unitaries, employed within Eq.~\ref{eq:it-app}, yield unequivocally satisfactory results for the cases investigated, as illustrated in Fig.~\ref{fig:repeater-twirl}. Hence, the results of our investigation establish the remarkable efficiency and effectiveness of approximating the isotropic twirling $\mathcal{T}_{\text{iso}}$ through the utilization of $\mathcal{T}_{\text{iso}, \text{app}}$.

\begin{figure}[h]
    \centering
    \includegraphics[width=0.8\textwidth]{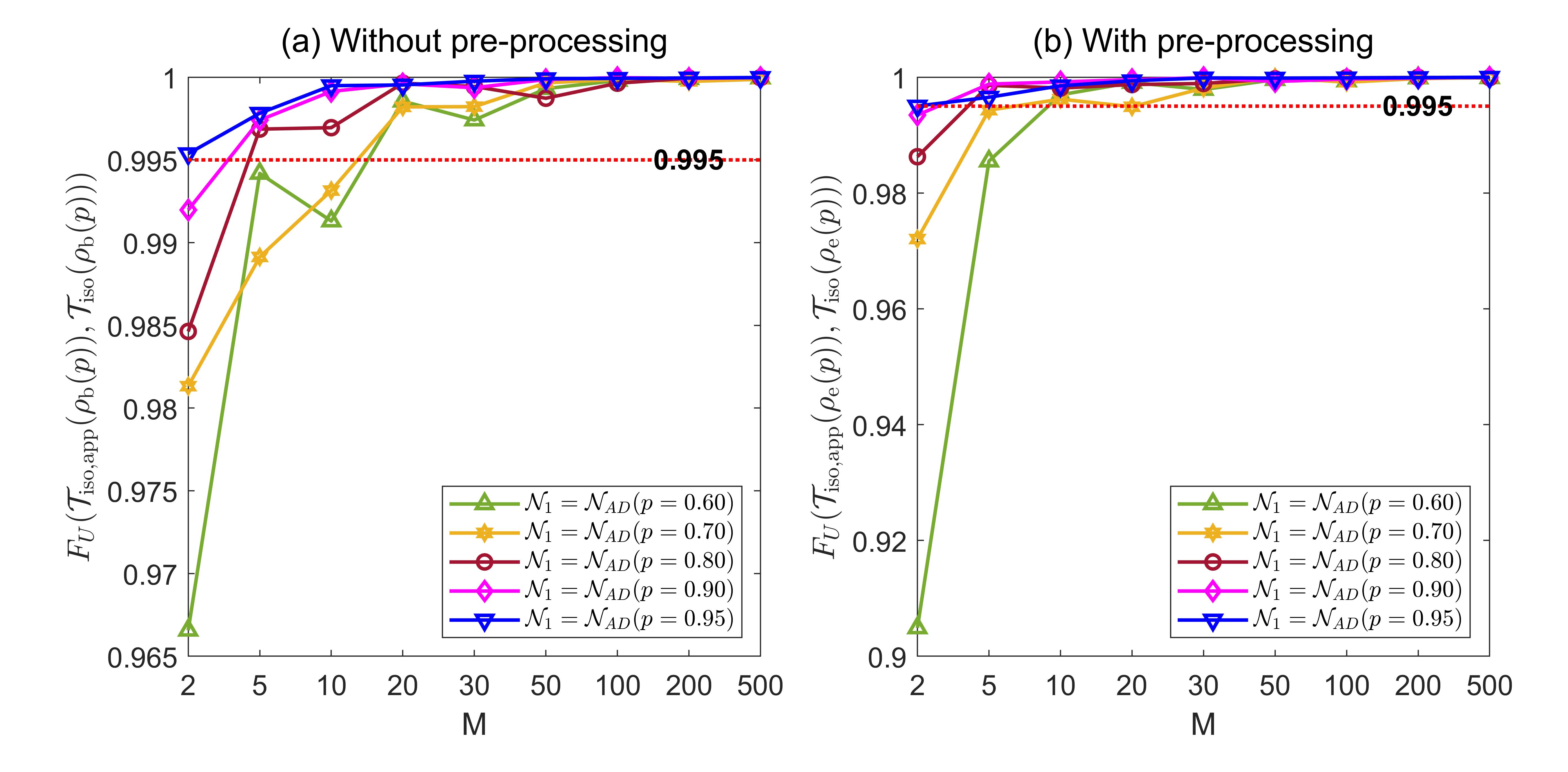}
       \caption{(Color online) Comparison of $F_{\text{U}}(\mathcal{T}{\text{iso}, \text{app}}(\rho), \mathcal{T}{\text{iso}}(\rho))$ (see Eq.~\ref{eq:FU_mix}) with respect to varying numbers of random unitaries $M$ (see Eq.~\ref{eq:it-app}): Figure (a) illustrates the comparison for different noisy parameter $p$ when considering the case $\rho=\rho_{\text{b}}$. Figure (b) explores the analogous scenario for the case $\rho= \rho_{\text{e}}(\alpha, \beta)$. The amplitude damping channel $\mN_1(p)= \mN_{\text{AD}}(p)$ (with $p\in\{0.60, 0.70, 0.80, 0.90, 0.95\}$) is employed, while $\mN_2$ is set as the identity channel $\mN_2= \id$ (see Fig.~\ref{fig:repeater-iso}).} 
    \label{fig:repeater-twirl}
\end{figure} 

Furthermore, the comprehensive analysis depicted in Fig.~\ref{fig:repeater-twirl} unveils a remarkable correlation between the accuracy of the numerical simulations, i.e., $F_{\text{U}}(\mT_{\text{iso}, \text{app}}(\rho), \mT_{\text{iso}}(\rho))$, and the parameter $p$ in the noise models. Notably, as the value of $p$ increases, the simulation accuracy consistently improves, while still allowing for appropriate fluctuations. This observation bears profound significance, as the parameter $p$ directly influences the extent to which the noise channel impacts the entanglement shared between the communication nodes in quantum repeater protocols. By selecting a higher value of $p$, we effectively enhance the entanglement fidelity of the resulting quantum state. Based on this crucial finding, we put forth a compelling conjecture: the entanglement fidelity of the bipartite state shared between the sender and receiver plays a pivotal role in determining the number of random unitary samples necessary to achieve a desired level of simulation accuracy. Intriguingly, our conjecture suggests an inverse relationship between the entanglement fidelity of the bipartite state shared between the sender and receiver and the computational demands of accurate simulations. In essence, when the bipartite state shared between the sender and receiver exhibits higher entanglement fidelity, fewer random matrix samples are required to attain the desired level of simulation accuracy. Ultimately, these findings contribute to a deeper understanding of entanglement dynamics in the presence of noise.
\end{remark}

In our previous discussions, we focused on the cases where the bipartite state shared between parties is subject to an isotropic twirling (colored pink in Fig.~\ref{fig:repeater-iso}), as discussed in Thm.~\ref{thm:repeater} and our first numerical experiment (see Fig.~\ref{fig:Network-Twirl-AD}). Although this assumption has simplified our analysis and provided a general formula for $N$ copies in quantum repeater protocols (see Fig.~\ref{fig:repeater}), it's important to note that the benefits of using pre-processing are not limited to this scenario. We now move on to investigating a different noise model and exploring the advantages of pre-processing. In particular, we direct our attention towards the state acting on the subspace that is spanned by six operators; that are $\ketbra{00}{00}$, $\ketbra{00}{11}$, $\ketbra{11}{00}$, $\ketbra{11}{11}$, $\ketbra{01}{01}$, and $\ketbra{10}{10}$. Our aim is to present a general formula for the entanglement fidelity between two distinct quantum states subjected to this specific noise model. Additionally, we provide numerical evidences that pre-processing techniques confer benefits. To begin with, let us introduce the following lemma.

\begin{mylem}{Concentrating States of Eq.~\ref{eq:6-rho} Through Entanglement Swapping}{6-rho-repeater}
Consider two distinct quantum systems $AC$ and $CB$ and let $\rho_{AC}$ and $\sigma_{CB}$ be two special quantum states that act on these systems, respectively. In particular, we make the assumption that $\rho$ and $\sigma$ have the following mathematical form:
\begin{align}
    \rho_{AC}&=x_{1}\ketbra{00}{00}+\hspace{-0.07em}x_{2}\hspace{0.07em}\ketbra{00}{11}
    +\hspace{-0.07em}x_{3}\hspace{0.07em}\ketbra{11}{00}+\hspace{-0.07em}x_{4}\hspace{0.07em}\ketbra{11}{11}+\hspace{-0.07em}x_{5}\hspace{0.07em}\ketbra{01}{01}
    +\hspace{-0.07em}x_{6}\hspace{0.07em}\ketbra{10}{10},
    \label{eq:6-rho}\\
    \sigma_{CB}&=\hspace{0.05em}y_{1}\hspace{0.045em}\ketbra{00}{00}+y_{2}\hspace{0.08em}\ketbra{00}{11}
    +y_{3}\hspace{0.08em}\ketbra{11}{00}+y_{4}\hspace{0.08em}\ketbra{11}{11}+y_{5}\hspace{0.08em}\ketbra{01}{01}
    +y_{6}\hspace{0.08em}\ketbra{10}{10}.\label{eq:6-sigma}
\end{align}
By performing entanglement swapping (as depicted in Fig.~\ref{fig:repeater}) on system $C$, we can create an entangled state $\tau$ between the nodes $A$ and $B$ that is a composite of the two initial states $\rho$ and $\sigma$. Denote $\tau$ as
\begin{align}\label{eq:6-tau}
    \tau=z_{1}\ketbra{00}{00}+z_{2}\ketbra{00}{11}
    +z_{3}\ketbra{11}{00}+z_{4}\ketbra{11}{11}+z_{5}\ketbra{01}{01}
    +z_{6}\ketbra{10}{10}.
\end{align}
Then, the coefficients connecting these three states, i.e., $\rho$, $\sigma$, and $\tau$, can be determined using the following equations
\begin{align}
    z_1&=x_1(y_1+y_4)+x_5(y_5+y_6),\label{eq:6-tau-z1}\\ 
    z_2&=x_2(y_2+y_3),\\
    z_3&=x_3(y_2+y_3),\\ 
    z_4&=x_4(y_1+y_4)+x_6(y_5+y_6),\\
    z_5&=x_1(y_5+y_6)+x_5(y_1+y_4),\\
    z_6&=x_4(y_5+y_6)+x_6(y_1+y_4).\label{eq:6-tau-z6}
\end{align}
In repeater-based communication protocols, the quantum systems $A$ and $B$ typically correspond to the sender and receiver of quantum information, respectively, while the system $C$ represents an intermediate node that serves as a relay for the quantum information. 
\end{mylem}

\begin{proof}
Entanglement swapping, a critical technique for achieving long-range quantum communication, involves the transfer of entanglement between two physically separated particles without direct interaction between them. As shown in Fig.~\ref{fig:repeater}, the protocol typically comprises two steps: First, a Bell measurement is performed on system $C$ of the joint state $\rho_{AC}\otimes\sigma_{CB}$; Second, a unitary operation is applied to the system $B$ to correct its quantum state, resulting in the creation of an entangled pair $\tau_{AB}$ shared by particles $A$ and $B$.   

To describe these steps mathematically, we consider the following decomposition of the initial state $\rho\otimes\sigma$. 
\begin{align}
    \rho\otimes\sigma
    =
    &x_1\bigg(y_{1}\ketbra{0000}{0000}+y_{2}\ketbra{0000}{0011}+y_{3}\ketbra{0011}{0000}+y_{4}\ketbra{0011}{0011}
    +y_{5}\ketbra{0001}{0001}+y_{6}\ketbra{0010}{0010}\bigg)\\
    +
    &x_2\bigg(y_{1}\ketbra{0000}{1100}+y_{2}\ketbra{0000}{1111}+y_{3}\ketbra{0011}{1100}+y_{4}\ketbra{0011}{1111}
    +y_{5}\ketbra{0001}{1101}+y_{6}\ketbra{0010}{1110}\bigg)\\
    +
    &x_3\bigg(y_{1}\ketbra{1100}{0000}+y_{2}\ketbra{1100}{0011}+y_{3}\ketbra{1111}{0000}+y_{4}\ketbra{1111}{0011}
    +y_{5}\ketbra{1101}{0001}+y_{6}\ketbra{1110}{0010}\bigg)\\
    +
    &x_4\bigg(y_{1}\ketbra{1100}{1100}+y_{2}\ketbra{1100}{1111}+y_{3}\ketbra{1111}{1100}+y_{4}\ketbra{1111}{1111}
    +y_{5}\ketbra{1101}{1101}+y_{6}\ketbra{1110}{1110}\bigg)\\
    +
    &x_5\bigg(y_{1}\ketbra{0100}{0100}+y_{2}\ketbra{0100}{0111}+y_{3}\ketbra{0111}{0100}+y_{4}\ketbra{0111}{0111}
    +y_{5}\ketbra{0101}{0101}+y_{6}\ketbra{0110}{0110}\bigg)\\
    +
    &x_6\bigg(y_{1}\ketbra{1000}{1000}+y_{2}\ketbra{1000}{1011}+y_{3}\ketbra{1011}{1000}+y_{4}\ketbra{1011}{1011}
    +y_{5}\ketbra{1001}{1001}+y_{6}\ketbra{1010}{1010}\bigg).
\end{align}
Here, we safely omit the subscripts since they are self-evident from the context. Upon performing a Bell measurement $\ket{\phi_{ij}}:= \1\otimes X^j Z^i\ket{\phi^+}$ with $i,j \in \{0,1\}$, the joint state transforms into
\begin{align}\label{eq:gen-ij} 
    \frac{
    \left(\1_{A}\otimes \ketbra{\phi_{ij}}{\phi_{ij}}_{CC}\otimes\1_{B}\right)\left(\rho_{AC}\otimes \sigma_{CB}\right)\left(\1_{A}\otimes \ketbra{\phi_{ij}}{\phi_{ij}}_{CC}\otimes \1_{B}\right)
    }
    {p_{ij}}.
\end{align}
In this expression, $p_{ij}:= \Tr[\left(\1_{A}\otimes \ketbra{\phi_{ij}}{\phi_{ij}}_{CC}\otimes\1_{B}\right)\left(\rho_{AC}\otimes \sigma_{CB}\right)\left(\1_{A}\otimes \ketbra{\phi_{ij}}{\phi_{ij}}_{CC}\otimes \1_{B}\right)]$ represents the probability of obtaining classical outcomes $(i, j)$. After tracing out system $C$ from the numerator of Eq.~\ref{eq:gen-ij}, we denote the resulting (unnormalized) state as $\tau_{ij}$, giving us
\begin{align}
    \tau_{00}=
    \frac{1}{2}\bigg[
    &(x_1y_1+x_5y_6)\ketbra{00}{00}+x_2y_2\ketbra{00}{11}+x_3y_3\ketbra{11}{00}+(x_4y_4+x_6y_5)\ketbra{11}{11}
    +(x_1y_5+x_5y_4)\ketbra{01}{01}\notag\\
    &\quad\quad\quad\quad\quad\quad\quad\quad
    \quad\quad\quad\quad\quad\quad\quad\quad
    \quad\quad\quad\quad\quad\quad\quad\quad
    \quad\quad\quad\quad\quad\quad\quad\quad+
    (x_4y_6+x_6y_1)\ketbra{10}{10}\bigg],\label{eq:tau-00}\\
    \tau_{01}=
    \frac{1}{2}\bigg[
    &(x_1y_4+x_5y_5)\ketbra{01}{01}+x_2y_3\ketbra{01}{10}+x_3y_2\ketbra{10}{01}+(x_4y_1+x_6y_6)\ketbra{10}{10} 
    +(x_1y_6+x_5y_1)\ketbra{00}{00}\notag\\
    &\quad\quad\quad\quad\quad\quad\quad\quad
    \quad\quad\quad\quad\quad\quad\quad\quad
    \quad\quad\quad\quad\quad\quad\quad\quad
    \quad\quad\quad\quad\quad\quad\quad\quad+
    (x_4y_5+x_6y_4)\ketbra{11}{11}\bigg],\label{eq:tau-01}\\
    \tau_{10}=
    \frac{1}{2}\bigg[
    &(x_1y_1+x_5y_6)\ketbra{00}{00}-x_2y_2\ketbra{00}{11}-x_3y_3\ketbra{11}{00}+(x_4y_4+x_6y_5)\ketbra{11}{11} 
    +(x_1y_5+x_5y_4)\ketbra{01}{01}\notag\\
    &\quad\quad\quad\quad\quad\quad\quad\quad
    \quad\quad\quad\quad\quad\quad\quad\quad
    \quad\quad\quad\quad\quad\quad\quad\quad
    \quad\quad\quad\quad\quad\quad\quad\quad+
    (x_4y_6+x_6y_1)\ketbra{10}{10}\bigg],\label{eq:tau-10}\\
    \tau_{11}=
    \frac{1}{2}\bigg[
    &(x_1y_4+x_5y_5)\ketbra{01}{01}-x_2y_3\ketbra{01}{10}-x_3y_2\ketbra{10}{01}+(x_4y_1+x_6y_6)\ketbra{10}{10} 
    +(x_1y_6+x_5y_1)\ketbra{00}{00}\notag\\
    &\quad\quad\quad\quad\quad\quad\quad\quad
    \quad\quad\quad\quad\quad\quad\quad\quad
    \quad\quad\quad\quad\quad\quad\quad\quad
    \quad\quad\quad\quad\quad\quad\quad\quad+
    (x_4y_5+x_6y_4)\ketbra{11}{11}\bigg]\label{eq:tau-11}.
\end{align}
Applying the unitary operation $X^jZ^i$ to system $B$ results in the following expression for the entangled state acting on systems $A$ and $B$.
\begin{align}
    (\1\otimes X^j Z^i)
    \cdot\frac{\tau_{ij}}{p_{ij}}\cdot
    (\1\otimes Z^i X^j).
\end{align}
Accordingly, the entangled state $\tau_{AB}$ obtained by performing entanglement swapping becomes
\begin{align}
    \tau
    =&
    \sum_{ij}p_{ij}
    (\1\otimes X^j Z^i)
    \cdot\frac{\tau_{ij}}{p_{ij}}\cdot
    (\1\otimes Z^i X^j)\\
    =&
    \sum_{ij}
    (\1\otimes X^j Z^i)
    \cdot\tau_{ij}\cdot
    (\1\otimes Z^i X^j)\\
    =&
    \left(x_1(y_1+y_4)+x_5(y_5+y_6)\right)
    \ketbra{00}{00}+\left(x_4(y_1+y_4)
    +x_6(y_5+y_6)\right)\ketbra{11}{11}
    +x_2(y_2+y_3)\ketbra{00}{11}\\
    +&    \left(x_1(y_5+y_6)+x_5(y_1+y_4)\right)\ketbra{01}{01}
+\left(x_4(y_5+y_6)+x_6(y_1+y_4)\right)\ketbra{10}{10}+x_3(y_2+y_3)\ketbra{11}{00},
\end{align}
which concludes the proof.
\end{proof}

We will focus specifically on the case where the resourceful state shared between nodes is represented as $\rho_{\text{b}}$ (that is, in Fig.~\ref{fig:repeater}, $\rho$ has been substituted with $\rho_{\text{b}}$), which is defined in Eq.~\ref{eq:rho-b} with $\mN_1(p)$ being the amplitude damping channel (i.e., $\mN_1(p)= \mN_{\text{AD}}(p)$), and where we assume that $\mN_2(q=1)=\id$. Namely, we have eliminated the isotropic twirling in $\mu$ and $\nu(\alpha, \beta)$ (see Fig.~\ref{fig:repeater-iso}). For this scenario, the initial state (see Eq.~\ref{eq:rho-p}) shared between the nodes is given by
\begin{align}
    \rho(p)
    := 
    &\mN_{\text{AD}}(p)_{A\to A}\otimes \id_{B\to B}\circ \text{C}_{\text{NOT}, AB\to AB}\circ \text{H}_{A\to A}\otimes\id_{B\to B}
    (\ketbra{0}{0}_{A}\otimes\ketbra{0}{0}_{B})\\
    =
    &\frac{1}{2}
    \left(
    \ketbra{00}{00}+\sqrt{p}\ketbra{00}{11}+\sqrt{p}\ketbra{11}{00}+p\ketbra{11}{11}+(1-p)\ketbra{01}{01}
    \right).\label{eq:rho-p-AD}
\end{align}
For a single-copy of the initial state $\rho(p)$, the relationship between the noisy parameter $p$ and entanglement fidelity $F_1$, defined as $F$ (i.e., $F_1:= F$), can be accurately characterized by the following formula.
\begin{align}\label{eq:F1_rho-p-AD}
    F_1(\rho(p)):= F(\rho(p)) =\frac{1}{4}(1+\sqrt{p})^2.
\end{align}

Our foremost objective is to construct a quantum network that employs $\rho(p)$ considered in Eq.~\ref{eq:rho-p-AD} as the entangled resource, with the aim of demonstrating the benefits of pre-processing in quantum network communication. To assess the effectiveness of these resources, we will measure the entanglement fidelity $F_N(\rho(p))$ of the resultant state that arises from concentrating $N$ copies of $\rho(p)$ under entanglement swapping, as shown in Fig.~\ref{fig:repeater}. Mathematically, the performance of employing $\rho(p)$ in quantum network communication is characterized by the following theorem.

\begin{mythm}{Performance of Quantum Network with State of Eq.~\ref{eq:rho-p-AD}}{AD-repeater}
By entanglement swapping $N$ adjacent pairs of quantum states $\rho(p)$ (defined in Eq.~\ref{eq:rho-p-AD}) in a quantum network (illustrated in Fig.~\ref{fig:repeater}), a new pair, denoted by $\rho_N(p)$, can be generated in the form of
\begin{align}\label{eq:rho-p-AD-N}
    \rho_N(p)=
    \frac{1}{4}\bigg(
    &(2-p+p^{N})\ketbra{00}{00}
    +2p^{\frac{N}{2}}\ketbra{00}{11}
    +2p^{\frac{N}{2}}\ketbra{11}{00}
    +(p+p^{N})\ketbra{11}{11}
    +(2-p-p^{N})\ketbra{01}{01}\notag\\
    &\quad\quad\quad\quad\quad\quad\quad\quad
    \quad\quad\quad\quad\quad\quad\quad\quad
    \quad\quad\quad\quad\quad\quad\quad\quad
    \quad\quad\quad\quad\quad\quad
    +(p-p^{N})\ketbra{10}{10}\bigg),
\end{align}
whose performance, as measured by its entanglement fidelity, is determined by
\begin{align}\label{eq:FN_rho-p-AD-0}
    F_{N}(\rho(p))
    := F(\rho_N(p))
    =\frac{1}{4}(1+p^{\frac{N}{2}})^2.
\end{align}
Utilizing the relationship established in Eq.~\ref{eq:F1_rho-p-AD}, the expression linking the entanglement fidelity of $\rho_N(p)$ to the initial state $\rho(p)$ shared between parties can be derived as
\begin{align}\label{eq:FN_rho-p-AD}
    F_N(\rho(p))=
    \frac{1}{4}\left(1+\left(2\sqrt{F_1(\rho(p))}-1\right)^N\right)^2.
\end{align}
\end{mythm}

\begin{proof}
By taking $\rho= \sigma= \rho(p)$ in Lem.~\ref{lem:6-rho-repeater}, we can readily obtain $\rho_2(p)$, which is given by 
\begin{align}
    \rho_2(p)
    =
    \frac{1}{4}
    \bigg(
    (2-p+p^{2})\ketbra{00}{00}+2p\ketbra{00}{11}+2p\ketbra{11}{00}
    +(p+p^{2})\ketbra{11}{11}+&(2-p-p^{2})\ketbra{01}{01}\notag\\
    &\quad\quad\quad\quad
    +(p-p^{2})\ketbra{10}{10}
    \bigg).
\end{align}
The following expression for the entanglement fidelity can be obtained by explicitly writing out its form
\begin{align}
    F_2(\rho(p))=\frac{1}{4}(1+p)^2.
\end{align}
We will use mathematical induction to prove the validity of the expression for the general case of $N$. Assuming the validity of Eq.~\ref{eq:rho-p-AD-N} for $N-1$, which states that
\begin{align}\label{eq:rho-p-AD-N-1}
    \rho_{N-1}(p)=
    \frac{1}{4}\bigg(
    &(2-p+p^{N-1})\ketbra{00}{00}
    +2p^{\frac{N-1}{2}}\ketbra{00}{11}
    +2p^{\frac{N-1}{2}}\ketbra{11}{00}
    +(p+p^{N-1})\ketbra{11}{11}\notag\\
    &\quad\quad\quad\quad\quad\quad\quad\quad
    \quad\quad\quad\quad\quad\quad\quad\quad
    \quad\quad\quad
     +(2-p-p^{N-1})\ketbra{01}{01}
    +(p-p^{N-1})\ketbra{10}{10}\bigg).
\end{align}
Applying Lem.~\ref{lem:6-rho-repeater} once again by setting $\rho=\rho_{N-1}(p)$ and $\sigma=\rho(p)$, we can derive the general form of $\rho_{N}(p)$
\begin{align}
    \rho_{N}(p)=
    z_{1}\ketbra{00}{00}+z_{2}\ketbra{00}{11}
    +z_{3}\ketbra{11}{00}+z_{4}\ketbra{11}{11}+z_{5}\ketbra{01}{01}
    +z_{6}\ketbra{10}{10}
\end{align}
with the following coefficients
\begin{align}
    z_1&=\frac{1}{4}(2-p+p^{N-1}), \quad  
    z_2=\frac{1}{2}p^{\frac{N-1}{2}}, \quad 
    \hspace{4.2em}z_3=\frac{1}{2}p^{\frac{N-1}{2}},\\
    z_4&=\frac{1}{4}(p+p^{N-1}),\quad\hspace{1.8em} 
    z_5=\frac{1}{4}(2-p-p^{N-1}),
    \quad z_6=\frac{1}{4}(p-p^{N-1}),
\end{align}
as required. The expressions in Eq.~\ref{eq:FN_rho-p-AD-0} and Eq.~\ref{eq:FN_rho-p-AD} follow straightforwardly from Eq.~\ref{eq:rho-p-AD-N}, thus concluding the proof.
\end{proof}

Turning to the examination of pre-processing effects, it is crucial to emphasize that our objective is not to provide an analytic formula for pre-processing in general. In other words, we are not going to demonstrate the all-encompassing formula of $F_N$ for the state $\rho(p, \alpha, \beta)$ obtained by applying $U_1(\alpha)\otimes 
 U_2(\beta)$, as defined in Eq.~\ref{eq:u12}.
\begin{align}\label{eq:rho-p-ab-AD}
    \rho(p, \alpha, \beta)
    :=
    &\mN_{\text{AD}}(p)_{A\to A}\otimes \id_{B\to B}\circ \text{C}_{\text{NOT}, AB\to AB}\circ \text{H}_{A\to A}\otimes\id_{B\to B}\notag\\
    &\quad\quad\quad\quad\quad\quad\quad
    \quad\quad\quad\quad\quad\quad\quad
    \quad\quad\quad\quad\quad\quad
    \left(\left(U_1(\alpha)\ketbra{0}{0} U_1^{\dag}(\alpha)\right)_A\otimes
    \left(U_2(\beta)\ketbra{0}{0} U_2^{\dag}(\beta)\right)_B\right).
\end{align}
Rather, we center our attention on a particular form of $U_1(\alpha)\otimes U_2(\beta)$ with $\beta=0$, namely $U_1(\alpha)\otimes \1$, that produces relatively straightforward outcomes. This approach proves sufficient for observing the influence of pre-processing and deriving critical insights. In this case, the bipartite state that is shared between the parties is given by 
\begin{align}\label{eq:rho-p-ab-AD-sim}
    \rho(p, \alpha)
    :=
    \rho(p, \alpha, 0)
    =
    \frac{1}{2}
    \bigg(
    a^2\ketbra{00}{00}+ab\sqrt{p}\ketbra{00}{11}+ab\sqrt{p}\ketbra{11}{00}+b^2p\ketbra{11}{11}+b^2(1-p)\ketbra{01}{01}
    \bigg)
    .
\end{align}
Here, the parameters $a$ and $b$ within the coefficients take on the following forms:
\begin{align}
    a:=&\cos{\frac{\alpha}{2}}+\sin{\frac{\alpha}{2}},\\  b:=&\cos{\frac{\alpha}{2}}-\sin{\frac{\alpha}{2}}.
\end{align}

The entanglement fidelity $F_N(\rho(p, \alpha))$ of the resulting state that emerges from concentrating $N$ copies of $\rho(p, \alpha)$ (see Eq.~\ref{eq:rho-p-ab-AD-sim}) via entanglement swapping, depicted in Fig.~\ref{fig:repeater}, is determined by the following theorem.

\begin{mythm}{Performance of Quantum Network with State of Eq.~\ref{eq:rho-p-ab-AD-sim}}{AD-repeater-pre}
Through the process of entanglement swapping, $N$ neighboring pairs of states $\rho(p, \alpha)$ (as defined in Eq.\ref{eq:rho-p-ab-AD-sim}) within a quantum network (depicted in Fig.~\ref{fig:repeater}) can produce a new pair, denoted as $\rho_N(p, \alpha)$, in the following form
\begin{align}
    \rho_N(p, \alpha)=
    \frac{1}{2^N}
    \bigg(
    w_{N, 1}\ketbra{00}{00}+
    w_{N, 2}\ketbra{00}{11}+
    w_{N, 3}\ketbra{11}{00}+
    w_{N, 4}\ketbra{11}{11}+
    w_{N, 5}\ketbra{01}{01}+
    w_{N, 6}\ketbra{10}{10}
    \bigg),
\end{align}
where the coefficients $w_{N, i}$ ($N\geqslant2$, $i\in\{1, \ldots, 6\}$) are described by the following recursive formula
\begin{align}
    w_{N, 1}
    :=&
    (a^2+b^2p) w_{N-1,1} + b^2(1-p) w_{N-1,5},
    \\
    w_{N, 2}
    :=&
    (2ab\sqrt{p}) w_{N-1,2},
    \\
    w_{N, 3}
    :=&
    (2ab\sqrt{p}) w_{N-1,3},
    \\
    w_{N, 4}
    :=&
    (a^2+b^2p) w_{N-1,4} + b^2(1-p) w_{N-1,6},
    \\
    w_{N, 5}
    :=&
    b^2(1-p) w_{N-1,1} + (a^2+b^2p) w_{N-1,5},
    \\
    w_{N, 6}
    :=&
    b^2(1-p) w_{N-1,4} + (a^2+b^2p) w_{N-1,6}.
\end{align}
In the case where $N=1$, we have
\begin{align}
    w_{1, 1}&= a^2,\\
    w_{1, 2}&= ab\sqrt{p},\\
    w_{1, 3}&= ab\sqrt{p},\\
    w_{1, 4}&= b^2p,\\
    w_{1, 5}&= b^2(1-p),\\
    w_{1, 6}&= 0,
\end{align}
which denote the coefficients (up to $1/2$) of $\rho_N(p, \alpha)$. The performance of $\rho_N(p, \alpha)$ in quantum network communication, as evaluated by its entanglement fidelity, is determined using the following equation
\begin{align}
    F_N(\rho(p, \alpha)):= 
    F(\rho_N(p, \alpha))=
    \frac{1}{2^{N+1}}(w_{N, 1}+ w_{N, 2}+ w_{N, 3}+ w_{N,4}).
\end{align}
\end{mythm}

\begin{figure}[h]
    \centering
    \includegraphics[width=1\textwidth]{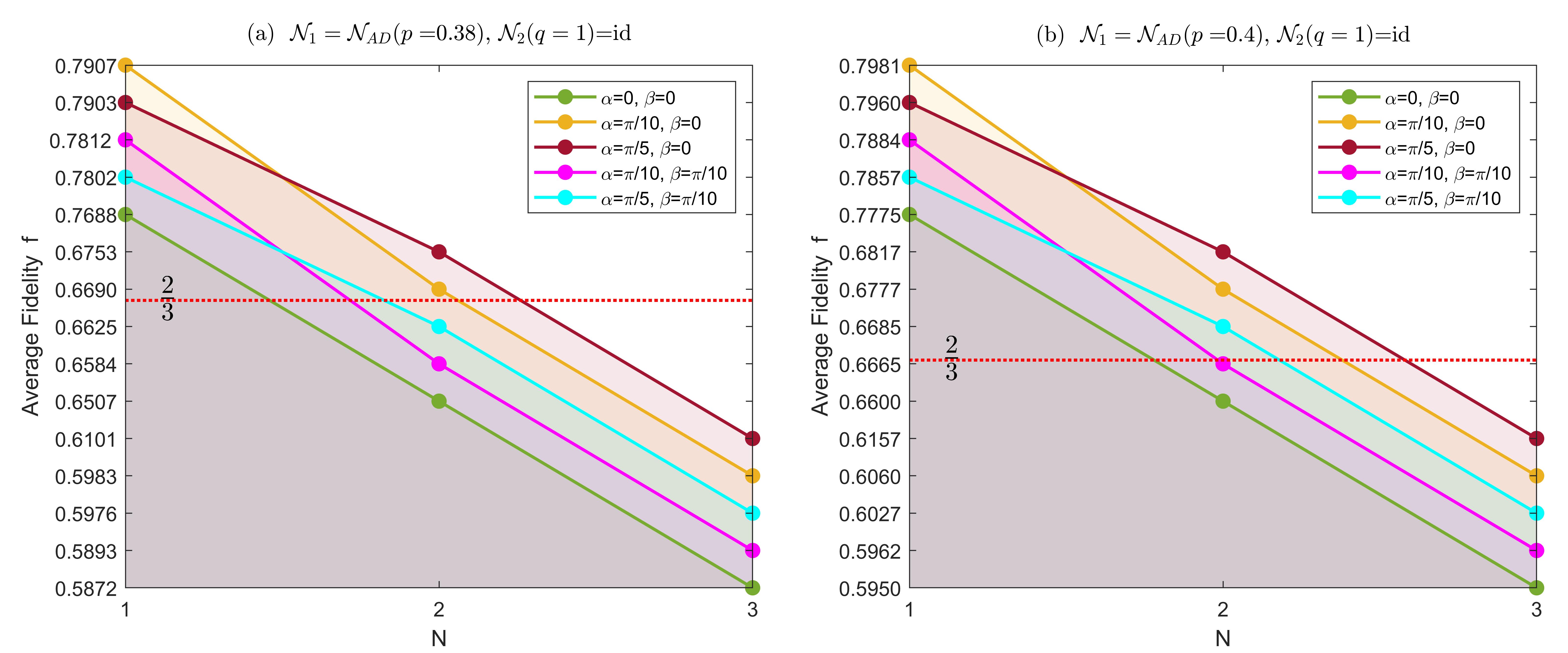}
    \caption{(Color online) Comparison of the average fidelities $f_N$ obtained by connecting $N$ copies of $\rho(p, \alpha, \beta)$, defined in Eq.~\ref{eq:rho-p-ab-AD}, through entanglement swapping with $\mN_1= \mN_{\text{AD}}(p)$ and $\mN_2= \id$. Specifically, figure (a) demonstrates the obtained results for $p=0.38$, while figure (b) illustrates the case of $p=0.40$.}
    \label{fig:Network-AD-Comparison}
\end{figure}

\begin{remark}
Given that the proof of the aforementioned Theorem~\ref{thm:AD-repeater-pre} is an immediate consequence of Lemma~\ref{lem:6-rho-repeater}, a detailed proof is omitted here. The entanglement fidelity $F_N$ of the state $\rho(p, \alpha)$ for $N=1,2,3$ is expressed by the following formulas. 
\begin{align}
    F_1(\rho(p, \alpha))&:= 
    F(\rho(p, \alpha))= 
    \frac{1}{4}(a+b\sqrt{p})^2,\\
    F_2(\rho(p, \alpha))&:= 
    F(\rho_2(p, \alpha))= 
    \frac{1}{8}\left(a^4+b^4+(6a^2b^2-2b^4)p+2b^4p^2\right),\\
    F_3(\rho(p, \alpha))&:= 
    F(\rho_3(p, \alpha))= 
    \frac{1}{16}
    \left(
    a^6+3a^2b^4+(3a^4b^2-6a^2b^4+3b^6)p+8a^3b^3p\sqrt{p}+(6a^2b^4-6b^6)p^2+4b^6p^3
    \right).
\end{align}
These formulas can be used to plot the yellow, currant, and green lines in Fig.~\ref{fig:Network-AD-Comparison}. 
\end{remark}

\begin{remark}
    Even for bipartite qubit systems, calculating the entanglement fidelity $F_N$ of the state obtained by entanglement swapping $N$ adjacent pairs of an arbitrary state $\rho$ in a quantum network, as depicted in Fig.~\ref{fig:repeater}, presents a challenging problem that goes beyond the scope of this work. This calculation requires some techniques and a deep understanding of the underlying structures, and as such, it remains an open problem that warrants further investigation in future researchs.
\end{remark}

\begin{figure}[h]
    \centering   \includegraphics[width=1\textwidth]{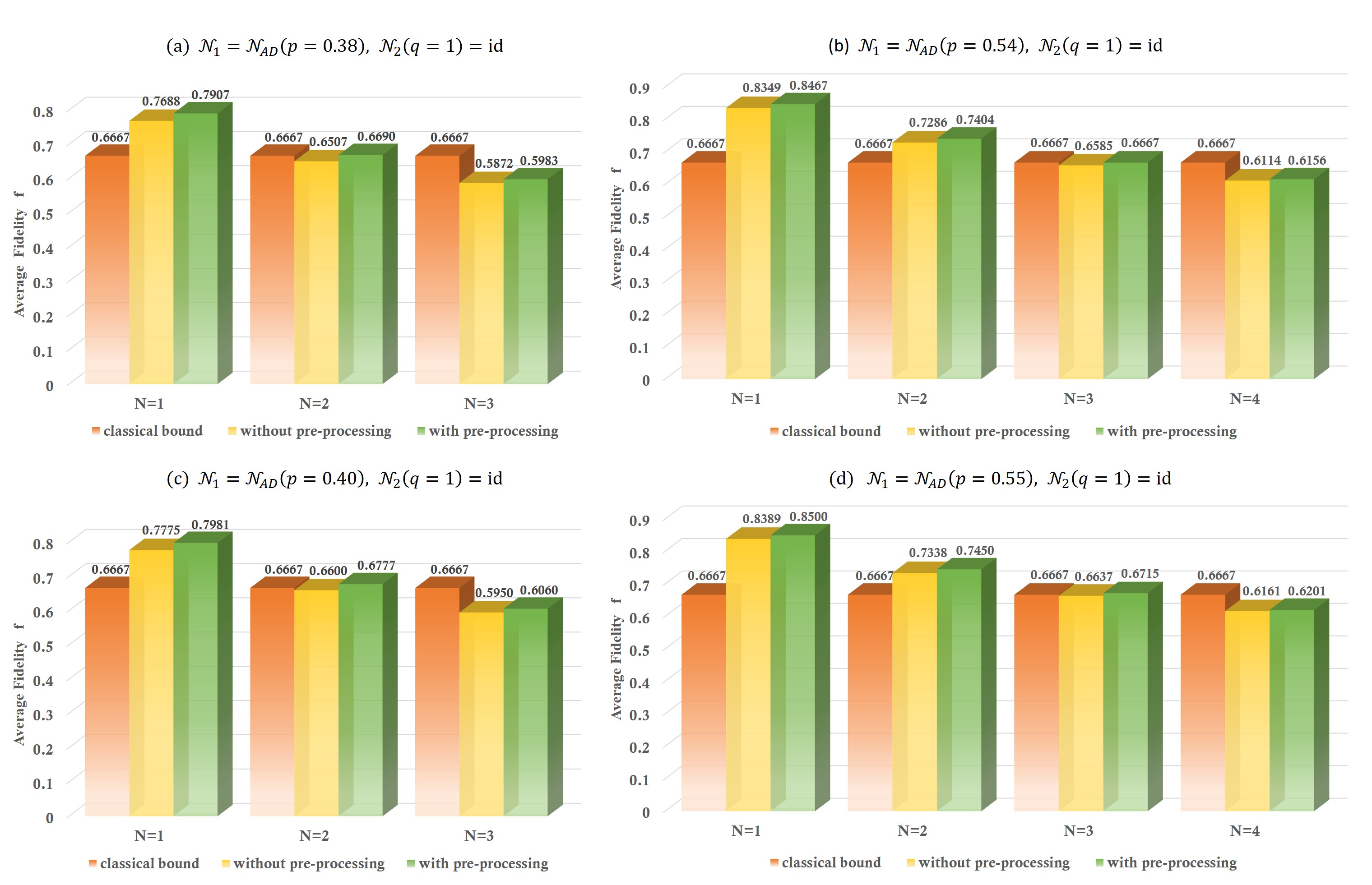}
    \caption{(Color online) Enhancing the efficiency of quantum network communication by employing pre-processing techniques. Specifically, figures (a) and (c) show a doubling effect in the communication distance achieved with pre-processing compared to without pre-processing, as evidenced by the comparison between states $\rho(p)$ (defined in Eq.~\ref{eq:rho-p-AD}) and $\rho(p, \alpha, \beta)$ (defined in Eq.\ref{eq:rho-p-ab-AD}). Moreover, our numerical experiments indicate that pre-processing techniques enable the extension of connective copies of states in quantum networks from $N=2$ to $N=3$, as illustrated in figures (b) and (d) for different noise parameters. 
   }
    \label{fig:Network-AD}
\end{figure}

In addition to theoretical developments, we turn our attention to numerical experiments related to the state $\rho(p, \alpha, \beta)$ (defined in Eq.~\ref{eq:rho-p-ab-AD}) in the context of quantum network communication. Specifically, we perform numerical experiments by scanning over all possible values of $\alpha$ and $\beta$ in the pre-processing of form $U_1(\alpha)\otimes U_2(\beta)$. In Fig.~\ref{fig:Network-AD}(a) and Fig.~\ref{fig:Network-AD}(c), we demonstrate how our pre-processing techniques can double the communication distance when transmitting quantum information, similar to the data investigated in Fig.~\ref{fig:Network-Twirl-AD}. However, in Fig.~\ref{fig:Network-AD}(b) and Fig.~\ref{fig:Network-AD}(d), we show that doubling the communication distance is not always feasible. Nevertheless, the pre-processing techniques can still increase the number of connective copies of states in the quantum network of Fig.~\ref{fig:repeater} from $2$ to $3$.

Through these numerical experiments (see Fig.~\ref{fig:Network-Twirl-AD} and Fig.~\ref{fig:Network-AD}), our findings have exhibited the capability of pre-processing-assisted quantum network communication to increase the number of connective copies of states, thereby enhancing the distance of communication. Building on this, we now move on to proposing a comprehensive investigation of the efficacy of pre-processing techniques. Specifically, we will explore the impact of various parametric values of pre-processing, namely $\alpha$ and $\beta$, on the performance of quantum network communication for the noise model considered in Fig.~\ref{fig:Network-AD}(c) -- $\mN_1= \mN_{\text{AD}}(p= 0.4)$ and $\mN_2= \id$. Fig.~\ref{fig:Network-AD-Comparison} showcases a detailed comparison between various parametric values of $U_1(\alpha)\otimes U_2(\beta)$ in pre-processing, shedding light on their respective impacts on the performance of the system.

The quantum network communication model considered in Lem.~\ref{lem:EF-repeater} tells us that higher entanglement fidelity of the initial state shared between parties, denoted by $F_1$, leads to better performance in network communication when connecting $N$ copies. Mathematically, it is equivalent to say that if we have two noisy singlets $\rho$ and $\sigma$ satisfying
\begin{align}\label{eq:mono-F1}
    F_1(\rho):= F(\rho)\geqslant F_1(\sigma):= F(\sigma),
\end{align}
then the entanglement fidelities obtained by connecting $N$ copies of $\rho$ and $\sigma$ through entanglement swapping (see Fig.~\ref{fig:repeater}) meet the following inequality
\begin{align}\label{eq:mono-FN}
    F_N(\rho):= F(\rho_N)\geqslant F_N(\sigma):= F(\sigma_N),
\end{align}
due to the result of Eq.~\ref{eq:EF-repeater}. Here, $\rho_N$ denotes the entangled state resulting from the tensor product of $N$ identical copies of $\rho$, followed by entanglement swapping. A similar definition holds for $\sigma_N$. Subsec.~\ref{subsec:FB} clarifies that the entangled states $\rho_N$ and $\sigma_N$ can be employed to construct teleportation channels, namely $\mF(\rho_N)$ and $\mF(\sigma_N)$. The process of constructing these channels is depicted in Figure~\ref{fig:channel-rho}, and their mathematical definition is given by Equation~\ref{eq:channel-rho}. Thanks to Lem.~\ref{lem:AFEF} and Lem.~\ref{lem:EFUF}, we have 
\begin{align}
    f_N(\rho):= f(\mF(\rho_N))= \frac{d F_N(\rho)+ 1}{d+1},
\end{align}
where $f_N(\rho)$ denotes the average fidelity of teleportation channel $\mF(\rho_N)$ obtained by connecting $N$ copies of bipartite state $\rho$ through entanglement swapping and implementing $\Theta^{\text{post}}_{0}$ on the message, sender, and receiver system. Then the monotonicity property of $F_N$ suggests that there is a direct correlation between the quality of the shared entanglement and the efficiency of quantum communication. Expressing the results in terms of average fidelity, Eq.~\ref{eq:mono-FN} indicates that
\begin{align}\label{eq:mono-fN}
    f_N(\rho)\geqslant f_N(\sigma).
\end{align}

It is natural to inquire whether the monotonicity property of entanglement fidelity holds for general quantum states beyond the noisy singlet? In other words, can we expect that higher entanglement fidelity of the initial state shared between parties always leads to better performance in quantum network communication when connecting multiple copies, regardless of the specific structure of quantum state in question? This question is of great interest and importance for the field of quantum communication, as it has significant implications for the development and optimization of repeater-based quantum network communication protocols. Based on the data presented in Fig.~\ref{fig:Network-AD-Comparison}, we can confidently conclude that the answer to the aforementioned question is simply negative! For state $\rho(p, \alpha, \beta)$ (see Eq.~\ref{eq:rho-p-ab-AD}) with $\mN_1= \mN_{\text{AD}}(p= 0.38)$ and $\mN_2= \id$, we have 
\begin{align}\label{eq:0.4->}
    f_1(\rho(0.38, \frac{\pi}{10}, 0))
    \approx
    0.7907
    >
    f_1(\rho(0.38, \frac{\pi}{5}, 0))
    \approx
    0.7903
    .
\end{align}
Upon connecting two copies of the quantum states, we observe a reverse direction of Eq.~\ref{eq:0.4->}. Specifically, we find that the entanglement fidelity of the resulting state after connecting two copies in one case can be lower than the entanglement fidelity of the other case, even though the opposite was true for the initial states; that is
\begin{align}\label{eq:0.4-<}
    f_2(\rho(0.38, \frac{\pi}{10}, 0))
    \approx
    0.6690
    <
    f_2(\rho(0.38, \frac{\pi}{5}, 0))
    \approx
    0.6753
    .
\end{align}
In Fig.~\ref{fig:Network-AD-Comparison}, the data related to states with $(\alpha, \beta)= (\frac{\pi}{10}, 0)$ and $(\alpha, \beta)= (\frac{\pi}{5}, 0)$ are represented by yellow and currant lines, respectively. The comparison of Eq.~\ref{eq:0.4->} and Eq.~\ref{eq:0.4-<} reveals that the relationship between a single copy and the multiple copies of quantum states in quantum network communication is complex and non-monotonic, which poses a significant challenge to the development of quantum communication technologies. This observation underscores the urgent need for further theoretical investigations to establish a comprehensive understanding of the underlying principles governing quantum communication networks. Such efforts will enable the design of more efficient and robust quantum communication protocols and pave the way for the realization of a global quantum network. 

In concluding this subsection, we shall undertake a more comprehensive comparison of the datasets presented in Fig.~\ref{fig:Network-Twirl-AD} and Fig.~\ref{fig:Network-AD} by tabulating their respective attributes in Tab.~\ref{tab:comparison}.
It is worth mentioning that our analysis extends beyond the noise model discussed in Fig.~\ref{fig:Network-Twirl-AD} and Fig.~\ref{fig:Network-AD}, as evidenced by the additional data presented in Tab.~\ref{tab:comparison}. With the addition of this expanded dataset, we can conduct a more thorough and in-depth analysis of the quantum network's performance, offering a more nuanced understanding of its capabilities and limitations under a broader range of noise conditions.

\begin{table}[h]
    \centering
   \includegraphics[width=1\textwidth]{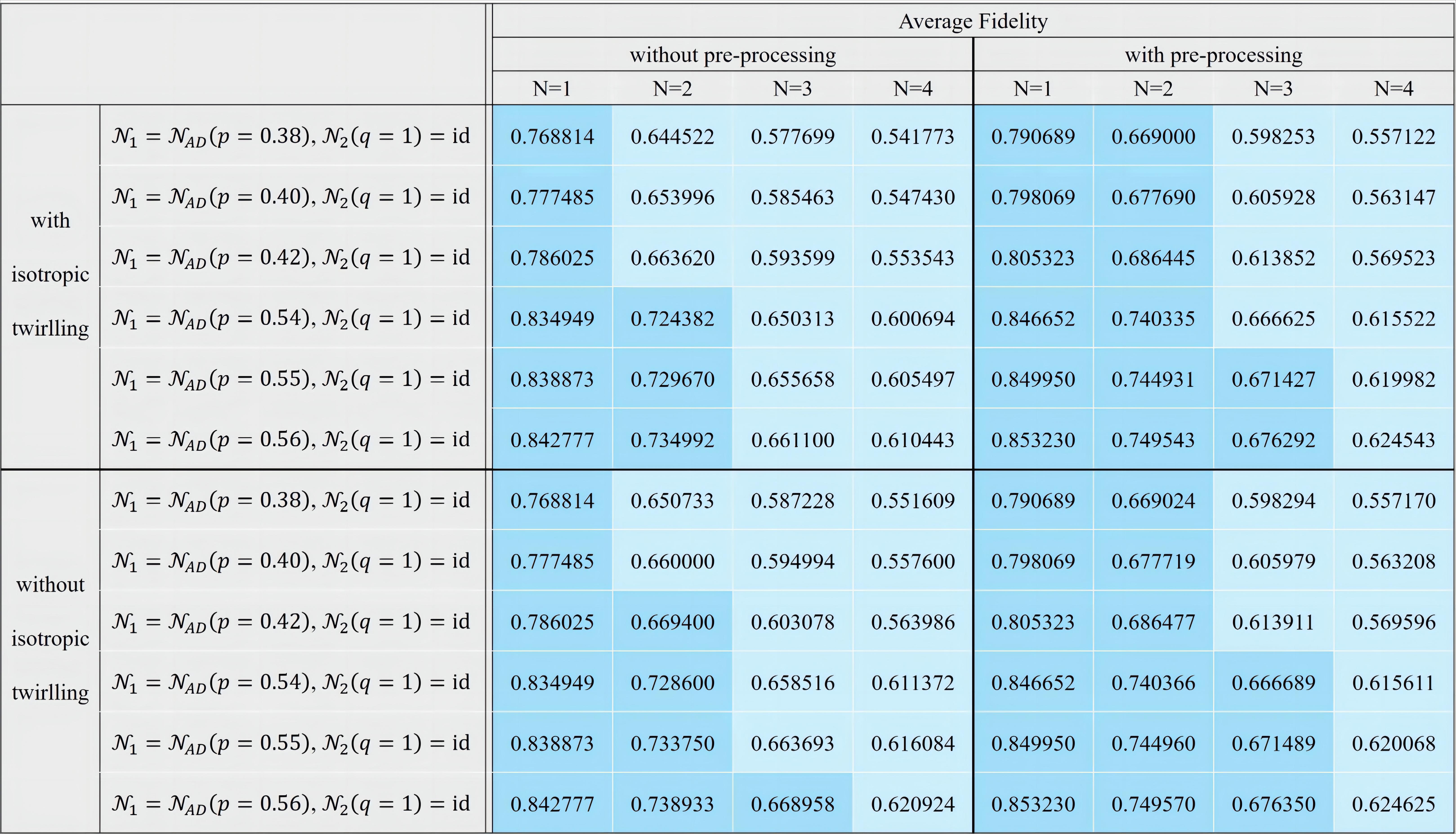}
\caption{Investigating the effectiveness of pre-processing techniques: analyzing the average fidelity $f_N$ of a state obtained through entanglement swapping (see Fig.~\ref{fig:repeater}) under various noise models, with and without pre-processing. More precisely, we analyze the noise models of amplitude damping channel $\mN_1(p)= \mN_{\text{AD}}(p)$ with $p\in\{0.38, 0.40, 0.42, 0.54, 0.55, 0.56\}$, while employing the identity channel $\mN_2= \id$ to represent a noiseless evolution of the system (see Fig.~\ref{fig:repeater-iso}). Here, we use dark blue shading to indicate that the corresponding average fidelity exceeds the classical limit of $2/3$. Conversely, light blue shading denotes cases where the performance falls below this limit.}
\label{tab:comparison}
\end{table}

Notably, the primary variance between the two figures (namely Fig.~\ref{fig:Network-Twirl-AD} and Fig.~\ref{fig:Network-AD}) pertains to the adoption of isotropic twirling, which involves randomly applying a set of unitary operations to a quantum state, in a way that preserves the overall symmetry of the system. As explained in Subsec.~\ref{subsec:QQC}, isotropic twirling $\mT_{\text{iso}}$ does not affect the entanglement fidelity of a state. Therefore, if we have a state $\rho$, its performance and that of $\mT_{\text{iso}}(\rho)$ in teleportation remains unchanged, since for a single copy of
resource shared between the sender and the receiver, we have
\begin{align}
    f_1(\rho):= f(\rho)
    = 
    f_1(\mT_{\text{iso}}(\rho)):= f(\mT_{\text{iso}}(\rho)).
\end{align}
However, when multiple copies of bipartite states are connected through entanglement swapping, isotropic twirling might decrease the fidelity of the resulting entangled state. This is in contrast to the case where single bipartite states are considered. The data listed in Tab.~\ref{tab:comparison} supports the following inequality,
\begin{align}
    f_N(\rho)\geqslant f_N(\mT_{\text{iso}}(\rho)),
\end{align}
demonstrating that the average fidelity of quantum network communication generally decreases as a result of isotropic twirling. Here $f_N(\rho)$ stands for the average fidelity of a state obtained by connecting $N$ copies of a given bipartite state $\rho$ through entanglement swapping (as illustrated in Fig.~\ref{fig:repeater}). Moreover, Tab.~\ref{tab:comparison} reveals an intriguing trend in the performance of quantum network communication. Despite a noticeable decrease in average fidelity, a striking result emerges -- pre-processing-assisted network communication exhibits remarkable noise resilience against isotropic twirling. Quite remarkable, the performance remains almost unaffected whether the bipartite channel is subjected to isotropic twirling or not. Although our numerical investigations provide important insights into the robustness of pre-processing-assisted network communication under isotropic twirling, the generality of these results beyond the specific models considered in our work remains an open question. Further exploration of this issue is necessary to establish the extent to which our findings hold more generally. However, this is a challenging and multi-faceted problem that extends beyond the scope of our present work, and will be the subject of future research.


\subsection{\label{subsec:AIO}All-In-One Communication Framework: That's all you need}

The quantum repeater is a promising approach for building a quantum network that allows for the reliable transfer of unknown quantum states over long distances, as explained in Subsec.~\ref{subsec:QR}. However, the assumption of identical Bell measurements on independent and identically distributed (i.i.d.) quantum states in each protocol round, commonly used in this approach, cannot be justified a priori despite its analytical convenience. In reality, error rates in quantum network communication can fluctuate due to variations in nodes or environmental factors, thus necessitating the introduction of an adaptive communication protocol that can dynamically adjust its behavior in response to such changes. Imperfect devices and transmission channels also introduce noise into quantum networks, making continuous monitoring and adjustment of the network critical for reliable and efficient quantum communication. As in classical wireless communication, adaptive communication protocols~\cite{10.1145/313451.313529,10.1145/1322263.1322295,10.1145/1754414.1754423} optimize communication parameters for successful transmission by accounting for variations in network conditions such as distance, obstacles, and interference. In this subsection, we will introduce the most general framework of adaptive quantum communication protocols, analyse their performance, determine their limitations in terms of average fidelities, and explore the relationship between their communication capability and inherent temporal entanglement (see Sec.~\ref{sec:TE}).

Assuming, without loss of generality, that the adaptive quantum communication protocol is designed to transmit information from the sender, Alice, to the receiver, Bob, denoted by systems $A$ and $B$, respectively, we consider a communication channel with $N-1$ intermediate nodes. These nodes, denoted by $C_i$ with $i \in \{1, \ldots, N-1\}$, are controlled by agents $1$ to $N-1$. The agents serve as intermediaries between the sender and receiver, enabling the transfer of information in a quantum network. Traditional quantum communication protocols have focused on the quantum channel between Alice, Bob, and agents in a single round. However, our approach goes beyond this limitation and considers the use of multiple rounds of quantum channels shared between them. This enables the creation of an adaptive quantum communication protocol that can more effectively transmit quantum information. Key to our approach is the assumption that the sender, receiver, and agents have access to a quantum memory that connects the different quantum channels together. By utilizing this quantum memory, we can optimize the transmission of quantum information, resulting in a more reliable and efficient communication protocol. 

\begin{figure}[h]
    \centering
    \includegraphics[width=1\textwidth]{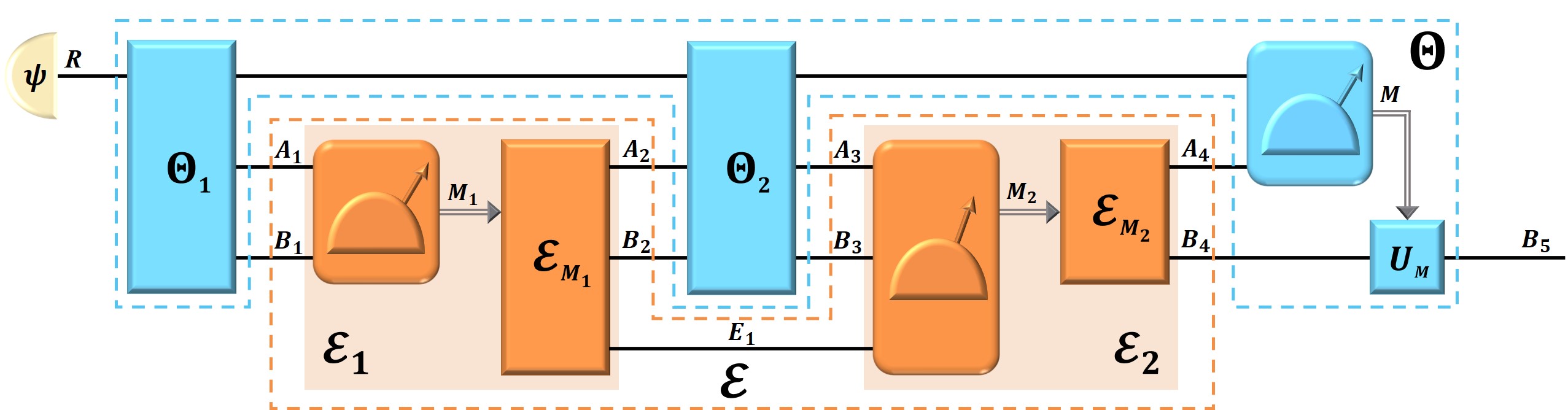}
    \caption{(Color online) A specific adaptive quantum communication protocol $\mE$ works in action, using real-time feedback to ensure secure transmission of quantum information encoded by $\psi$ on system $R$. The present protocol $\mE$ relies on two highly versatile bipartite operations, denoted as $\mE_1$ and $\mE_2$, which are implemented through a learning machine that employs a joint positive operator-valued measure (POVM) on multiple quantum systems. The output of each operation, denoted by $M_1$ and $M_2$ respectively, is then transmitted to a retrieving device, which is realized by quantum channels $\mE_{M_1}$ and $\mE_{M_2}$ respectively. Notably, these operations are connected by a quantum memory $E_1$, which enables their communication to be dynamically adjusted in response to changes in the channel conditions.}
    \label{fig:adaptive}
\end{figure}

Before elucidating the mathematical underpinnings of adaptive quantum communication protocols, we focus on a specific instance illustrated in Fig.~\ref{fig:adaptive} to develop a more intuitive grasp. This protocol consists of two resourceful bipartite operations, denoted as $\mE_1$ and $\mE_2$, each comprising a learning machine implemented through a joint POVM on multiple systems, and a retrieving device realized using a quantum channel. These operations are linked by a quantum memory, which enables their communication to be dynamically adjusted based on changing conditions of the channel. Here, the ability to adaptively optimize the communication in this way arises from the use of a memory-assisted quantum feedback control scheme. While this special scheme does not account for intermediate nodes between sender and receiver, such nodes can greatly enhance long-distance quantum communication.

From a resource-theoretical perspective, the bipartite superchannel $\mE$ (refer to the orange dashed box in Fig.~\ref{fig:adaptive}) shared between the sender and receiver is a critical resource for the transmission of quantum information. To fully leverage its power, we employ a free morphism $\Theta\in\mathfrak{F}_{3}(\mS)$ (as detailed in Subsec.~\ref{subsec:Fragment} and Sec.~\ref{sec:TE}), where $\mS$ is a set of permissible operations from $\{\text{LOCC}_{1}(\text{poly}(d)), \text{LOCC}_{k}, \text{LOCC}_{\mathds{N}}, \text{LOCC}, \overline{\text{LOCC}_{\mathds{N}}}, \text{SEP}, \text{SEPP}, \text{PPT}\}$. By applying this free morphism, we transform $\mE$ into a quantum channel $\Theta(\mE)_{R\to B_{5}}$ between the message system $R$ and the receiver system $B_5$.

\begin{figure}[h]
    \centering
    \includegraphics[width=1\textwidth]{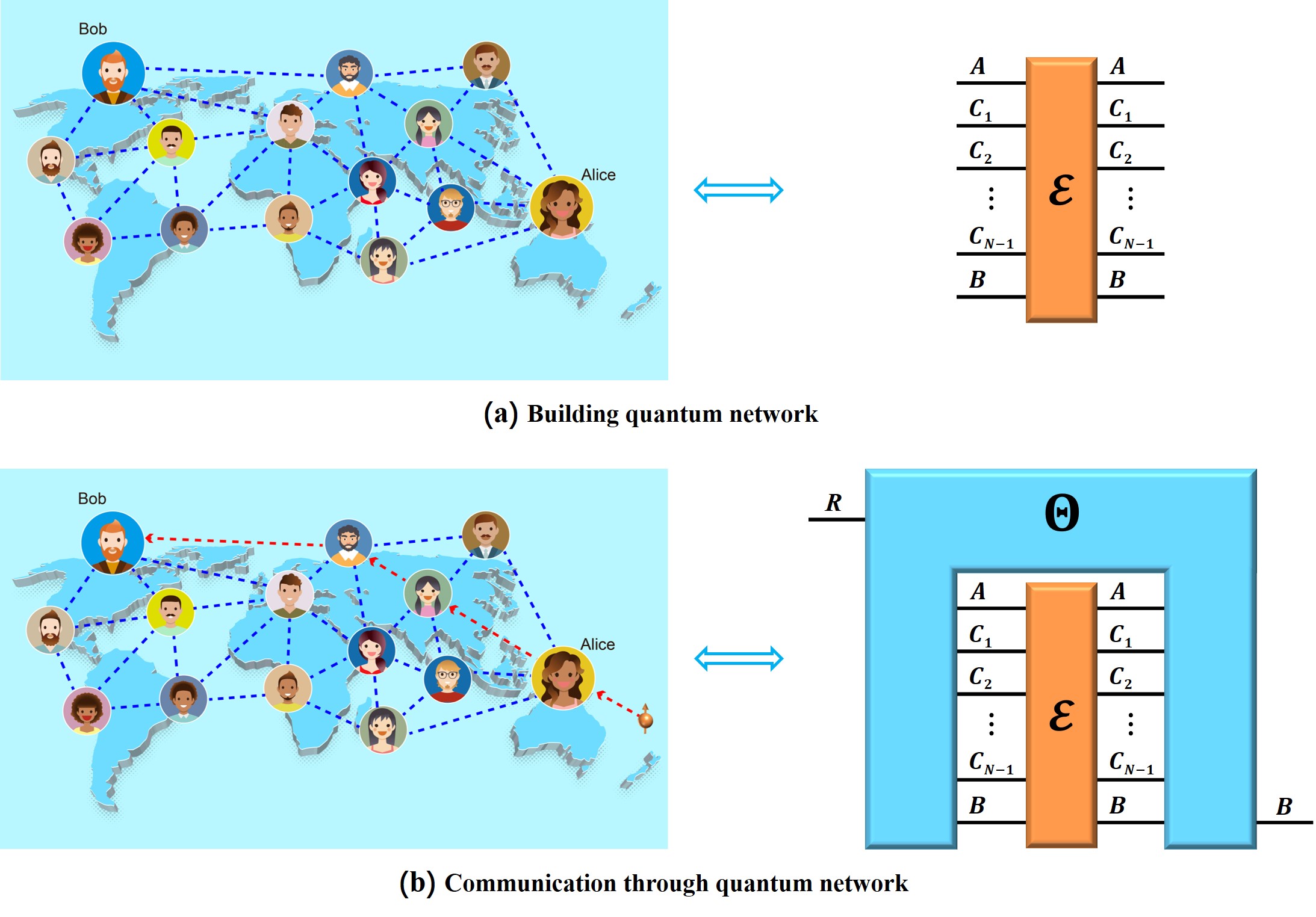}
    \caption{(Color online) A schematic diagram of single-round communication through a quantum network: In order to establish a reliable and secure physical quantum network between the sender Alice, receiver Bob, and third-party agents, we employ a multi-input and multi-output quantum channel denoted as $\mE$, as described in the orange box of (a). However, to fully exploit the capabilities of multipartite quantum channel $\mE$ in quantum network communications, it is crucial to integrate free operations like $\text{LOCC}_{1}(\text{poly}(d))$ (see Subsec.~\ref{subsec:ECF}), and utilize local memory to construct a quantum superchannel $\Theta$, highlighted in the blue ``$2$-comb'' of (b). By utilizing this quantum superchannel, we can transform the quantum network into a direct point-to-point channel between the message system $R$ and the receiver's system $B$, thereby providing a secure communication pathway.}
    \label{fig:Network-single-round}
\end{figure}

As a fundamental building block of quantum communication technologies, quantum teleportation relies on the creation (represented by the brown box of Fig.~\ref{fig:qt}(f)) and manipulation (identified by the gradient superchannel in Fig.~\ref{fig:qt}(f)) of entangled states between distant parties. In particular, the success of the protocol is determined by the quality of the bipartite channel, which carries the resources of temporal entanglement. Moving beyond teleportation to quantum network communication, the challenge becomes more complex due to the involvement of multiple parties and channels~\cite{Kimble2008,Pirandola2016,Simon2017,doi:10.1126/science.aam9288,Pant2019,PRXQuantum.1.020317,PRXQuantum.2.017002,Hermans2022,doi:10.1116/5.0051881,doi:10.1116/5.0092069,doi:10.1116/5.0118569,fang2022quantum,azuma2022quantum}. Specifically, the capacity for transmitting quantum information through a quantum network is limited by multipartite quantum channels that generate and preserve entanglement among the sender, the receiver, and the intermediaries, as outlined in Fig.~\ref{fig:Network-single-round}(a). To harness the full potential of multipartite quantum channels, it is essential to employ free morphism for manipulating the channels, as illustrated in Fig.~\ref{fig:Network-single-round}(b). A case in point is the $\text{LOCC}_{1}(\text{poly}(d))$-superchannel, namely $\mathfrak{F}_{2}(\text{LOCC}_{1}(\text{poly}(d)))$ (see Subsec.~\ref{subsec:ECF}), which uses $\text{LOCC}_{1}(\text{poly}(d))$ for both the pre-processing and post-processing channels. The network described here is designed for single-use only, which makes it ``disposable''. As a result, the practicality of a disposable quantum network is somewhat limited, as it cannot be reused after its initial use.

\begin{figure}[h]
    \centering
    \includegraphics[width=1\textwidth]{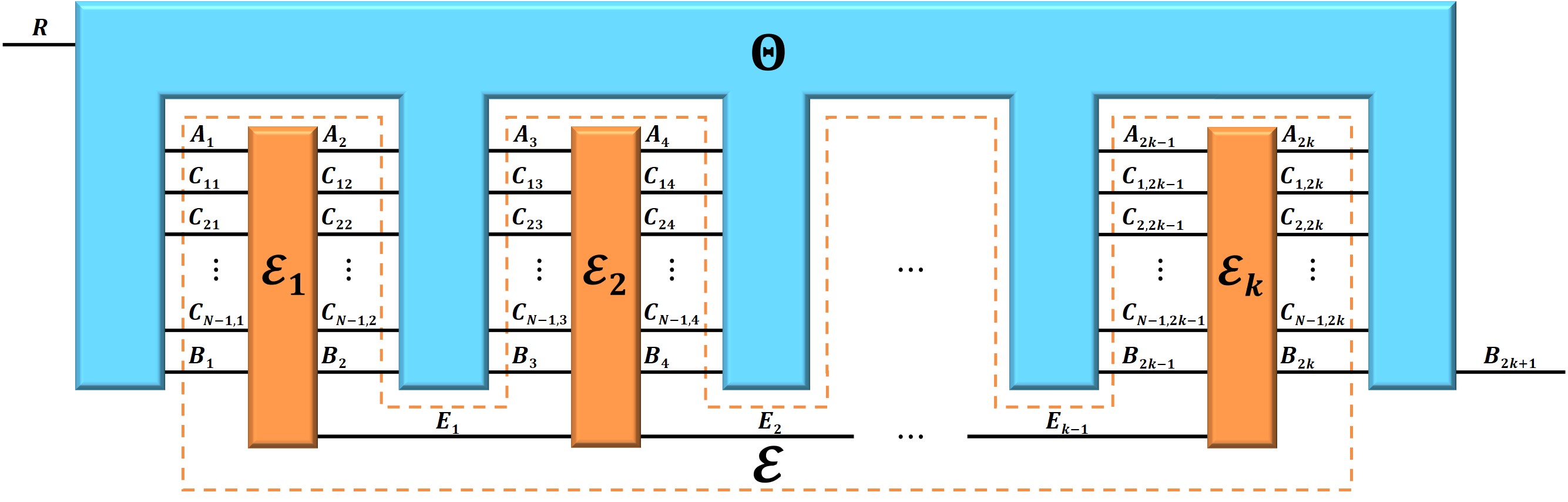}
    \caption{(Color online) The architecture of general adaptive quantum communication protocol $\mE$, utilizing non-Markovian quantum feedback to ensure secure transmission of quantum information encoded by state $\psi$ on system $R$. The orange dashed box represents the protocol $\mE$ itself, while the blue ``comb'' represents a free morphism $\Theta\in\mathfrak{F}_{k+1}(\mS)$ (see Subsec.~\ref{subsec:Fragment} and Sec.~\ref{sec:TE}) that transforms $\mE$ into a quantum channel from the message system, denoted as $R$, to the receiver's system, denoted as $B_{2k+1}$. The construction of $\mE$ (see Eq.~\ref{eq:qcf-network}) involves $k$ rounds of multipartite quantum operations denoted as $\mE_i$, with $1\leqslant i\leqslant k$. These operations are shared among the sender, intermediary agents, and the receiver, and are used to manipulate quantum systems. This manipulation includes processes such as learning and retrieving information. Here, the sender holds the message system $R$ and the quantum systems related to $A$, while the intermediary agents hold those related to $C$, and the receiver holds those related to $B$. To connect the quantum operations together, quantum memory systems denoted as $E_i$, with $1\leqslant i\leqslant k-1$, are employed. These memory systems are used to store the quantum state of the system at the end of each round, preserving the information for use in the subsequent rounds.}
    \label{fig:qcf-network}
\end{figure}

To address this limitation and establish a practical and general framework for quantum networks, we propose the adaptive quantum communication protocol. This protocol involves $k$ rounds of multipartite quantum operations, where the $i$-th round of quantum operations, denoted as $\mE_i$, is shared between the sender, receiver, and agents, with $1 \leqslant i \leqslant k$. This linear map transforms the quantum states of the composite system 
\begin{align}
    A_{2i-1}C_{1,2i-1}C_{2,2i-1}\cdots C_{N-1,2i-1}B_{2i-1}E_{i-1}
\end{align} 
to those of the system 
\begin{align}
    A_{2i}C_{1,2i}C_{2,2i}\cdots C_{N-1,2i}B_{2i}E_{i}.
\end{align} 
As a simplification, we have assumed that the memory systems of the boundary channels that are not connected to other channels are trivial. Namely, $E_0= E_k=\mathds{C}$. The quantum dynamics of $k$ connected quantum channels can be represented as a quantum circuit fragment (see Subsec.~\ref{subsec:Fragment}). We denote this protocol as $\mE$, and its behavior is given by
\begin{align}\label{eq:qcf-network}
\mE:=\mE_k\circ\mE_{k-1}\circ\cdots\circ\mE_2\circ\mE_1.
\end{align}
Fig.~\ref{fig:qcf-network} provides a visual representation of the adaptive quantum communication protocol $\mE$, namely Eq.~\ref{eq:qcf-network}. It is worth noting that as the learning machine and retrieving device form a multipartite quantum channel (highlighted by the orange dashed box in Fig.~\ref{fig:adaptive}), Fig.~\ref{fig:adaptive} represents a specific instance of the more general adaptive quantum communication protocol shown in Fig.~\ref{fig:qcf-network}. Our framework for adaptive quantum communication protocols is highly versatile, offering broad applicability across a range of communication scenarios. Unlike many existing models, our framework does not rely on any specific assumptions or structures about the protocol, including the number of communication rounds and the spatial correlations shared between quantum systems. The only fundamental constraint in our model is the principle of causality, which ensures that information cannot be transmitted from the future to the past. This allows our framework to be readily adapted to arbitrary quantum causal networks, enabling dynamic adjustments that can help to optimize communication performance in real-world settings.

The optimization of adaptive quantum communication protocols represents a crucial research challenge with far-reaching implications for the development of quantum communication technologies. In this context, a fundamental question arises: what is the maximum information transmission performance of a given protocol $\mE$ (refer to the orange dashed box highlighted in Fig.~\ref{fig:qcf-network}) that adapts to dynamic network conditions? To address this question, we propose a novel approach based on the use of a high-level free morphism $\Theta\in\mathfrak{F}_{k+1}(\mS)$ (examine the blue ``comb'' highlighted in Fig.~\ref{fig:qcf-network}) with $\mS$ being a set of permissible operations from
\begin{align}
    \{\text{LOCC}_{1}(\text{poly}(d)), \text{LOCC}_{k}, \text{LOCC}_{\mathds{N}}, \text{LOCC}, \overline{\text{LOCC}_{\mathds{N}}}, \text{SEP}, \text{SEPP}, \text{PPT}\},
\end{align} 
which acts as a transformative tool to convert the dynamical resource $\mE$ into a functional channel $\Theta(\mE)_{R\to B_{2k+1}}$ for transmitting message encoded by $\psi$ on system $R$ (see Subsec.~\ref{subsec:Fragment} and Sec.~\ref{sec:TE}). As $\mE$ lacks any direct input port for $\psi$ (see also the role of $\mE(p,q)$ in quantum teleportation), the use of $\Theta$ is essential to enable the information transmission. Here both $R$ and $A$-related systems are under the control of the sender.

From a quantum communication perspective, the success of quantum communication relies on the ability to preserve and transmit quantum states accurately and reliably. In this context, the performance of a quantum channel, denoted by $\Theta(\mE)$, is characterized by the maximal average fidelity $f(\Theta(\mE))$ (see Def.~\ref{def:AF}), which measures the degree of similarity between the transmitted quantum state and the received one. The goal of quantum network communication is to maximize this fidelity under all possible free morphisms $\Theta\in\mathfrak{F}_{k+1}(\mS)$, which captures different physical restrictions in the laboratory. From an entanglement-theoretic perspective, especially in the single-shot setting, where the transmission of a single copy of adaptive quantum communication protocol $\mE$ is considered, the quality of the channel is captured by the $d$-fidelity of $\mS$ distillation, denoted by $F_{d, \mS}(\mE)$ (see Def.~\ref{def:FD}). This fidelity measures the ability to extract the maximum amount of entanglement from the initial adaptive quantum communication protocol $\mS$. In what follows, we will show that the temporal entanglement associated with the adaptive quantum communication protocol $\mE$ under free morphisms $\mathfrak{F}_{k+1}(\mS)$ determines its capability of communication in a quantum network. Specifically, the higher the $d$-fidelity of $\mS$ distillation, the better the performance of the resulting channel, as it implies a higher degree of entanglement between the sender and receiver. This result highlights the importance of understanding the underlying temporal entanglement properties of quantum communication protocols and the role of free morphisms in shaping their performance.

\begin{mythm}
{Temporal Entanglement Determines Quantum Network Communications}{All-in-One}
Given an adaptive quantum communication protocol $\mE$ (see Eq.~\ref{eq:qcf-network}), a free morphism $\Theta\in\mathfrak{F}_{k+1}(\mS)$ will transform the protocol $\mE$ into a point-to-point channel $\Theta(\mE)$, as illustrated in Fig.~\ref{fig:qcf-network}. Under all free morphisms, the resulting channel's highest efficacy is determined by the equation provided below
\begin{align}\label{eq:network-fidelity-thm}
\max_{\Theta\in\mathfrak{F}_{k+1}(\mS)}f(\Theta(\mE))
=
\frac{d F_{d, \mS}(\mE)+ 1}{d+1},
\end{align}
Here, we use $f$ to denote the average fidelity, as defined in Def.~\ref{def:AF}, and $F_{d, \mS}$ to represent the $d$-fidelity of $\mS$ distillation, as defined in Def.~\ref{def:FD}. The set $\mS$ encompasses permissible operations, including $\text{LOCC}_{1}(\text{poly}(d))$, $\text{LOCC}_{k}$, $\text{LOCC}_{\mathds{N}}$, $\text{LOCC}$, $\overline{\text{LOCC}_{\mathds{N}}}$, $\text{SEP}$, $\text{SEPP}$, and $\text{PPT}$. For ease of exposition, we assume that all systems have the same dimension $d$.
\end{mythm}

\begin{proof}
To establish the validity of Eq.~\ref{eq:network-fidelity-thm}, we first demonstrate that the left-hand side is upper-bounded by the right-hand side. Let us suppose that there exists a free morphism $\Xi\in\mathfrak{F}_{k+1}(\mS)$ which attains the maximal performance of adaptive quantum communication protocol $\mE$ in terms of average fidelity $f$. Then we have 
\begin{align}\label{eq:network-fidelity-thm-left}
\max_{\Theta\in\mathfrak{F}_{k+1}(\mS)}f(\Theta(\mE))
=
f(\Xi(\mE))
=
\frac{d F(\Xi(\mE))+ 1}{d+1}
=
\frac{d \Tr[\Xi(\mE)(\phi^{+}_{d})\cdot\phi^{+}_{d}]+ 1}{d+1}
\leqslant
\frac{d F_{d, \mS}(\mE)+ 1}{d+1}.
\end{align}
The second equation of Eq.~\ref{eq:network-fidelity-thm-left} follows directly from Lem.~\ref{lem:AFEF}. The third equation of Eq.~\ref{eq:network-fidelity-thm-left} is based on the standard definition of entanglement fidelity, defined in Def.~\ref{def:EF}. Without loss of generality, we can assume that $\phi^{+}_{d}$ acts on the bipartite system $A_{2k+1}B_{2k+1}$. To prove the inequality of Eq.~\ref{eq:network-fidelity-thm-left}, we apply a local operation $\id_{R\to A_{2k+1}}$ to $\Xi(\mE)(\phi^{+}_{d})$ that maps the subsystem $R$ to $A_{2k+1}$ and leaves the subsystem $B_{2k+1}$ unchanged, as illustrated in Fig.~\ref{fig:network-fidelity-thm}(a). The transformation $\id_{R\to A_{2k+1}}$ corresponds to a local operation, given that the sender controls the message system $R$ and all systems related to $A$. Consequently, the overall operation $\id_{R\to A_{2k+1}}\circ\Xi(\phi^{+}_{d, RR^{'}})$ constitutes a free morphism in $\mathfrak{F}_{k+1}(\mS)$, leading to the conversion of the adaptive quantum communication protocol $\mE$ into a state acting on system $A_{2k+1}B_{2k+1}$.

\begin{figure}[h]
    \centering
    \includegraphics[width=1\textwidth]{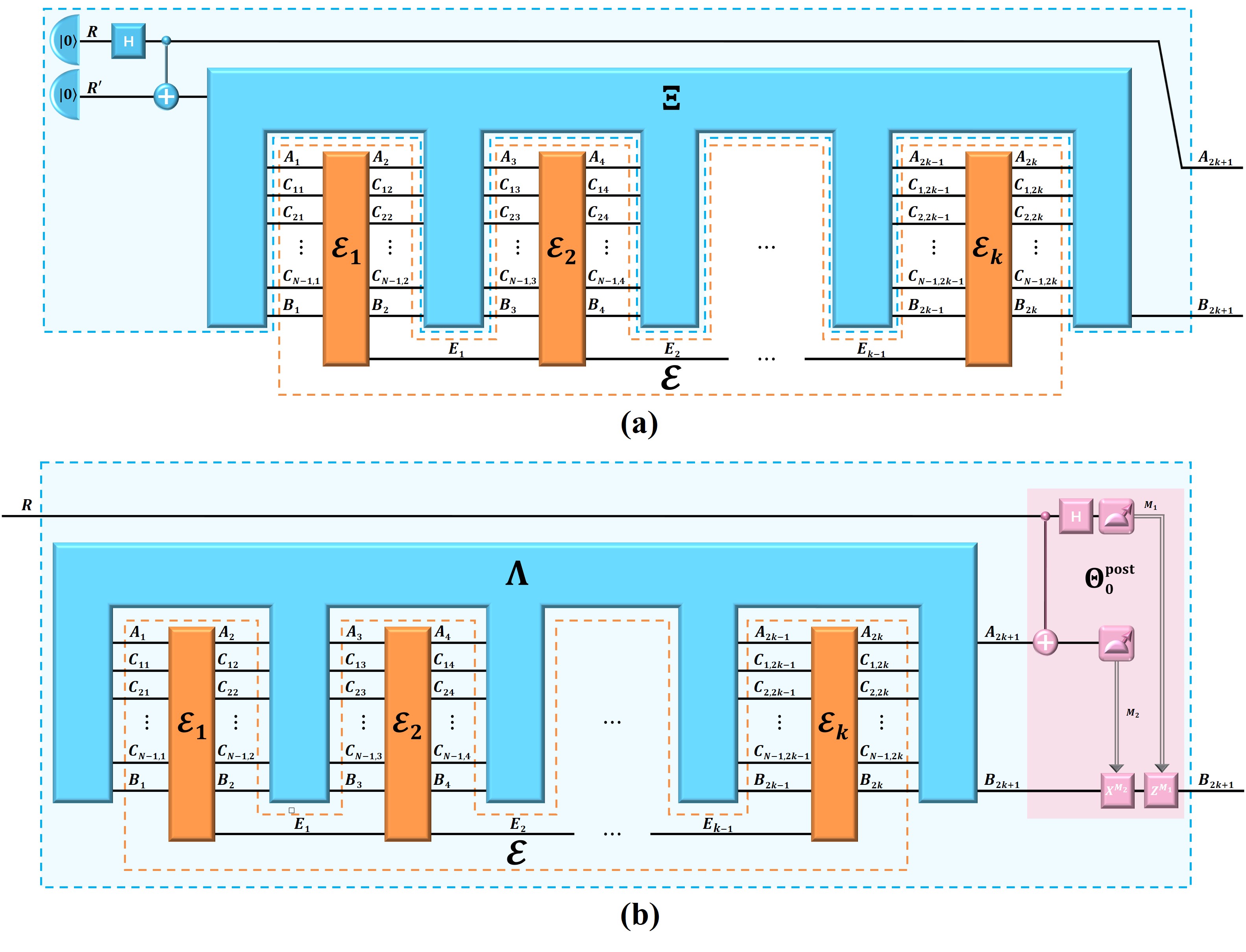}
    \caption{(Color online) Graphical representation of the proof of Thm.~\ref{thm:All-in-One}, showcasing the fundamental mechanisms underlying Eq.~\ref{eq:network-fidelity-thm}. In figure (a), we illustrate how the free morphism $\Xi\in\mathfrak{F}_{k+1}(\mS)$ supports Eq.~\ref{eq:network-fidelity-thm-left}, a key component of our proof. Figure (b) shows the construction of the network communication channel using $\Lambda$ and $\Theta^{\text{post}}_{0}$, as described in Eq.~\ref{eq:network-fidelity-thm-right}. The dashed orange box represents the adaptive quantum communication protocol $\mE$. While we have presented qubit cases for simplicity, it is important to emphasize that our proof for Thm.~\ref{thm:All-in-One} is applicable to quantum systems of any finite dimension $d$. 
    }
    \label{fig:network-fidelity-thm}
\end{figure}

We now proceed to prove the converse direction by demonstrating that the right-hand side is smaller than or equal to the left-hand side of Eq.~\ref{eq:network-fidelity-thm}. To this end, we assume that the right-hand side value of $F_{d, \mS}(\mE)$ is attained by a free morphism $\Lambda\in\mathfrak{F}_{k+1}(\mathcal{S})$, namely
\begin{align}\label{eq:network-fidelity-thm-FD}
F_{d, \mS}(\mE)
=
\Tr[\Lambda(\mE)\cdot\phi^{+}_{d}]
=
F(\mF(\Lambda(\mE)))
=
F(\Theta^{\text{post}}_{0}\circ\Lambda(\mE)),
\end{align}
where $\mF$ stands for the teleportation channel created by consuming resourceful state $\Lambda(\mE)$ (see Fig.~\ref{fig:channel-rho}), and $\Theta^{\text{post}}_{0}$ is the standard teleportation operation (see the pink box of Fig.~\ref{fig:qt}(a)).
The second equation in Eq.~\ref{eq:network-fidelity-thm-FD} directly follows from Lem.~\ref{lem:EFUF}, while the last equation is obtained by using the definition of $\mF$ (See Eq.~\ref{eq:channel-rho}). It is important to note that in this context, the quantum state $\Lambda(\mE)$ operates on the composite system $A_{2k+1}B_{2k+1}$, whereas the quantum channel $\Theta^{\text{post}}_{0}\circ\Lambda(\mE)$ maps the message system to receiver's system. With this defined, we see that
\begin{align}\label{eq:network-fidelity-thm-right}
\frac{d F_{d, \mS}(\mE)+ 1}{d+1}
=
\frac{d F(\Theta^{\text{post}}_{0}\circ\Lambda(\mE))+ 1}{d+1}
=
f(\Theta^{\text{post}}_{0}\circ\Lambda(\mE))
\leqslant
\max_{\Theta\in\mathfrak{F}_{k+1}(\mS)}f(\Theta(\mE)).
\end{align}
The standard teleportation operation $\Theta^{\text{post}}_{0}$ is protocol that can be implemented using only local operations and one-way classical communication with the communication complexity polynomially dependent on the dimension of the message system. As such, it is an element of the set $\text{LOCC}_{1}(\text{poly}(d))$. When we compose the free morphism $\Lambda$ with the standard teleportation operation $\Theta^{\text{post}}_{0}$, the resulting quantum circuit fragment $\Theta^{\text{post}}_{0}\circ\Lambda$ still belongs to the set $\mathfrak{F}_{k+1}(\mS)$ for any $\mS\in\{\text{LOCC}_{1}(\text{poly}(d)), \text{LOCC}_{k}, \text{LOCC}_{\mathds{N}}, \text{LOCC}, \overline{\text{LOCC}_{\mathds{N}}}, \text{SEP}, \text{SEPP}, \text{PPT}\}$. This implies the inequality shown in Eq.~\ref{eq:network-fidelity-thm-right}. Combining Eq.~\ref{eq:network-fidelity-thm-left} with Eq.~\ref{eq:network-fidelity-thm-right}, we have demonstrated that the two sides of Eq.~\ref{eq:network-fidelity-thm} are equal, as required by the theorem.
\end{proof}

The advent of the worldwide network has revolutionized the way we live our lives, connecting people and information in ways that were once unimaginable. The potential of a global quantum network is even more profound, promising to usher in a new era of communication by enabling quantum information exchange between any two points on Earth. By harnessing the power of quantum mechanics, a quantum network has the potential to achieve unparalleled capabilities that are provably impossible with classical information processing alone, when combined with the classical network we have today. Quantum information comes in various forms within a quantum network, such as the polarization state of a photon, the spin of an electron, or the excitation state of an atom. While numerous technologies have been developed to teleport these states, optimizing the performance of quantum communication remains a challenge. Despite the availability of sequential resourceful operations and quantum memory, which form an adaptive quantum communication protocol (highlighted in the orange dashed box of Fig.~\ref{fig:qcf-network}), the fundamental limitation (in terms of average fidelity) associated with this resource remains unknown.

Here, we have successfully addressed the question of optimal quantum communication performance by establishing a connection with the concept of temporal entanglement (introduced in Sec.~\ref{sec:TE}), as demonstrated by our Thm.~\ref{thm:All-in-One}. This achievement paves the way for the future development of a global quantum network and brings us closer to realizing the full potential of this revolutionary technology. Despite such a progress in the field of quantum network communications, there are still several questions that remain unanswered. For instance, given an adaptive quantum communication protocol $\mE$, how can we efficiently compute its $d$-fidelity of $\mS$ distillation, i.e., $F_{d, \mS}(\mE)$ defined in Def.~\ref{def:FD}, which involves determining the fidelity of a set of permissible operations from $\{\text{LOCC}_{1}(\text{poly}(d)), \text{LOCC}_{k}, \text{LOCC}_{\mathds{N}}, \text{LOCC}, \overline{\text{LOCC}_{\mathds{N}}}, \text{SEP}, \text{SEPP}, \text{PPT}\}$? As a free morphism (see the blue ``comb'' of Fig.~\ref{fig:qcf-network}) can involve multiple rounds of quantum channels, and quantum memory may exist between them, finding an exact solution is extremely challenging. Therefore, one possible approach is to relax the free morphisms considered in this work and develop upper bounds for the performance of the adaptive quantum communication protocol $\mE$. This strategy can help us gain a better understanding of the capabilities and limitations of quantum network communication protocols and guide the development of future protocols with enhanced performance. Readers who are interested in exploring the relaxation of free morphisms are encouraged to refer to Ref.~\cite{PhysRevLett.123.070502,PhysRevA.104.022401,Berta2022,PhysRevA.107.012428,doi:10.1116/5.0135467} and references therein, which offer in-depth investigations on the relaxation of free morphisms in the context of quantum channels, a special case of quantum circuit fragment (see Subsec.~\ref{subsec:Fragment}) considered here.


\bibliography{Bib}